\documentclass[11pt,oneside]{article}
\usepackage{epsfig,lscape,amsmath, amssymb,amsfonts, amsthm, bm, bbm, graphicx,geometry,float, subcaption, enumerate, color, verbatim ,mathtools,algorithm, algorithmic, url, amssymb, bm,setspace, natbib}
\usepackage[svgnames,x11names,table]{xcolor}
\usepackage[colorlinks=true,citecolor=Navy,linkcolor=.,urlcolor=black]{hyperref}
\mathtoolsset{showonlyrefs}

\usepackage{tikz}
\usetikzlibrary{arrows.meta, positioning}
\usepackage{setspace}
\usepackage[utf8]{inputenc}
\usepackage[english]{babel}
\usepackage{framed}
\usepackage{multibib}
\usepackage[symbol]{footmisc}
\newcites{append}{Supplement References}

\newenvironment{prf}
{\noindent \textbf{Proof.}}{\hfill $\Box$ \vspace{.1in}}

\newtheorem{lem}{Lemma}
\newtheorem{pro}{Proposition}
\newtheorem{cor}{Corollary}

\theoremstyle{definition}

\theoremstyle{definition}

\usepackage[medium,compact]{titlesec}

\titlespacing*{\section} {0pt}{1.5ex}{1ex}
\titlespacing*{\subsection} {0pt}{1.5ex}{1ex}
\titlespacing*{\subsubsection} {0pt}{1ex}{1ex}

\graphicspath{{figures/}}
\allowdisplaybreaks

\newcommand{\D}{\mathcal{D}}

\newcommand{\ud}{\mathrm{d}}

\newcommand{\tE}{\tilde{E}}

\newcommand{\halpha}{\hat{\alpha}}
\newcommand{\hbeta}{\hat{\beta}}
\newcommand{\hgamma}{\hat{\gamma}}
\newcommand{\RR}{\mathrm{RR}}
\newcommand{\ORR}{\mathrm{ORR}}
\newcommand{\CRR}{\mathrm{CRR}}
\newcommand{\odds}{\mathrm{odds}}

\theoremstyle{definition}

\def\mg{\eta}

\def\bbR{{\mathbb{R}}}

\newcommand\mytext[1]{\text{\scriptsize{#1}}}

\allowdisplaybreaks

\begin{document}
\setlength{\abovedisplayskip}{5pt}
\setlength{\belowdisplayskip}{5pt}

\begin{titlepage}
\begin{center}
{\LARGE\bf Enhanced Marginal Sensitivity Model and Bounds}\\
\vspace{.1in}
Yi Zhang\footnotemark[1]\footnotemark[2], Wenfu Xu\footnotemark[1]\footnotemark[3], Zhiqiang Tan\footnotemark[2]\\
\footnotetext[1]{These two authors contributed equally to this work}
\footnotetext[2]{Department of Statistics, Rutgers University}
\footnotetext[3]{College of Economics and Management, China Jiliang University}
\vspace{.1in}
\today
\end{center}


\paragraph{Abstract.}
Sensitivity analysis is important to assess the impact of unmeasured confounding in causal inference from observational studies. 
The marginal sensitivity model (MSM) provides a useful approach in quantifying the influence of unmeasured confounders on treatment assignment
and leading to tractable sharp bounds of common causal parameters. In this paper, to tighten MSM sharp bounds,
we propose the enhanced MSM (eMSM) by incorporating another sensitivity constraint which quantifies the influence of unmeasured confounders on outcomes.
We derive sharp population bounds of expected potential outcomes under eMSM, which are always narrower than the MSM sharp bounds in a simple and interpretable way. We further discuss desirable specifications of sensitivity parameters related to the outcome sensitivity constraint, and
obtain both doubly robust point estimation and confidence intervals for the eMSM population bounds. 
The effectiveness of eMSM is also demonstrated numerically through two real-data applications. 
Our development represents for the first time a satisfactory extension of MSM to exploit both treatment and outcome sensitivity constraints on unmeasured confounding.  


\end{titlepage}







\section{Introduction}
Without randomization, causal inference in observational studies is challenging due to possible confounding. When all confounding is assumed to be accounted for by measured covariates $X$, causal parameters can be point identified from observed data.
However, the assumption of no unmeasured confounding is untestable and may often be violated in practice.
Assessing the sensitivity of causal estimates against possible unmeasured confounding is known as sensitivity analysis. Typically, a sensitivity model
with a sensitivity parameter is employed to quantify unmeasured confounding, and then the causal parameter of interest
can be partially identified or bounded.
There is an extensive and growing literature on sensitivity analysis. See, for example, \citet{PRosenbaum2002}, \citet{tan2006distributional, tan2024model}, and references therein.

A popular approach is the marginal sensitivity model (MSM) of \citet{tan2006distributional}. Given measured covariates $X$, an MSM assumes a uniform bound $\Lambda$ on the density ratio of each potential outcome between the treated and untreated groups.
By introducing an unmeasured confounder $U$ which together with $X$ accounts for all confounding,
an MSM is equivalent to a constraint on the treatment assignment, which uniformly bounds the odds ratio of receiving the treatment given $(X,U)$ and that given $X$ by the same $\Lambda$.
Hence the MSM is related to but distinct from latent sensitivity model of \citet{PRosenbaum2002}, which uniformly bounds the odds ratio of receiving the treatment given the same level of $X$ but any two different levels of $U$. For inference about the average treatment effect (ATE) under MSM,
there has been a notable line of works which provide tractable formulas for sharp population bounds and estimation methods for obtaining sample bounds
which incorporate both partial identification and sampling uncertainty.
See \citet{zhao2019sensitivity}, \citet{dorn2023sharp}, \citet{dorn2024doubly}, and \citet{tan2024model}.

Various MSM-related sensitivity models have also been proposed to address different needs.
For example, \citet{jesson2022scalable} extended MSM to continuous treatments, while \citet{bonvini2022sensitivity}, \citet{frauen2023sharp} and \citet{tan2024sensitivity} considered time-varying treatments in longitudinal settings. Another collection of works replace the uniform bounds in MSMs with bounds on average quantities related to the probabilities of treatment assignment given $X$ or given $(X,U)$. See, for example, \citet{jin2022sensitivity}, \citet{zhang2022infty} and \citet{huang2025variance}. While average-based constraints may be more realistic,
these models often result in wider population bounds.

To tighten MSM sensitivity bounds, we propose the enhanced marginal sensitivity model (eMSM). 
While the MSM constraint concerns the effect of unmeasured confounders on treatment assignment, 
another sensitivity constraint that captures the effect of unmeasured confounders on the outcome is incorporated. 
Specifically, the outcome sensitivity constraint bounds how much the expected outcome given $(X,U)$ differs from that given $X$,
by a pair of sensitivity parameters $(\Delta_1,\Delta_2)$. All sensitivity parameters are made covariate--dependent for generality of the discussion.

We further develop and support sensitivity analysis based on eMSM in the following three aspects.
First, we derive sharp bounds for the expected potential outcomes, which can be simultaneously achieved across treatment levels and, hence, lead to sharp bounds of other causal parameters such as ATE. The sharp bounds for the potential outcome means are simple and readily interpretable in terms of the two sensitivity constraints in eMSM: the treatment sensitivity constraint, as in the MSM sharp bounds, indicates the quantile level at which the mean parameter may deviate from its reference value under unconfoundedness,
and the outcome sensitivity constraint indicates the shrinkage level at which the MSM sharp bounds are pulled toward the reference value. In addition to the sharp bounds, we also characterize the worst-case distributions that attain the sharp bounds along with their associated unmeasured confounders.

Second, we identify a particular specification of the sensitivity parameters $(\Delta_1,\Delta_2)$ for practical interpretation and implementation, under which the eMSM sharp bounds can be further simplified. Due to this specification, we obtain both doubly robust point estimation and confidence intervals for the eMSM population bounds through calibrated estimation similarly as in \citet{tan2024model}. Such an extension seems infeasible for other general eMSM specifications.

Third, we establish an interesting relation between eMSM and the sensitivity approach of \citet{ding2016sensitivity} (DV) for binary outcomes. DV imposes a treatment sensitivity constraint as in MSM and an outcome sensitivity constraint on the ratio of expected outcomes between any two levels of $U$.
We show that a DV model is a collection of eMSMs with compatible sensitivity parameters
and then derive sharp bounds for the DV model, which are previously unrealized. A special case of these sharp bounds recovers those in \cite{sjolander2024sharp}.
Moreover, similar relations can be established between eMSM and two other related sensitivity models that allow non-binary outcomes. These findings attest to the fundamental value of eMSM.

The remaining of the paper is organized as follows. In Section \ref{sec:setup}, we review the marginal sensitivity model. In Section \ref{sec:propsens}, we develop the proposed sensitivity model and its sharp population bounds, and then discuss specification and interpretation of sensitivity parameters. In Section \ref{sec:est}, we use calibrated estimation to obtain both doubly robust point estimation and confidence intervals for eMSM population bounds. In Sections \ref{sec:compare}, we compare eMSM with DV. Two real data applications are provided in Section \ref{sec:numerical}. Finally, we conclude the paper and discuss two alternative sensitivity models in Section \ref{sec:conclusion}. All proofs are provided in the Supplement Material. 
\section{Setup and marginal sensitivity model}\label{sec:setup}
Suppose that the observed data $ \{(Y_i,T_i,X_i):i=1,\dots,n\}$ is an independent and identically distributed sample from a joint distribution of $(Y,T,X)$,
where $Y$ is an outcome variable, $T$ a binary treatment (taking value 0 or 1), and $X$ a vector of measured covariates.
Following the potential-outcome framework \citep{rubin1974estimating, Neyman1923}, let $Y^0$ or $Y^1$ be the potential outcomes that would be observed under treatment $0$ or $1$ respectively.
The observed outcome $Y$ is determined from the potential outcomes through a consistency assumption,
$Y=TY^1+(1-T)Y^0$, i.e., $Y=Y^1$ if $T=1$ and $Y=Y^0$ if $T=0$.
In other words, only one of the two potential outcomes can be identified from the observed outcome $Y$ at the unit level, depending on $T$.

Point identification of the means, $\mu^0 = E(Y^0)$ and $\mu^1 = E(Y^1)$, and the average treatment effect (ATE), defined as $\mu^1-\mu^0$, typically relies on the following assumptions:
(i) unconfoundedness, $(Y^0,Y^1) \perp T | X$, i.e., $(Y^0,Y^1)$ and $T$ are conditionally independent given $X$,
and (ii) overlap, $0 < \pi^*(X) < 1$ almost surely, where $\pi^*(X) = P(T=1|X)$ is the propensity score \citep{rosenbaum1983central}.
Assumption (i) is also known as the assumption of no unmeasured confounding:
adjustment for measured covariates $X$ removes all confounding between treatment selection and potential outcomes.
In the following, we discuss inference about $\mu^1$ when unmeasured confounding may exist, and defer to Supplement Section \ref{sec:additionalbounds}
the discussion of $\mu^0$ and ATE.

To allow unmeasured confounding, a marginal sensitivity model (MSM) of \cite{tan2006distributional}
assumes that there exists an unmeasured variable $U$ such that
\begin{subequations} \label{eq:msm}
\begin{align}
 &  (Y^0,Y^1)\perp T | X,U, \label{eq:msm-indep} \\
 &  \lambda^*(X,U) \in [\Lambda_1(X),\Lambda_2(X)]. \label{eq:msm-bound}
\end{align}
\end{subequations}
Constraint \eqref{eq:msm-indep} indicates that $U$ and $X$ together account for all confounding.
For constraint \eqref{eq:msm-bound},
$\Lambda_2 (X) \ge 1 \ge \Lambda_1(X) \ge 0$ are some \textit{pre-specified} covariate functions, and
$\lambda^*(X,u)$ is the Radon--Nikodym derivative (or density ratio) at $U=u$ between the two conditional distributions $P_U(\cdot|T=0,X)$ and $P_U(\cdot|T=1,X)$, or equivalently,
by Bayes' rule, the odds ratio associated with the two conditional probabilities $P(T=1|X)$ and $P(T=1|X,U=u)$:
\begin{equation}
    \lambda^*(X,u) = \frac{\ud P_U(u|T=0,X)}{\ud P_U(u|T=1,X)} =\frac{P(T=1|X)}{P(T=0|X)}\Big/\frac{P(T=1|X,U=u)}{P(T=0|X,U=u)} .  \label{eq:w-def}
\end{equation}
The extreme case $\Lambda_1(X) = \Lambda_2(X) \equiv 1$ indicates $U \perp T | X$, which
together with $(Y^0,Y^1)\perp T|X,U$ implies unconfoundedness, $(Y^0,Y^1) \perp T | X$.
The further $\Lambda_1(X)$ and $\Lambda_2(X)$ are away from $1$, the greater unmeasured confounding is allowed.
See Supplement Section~\ref{sec:addonMSM} for a discussion related to the original MSM in \cite{tan2006distributional} without introducing $U$.

To clarify our notation and perspective, we view the potential-outcome framework as a latent-variable model.
At the unit level, the observed data consist of $(Y,T,X)$, and
the full data consist of
$(Y^0,Y^1,T,X,U)$ with latent variables
$(Y^0,Y^1,U)$ in the case of MSM.
The observed outcome $Y$ is determined from $(Y^0,Y^1)$ through the consistency assumption. The unmeasured variable $U$ can be arbitrary in the sense that no knowledge is assumed, except \eqref{eq:msm-indep}. We take a frequentist perspective and denote by $P$ the true distribution of the full data. The
induced distribution of $P$ on $(Y,T,X)$ is the true distribution of the observed data.
The expectation with respect to $P$ is denoted by $E(\cdot)$. We use an asterisk to signify a quantity
defined from the true distribution of the observed or full data, for example, $\pi^*(\cdot)$ and $\lambda^*(\cdot)$ above.

From this perspective, the sharp population upper bound of $\mu^1$ under MSM is defined as
\begin{align}
\mu^{1+}_{\mytext{MSM}} = \sup_Q \; E_Q (Y^1) ,  \label{eq:msm-bound-mu1}
\end{align}
over any joint distribution $Q$ of the full data $(Y^0,Y^1,T,X,U)$ such that
\begin{itemize}\addtolength{\itemsep}{-.1in}
\item[(i)] $Q$ is compatible with the observed-data distribution:
the induced distrbution of $Q$ on $(Y,T,X)$ coincides with the true distribution, and

\item[(ii)] $Q$ satisfies MSM: properties \eqref{eq:msm-indep} and \eqref{eq:msm-bound} hold under $Q$, with $\lambda^*(X,U)$ replaced by its $Q$-analogue, $\lambda_Q (X,U)$.
\end{itemize}
Here $E_Q(\cdot)$ denotes the expectation with respect to $Q$. 
A subtle point is that the specification of $U$ may vary across different distributions $Q$, especially in comparison to the true distribution $P$.
For example, in a given study, even if the full data are generated from $P$ with a continuous unmeasured variable $U$, 
a joint distribution $Q$ is allowed with a binary unmeasured variable $U$ provided the preceding conditions (i) and (ii) are satisfied.

The sharp bound $\mu^{1+}_{\mytext{MSM}}$ has been solved as follows \citep{dorn2023sharp, tan2024model}:
\begin{subequations}\label{eq:msm-sol-mu1}
\begin{align}
 \mu^{1+}_{\mytext{MSM}} & = E \{ TY + (1-T) \nu^{1+}_{\mytext{MSM}} (X)\},  \label{eq:MSM-bound-mu} \\
 \nu^{1+}_{\mytext{MSM}} (X) &=
E\left[ Y +\{\Lambda_2(X)-\Lambda_1(X)\}  \rho_{\tau}(Y,q^*_{1,\tau}(X) )|T=1,X \right],  \label{eq:MSM-bound-nu}
\end{align}
\end{subequations}
where
$\tau(X)=\{\Lambda_2(X)-1\}/\{\Lambda_2(X)-\Lambda_1(X)\}$ and is set to 1/2 when $\Lambda_2(X)=\Lambda_1(X)=1$,
 $\rho_{\tau} (y,q)= \tau (y-q)_{+}+(1-\tau) (q-y)_{+}$ is the ``check'' function,
 $q^*_{1,\tau}(X)$ is the $\tau(X)$-quantile of $P_Y(\cdot| T=1,X)$,
 defined as a minimizer of the quantile loss $E \{\rho_{\tau} (Y, q) | T=1,X\}$ over $q\in\bbR$ \citep{koenker1978regression},
 and $c_+=\max\{c,0\}$ for $c \in \bbR$.
In fact, $\nu^{1+}_{\mytext{MSM}} (X)$ can be shown to be the sharp upper bound of $\nu^1(X)=E(Y^1|T=0,X)$, i.e.,
$ \nu^{1+}_{\mytext{MSM}} (X) = \sup_Q E_Q (Y^1 | T=0,X)$ over all possible distributions $Q$ as in \eqref{eq:msm-bound-mu1}.
\section{Proposed sensitivity model and population bounds}\label{sec:propsens}
\subsection{Enhanced MSM} \label{sec:eMSMmodel}
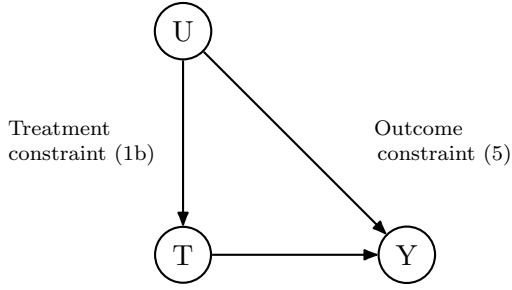
\begin{figure}
    \centering
    \begin{tikzpicture}[
    roundnode/.style={circle, draw=black, thick, minimum size=.6cm},
    directed edge/.style={-{Latex[round]} , thick},
    node distance=2.2cm
]
    \node[roundnode] (T) {T};
    \node[roundnode,right=of T] (Y) {Y};
    \node[roundnode,above=of T] (U) {U};

    \draw[directed edge] (T) -- (Y);
    \draw[directed edge] (U) -- (Y) node[midway, right] {\parbox{3cm}{\centering \scriptsize Outcome \\\hspace{2em} constraint \eqref{eq:UYconstr}}};
    \draw[directed edge] (U) -- (T) node[midway, left] {\parbox{3cm}{\centering \scriptsize Treatment\\ \hspace{2em}constraint \eqref{eq:msm-bound}}};
\end{tikzpicture}
    \caption{Graphical representation of eMSM constraints conditional on $X$.}
    \label{fig:sensitivity-assumption-dag}
\end{figure}

As a motivation for the proposed model, we observe that to produce confounding beyond $X$,
an unmeasured variable $U$ needs to be associated with \textit{both} $T$ \textit{and} $(Y^0,Y^1)$ given $T$, within some level of $X$.
Stated in reverse, if \eqref{eq:msm-indep} is satisfied, then
\textit{either} $U\perp T|X$ \textit{or} $U\perp (Y^0,Y^1)|T,X$ implies no unmeasured confounding, $(Y^0, Y^1) \perp T | X$.
Such results are known when characterizing confounders in general (e.g., \citeauthor{greenland1986identifiability}, \citeyear{greenland1986identifiability}).
See Supplement Lemma~\ref{lem:U-confounding} for mathematical details.
Therefore, both the $U$-$T$ association and $U$-$(Y^0,Y^1)$ association, conditional on $X$, can be used to measure confounding strength.

From the preceding perspective, MSM constraint \eqref{eq:msm-bound} posits a bound on the conditional association between $U$ and $T$ in the odds ratio scale as in \eqref{eq:w-def}, but leaves the association between $U$ and $(Y^0,Y^1)$ arbitrary. This may often lead to wide population sharp bounds under MSMs.
To address this limitation when inferring $\mu^1$, we postulate, in addition to \eqref{eq:msm}, the following constraint on the the deviation from $U\perp Y^1|T=1,X$:
\begin{equation}\label{eq:UYconstr}
 \mg^*(X,U) - E(Y^1|T=1,X)\in[-\Delta_1(X), \Delta_2(X)],
\end{equation}
where $\mg^*(X,U)=E (Y^1|T=1,X,U)$, and $\Delta_1(X), \Delta_2(X) \ge 0$ are some \textit{pre-specified} covariate functions.
See Figure \ref{fig:sensitivity-assumption-dag} for an illustration in a directed cyclic graph.
The larger $\Delta_1(X)$ and $\Delta_2(X)$ are, the more $U$ may influence $Y^1$ given $T=1$ and hence the greater unmeasured confounding is allowed. Note that, due to $E \{ E(Y^1|T=1,X,U)|T=1,X\} = E(Y^1 |T=1,X)$, the interval $[-\Delta_1(X), \Delta_2(X)]$ should always contain 0 and hence $\Delta_1(X), \Delta_2(X) \ge 0$. By the consistency assumption, \eqref{eq:UYconstr} can also be expressed in terms of the observed outcome:
$E (Y|T=1,X,U)- E(Y|T=1,X)\in[-\Delta_1(X), \Delta_2(X)]$.
The constraint \eqref{eq:UYconstr} can be considered ``marginal'', concerning how the mean of $Y^1$ given $T=1,X,U$ can differ from that with $U$ marginalized out,
which is in a similar sense of constraint \eqref{eq:msm-bound} concerning how the odds ratio of $T$  given $X,U$ can differ from that with $U$ marginalized out.

We refer to as an enhanced MSM (eMSM) the model defined by the constraints \eqref{eq:msm-indep}, \eqref{eq:msm-bound}, and \eqref{eq:UYconstr},
where $(\Lambda_1(X), \Lambda_2(X))$ and $(\Delta_1(X), \Delta_2(X))$ are sensitivity parameters,
controlling the influence of $U$ on $T$ and, respectively, that of $U$ on $Y^1$ given $T=1$.
For convenience, the constraint \eqref{eq:msm-bound} can be called a treatment sensitivity constraint,
and \eqref{eq:UYconstr} an outcome sensitivity constraint,
even though the constraint \eqref{eq:msm-bound} is also relevant in terms of the density ratio of potential outcomes
as shown by the equivalence between model \eqref{eq:msm} and model \eqref{eq:msm-Y0Y1} of Supplement.
By design, MSM \eqref{eq:msm} is recovered when $\Delta_1(X)=\Delta_2(X) \equiv \infty$, i.e., $E (Y^1|T=1,X,U)$ is unconstrained.
If $\Lambda_1(X) = \Lambda_2(X) \equiv 1 $, unconfoundedness holds as discussed earlier. Alternatively, if $\Delta_1 (X) = \Delta_2 (X) \equiv 0$,
then unconfoundedness holds in mean, i.e., $E(Y^1 | T=0,X) = E(Y^1 | T=1,X)$.
In either case, $\mu^1 = E(Y^1)$ is identified from the observed data as $\mu^1=  E \{ E (Y | T=1,X)\}$. See Supplement Section~\ref{sec:Characterize-U} for details.

\subsection{Population sharp bounds} \label{sec:sharpbounds}
For sensitivity analysis with eMSM defined by \eqref{eq:msm} and \eqref{eq:UYconstr},
we define the sharp lower and upper bounds of $\mu^1$ as
\begin{align}
& \mu^{1-}_{\mytext{eMSM}} = \inf_Q \; E_Q (Y^1) , \quad
\mu^{1+}_{\mytext{eMSM}} = \sup_Q \; E_Q (Y^1) ,  \label{eq:emsm-bound-mu1}
\end{align}
over any joint distribution $Q$ on any choice of the full data $(Y^0,Y^1,T,X,U)$ such that
\begin{itemize}\addtolength{\itemsep}{-.1in}
\item[(i)] $Q$ is compatible with the observed-data distribution, in the same manner as in the constraint~(i) in
the MSM sharp bound \eqref{eq:msm-bound-mu1}, and

\item[(ii)] $Q$ satisfies eMSM: properties \eqref{eq:msm-indep}, \eqref{eq:msm-bound}, and \eqref{eq:UYconstr} hold under $Q$, i.e.,
\begin{subequations}
\begin{align}
 &  (Y^0,Y^1)\perp T | X,U \quad \text{under $Q$}, \label{eq:Q-indep} \\
 &  \lambda_Q(X,U) \in [\Lambda_1(X),\Lambda_2(X)], \label{eq:Q-lam-bound}\\
 & \mg_Q(X,U) - E(Y^1|T=1,X)\in[-\Delta_1(X), \Delta_2(X)], \label{eq:Q-g-bound}
\end{align}
\end{subequations}
where $\lambda_Q (X,u) = \frac{\ud Q_U (u|T=0,X)}{\ud Q_U (u|T=1,X)}$
and $\mg_Q (X,u) =E_Q ( Y^1 |T=1, X,U=u)$, satisfying $\mg_Q (X,u) =E_Q ( Y^1 |T=0, X,U=u) = E_Q (Y^1 | X,U=u)$ by \eqref{eq:Q-indep}.
\end{itemize}
The central question of this section is then to derive implementable representations of the sharp bounds
$\mu^{1-}_{\mytext{eMSM}}$ and $ \mu^{1+}_{\mytext{eMSM}} $. By symmetry, we mainly discuss the sharp upper bound.

To start, for any joint distribution $Q$ allowed in \eqref{eq:emsm-bound-mu1}, we note that the mean
$E_Q(Y^1)$ can be expressed as follows:
\begin{subequations}\label{eq:Expected-Y}
\begin{align}
E_Q(Y^1) & =E \left\{TY+(1-T) E_Q (Y^1|T=0,X) \right\}, \label{eq:mu1-formula}\\
E_Q (Y^1|T=0,X) & = \int \mg_Q(X,u) \lambda_Q(X,u) \ud Q_U (u|T=1,X) . \label{eq:nu1-formula}
\end{align}
\end{subequations}
See Supplement Section \ref{sec:proof-eq-EQY} for a proof.
From the expressions \eqref{eq:mu1-formula}--\eqref{eq:nu1-formula}, the mean $E_Q(Y^1)$ depends on $Q$ through
the conditional expectation $E_Q (Y^1|T=0,X=x)$ separately in $x$. Hence the sharp upper bound can be determined as
$\mu^{1+}_{\mytext{eMSM}} = E \{TY+(1-T) \nu^{1+}_{\mytext{eMSM}} (X) \}$, where
$\nu^{1+}_{\mytext{eMSM}} (X)$ is the sharp upper bound of $\nu^1(X) = E(Y^1 | T=0,X)$ defined as
\begin{align}
\nu^{1+}_{\mytext{eMSM}} (X) 
=\sup_Q\; \int \mg_Q (X,u)\lambda_Q (X,u)\ud Q_U (u|T=1,X),\label{eq:condtightup}
\end{align}
over any joint distribution $Q$ allowed in \eqref{eq:emsm-bound-mu1} conditionally on $X$.
Therefore, it suffices to solve \eqref{eq:condtightup}, i.e., derive an implementable representation of $\nu^{1+}_{\mytext{eMSM}} (X)$.

A major challenge in solving \eqref{eq:condtightup} is that the optimization is over $Q$, whose effect on the objective function in \eqref{eq:condtightup} is mediated by the interdependent triple $\{\lambda_Q(X,\cdot),\mg_Q(X,\cdot),Q_U(\cdot|T=1,X)\}$. In addition to the range constraints \eqref{eq:Q-lam-bound}--\eqref{eq:Q-g-bound}, $\lambda_Q(X,\cdot)$ and $\mg_Q(X,\cdot)$, by definition, are subject to \textit{inherent} constraints such as
\begin{subequations}
\begin{align}
 & \int \lambda_Q(X,u) \ud Q_U (u|T=1,X) \equiv 1,  \label{eq:lam-inherent} \\
 & \int \mg_Q(X,u) \ud Q_U (u|T=1,X) = E ( Y | T=1, X).  \label{eq:g-inherent}
\end{align}
\end{subequations}
It can be shown that constraint~\eqref{eq:lam-inherent} is sufficient for $\lambda_Q(X,\cdot)$, in the sense that for any $Q$ allowed in \eqref{eq:condtightup} and any nonnegative function $\lambda (X,\cdot)$ satisfying \eqref{eq:lam-inherent}, replacing
$\ud Q_{Y^0,Y^1,U}(\cdot|T=0,X)$ with $\ud Q_{Y^0,Y^1}(\cdot|T=0,X,U)\lambda(X,\cdot)\ud Q_U(\cdot|T=1,X)$
defines another joint distribution allowed in \eqref{eq:condtightup}.
See \cite{tan2024sensitivity} Lemma~3 for a similar result. However, constraint~\eqref{eq:g-inherent} is generally not sufficient for $\mg_Q(X,\cdot)$. For any $u$ satisfying $c_Q (X,u) = Q(u | T=1,X)>0$, the following constraints hold:
\begin{subequations}\label{eq:g-inherentNP}
\begin{align}
 &  \mg_Q(X,u) \le E \left\{ Y + c_Q^{-1} \rho_{1-c_Q} (Y, q^*_{1,1-c_Q}(X)) |T=1,X \right\},  \label{eq:g-inherentNP-up} \\
 &  \mg_Q(X,u) \ge E \left\{ Y - c_Q^{-1} \rho_{c_Q} (Y, q^*_{1,c_Q}(X)) |T=1,X \right\},\label{eq:g-inherentNP-low}
\end{align}
\end{subequations}
where $c_Q = c_Q (X,u)$, $\rho_{c_Q}(y,q)$ is the check function and $q^*_{1,c_Q}(X)$ is the $c_Q$-quantile of $P_Y(\cdot|T=1,X)$ as in Section~\ref{sec:setup}.
This result is a variation of the optimization for MSM sharp bounds, and hence resembles \eqref{eq:MSM-bound-nu} and its lower-bound counterpart with sensitivity parameters $(\Lambda_1,\Lambda_2)=(0,c_Q^{-1})$. The proof is provided in Supplement Section~\ref{sec:proof-g-inherentNP}.

In spite of the challenge, we find transparent, implementable representations of the eMSM sharp bounds,
as stated in the following result.

\begin{pro}[Sharp upper bounds]\label{pro:pop-sol}
The sharp upper bound $\nu^{1+}_{\mytext{eMSM}} (X)$ in \eqref{eq:condtightup} for
$\nu^1(X) = E(Y^1 | T=0,X)$ can be determined as
\begin{align}\label{eq:asymomax}
\begin{split}
& \nu^{1+}_{\mytext{eMSM}} (X) = E(Y|T=1,X) +\\
& \quad \{\Lambda_2(X)-\Lambda_1(X)\} \min\big[ \tau(X) \Delta_1(X), \{1-\tau(X)\} \Delta_2 (X), E \{\rho_{\tau}(Y,q^*_{1,\tau})|T=1,X \} \big],
\end{split}
\end{align}
where $\tau(X)=\{\Lambda_2(X)-1\}/\{\Lambda_2(X)-\Lambda_1(X)\}$,
 $\rho_{\tau} (y,q)$ is the check function, and $q^*_{1,\tau}=q^*_{1,\tau}(X)$ is the $\tau(X)$-quantile of $P_Y(\cdot| T=1,X)$
 as in \eqref{eq:MSM-bound-mu}--\eqref{eq:MSM-bound-nu}.
The sharp upper bound $\mu^{1+}_{\mytext{eMSM}} $ in \eqref{eq:emsm-bound-mu1} for $\mu^1=E(Y^1)$ can be obtained as
$\mu^{1+}_{\mytext{eMSM}} = E \{TY+(1-T) \nu^{1+}_{\mytext{eMSM}} (X) \}$.
\end{pro}

\begin{figure}
    \centering
    \includegraphics[width=.7\linewidth]{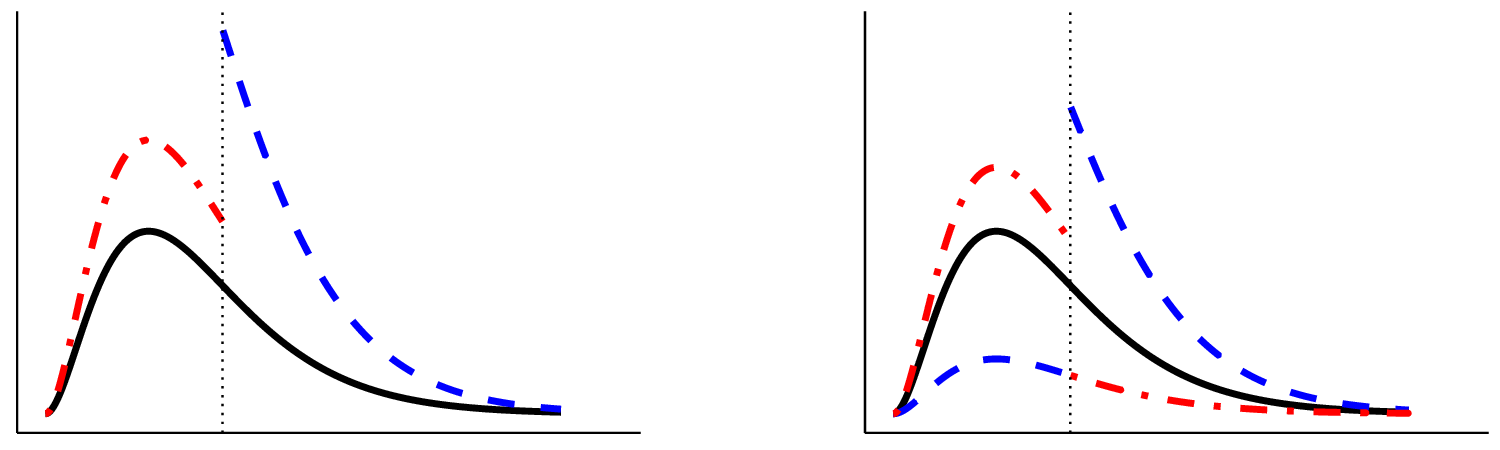}
    \caption{Effect of worst-case confounder $U$ on $Y^1$, conditioned on $X$, according to Corollary~\ref{cor:eMSM-Q}.
    Density functions of $Y|T=1$ (solid black), $Y^1|U=1$ (dashed blue), and $Y^1|U=0$ (dotdash red) are plotted.
    $\Lambda_2=\Lambda_1^{-1}=2$, and vertical dotted line marks the $2/3$-quantile of $Y|T=1$.
    The parameters $(\Delta_1,\Delta_2)$ are set such that $\psi_{1+}=1$ (left) or $\psi_{1+}=0.7$ (right).}
    \label{fig:emsmU}
\end{figure}

For later discussion, it is helpful to rewrite the representation of $\nu^{1+}_{\mytext{eMSM}} (X)$ in \eqref{eq:asymomax} as
\begin{align}
& \nu^{1+}_{\mytext{eMSM}} (X) = E(Y|T=1,X) + \{\Lambda_2(X)-\Lambda_1(X)\}\psi_{1+}(X)E\{ \rho_{\tau}(Y,q^*_{1,\tau})|T=1,X\}, \label{eq:asymomax-b}
\end{align}
where $\psi_{1+}(X)=\min \{ \delta_{1+}(X), \delta_{2+}(X), 1\}$ with
\begin{align*}
 \delta_{1+} (X) = \frac{\tau(X) \Delta_1(X)}{E\{\rho_{\tau}(Y,q^*_{1,\tau})|T=1,X\}},\quad
 \delta_{2+}(X) = \frac{\{1-\tau(X)\}\Delta_2(X)}{E\{\rho_{\tau}(Y,q^*_{1,\tau})|T=1,X\}}.
\end{align*}
The eMSM bound in \eqref{eq:asymomax-b} resembles the MSM bound in \eqref{eq:MSM-bound-nu}, except for the factor $\psi_{1+}(X)$.
Moreover, the sharp bounds in Proposition \ref{pro:pop-sol} can be achieved by a joint distribution $Q$ as in Corollary~\ref{cor:eMSM-Q}, with $Y^0$ omitted for simplicity. See Figure~\ref{fig:emsmU} for an illustration.

\begin{cor}[Worst-case unmeasured confounding]  \label{cor:eMSM-Q}
A joint distribution $Q$ on $(Y^0,Y^1,T,X,U)$ which achieves
the sharp upper bound $\nu^{1+}_{\mytext{eMSM}} (X)$ given $X$
and the sharp upper bound $\mu^{1+}_{\mytext{eMSM}} $ can be specified with the following properties:
(i) the unmeasured confounder $U$ is binary, (ii)
$Q_ U (u | T=0, X)$ is re-weighted from $Q_ U (u | T=1, X)$ by $\Lambda_1(X)$ or $\Lambda_2(X)$ for $u=0$ or $1$:
\begin{align*}
 Q_U ( 1 | T=1, X) & = 1-Q_U ( 0 | T=1, X) = 1-\tau(X),\\
 \lambda_Q (X,u) &=\begin{cases}
        \Lambda_1(X)&\text{if}\quad u=0, \\
        \Lambda_2(X)&\text{if}\quad u=1, \\
    \end{cases}
\end{align*}
and (iii) for $u=0,1$, $Q_{Y^1}(\cdot|X,U=u)$ is a re-mix of $P_Y(\cdot|T=1,X)$:
\begin{align*}
 \ud Q_{Y^1} (y | X, U=1) & = \begin{cases}
        \left(1+\frac{\tau(X)}{1-\tau(X)}\psi_{1+}\right)\ud P_Y (y | T=1,X) &\text{if}\quad y>q^*_{1,\tau}, \\
        \left(1-\psi_{1+}\right)\ud P_Y (y | T=1,X) &\text{if}\quad y<q^*_{1,\tau}, \\
   \end{cases} \\
 \ud Q_{Y^1} (y | X, U=0) & = \begin{cases}
        \left(1-\psi_{1+}\right)\ud P_Y (y | T=1,X)&\text{if}\quad y>q^*_{1,\tau}, \\
        \left(1+\frac{1-\tau(X)}{\tau(X)}\psi_{1+}\right)\ud P_Y (y | T=1,X) &\text{if}\quad y<q^*_{1,\tau}. \\
   \end{cases}
\end{align*}
Wherever $P_Y(q^*_{1,\tau}|T=1,X)>0$,
\begin{align*}
    &Q_{Y^1} (q^*_{1,\tau}| X, U=1)=\left(1-\psi_{1+}\right) P_Y(q^*_{1,\tau} | T=1,X)+\psi_{1+} \left(1-\frac{P(Y>q^*_{1,\tau}|T=1,X)}{1-\tau(X)}\right) ,\\
    &Q_{Y^1} (q^*_{1,\tau} | X, U=0)=\left(1-\psi_{1+}\right) P_Y(q^*_{1,\tau} | T=1,X)+ \psi_{1+} \left(1-\frac{P(Y< q^*_{1,\tau}|T=1,X)}{\tau(X)}\right).
\end{align*}
The worst-case $Q$ may also be specified differently from the above.
\end{cor}

We describe our strategy in solving the distributional optimization \eqref{eq:condtightup} that leads to Proposition \ref{pro:pop-sol} and Corollary \ref{cor:eMSM-Q}.
The eMSM sharp upper bound of $\nu^1$ in \eqref{eq:asymomax} can be represented as the minimum of two relaxed upper bounds:
$$ \nu^{1+}_{\mytext{eMSM}} (X) = \min \{\nu^{1+}_{\mytext{MSM}} (X), \nu^{1+}_{\mytext{e}} (X)\},$$
where $\nu^{1+}_{\mytext{MSM}} (X)$ is the MSM sharp upper bound of $\nu^1(X)$ in \eqref{eq:MSM-bound-nu} and
\begin{align*}
& \nu^{1+}_{\mytext{e}} (X) = E(Y|T=1,X) +\{\Lambda_2(X)-\Lambda_1(X)\} \min\big[ \tau(X) \Delta_1(X), \{1-\tau(X)\} \Delta_2 (X) \big].
\end{align*}
To prove the preceding result, we first note that $\nu^{1+}_{\mytext{MSM}} (X)$
remains a valid but possibly relaxed upper bound of $\nu^1(X)$ under eMSM.
Second, as shown in Supplement Section \ref{sec:proof-prop1-cor1},
we obtain $\nu^{1+}_{\mytext{e}} (X)$ by solving the following distributional-functional optimization:
\begin{align}
\nu^{1+}_{\mytext{e}} (X) =\sup_{Q_U(\cdot|T=1,X),\lambda(X,\cdot),\mg(X,\cdot)}\;
\int \mg(X,u) \lambda (X,u)\ud Q_U (u|T=1,X),  \label{eq:condtightup-b}
\end{align}
over any choice of distribution $Q_U(\cdot|T = 1,X)$, and two measurable functions $\lambda (X,\cdot)$ (nonnegative)
and $\mg(X,\cdot)$ satisfying the inherent constraints \eqref{eq:lam-inherent}--\eqref{eq:g-inherent} and the range constraints \eqref{eq:Q-lam-bound}--\eqref{eq:Q-g-bound} as $\lambda_Q (X,\cdot)$ and $\mg_Q (X,\cdot)$. 
Compared with \eqref{eq:condtightup} for $ \nu^{1+}_{\mytext{eMSM}} (X)$, the optimization in \eqref{eq:condtightup-b} is much simpler:
$Q$ is defined only for $U$ given $T=1$ and $X$,
and both $ \lambda (X,\cdot)$ and $\mg(X,\cdot)$ are related to $Q$ only through the constraints corresponding to \eqref{eq:lam-inherent}--\eqref{eq:g-inherent},
but without accounting for further constraints such as \eqref{eq:g-inherentNP-up}--\eqref{eq:g-inherentNP-low} on $\mg_Q(X,\cdot)$.
Therefore, $\nu^{1+}_{\mytext{e}} (X)$ is a valid but possibly relaxed upper bound of $\nu^1(X)$ under eMSM,
and so is $\min \{\nu^{1+}_{\mytext{MSM}} (X), \nu^{1+}_{\mytext{e}} (X)\}$.
Finally, sharpness is proved by constructing a joint distribution $Q$ that, as described in Corollary \ref{cor:eMSM-Q}, attains the bound.

We provide additional remarks to help understand Proposition \ref{pro:pop-sol} and Corollary \ref{cor:eMSM-Q}.
For brevity, the discussion mainly deals with the conditional sharp bound $\nu^{1+}_{\mytext{eMSM}} (X)$, and a similar discussion
holds for the unconditional sharp bound $\mu^{1+}_{\mytext{eMSM}} $.

First, as a sanity check, if $\Lambda_1(X) = \Lambda_2(X) = 1$ but $(\Delta_1(X), \Delta_2(X))$ may be arbitrary, then
$\nu^{1+}_{\mytext{eMSM}} (X)$ reduces to $E(Y|T=1,X)$, which identifies $\nu^1 (X)$ under ignobility.
If $\Delta_1(X) = \Delta_2(X) =0$ but $(\Lambda_1(X), \Lambda_2(X))$ may be arbitrary, then
$\nu^{1+}_{\mytext{eMSM}} (X)$ also reduces to $E(Y|T=1,X)$.
These ``degenerate'' cases are consistent with our earlier discussions in Section \ref{sec:eMSMmodel}.

Second, Proposition \ref{pro:pop-sol} demonstrates in a transparent manner how the eMSM bound is compared with the MSM bound, depending on
the factor $\psi_+ \,(\le 1)$.
In fact, the eMSM upper bound $\nu^{1+}_{\mytext{eMSM}} (X)$ \textit{either} coincides with the MSM upper bound $\nu^{1+}_{\mytext{MSM}} (X)$
when $\psi_{1+}(X) = 1$, i.e., $\min[\tau(X) \Delta_1(X), \{1-\tau(X)\} \Delta_2 (X)] \ge E\{\rho_{\tau}(Y,q^*_{1,\tau})|T=1,X\}$,
\textit{or} is smaller (i.e., tighter) than the MSM upper bound
when $\psi_{1+}(X)<1$, i.e., $\min[\tau(X) \Delta_1(X), \{1-\tau(X)\} \Delta_2 (X)] < E\{\rho_{\tau}(Y,q^*_{1,\tau})|T=1,X\}$.
Therefore, incorporating the outcome sensitivity constraint \eqref{eq:UYconstr} with relatively large $(\Delta_1(X), \Delta_2(X))$
leads to the same sensitivity bounds as under MSM, but that with relatively small $(\Delta_1(X), \Delta_2(X))$
leads to improved (narrower) sensitivity bounds. More importantly,
Proposition \ref{pro:pop-sol} makes it clear that for such an improvement to happen,
$\min[\tau(X) \Delta_1(X), \{1-\tau(X)\} \Delta_2 (X)]$ needs to be smaller than the optimized quantile loss $E\{\rho_{\tau}(Y,q^*_{1,\tau})|T=1,X\}$, which is data dependent.
From this observation, $E\{\rho_{\tau}(Y,q^*_{1,\tau})|T=1,X\}$ serves as a natural yardstick for the specification
of sensitivity parameters $(\Delta_1(X), \Delta_2(X))$. This topic is further investigated in Section~\ref{sec:ParaSpec}.

Third, from Corollary~\ref{cor:eMSM-Q}, the worst-case unmeasured confounding under eMSM can be induced by
a binary unmeasured confounder $U$, which takes value 1 more likely (and hence value 0 less likely) in the untreated ($T=0$) than in the treated ($T=1$),
while $U$ taking value 1 is, on average, associated with larger values of $Y^1$ than $U$ taking value 0, conditionally on $X$. It is informative to distinguish two cases as illustrated in Figure~\ref{fig:emsmU}.
\vspace{-.05in}
\begin{itemize}\addtolength{\itemsep}{-.1in}
\item In the case of $\psi_{1+}(X) = 1$, indicating $\nu^{1+}_{\mytext{eMSM}} (X) = \nu^{1+}_{\mytext{MSM}} (X)\le \nu^{1+}_{\mytext{e}} (X)$,
the association of $U$ with $Y^1$ is simple:
$U$ taking value $1$ is \textit{deterministically} (not just on average) associated with
larger values of $Y^1$ than $U$ taking value 0, such that
$Q_{Y^1} (\cdot| X,U=1)$ or $Q_{Y^1} (\cdot| X,U=0)$ corresponds to, respectively,
the higher or lower segment of the distribution $P_Y (\cdot | T=1,X)$ divided at the $\tau(X)$-quantile $q^*_{1,\tau}(X)$.
This worst-case construction agrees with that in \citet{dorn2023sharp} obtained under the MSM.

\item In the case of $\psi_{1+}(X) <1$, indicating $\nu^{1+}_{\mytext{eMSM}} (X) = \nu^{1+}_{\mytext{e}} (X)< \nu^{1+}_{\mytext{MSM}} (X)$,
the association of $U$ with $Y^1$ is more complicated but remains interpretable:
$Q_{Y^1} (\cdot| X,U=1)$ or $Q_{Y^1} (\cdot| X,U=0)$ each corresponds to a non-degenerate mixture of
the two segments of the distribution $P_Y (\cdot | T=1,X)$ mentioned above,
but with different mixture proportions, such that the relaxed bound $\nu^{1+}_{\mytext{e}} (X)$ is realized and hence becomes a sharp bound.
\end{itemize}
In either case, the binary unmeasured confounder $U$ acts as a selection indicator deciding how $Y^1$ is related to $Y$ in the two segments of the distribution $P_Y (\cdot | T=1,X)$.

Fourth, by the definition of $q^*_{1,\tau}(X)$ as a minimizer of $E \{\rho_{\tau} (Y, q) | T=1,X\}$ over $q\in\bbR$,
the formula \eqref{eq:asymomax} shows that conditionally on $X$,
\begin{align*}
\nu^{1+}_{\mytext{eMSM}} (X) & = \inf_{q\in\bbR} \; \nu^{1+}_{\mytext{eMSM}} (X; q ) ,
\end{align*}
where $\nu^{1+}_{\mytext{eMSM}} (X; q)$ is defined as the right-hand side of \eqref{eq:asymomax} with $q^*_{1,\tau}(X)$ replaced by $q$.
The minimum above may be achieved by $q$ different from $q^*_{1,\tau}(X)$ in the case of $\psi_{1+}(X) <1$.
This provides a dual representation for $\nu^{1+}_{\mytext{eMSM}} (X)$ defined via maximization in \eqref{eq:condtightup},
similarly as in the duality relationships established for the MSM sharp bounds \citep{tan2024model}.
Moreover, due to the separability in different values of $X$, a dual representation also holds for $\mu^{1+}_{\mytext{eMSM}}$:
\begin{align*}
& \mu^{1+}_{\mytext{eMSM}} = \inf_{q(\cdot)} \; E \Big\{ T Y + (1-T) \nu^{1+}_{\mytext{eMSM}} (X; q(X)) \Big\},
\end{align*}
over any covariate function $q(X)$.
A simple but important implication from these duality relationships is that
$E \{ T Y + (1-T) \nu^{1+}_{\mytext{eMSM}} (X; q(X)) \}$ yields a valid (relaxed) upper bound on $\mu^1$
for any covariate function $q(X)$, and becomes sharp when $q(X) = q^*_{1,\tau}(X)$.

Next, we present corresponding representations for eMSM sharp lower bounds.
The representation for $\nu^{1-}_{\mytext{eMSM}} (X)$ below can be derived by negating the representation for
the sharp upper bound $\nu^{1+}_{\mytext{eMSM}} (X)$ in \eqref{eq:asymomax} with $Y$ and $(\Delta_1(X),\Delta_2(X))$ replaced by
$-Y$ and $(\Delta_2(X), \Delta_1(X))$.
The worst-case unmeasured confounding can also be characterized similarly as in Corollary~\ref{cor:eMSM-Q}.

\begin{pro}[Sharp lower bounds]\label{pro:pop-sol-lower}
The sharp lower bound $\nu^{1-}_{\mytext{eMSM}} (X)$ for $\nu^1(X) = E(Y^1 | T=0,X)$, defined as $\nu^{1+}_{\mytext{eMSM}} (X)$ in \eqref{eq:condtightup}
with supremum replaced by infimum, can be determined as
\begin{align*}
& \nu^{1-}_{\mytext{eMSM}} (X) = E(Y|T=1,X)- \\
& \quad \{\Lambda_2(X)-\Lambda_1(X)\} \min\big[ \tau(X) \Delta_2(X), \{1-\tau(X)\} \Delta_1 (X), E \{\rho_{1-\tau}(Y,q^*_{1,1-\tau})|T=1,X \} \big] \\
& = E(Y|T=1,X)-\{\Lambda_2(X)-\Lambda_1(X)\}\psi_{1-}(X)E\{\rho_{1-\tau}(Y,q^*_{1,1-\tau})|T=1,X\},
\end{align*}
where
$\psi_{1-}(X)=\min \{\delta_{1-}(X), \delta_{2-}(X), 1\}$ with
\begin{align*}
\delta_{1-}(X) = \frac{\{1-\tau(X)\}\Delta_1(X)}{E\{\rho_{1-\tau}(Y,q^*_{1,1-\tau})|T=1,X\}} , \quad
\delta_{2-}(X) = \frac{\tau(X) \Delta_2(X)}{E\{\rho_{1-\tau}(Y,q^*_{1,1-\tau})|T=1,X\}}.
\end{align*}
The sharp lower bound $\mu^{1-}_{\mytext{eMSM}} $ in \eqref{eq:emsm-bound-mu1} for $\mu^1=E(Y^1)$ can be obtained as
$\mu^{1-}_{\mytext{eMSM}} = E \{TY+(1-T) \nu^{1-}_{\mytext{eMSM}} (X) \}$.
\end{pro}

\subsection{Recommended specification of sensitivity parameters}\label{sec:ParaSpec}
Our development in Section~\ref{sec:eMSMmodel} is general, allowing specification of sensitivity parameters
$(\Lambda_1(X), \Lambda_2(X))$ and $(\Delta_1(X), \Delta_2(X))$ to be covariate-dependent.
Typically, sensitivity analysis involves varying the magnitudes of such sensitivity parameters
and examining the resulting bounds on the causal parameters of interest.
For practical interpretation and implementation, we discuss suitable specification of sensitivity parameters
which can be varied in a covariate-independent manner, possibly after some re-parametrization.

For the treatment sensitivity parameters $(\Lambda_1(X), \Lambda_2(X))$, a simple specification as in \cite{tan2006distributional} is to
take $\Lambda_1(X) \equiv \Lambda^{-1}$ and
$ \Lambda_2(X) \equiv \Lambda$ for a constant $\Lambda \ge 1$. In this case, $\tau(X) \equiv \Lambda / (\Lambda+1) \ge 1/2$.
In the subsequent discussion, to accommodate asymmetric specifications, we assume
that $\Lambda_1(X)$ and $ \Lambda_2(X)$ are specified as two separate constants, $\Lambda_1\,(\le 1)$ and $\Lambda_2\,(\ge 1)$.
Then $\tau(X)\equiv\tau = (\Lambda_2-1)/(\Lambda_2-\Lambda_1)$ satisfies $\tau \ge 1/2$ if and only if $\Lambda_1 + \Lambda_2 \ge 2$.

For the outcome sensitivity parameters $(\Delta_1(X), \Delta_2(X))$, more care is needed to find a simple and meaningful specification.
Instead of directly specifying $(\Delta_1(X), \Delta_2(X))$ as constants, we see from the second remark after Corollary~\ref{cor:eMSM-Q} that
it is more natural to parameterize the relative magnitudes of
$\Delta_1(X)$ and $\Delta_2(X)$ against the optimized quantile loss $E\{ \rho_{\tau}(Y,q^*_{1,\tau})|T=1,X\}$,
when evaluating the sharp upper bound of $\mu^1$.
Taking $\delta_{1+}(X) =\delta_{2+}(X) \equiv\delta $ would lead to the specification
\begin{align}\label{eq:upspec}
    [-\Delta_{1,\mytext{U}} (X),\Delta_{2,\mytext{U}} (X)]&=[ -\delta /\tau, \delta /(1-\tau)]\cdot E\{ \rho_{\tau}(Y,q^*_{1,\tau})|T=1,X\},
\end{align}
where $\delta \in  [0,1]$ is a constant.
The sharp upper bound $\nu^{1+}_{\mytext{eMSM}} (X)$ would become
\begin{align}  \label{eq:nu-upper-r}
 \nu^{1+}_{\mytext{eMSM,U}} (X) = E(Y|T=1,X)
  +(\Lambda_2-\Lambda_1) \delta E \{\rho_{\tau}(Y,q^*_{1,\tau})|T=1,X \} .
\end{align}
Similarly, when evaluating the sharp lower bound of $\mu^1$ based on Proposition \ref{pro:pop-sol-lower},
taking $\delta_{1-}(X) = \delta_{2-}(X) \equiv\delta $ would give the specification
\begin{align}\label{eq:lowspec}
[-\Delta_{1,\mytext{L}}(X),\Delta_{2,\mytext{L}} (X)]&=[ -\delta /(1-\tau),\delta /\tau]\cdot E\{ \rho_{1-\tau}(Y,q^*_{1,1-\tau})|T=1,X\}.
\end{align}
The sharp lower bound $\nu^{1-}_{\mytext{eMSM}} (X)$ would become
\begin{align}  \label{eq:nu-lower-r}
 \nu^{1-}_{\mytext{eMSM,L}} (X) = E(Y|T=1,X)
  -(\Lambda_2-\Lambda_1) \delta E \{\rho_{1-\tau}(Y,q^*_{1,1-\tau})|T=1,X \} .
\end{align}
However, the two specifications \eqref{eq:upspec} and \eqref{eq:lowspec} are in general incompatible with each other:
if \eqref{eq:upspec} gives $(\Delta_1(X), \Delta_2(X))$, then \eqref{eq:lowspec} would not, and vice versa.

To address the above issue, in the case of $\tau \ge 1/2$, we recommend taking $\Delta_2(X)$  and $\Delta_1(X)$ from \eqref{eq:upspec} and \eqref{eq:lowspec} \textit{respectively}
with a sensitivity parameter $\delta \in  [0,1]$:
\begin{align} \label{eq:recspec}
\begin{split}
& \Delta_2(X)=\Delta_{2,\mytext{U}}= \delta /(1-\tau) \cdot E\{ \rho_{\tau}(Y,q^*_{1,\tau})|T=1,X\}, \\
& \Delta_1(X)=\Delta_{1,\mytext{L}} = \delta /(1-\tau) \cdot E\{ \rho_{1-\tau}(Y,q^*_{1,1-\tau})|T=1,X\},
\end{split}
\end{align}
i.e., $\Delta_2(X)$ and $\Delta_1(X)$ are specified as the products of
a constant $\delta /(1-\tau)$ and the optimized quantile losses $E\{ \rho_{\tau}(Y,q^*_{1,\tau})|T=1,X\}$
and $E\{ \rho_{1-\tau}(Y,q^*_{1,1-\tau})|T=1,X\}$ \textit{respectively}. The resulting sharp upper and lower bounds of $\nu^1(X)$ from Propositions \ref{pro:pop-sol} and \ref{pro:pop-sol-lower} are
\begin{subequations}
\begin{align}
 & \nu^{1+}_{\mytext{eMSM}} (X) = E(Y|T=1,X)\nonumber \\
 &\quad +(\Lambda_2-\Lambda_1) \min\left[\delta \frac{\tau}{1-\tau} E\{ \rho_{1-\tau}(Y,q^*_{1,1-\tau})|T=1,X\}, \delta E \{\rho_{\tau}(Y,q^*_{1,\tau})|T=1,X \} \right] ,
    \label{eq:nu-upper-complex} \\
 & \nu^{1-}_{\mytext{eMSM}} (X) = E(Y|T=1,X)\nonumber \\
 &\quad -(\Lambda_2-\Lambda_1) \min \left[\delta E\{ \rho_{1-\tau}(Y,q^*_{1,1-\tau})|T=1,X\}, \delta \frac{\tau}{1-\tau} E \{\rho_{\tau}(Y,q^*_{1,\tau})|T=1,X \} \right] .
   \label{eq:nu-lower-complex}
\end{align}
\end{subequations}
Interestingly, these more complex expressions, \eqref{eq:nu-upper-complex} and \eqref{eq:nu-lower-complex},
can be shown to reduce back to \eqref{eq:nu-upper-r} and \eqref{eq:nu-lower-r} provided
$ \tau \ge 1/2$, which automatically holds for the typical choice $\Lambda_1^{-1}=\Lambda_2=\Lambda$.
Another important advantage of removing the minimum operator from \eqref{eq:nu-upper-complex}--\eqref{eq:nu-lower-complex}
is that it enables identification of the sharp bounds of $\mu^1$ by inverse probability weighting in Proposition~\ref{prop:propspecbd}
and by doubly robust functionals in Proposition~\ref{prop:doublyrobust} later.
For completeness, additional discussion in the case of $ \tau \le 1/2$ is provided in Supplement Section \ref{sec:proof-propspecbd}.

\begin{pro}[Sharp bounds under recommended specification] \label{prop:propspecbd}
Assume that $\tau \ge 1/2$ or equivalently $\Lambda_1 + \Lambda_2 \ge 2$.
For the recommended specification \eqref{eq:recspec} with a constant $\delta \in  [0,1]$,
the eMSM sharp upper and lower bounds of $\nu^1(X)$ can be simplified to
\eqref{eq:nu-upper-r} and \eqref{eq:nu-lower-r} respectively.
Then the eMSM sharp bounds of $\mu^1$ can be expressed as
\begin{subequations}
\begin{align}
    \mu^{1+}_{\mytext{eMSM}} =\mu^{1+}_{\mytext{eMSM}} (q^*_{1,\tau})
    &=E\left[E (Y|T=1,X) +(1-T)(\Lambda_2-\Lambda_1)\delta E\left\{\rho_{\tau}(Y,q^*_{1,\tau})|T=1,X\right\}\right]\nonumber\\
    &=E\left\{\frac{T}{\pi^*(X)}Y+T\frac{1-\pi^*(X)}{\pi^*(X)}(\Lambda_2-\Lambda_1)\delta\rho_{\tau}(Y,q^*_{1,\tau})\right\}  ,  \label{eq:uncondup}\\
    \mu^{1-}_{\mytext{eMSM}} =\mu^{1-}_{\mytext{eMSM}} (q^*_{1,1-\tau})
    &=E\left[E (Y|T=1,X) -(1-T)(\Lambda_2-\Lambda_1)\delta E\left\{\rho_{1-\tau}(Y,q^*_{1,1-\tau})|T=1,X\right\}\right]\nonumber\\
    &=E\left\{\frac{T}{\pi^*(X)}Y-T\frac{1-\pi^*(X)}{\pi^*(X)}(\Lambda_2-\Lambda_1)\delta\rho_{1-\tau}(Y,q^*_{1,1-\tau})\right\},\label{eq:uncondlow}
\end{align}
\end{subequations}
where $\mu^{1+}_{\mytext{eMSM}} (q)$ and $\mu^{1-}_{\mytext{eMSM}} (q)$ denote the right-hand sides of \eqref{eq:uncondup} and \eqref{eq:uncondlow}
respectively, with $q^*_{1,\tau}$ and $q^*_{1,1-\tau}$ replaced by any covariate function $q(X)$.
\end{pro}

We provide some remarks related to Proposition \ref{prop:propspecbd}. First, the simplification of
the expressions of sharp bounds in \eqref{eq:nu-upper-complex}--\eqref{eq:nu-lower-complex}  to  \eqref{eq:uncondup}--\eqref{eq:uncondlow} follows from
the fact that with $\tau\ge 1/2$,
\begin{align}  \label{eq:quantile-loss-ineq}
 \frac{1-\tau}{\tau} \le \frac { E\{ \rho_{1-\tau}(Y,q^*_{1,1-\tau})|T=1,X\} }{ E \{\rho_{\tau}(Y,q^*_{1,\tau})|T=1,X \} }
 \le  \frac{\tau} {1-\tau} ,
\end{align}
for any conditional distribution of $Y$ given $T=1$ and $X$, where $0/0$ is treated as $1$.
If the conditional distribution of $Y$ is symmetric around its median, then \eqref{eq:quantile-loss-ineq} is trivial
because the ratio of the two optimized quantile losses in \eqref{eq:quantile-loss-ineq} is $1$.
For an asymmetric conditional distribution, inequality \eqref{eq:quantile-loss-ineq} appears to be new.
See Supplement Section~\ref{sec:proof-propspecbd} for a proof.

Second, inequality \eqref{eq:quantile-loss-ineq} also helps to shed additional light on the comparison between
the specifications \eqref{eq:upspec}, \eqref{eq:lowspec}, and \eqref{eq:recspec}. In fact, it can be directly shown using \eqref{eq:quantile-loss-ineq} with $\tau\ge 1/2$ that
$\Delta_{1,\mytext{U}} (X) \le \Delta_{1,\mytext{L}} (X)$ and
$\Delta_{2,\mytext{L}} (X) \le \Delta_{2,\mytext{U}} (X)$, and hence as intervals
\begin{align*}
 [-\Delta_{1,\mytext{U}} (X),\Delta_{2,\mytext{U}} (X)]
 \subset  [-\Delta_{1,\mytext{L}} (X),\Delta_{2,\mytext{U}} (X)], \quad
  [-\Delta_{1,\mytext{L}} (X),\Delta_{2,\mytext{L}} (X)]
 \subset  [-\Delta_{1,\mytext{L}} (X),\Delta_{2,\mytext{U}} (X)].
\end{align*}
Therefore, the recommended specification \eqref{eq:recspec} is less restrictive than
the naive specifications \eqref{eq:upspec} and \eqref{eq:lowspec}, while leads to both the same sharp upper bound of $\nu^1(X)$ as \eqref{eq:upspec}
and the same sharp lower bound as \eqref{eq:lowspec}.
In this sense, the recommended specification \eqref{eq:recspec}
serves as a satisfactory combination of \eqref{eq:upspec} and \eqref{eq:lowspec}.
For completeness, we note that the eMSM sharp lower bound of $\nu^1(X)$ with the specification \eqref{eq:upspec} can be shown
using \eqref{eq:quantile-loss-ineq} as
\begin{align*}
 \nu^{1-}_{\mytext{eMSM,U}} (X) & = E(Y|T=1,X)
  - (\Lambda_2-\Lambda_1) \delta \frac{1-\tau}{\tau} E \{\rho_{\tau}(Y,q^*_{1,\tau})|T=1,X \}
 \ge \nu^{1-}_{\mytext{eMSM,L}} (X) ,  
\end{align*}
where $\nu^{1-}_{\mytext{eMSM,L}} (X)$ in \eqref{eq:nu-lower-r} is
the sharp lower bound with specification \eqref{eq:recspec} as well as \eqref{eq:lowspec}.
Compared with $\nu^{1-}_{\mytext{eMSM,L}} (X)$, the lower bound $\nu^{1-}_{\mytext{eMSM,U}} (X)$ is also of a simple form, but
involves the optimized $\tau$-quantile loss instead of $1-\tau$ and hence does not recover the MSM sharp lower bound $\nu^{1-}_{\mytext{MSM}} (X)$ when $\delta=1$.
Similar calculation holds for the sharp upper bound of $\nu^1(X)$ with specification \eqref{eq:lowspec}.

Third, the interpretation of sensitivity analysis using eMSMs with constant $(\Lambda_1,\Lambda_2)$ and recommended $(\Delta_1,\Delta_2)$-specification \eqref{eq:recspec} is simple and informative. As in \citet{tan2024model,tan2024sensitivity}, the sensitivity parameters $(\Lambda_1, \Lambda_2)$ indicate the quantile level of sensitivity for the associated MSM bounds,
for example, $(\Lambda_1,\Lambda_2)=(1/2,2)$ indicates 67\%-quantile level for upper bounds or 33\%-quantile level for lower bounds.
This approach is similar to the use of quantile levels in defining the conditional value-at-risk in finance and operations research \citep{rockafellar2000optimization}.
The sensitivity parameter $\delta$, as demonstrated in Proposition \ref{prop:propspecbd}, indicates the shrinkage level at which the MSM bounds (when $\delta=1$) are pulled toward the values that would be identified
under unconfoundedness (when $\delta=0$).
Hence the interpretations of $(\Lambda_1, \Lambda_2)$ and $\delta$ are data-independent (or study-independent), i.e.,
the meanings of fixed choices of $(\Lambda_1, \Lambda_2)$ and $\delta$ are the same across studies.
Note that the interpretation of $(\Delta_1(X), \Delta_2(X))$ under \eqref{eq:recspec} are data-dependent, as the product of
data-independent $\delta/(1-\tau)$ and 
the optimized quantile losses. For fixed $(\Lambda_1, \Lambda_2)$ and $\delta$, the width of the sensitivity interval for $\mu^1$, with end points \eqref{eq:uncondup} and \eqref{eq:uncondlow}, depends on the observed data through the magnitudes of
\begin{align*}
 E [(1-T)E \{\rho_{\tau}(Y,q^*_{1,\tau})|T=1,X \} ] &= E [(1-\pi^*(X) )E \{\rho_{\tau}(Y,q^*_{1,\tau})|T=1,X \} ], \\
 E [(1-T)E \{\rho_{1-\tau}(Y,q^*_{1,1-\tau})|T=1,X \} ] &= E [(1-\pi^*(X))E \{\rho_{1-\tau}(Y,q^*_{1,1-\tau})|T=1,X \} ].
\end{align*}
These two quantities decrease (hence sensitivity intervals become narrower) as $\pi^*(X)$ increases or as
the optimized quantile losses
$ E \{\rho_{\tau}(Y,q^*_{1,\tau})|T=1,X \} $ and $E \{\rho_{1-\tau}(Y,q^*_{1,1-\tau})|T=1,X \} $ decrease
(i.e., $Y$ can be predicted more accurately given $T=1$ and $X$).
Both properties of monotonicity seem to be sensible and reflect important aspects of uncertainty to infer about $\mu^1$ as discussed in \citet{tan2024model,tan2024sensitivity}.
In short, our approach treats the sensitivity parameters $(\Lambda_1, \Lambda_2)$ and $\delta$ data-independently
and then evaluate the sensitivity bounds which are data-dependent.

The preceding discussion leaves open the question of how to decide plausible values of the sensitivity parameters $(\Lambda_1, \Lambda_2)$ and $\delta$. A possible strategy is to empirically calibrate the ranges of sensitivity parameters by holding out each measured covariate as $U$
and estimating feasible values of $(\Lambda_1, \Lambda_2)$ and $\delta$, similarly as in \citet{hsu2013calibrating}.
However, this approach may be only speculative, because how the measured
covariates are related to the treatment and outcome may not be extrapolatable to unmeasured convariates.
Alternatively, given our data-independent interpretation of $(\Lambda_1, \Lambda_2)$ and $\delta$,
it seems amenable to commonly categorize certain ranges of the quantile level $\tau$ and shrinkage level $\delta$ as being small, moderate or large,
in a similar spirit to categorizing $p$-values below $5\%$ as being statistically significant.
For example, $\tau$ from 67\% to 75\%, corresponding to $\Lambda$ from 2 to 3 for $(\Lambda_1,\Lambda_2)=(\Lambda^{-1},\Lambda)$,
and $\delta$ from $0.5$ to $0.67$ may be considered being moderate.
Further investigation and experience are needed on this issue.

Finally, to prepare for sample estimation in Section~\ref{sec:est}, we describe doubly robust, relaxed bounds for $\mu^1$ at the population level
with constant $(\Lambda_1,\Lambda_2)$ and recommended $(\Delta_1,\Delta_2)$-specification \eqref{eq:recspec}.
For the sharp upper bound of $\mu^1$, the two expressions in \eqref{eq:uncondup} in general require different sets of unknown functions of covariates.
The first expression can be rewritten as
$ \mu^{1+}_{\mytext{eMSM}} = E [TY + (1-T) E \{ \tilde{Y}_+ (q^*_{1,\tau} ) | T=1, X \} ]$
and hence requires evaluation of the conditional expectation $E \{ \tilde{Y}_+ (q^*_{1,\tau} ) | T=1, X \}$, in addition to $q^*_{1,\tau} (X)$,
where $ \tilde{Y}_+ (q)=Y+(\Lambda_2-\Lambda_1)\delta\rho_\tau(Y,q)$ is a transformed response.
The second expression requires knowledge of the propensity score $\pi^*(X)$, as well as $q^*_{1,\tau} (X)$.
By extending related results in \citet{dorn2024doubly} and \citet{tan2024model}, consider the augmented IPW estimating function
\begin{align}\label{eq:doublyrobustest}
    \varphi_{1+} (\pi,q, m_+ )&=\frac{T}{\pi(X)}Y+T\frac{1-\pi(X)}{\pi(X)}(\Lambda_2-\Lambda_1)\delta\rho_{\tau}(Y,q)
    -\left(\frac{T}{\pi(X)}-1\right)  m_+(X;q),
\end{align}
where $\pi(X)$, $q(X)$, and $m_+ (X;q)$ are covariate functions expected to approximate $\pi^*(X)$, $q^*_{1,\tau}(X)$, and
$ m^*_+ (X; q) = E\{\tilde{Y}_+ (q)|T=1,X\}$ respectively.

\begin{pro}[Doubly robust, relaxed upper bound]\label{prop:doublyrobust}
For any covariate function $q(X)$, if either $\pi=\pi^*$ or $m_+(\cdot;q)= m_+^*(\cdot;q)$, then
$E \{ \varphi_{1+}(\pi,q,m_+) \} =\mu^{1+}_{\mytext{eMSM}} (q) \ge\mu^{1+}_{\mytext{eMSM}} $,
where $\mu^{1+}_{\mytext{eMSM}} (q) $ is defined as in Proposition \ref{prop:propspecbd}.
If further $q=q^*_{1,\tau}$, then the inequality reduces to equality.
\end{pro}

From Proposition \ref{prop:doublyrobust}, $E \{ \varphi_{1+}(\pi,q,m_+) \}$
is, at the population level, doubly robust for identifying $\mu^{1+}_{\mytext{eMSM}} (q) $,
which in general provides a relaxed upper bound of $\mu^{1+}$
and, if further $q=q^*_{1,\tau}$, coincides with the sharp upper bound $\mu^{1+}_{\mytext{eMSM}} $.
A caveat is that $\mu^{1+}_{\mytext{eMSM}} (q)$ is a relaxed upper bound of $\mu^{1+}$ in eMSM
with constant $(\Lambda_1,\Lambda_2)$ and specification \eqref{eq:recspec},
where $(\Delta_1(X), \Delta_2(X))$ are specified in terms of the optimized quantile losses depending on the true quantile function $q^*_{1,\tau}$, not $q$.

For lower bounds of $\mu^1$, the augmented IPW estimating function is
\begin{align}\label{eq:doublyrobustest-low}
    \varphi_{1-} (\pi,q,m_- )&=\frac{T}{\pi(X)}Y - T\frac{1-\pi(X)}{\pi(X)}(\Lambda_2-\Lambda_1)\delta\rho_{1-\tau}(Y,q)
    -\left(\frac{T}{\pi(X)}-1\right) m_- (X;q),
\end{align}
where $\pi(X)$, $q(X)$, and $m_- (X;q)$ are covariate functions expected to approximate $\pi^*(X)$, $q^*_{1,1-\tau}(X)$, and
$ m^*_- (X ; q) = E\{\tilde{Y}_- (q)|T=1,X\}$ respectively,
with the transformed response $ \tilde{Y}_- (q) =Y - (\Lambda_2-\Lambda_1)\delta\rho_{1-\tau}(Y,q)$. Similar results to Proposition~\ref{prop:doublyrobust} also hold.

\section{Sample estimation}\label{sec:est}
For sensitivity analysis using eMSMs,
the population bounds developed in Section~\ref{sec:propsens} can be estimated, with associated confidence intervals, from sample data.
For concreteness, we discuss estimation of doubly robust bounds of $\mu^1$ with constant $(\Lambda_1,\Lambda_2)$
and recommended $(\Delta_1,\Delta_2)$-specification \eqref{eq:recspec}, as stated in Proposition~\ref{prop:doublyrobust}.
From the reasoning in Section~\ref{sec:ParaSpec}, doubly robust identification of eMSM sharp bounds of $\mu^1$ in Propositions~\ref{pro:pop-sol} and \ref{pro:pop-sol-lower}
may not be feasible for general $(\Delta_1(X),\Delta_2(X))$.
The upper bound $E \{ \varphi_{1+}(\pi,q,m_+) \}$ for $\mu^1$ can be estimated in two stages.
\begin{itemize}\addtolength{\itemsep}{-.1in}
\item[(i)]Specify working models for unknown covariate functions involved in $\varphi_{1+}(\cdot)$:
    propensity score model $\pi(X;\gamma)$ for $\pi^*(X)$,
   outcome quantile regression model $ q(X;\beta)$ for $q_{1,\tau}(X)$,
 and outcome mean regression model $m_+(X; q(\cdot;\beta),\alpha)$ for  $E \{ \tilde{Y}_+ (q(\cdot;\beta) )|X,T=1 \}$,
where $\alpha$, $\beta$, and $\gamma$ are associated parameters. Obtain sample estimates $(\hgamma,\hbeta,\halpha)$.
    \item[(ii)] Compute the sample upper bound
    \begin{align*}
    \hat{\mu}^{1+} (\hgamma,\hbeta,\halpha) = \tE \{ \varphi_{1+} (\hgamma,\hbeta,\halpha) \},
    \end{align*}
where $\varphi_{1+} (\hgamma,\hbeta,\halpha) = \varphi_{1+} (\pi(\cdot;\hgamma),q(\cdot;\hbeta),m(\cdot; q(\cdot;\hbeta), \halpha))$ for abbreviation, and
$\tilde E (\cdot)$ denotes the sample average over the observed data $\{ (Y_i, T_i, X_i) :i=1,\ldots,n\}$.
\end{itemize}
To focus on main ideas, consider the case where the working regression models are linear in the original or logistic scale:
\begin{align}
    \pi(X;\gamma) &=\left\{1+\exp\left(-f^\mytext{T}(X)\gamma\right)\right\}^{-1}, \label{eq:propensity} \\
    q(X;\beta) &=h^\mytext{T}(X)\beta, \label{eq:quantile} \\
    m_+ (X; q(\cdot;\beta),\alpha) & =f^\mytext{T}(X)\alpha. \label{eq:outreg}
\end{align}
Here $f(x)$ and $h(x)$ are pre-specified vector of regression terms or basis functions.
See \cite{tan2024model}, Section 5.2, for related discussion with nonlinear outcome mean regression models in place of \eqref{eq:outreg}.

As in \cite{tan2024model}, we use calibrated estimation (CAL) and regularized CAL (RCAL) for $(\gamma,\beta,\alpha)$ respectively under low- and sparse high-dimensional settings. Under the low-dimensional settings, the CAL estimates $(\hgamma_{\mytext{CAL}},\hbeta_{\mytext{WQ},+},\halpha_{\mytext{WL},+})$ sequentially solve
\begin{equation}\label{eq:calest}
    \frac{\partial\hat{\mu}^{1+}(\gamma,\beta,\alpha)}{\partial \alpha}=0,\quad
  \frac{\partial\hat{\mu}^{1+}(\gamma,\beta,\alpha)}{\partial \beta}\Bigr|_{\gamma=\hgamma_{\rm{CAL}}}=0,\quad
  \frac{\partial\hat{\mu}^{1+}(\gamma,\beta,\alpha)}{\partial \gamma}\Bigr|_{\substack{\gamma=\hgamma_{\rm{CAL}}\\\beta=\hbeta_{\rm{WQ},+}}}=0 .
\end{equation}
With some algebra, $(\hgamma_{\mytext{CAL}},\hbeta_{\mytext{WQ},+},\halpha_{\mytext{WL},+})$ can be shown to solve the following optimizations in the given order:
\begin{align}
    \hgamma_{\mytext{CAL}}&=\arg\min_{\gamma}\tE \left\{Te^{-f^\mytext{T}(X)\gamma}+(1-T)f^\mytext{T}(X)\gamma\right\},\label{eq:gammaest} \\
    \hbeta_{\mytext{WQ},+}&=\arg\min_{\beta}\tE\left\{ T\frac{1-\pi(X;\hgamma_{\mytext{CAL}})}{\pi(X;\hgamma_{\mytext{CAL}})}\delta\rho_\tau\left(Y,h^\mytext{T}(X)\beta\right)\right\},\label{eq:betaest}\\
   \halpha_{\mytext{WL},+}&=\arg\min_{\alpha}\tE\left[ T\frac{1-\pi(X;\hgamma_{\mytext{CAL}})}{\pi(X;\hgamma_{\mytext{CAL}})}\left\{\tilde{Y}_+\left(h^\mytext{T}\hbeta_{\mytext{WQ},+} \right)-f^\mytext{T}(X)\alpha\right\}^2\right].\label{eq:alphaest}
\end{align}
All three optimizations are convex.
The objective function in \eqref{eq:gammaest} is the calibration loss for fitting logistic regression as studied in \citet{tan2020regularized},
whereas \eqref{eq:betaest} corresponds to weighted quantile (WQ) regression and \eqref{eq:alphaest} corresponds weighted least-squares (WL) regression.
In fact, the estimators $\hgamma_{\mytext{CAL}}$ and $\hbeta_{\mytext{WQ},+}$ are the same as calibrated estimation for MSM sensitivity analysis in \citet{tan2024model},
but $\halpha_{\mytext{WL},+}$ differs from MSM analysis due to the presence of $\delta$ in $\tilde{Y}_+ (\cdot)$.

By extending the reasoning in \cite{tan2024model}, $\hat{\mu}^{1+}_{\mytext{CAL}} = \hat{\mu}^{1+}(\hgamma_{\mytext{CAL}},\hbeta_{\mytext{WQ},+},\halpha_{\mytext{WL},+})$
can be shown to satisfy the following properties under suitable regularity conditions, with $p,m$ fixed and sample size $n\rightarrow\infty$.
Let $(\bar{\gamma}_{\mytext{CAL}},\bar{\beta}_{\mytext{WQ},+},\bar{\alpha}_{\mytext{WL},+})$ be the limiting value of $(\hgamma_{\mytext{CAL}},\hbeta_{\mytext{WQ},+},\halpha_{\mytext{WL},+})$ in probability.
\begin{itemize}\addtolength{\itemsep}{-.1in}
\item[(i)] $\hat{\mu}^{1+}_{\mytext{CAL}}$ is pointwise doubly robust for the relaxed bound $\mu^{1+}_{\mytext{eMSM}}(h^\mytext{T}\bar{\beta}_{\mytext{WQ},+})$, i.e., the limiting value $\hat{\mu}^{1+}(\bar{\gamma}_{\mytext{CAL}},\bar{\beta}_{\mytext{WQ},+},\bar{\alpha}_{\mytext{WL},+})$ gives $\mu^{1+}_{\mytext{eMSM}} (h^\mytext{T}\bar{\beta}_{\mytext{WQ},+})$
if either model \eqref{eq:propensity} or \eqref{eq:outreg} is correctly specified. If further model \eqref{eq:quantile} is correctly specified, then
the relaxed bound $\mu^{1+}_{\mytext{eMSM}}(h^\mytext{T}\bar{\beta}_{\mytext{WQ},+})$ reduces to the sharp bound $\mu^{1+}_{\mytext{eMSM}}$.

\item[(ii)] $\hat{\mu}^{1+}_{\mytext{CAL}}$ admits the asymptotic expansion
\begin{align}\label{eq:calexpansion}
    \hat{\mu}^{1+}_{\mytext{CAL}}=\hat{\mu}^{1+}(\bar{\gamma}_{\mytext{CAL}},\bar{\beta}_{\mytext{WQ},+},\bar{\alpha}_{\mytext{WL},+})+o_p(n^{-1/2}),
\end{align}
regardless of misspecification of model \eqref{eq:propensity}, \eqref{eq:quantile} or \eqref{eq:outreg}.
The expansion \eqref{eq:calexpansion} enables the construction of valid Wald confidence intervals
which contain $\hat{\mu}^{1+}(\bar{\gamma}_{\mytext{CAL}},\bar{\beta}_{\mytext{WQ},+},\bar{\alpha}_{\mytext{WL},+})$ with the associated probabilities asymptotically,
regardless of model misspecification.
\end{itemize}
Let
$\hat V_{\mytext{CAL}}^{1+}= \tE \{\varphi_{1+}(\hgamma_{\mytext{CAL}},\hbeta_{\mytext{WQ},+},\halpha_{\mytext{WL},+}) - \hat{\mu}^{1+}_{\mytext{CAL}}\}^2$.
By combining (i) and (ii), the Wald confidence interval
\begin{equation}\label{eq:CIupper}
    \left(-\infty,\hat{\mu}^{1+}_{\mytext{CAL}}+z_{c}\sqrt{ \hat V_{\mytext{CAL}}^{1+}} /n\right]
\end{equation}
is doubly robust with asymptotic size $1-c$ for $\mu^{1+}_{\mytext{eMSM}} (h^\mytext{T}\bar{\beta}_{\mytext{WQ},+})$,
where $z_c$ is the $1-c$ quantile of standard normal distribution.
In other words, if either model \eqref{eq:propensity} or \eqref{eq:outreg} is correctly specified,
then \eqref{eq:CIupper} achieves asymptotic size $1-c$ in containing $\mu^{1+}_{\mytext{eMSM}} (h^\mytext{T}\bar{\beta}_{\mytext{WQ},+})$
and hence also asymptotic level $1-c$ in containing the sharp bound $\mu^{1+}_{\mytext{eMSM}} $ and the true value $\mu^1$.
 If further model \eqref{eq:quantile} is correctly specified, then \eqref{eq:CIupper}
is doubly robust with asymptotic size $1-c$ for $\mu^{1+}_{\mytext{eMSM}}$.

Doubly robust inference is one of the advantages of CAL (and RCAL) estimation.
By comparison, conventional estimation methods based on maximum likelihood or least-squares
generally require both models \eqref{eq:propensity} and \eqref{eq:outreg}
to be correctly specified in order to achieve a similar expansion to \eqref{eq:calexpansion} and hence valid Wald confidence intervals.
As demonstrated by Corollary 2 in \citet{tan2024model} another advantage of CAL is that if propensity score model \eqref{eq:propensity} is correctly specified,
then the limiting value $\mu^{1+}_{\mytext{eMSM}} (h^\mytext{T}\bar{\beta}_{\mytext{WQ},+})$ of $\hat{\mu}^{1+}_{\mytext{CAL}}$
provides a tighter relaxed bound for $\mu^1$ than that would be obtained from conventional estimation methods.

Similarly, lower bounds of $\mu^1$ can be estimated using the augmented IPW estimating function \eqref{eq:doublyrobustest-low}
in place of \eqref{eq:doublyrobustest} for upper bounds. Consider propensity score model $\pi(X;\gamma)$ for $\pi^*(X)$,
outcome quantile regression $q(X;\beta)$ for $q_{1,1-\tau}(X)$,
 and outcome mean regression model $m_- (X; q(\cdot;\beta),\alpha)$ for $E \{ \tilde{Y}_- (q(\cdot;\beta) )|X,T=1 \}$,
 specified in the same forms as in \eqref{eq:propensity}--\eqref{eq:outreg}.
The calibrated estimators $(\hgamma_{\mytext{CAL}},\hbeta_{\mytext{WQ},-},\halpha_{\mytext{WL},-})$
are defined by keeping \eqref{eq:gammaest} while replacing $\tau$ by $1-\tau$ and $\tilde{Y}_{+}(q)$ by $\tilde{Y}_{-}(q)$
in \eqref{eq:betaest} and \eqref{eq:alphaest}.
Let
$\hat{\mu}^{1-}_{\mytext{CAL}}=\tE\{\varphi_{-}( \hgamma_{\mytext{CAL}},\hbeta_{\mytext{WQ},-},\halpha_{\mytext{WL},-})\}$ and
$\hat V_{\mytext{CAL}}^{1-}= \tilde E [\{\varphi_{-}( \hgamma_{\mytext{CAL}},\hbeta_{\mytext{WQ},-},\halpha_{\mytext{WL},-}) - \hat{\mu}^{1-}_{\mytext{CAL}}\}^2]$,
where $\varphi_{1-} (\hgamma,\hbeta,\halpha) = \varphi_{1-} (\pi(\cdot;\hgamma),q(\cdot;\hbeta),m_-(\cdot; q(\cdot;\hbeta),\halpha))$ for abbreviation.
Then a doubly robust, asymptotic level $1-c$ confidence interval for
$\mu^{1-}_{\mytext{eMSM}}$ and $\mu^1$ is
 $ [\hat{\mu}^{1-}_{\mytext{CAL}}-z_{c}\sqrt{\tilde V_{\mytext{CAL}}^{1-}/n},\;\infty]$.
By combining one-sided confidence intervals, a two-sided, doubly robust, asymptotic level $1-c$ confidence interval for $\mu^1$ is
\begin{align}\label{eq:Twosided-mu1-CI}
\left[\hat{\mu}^{1-}_{\mytext{CAL}}-z_{c/2}\sqrt{\tilde V_{\mytext{CAL}}^{1-}/n},\quad\hat{\mu}^{1+}_{\mytext{CAL}}+z_{c/2}\sqrt{\tilde V_{\mytext{CAL}}^{1+}/n}\right].
\end{align}

In high-dimensional settings with $p,m$ close to or greater than $n$, the RCAL estimators,
denoted as $(\hgamma_{\mytext{RCAL}},\hbeta_{\mytext{RWQ},+},\halpha_{\mytext{RWL},+})$, for parameters $(\gamma,\beta,\alpha)$ in working models \eqref{eq:propensity}--\eqref{eq:outreg}
are defined by incorporating Lasso penalties into \eqref{eq:gammaest}--\eqref{eq:alphaest}.
The resulting upper-bound estimator,
$\hat{\mu}^{1+}_{\mytext{RCAL}} = \hat{\mu}^{1+}(\hgamma_{\mytext{RCAL}},\hbeta_{\mytext{RWQ},+},\halpha_{\mytext{RWL},+})$,
can be shown to satisfy similar properties, and Wald confidence intervals can be obtained similarly as discussed above,
under suitable sparsity conditions on the limiting values $(\bar{\gamma}_{\mytext{CAL}},\bar{\beta}_{\mytext{WQ},+},\bar{\alpha}_{\mytext{WL},+})$.
See \citet{tan2020regularized, tan2024model} for further details.
\section{Comparison of bounds for binary outcomes}\label{sec:compare}
We compare the sensitivity bounds from eMSM with those from \citet{ding2016sensitivity}, hereafter referred to as DV.
Both approaches place constraints on the associations between an unmeasured confounder $U$ and treatment $T$, and between $U$ and outcome $Y$. We consider binary $Y$ and focus on the conditional bounds given $X$ as in the main results of DV. For simplicity, all probability expressions are implicitly conditional on $X$. See Supplement Section \ref{sec:dvest} for DV unconditional bounds and sample estimation.

With binary outcome, the eMSM sharp bounds of $\mu^1 =E(Y^1)$ in Propositions \ref{pro:pop-sol} and \ref{pro:pop-sol-lower} can be simplified as
\begin{subequations}\label{eq:binary-eMSM}
  \begin{align}
    \mu^{1+}_{\mytext{eMSM}} &= p_1+ P(T=0) (\Lambda_2-\Lambda_1) \min \left[ \tau \Delta_1, (1-\tau)\Delta_2, E\{\rho_{\tau}(Y,q^*_{1,\tau})|T=1\}\right],
    \label{eq:binary-emsm-up}\\
    \mu^{1-}_{\mytext{eMSM}}  &= p_1- P(T=0) (\Lambda_2-\Lambda_1)\min \left[(1- \tau) \Delta_1, \tau\Delta_2, E\{\rho_{1-\tau}(Y,q^*_{1,1-\tau})|T=1\}\right] , \label{eq:binary-emsm-low}
\end{align}
\end{subequations}
where conditioning on $X$ is implicit in the notation, $p_1=E(Y|T=1)$, and
\begin{subequations}\label{eq:Expectedrho}
\begin{align}
    E\{\rho_\tau(Y,q^*_{1,\tau})|T=1\}&=\min\{(1-\tau)(1-p_1),\tau p_1\},\label{eq:rho_tau}\\
    E\{\rho_{1-\tau}(Y,q^*_{1,1-\tau})|T=1\}&=\min\{(1-\tau)p_1,\tau(1-p_1)\}.\label{eq:rho_1-tau} 
\end{align}
\end{subequations}

\subsection{DV and related results}

DV quantifies the confounder-treatment and confounder-outcome associations by
\begin{equation*}
\RR_{\mytext{UT,t}}=\max_{u}\frac{\ud P_U(u\mid T=1-t)}{\ud P_U(u\mid T=t)},
\quad\RR_{\mytext{UY,t}}=\frac{\max_u E(Y\mid T=t,U=u)}{\min_u E(Y\mid T=t,U=u)}.
\end{equation*}
Let $p_0=E(Y|T=0)$, $\mu^1=E(Y^1)$ and $\mu^0=E(Y^0)$.
Their main results relate the observed relative risk $\ORR= p_1/p_0$, a population quantity rather than a sample estimate, and the causal relative risk $\CRR=\mu^1/\mu^0$ through
\begin{equation}\label{eq:DV2016}
    \ORR/B(\RR_{\mytext{UT,0}},\RR_{\mytext{UY}})\le\CRR\le \ORR\cdot B(\RR_{\mytext{UT,1}},\RR_{\mytext{UY}}),
\end{equation}
where $\RR_{\mytext{UY}}=\max\{\RR_{\mytext{UY,0}},\RR_{\mytext{UY,1}}\}$ and $B(x,y)=xy/(x+y-1)$.

To compare with eMSMs, consider a sensitivity model, hereafter referred to as a DV model, which postulates
that an unmeasured variable $U$ accounts for all confounding, $(Y^0,Y^1) \perp T |U$, and satisfies the following constraints:

\begin{subequations}\label{eq:UTbounds}
\noindent\begin{minipage}{0.45\textwidth}
\begin{equation}
\RR_{\mytext{UT,0}}\le1/\Lambda_1, \label{eq:UT0}
\end{equation}
    \end{minipage}\hfill
    \begin{minipage}{0.45\textwidth}
\begin{equation}
\RR_{\mytext{UT,1}}\le\Lambda_2, \label{eq:UT1}
\end{equation}
\end{minipage}
\end{subequations}
\begin{subequations}\label{eq:UYbounds}
\noindent\begin{minipage}{0.45\textwidth}
\begin{equation}\label{eq:UY0}
 \RR_{\mytext{UY,0}}\le \Theta,
\end{equation}
\end{minipage}\hfill
\begin{minipage}{0.45\textwidth}
\begin{equation}\label{eq:UY1}
    \RR_{\mytext{UY,1}}\le \Theta,
\end{equation}
\end{minipage}\vspace{1em}
\end{subequations}
where $\Lambda_2 \ge 1 \ge \Lambda_1 \ge 0$ and $\Theta \ge 1 $ are sensitivity parameters.
Note that \eqref{eq:UT0} and \eqref{eq:UT1} represent respectively the lower and upper bounds on the same ratio $\frac{dP(U=u\mid T=0)}{dP(U=u\mid T=1)}$.
By the monotonicity of $B(x,y)$, the DV bounds \eqref{eq:DV2016} can be stated under \eqref{eq:UT0}--\eqref{eq:UY1} as
\begin{align}
\ORR/B(1/\Lambda_1,\Theta)\le\CRR\le \ORR\cdot B(\Lambda_2,\Theta).\label{eq:DVmodel-ab}
\end{align}

The upper and lower bounds in \eqref{eq:DVmodel-ab}, however, depend on different subsets of constraints \eqref{eq:UT0}--\eqref{eq:UY1}.
The upper bound of $\CRR$ is derived
from an upper bound of $\mu^1$ under \eqref{eq:UT1} and \eqref{eq:UY1} only and a lower bound of $\mu^0$
under \eqref{eq:UT1} and \eqref{eq:UY0} only,
and the lower bound of $\CRR$ is derived from a lower bound of $\mu^1$ under \eqref{eq:UT0} and \eqref{eq:UY1} only and an upper bound of $\mu^0$
under \eqref{eq:UT0} and \eqref{eq:UY0} only. For example, the upper bound of $\mu^1$
under \eqref{eq:UT1} and \eqref{eq:UY1} and its lower bound under \eqref{eq:UT0} and \eqref{eq:UY1} in DV are
\begin{align}\label{eq:dv-ab}
 &  \mu^{1+}_{\mytext{DV}} =
 p_1\left\{P(T=1)+P(T=0) B(\Lambda_2,\Theta)\right\}, \quad
  \mu^{1-}_{\mytext{DV}} =
  p_1\left\{P(T=1)+\frac{P(T=0)}{B(1/\Lambda_1,\Theta)}\right\}.
\end{align}
For simplicity, we mainly discuss upper and lower bounds of $\mu^1$ for binary outcomes.
See Supplement Section~\ref{sec:dvest} for additional discussion about $\mu_0$, ATE, and $\CRR$.

\citet{sjolander2024sharp} pointed out that the above DV bounds \eqref{eq:DV2016}, hence \eqref{eq:DVmodel-ab} and \eqref{eq:dv-ab}, are not sharp. 
For example, the upper bound $\mu^{1+}_{\mytext{DV}}$ may exceed $1$ for large $\Lambda_2$ and $\Theta$.
\citet{sjolander2024sharp} showed that the sharp upper bound of $\mu^1$ under \eqref{eq:UT1} and \eqref{eq:UY1}
and the sharp lower bound of $\mu^1$ under \eqref{eq:UT0} and \eqref{eq:UY1} are
\begin{align}\label{eq:Sjosharp-ab}
  & \mu^{1+}_{\mytext{Sj}} =
  p_1\left[P(T=1)+P(T=0) \min\{B(\Lambda_2,\Theta),1/p_1\}\right]
  ,\quad
  \mu^{1-}_{\mytext{Sj}} = \mu^{1-}_{\mytext{DV}}.
\end{align}
The sharp bounds of $\mu^0$ are similar and obtained by switching $T=0$ with $T=1$, $p_1$ with $p_0$, and $\Lambda_2$ with $1/\Lambda_1$ in the above display.
The resulting improved bound of CRR under \eqref{eq:UT0}, \eqref{eq:UT1}, \eqref{eq:UY0}, and \eqref{eq:UY1} is
\begin{align}\label{eq:SjosharpCRR}
&\ORR\frac{P(T=1)+P(T=0)/B(1/\Lambda_1,\Theta)}{P(T=1)\min\{B(1/\Lambda_1,\Theta),1/p_0\}+P(T=0)}\nonumber\\
&  \hspace{8em}  \le\CRR\le\\
&\ORR\frac{P(T=1)+P(T=0)\min\{B(\Lambda_2,\Theta),1/p_1\}}{P(T=1)/B(\Lambda_2,\Theta)+P(T=0)} . \nonumber
\end{align}
Removing respectively the argument $1/p_0$ and $1/p_1$ in the min operators in \eqref{eq:SjosharpCRR} leads back to the DV bound \eqref{eq:DVmodel-ab}.
However, due to the separate use of \eqref{eq:UT0} or \eqref{eq:UT1},
the improved DV bounds \eqref{eq:SjosharpCRR} are not sharp in the DV model defined by
\eqref{eq:UT0}, \eqref{eq:UT1}, \eqref{eq:UY0}, and \eqref{eq:UY1} jointly.

\subsection{Comparison and improvement}\label{sec:modelcompare}

The eMSM bounds in \eqref{eq:binary-eMSM} and the DV bounds in \eqref{eq:dv-ab} or \eqref{eq:Sjosharp-ab} are not directly comparable,
because the underlying sensitivity models are different.
In fact, the eMSM model constraint \eqref{eq:msm-bound} is same as the DV model constraints \eqref{eq:UT0} and \eqref{eq:UT1} jointly, but the eMSM constraint \eqref{eq:UYconstr} differs from the DV constraint \eqref{eq:UY1}.
The DV constraint \eqref{eq:UY0} can be ignored, when considering bounds of $\mu^1$.
Nevertheless, we show that the DV model is equivalent to the union of eMSMs with compatible parameters,
and, from this relationship, we derive sharp bounds of $\mu^1$ under the DV model defined by \eqref{eq:UT0}, \eqref{eq:UT1}, and \eqref{eq:UY1} jointly. See Supplement Section \ref{sec:proof-dv} for a stronger result which includes simultaneous sharp bounds of $(\mu^0,\mu^1)$.

We introduce the following notations.
Denote as $\rm{eMSM} (\Delta_{1},\Delta_{2})$ the eMSM defined by \eqref{eq:msm} and \eqref{eq:UYconstr},
and denote as  $\mu^{1+}_{\mytext{eMSM}} (\Delta_{1},\Delta_{2})$ and
$\mu^{1-}_{\mytext{eMSM}} (\Delta_{1},\Delta_{2})$ the eMSM sharp bounds of $\mu^1$ in \eqref{eq:binary-emsm-up} and \eqref{eq:binary-emsm-low}.
Moreover, denote as $\rm{DV}(\Theta)$ the DV model defined by \eqref{eq:UT0}, \eqref{eq:UT1}, and \eqref{eq:UY1}.
A sensitivity model is treated as the collection of allowed distributions, i.e., distributions consistent with the observed-data distribution and satisfying model constraints.
The parameters $(\Lambda_1,\Lambda_2)$ are assumed to be common among different models and hence omitted in the notation.
Finally, let
$\mathcal{D} (\Theta) = \{(\Delta_{1},\Delta_{2}):\theta (\Delta_{1},\Delta_{2})\le \Theta\}$
where $\theta (x,y)=\min\{p_1 +y,1\}/\max\{p_1 -x,0\}$.

\begin{pro}[Sharp bounds for DV model]\label{prop:DVcompare}
For the model $\rm{DV}(\Theta)$, the following results hold.

(i) $\bigcup_{(\Delta_{1},\Delta_{2})\in\mathcal{D}(\Theta)}\rm{eMSM}(\Delta_{1},\Delta_{2})=\rm{DV}(\Theta)$.

(ii) The sharp upper bound of $\mu^{1}$ under $\rm{DV}(\Theta)$ is
\begin{align}
   \mu^{1+}_{\mytext{DV}} (\Theta) &= \max_{(\Delta_{1},\Delta_{2})\in\mathcal{D}(\Theta)} \; \mu^{1+}_{\mytext{eMSM}} (\Delta_{1},\Delta_{2})\nonumber\\
     &=p_1+P(T=0)(\Lambda_2-\Lambda_1)\min\left[\frac{\tau(\Theta-1)p_1}{\odds(\tau)+\Theta},E\{\rho_\tau(Y,q^*_{1,\tau})|T=1\}\right],\label{eq:dvsharp+}
\end{align}
with $\odds(x) = x/(1-x)$ and $E\{\rho_\tau(Y,q^*_{1,\tau})|T=1\}$ as in \eqref{eq:rho_tau}. The sharp upper bound $\mu^{1+}_{\mytext{DV}} (\Theta)$ is attained by $\rm{eMSM} (\Delta_{1,\mytext{U}},\Delta_{2,\mytext{U}})$, where
$$
\Delta_{1,\mytext{U}} =\frac{(\Theta-1)p_1}{\odds(\tau)+ \Theta}, \quad \Delta_{2,\mytext{U}}=\frac{(\Theta-1)p_1}{1+ \Theta/\odds(\tau)}.
$$

(iii) The sharp lower bound of $\mu^{1}$ under $\rm{DV}(\Theta)$ is
\begin{align}
     \mu^{1-}_{\mytext{DV}} (\Theta) &= \min_{(\Delta_{1},\Delta_{2})\in\mathcal{D}(\Theta)} \; \mu^{1-}_{\mytext{eMSM}}(\Delta_{1},\Delta_{2})\nonumber\\
    &=p_1-P(T=0)(\Lambda_2-\Lambda_1)\min\left[\frac{(1-\tau)(\Theta-1)p_1}{\odds(1-\tau)+\Theta},E\{\rho_{1-\tau}(Y,q^*_{1,1-\tau})|T=1\}\right],\label{eq:dvsharp-}
\end{align}
with $E\{\rho_{1-\tau}(Y,q^*_{1,1-\tau})|T=1\}$ as in \eqref{eq:rho_1-tau}. The sharp lower bound $\mu^{1-}_{\mytext{DV}} (\Theta)$ is attained by $\rm{eMSM} (\Delta_{1,\mytext{L}},\Delta_{2,\mytext{L}})$, where
$$
\Delta_{1,\mytext{L}}=\frac{(\Theta-1)p_1}{\odds(1-\tau)+ \Theta}, \quad \Delta_{2,\mytext{L}}=\frac{(\Theta-1)p_1}{1+ \odds(\tau)\Theta}.
$$
Note that $(\Delta_{1,\mytext{L}}, \Delta_{2,\mytext{L}})$ are same as $(\Delta_{1,\mytext{U}}, \Delta_{2,\mytext{U}})$
with $\tau$ replaced by $1-\tau$.
\end{pro}
We provide several remarks to help understand Proposition~\ref{prop:DVcompare}.
First, Proposition \ref{prop:DVcompare} (i) says that $\rm{DV}(\Theta)$ corresponds to a collection of
eMSMs with compatible parameters $(\Delta_1,\Delta_2)$ satisfying $\theta (\Delta_{1},\Delta_{2})\le \Theta$.
In fact, a simple observation is that for any $Q\in\rm{eMSM}(\Delta_{1},\Delta_{2})$,
$$
 \frac{\max_{u}Q(Y=1|U=u,T=1)}{\min_{u}Q(Y=1|U=u,T=1)}\le  \theta(\Delta_{1},\Delta_{2})
$$
by the definition of eMSM and $\theta(\cdot)$.
Hence if $Q\in\rm{eMSM}(\Delta_{1},\Delta_{2})$ with $\theta(\Delta_{1},\Delta_{2})\le \Theta$,
then $Q \in \rm{DV}(\Theta)$.
The converse statement also holds: if $Q \in \rm{DV}(\Theta)$,
then there exists $(\Delta_1,\Delta_2)$ such that $\theta(\Delta_{1},\Delta_{2})\le \Theta$ and $Q\in\rm{eMSM}(\Delta_{1},\Delta_{2})$.
See the proof in Supplement Section \ref{sec:proof-dv}.

\begin{figure}[t]
\noindent\begin{minipage}[t]{.57\textwidth}
  \centering
  \includegraphics[width=1\linewidth]{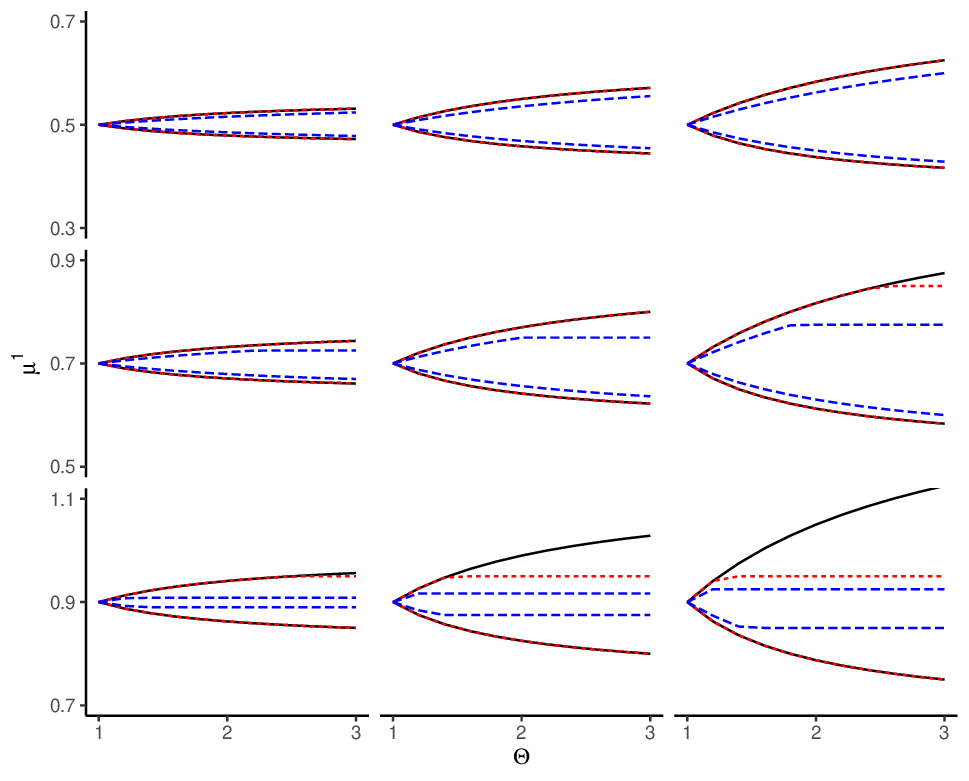}
\end{minipage}\hfill
\begin{minipage}[b]{.37\textwidth}
\captionof{figure}{Bounds of $\mu^1=E(Y^1)$ under DV model, conditioned on $X$. The original DV bounds \eqref{eq:dv-ab} (solid black), the improved bounds \eqref{eq:Sjosharp-ab} due to Sj{\"o}lander (dotted red), and the sharp bounds from Proposition \ref{prop:DVcompare} (dashed blue) are compared. By row, $p_1\in\{0.5,0.7,0.9\}$ increases from top to bottom. By column, $\Lambda_1^{-1}=\Lambda_2\in\{1.2,1.5,2\}$ increases from left to right. The treatment assignment is balanced with $P(T=1)=0.5$ in all plots.}
\label{fig:DVbounds}
\end{minipage}
\end{figure}

Second, the sharp bounds of $\mu^1$ in Proposition \ref{prop:DVcompare} (ii)--(iii) are derived by accounting for the constraints
\eqref{eq:UT0}, \eqref{eq:UT1}, and \eqref{eq:UY1} jointly, and hence are tighter than the DV bounds in \eqref{eq:dv-ab} or \eqref{eq:Sjosharp-ab}.
See Figure~\ref{fig:DVbounds} for a numerical demonstration. As a special case, setting $\Lambda_1=0$ or $\Lambda_2=\infty$ respectively recovers the upper or lower bound in \eqref{eq:Sjosharp-ab},
which confirms their sharpness under \eqref{eq:UT1} and \eqref{eq:UY1} or, separately,
under \eqref{eq:UT0} and \eqref{eq:UY1} \citep{sjolander2024sharp}.

Third, as shown in the proof, sharp bounds of $\mu^0$ under the DV model can be derived similarly as those of $\mu^1$,
and the sharp bounds of $\mu^0$ and $\mu^1$ can be simultaneously attained.
Therefore, the sharp bound on a contrast of $\mu^1$ and $\mu^0$, including CRR and risk difference, is determined by
the contrast of the sharp bounds on $\mu^1$ and $\mu^0$. For example,
the sharp bounds of $\CRR$ under the DV model with $\Lambda_1^{-1}=\Lambda_2=\Lambda$ are
\begin{align*}
     &\ORR\cdot\frac{P(T=1)+P(T=0)\max\{(\Lambda+\Theta)/(\Lambda\Theta+1),1-(\Lambda-1)/\odds(p_1)\}}{P(T=1)\min\{(\Lambda\Theta+1)/(\Lambda+\Theta),1+(1-\Lambda^{-1})/\odds(p_0)\}+P(T=0)}\nonumber\\
    &\hspace{12em}\le\CRR\le\\ &\ORR\cdot\frac{P(T=1)+P(T=0)\min\{(\Lambda\Theta+1)/(\Lambda+\Theta),1+(1-\Lambda^{-1})/\odds(p_1)\}}{P(T=1)\max\{(\Lambda+\Theta)/(\Lambda\Theta+1),1-(\Lambda-1)/\odds(p_0)\}+P(T=0)}\nonumber.
\end{align*}
Keeping only the first argument in the min and max operators above leads to a relaxed bound
\begin{align*}
&\ORR\Big/\frac{\Lambda\Theta+1}{\Lambda+\Theta}\le\CRR\le\ORR\cdot\frac{\Lambda\Theta+1}{\Lambda+\Theta},
\end{align*}
which is still strictly tighter than the DV bounds \eqref{eq:DVmodel-ab} with $\Lambda_1^{-1}=\Lambda_2=\Lambda$.

Fourth, for fixed $\Theta$, Proposition \ref{prop:DVcompare} (ii) identifies an eMSM $\subset\rm{DV}(\Theta)$
such that the sharp upper bound of $\mu^1$ under eMSM coincides with that under $\rm{DV}(\Theta)$.
The associated parameters $(\Delta_{1,\mytext{U}},\Delta_{2,\mytext{U}})$ corresponds to the specification \eqref{eq:upspec} with $\delta_{1+} = \delta_{2+} =\delta$ and
\begin{align}
 \delta =\frac{p_1(\Theta-1)  / \{1/(1-\tau)+\Theta/\tau\} }{E \{ \rho_{\tau}(Y,q^*_{1,\tau})|T=1 \} }.  \label{eq:DV-r-up}
\end{align}
Note that $\delta$ in \eqref{eq:DV-r-up} may be greater than 1, in which case the DV sharp upper bound reduces to the MSM sharp upper bound.
Similarly, Proposition \ref{prop:DVcompare} (iii) identifies an eMSM $\subset\rm{DV}(\Theta)$
such that the lower upper bound of $\mu^1$ under eMSM coincides with that under $\rm{DV}(\Theta)$.
The associated parameters $(\Delta_1,\Delta_2)$ corresponds to the specification \eqref{eq:lowspec} with $ \delta_{1-} = \delta_{2-} =\delta$ and
\begin{align}
\delta =\frac{p_1(\Theta-1)  / \{\Theta/(1-\tau)+1/\tau\} }{E \{ \rho_{1-\tau}(Y,q^*_{1,1-\tau})|T=1 \} }. \label{eq:DV-r-low}
\end{align}
When $\delta$ in \eqref{eq:DV-r-low} is greater than 1, the DV sharp lower bound reduces to the MSM sharp lower bound.
Moreover, the two values of $\delta$ in \eqref{eq:DV-r-up} and \eqref{eq:DV-r-low} are in general different.
Hence the sharp upper and lower bounds from $\rm{DV}(\Theta)$
are attained by two different MSMs respectively, but in general not by a single specification of MSM.
This situation differs from that of Section \ref{sec:ParaSpec}, where MSM with recommended specification \eqref{eq:recspec}
achieves the sharp upper and lower bounds from two MSMs respectively with specifications \eqref{eq:upspec} and \eqref{eq:lowspec} for the \textit{same} $\delta$.

Finally, the relationship \eqref{eq:DV-r-up} and \eqref{eq:DV-r-low} can be used in a converse direction to relate a specified eMSM to some DV model.
For fixed $\delta$, solving for $\Theta$ in \eqref{eq:DV-r-up} shows that the sharp upper bound of $\mu^1$ under eMSM with specification \eqref{eq:upspec} corresponds to that under $\rm{DV}(\Theta)$ with
\begin{align}
 \Theta & =\frac{p_1 + \delta/(1-\tau)E(\rho_\tau(Y,q^*_{1,\tau})|T=1) }{p_1 -\delta/\tau E(\rho_\tau(Y,q^*_{1,\tau})|T=1) }   \nonumber \\
 & = \frac{1+\delta\min\{\odds(\tau),\odds(1-p_1)\}}{1-\delta\cdot\odds(1-\tau)\min\{\odds(\tau),\odds(1-p_1)\}}.  \label{eq:DV-Theta-up}
\end{align}
Similarly, from \eqref{eq:DV-r-up}, the sharp lower bound of $\mu^1$ under eMSM with specification \eqref{eq:lowspec} corresponds to that under $\rm{DV}(\Theta)$ with
\begin{align}
 \Theta & = \frac{1+\delta\min\{\odds(1-\tau),\odds(1-p_1)\}}{1-\delta\cdot\odds(\tau)\min\{\odds(1-\tau),\odds(1-p_1)\}},  \label{eq:DV-Theta-low}
\end{align}
where the equality is same as \eqref{eq:DV-Theta-up} with $\tau$ replaced by $1-\tau$. For fixed $\delta$, the sharp upper or lower bound from eMSM with recommended specification \eqref{eq:recspec}, in the case of $\tau\ge 1/2$, matches that of
$\rm{DV}(\Theta)$ with $\Theta$ value in \eqref{eq:DV-Theta-up} or \eqref{eq:DV-Theta-low} respectively.
However, a subtle point is that each $\Theta$ value depends on the data quantity $p_1 = E(Y=1|T=1)$.
The adaptation to $p_1$ for the implied $\Theta$ by eMSM is reasonable because, for example, as $p_1$ approaches 1, $\RR_{\mytext{UY,1}}$ is expected to converge toward 1. Therefore, a sensible parameter $\Theta$ used in \eqref{eq:UY1} should converge to 1 instead of remaining constant.
Fundamentally, this data-dependency can be attributed
to the fact that $(\Delta_1,\Delta_2)$ are data-dependent under specification \eqref{eq:recspec}.

To facilitate comparisons of eMSM and DV models,
it is desirable to find a data-independent alignment between the respective parameters. Interestingly, $\Theta$ in \eqref{eq:DV-Theta-up} or \eqref{eq:DV-Theta-low} is at most respectively
\begin{align}
 & \Theta^+ = \{1+\delta\cdot\odds(\tau)\}/(1-\delta), \quad \Theta^- = \{1+\delta\cdot\odds(1-\tau)\}/(1-\delta), \label{eq:this}
\end{align}
independently of $p_1$, and both inequalities are exact for $p_1 \le \min\{1-\tau,\tau\}$.
Hence $\Theta^+$ or $\Theta^-$ can be considered to be roughly aligned with a fixed $\delta$.
When $\tau \ge 1/2$, we have $ \Theta^+ \ge \Theta^-$.
In numerical applications (Section \ref{sec:numerical}), we report eMSM bounds with recommended specification \eqref{eq:recspec}
and DV bounds with $\Theta \in \{\Theta^+/2, \Theta^+, 3 \Theta^+/2\}$, even though these bounds are not strictly comparable.
\section{Empirical applications}\label{sec:numerical}

We present two empirical applications to evaluate our proposed methods.
The first, based on a medical study by \cite{connors1996effectiveness},  
examines the effect of right heart catheterization (RHC) on a binary outcome, $30$-day survival status. 
For this study, sensitivity analysis of ATE is conducted by \cite{dorn2023sharp} and \cite{tan2024model} using MSMs.\
The second study, based on the 2013-2014 National Health and Nutrition Examination Survey (NHANES) as used in \cite{zhao2019sensitivity},
examines the effect of fish consumption on a continuous outcome, the total blood mercury. 
Sensitivity analysis for this study can be found in \cite{zhao2019sensitivity}, \cite{soriano2023interpretable}, and \cite{huang2025variance} using MSMs or related models.

\subsection{Application to RHC} \label{sec:rhc}
The study included $n = 5735$ critically ill patients admitted to the intensive care units of five medical centers. For each patient, the data consist of treatment status $T$ ($ = 1$ if RHC was used within 24 hours of admission and 0 otherwise), binary outcome $Y$ ($30$-day survival status), and a list of 75 covariates $X$ specified by medical specialists in critical care.

For sensitivity analysis, we apply \cite{tan2024model} method under MSM and the proposed method under eMSM in two settings: non-regularized estimation (CAL under MSM or eMSM) using only main-effect working models, and regularized estimation (RCAL under MSM or eMSM) using high-dimensional working models that incorporate main effects and interactions.
Calibrated estimation, CAL or RCAL, is implemented for $\gamma$ and $\alpha$ using the R package \texttt{RCAL} \citep{tan2020rcalpkg},
and for $\beta$ using the R package \texttt{quantreg} \citep{koenker2022quantregpkg}.
Moreover, we apply a sample version of the DV method with non-regularized estimation and bootstrapping as described in Supplement Section \ref{sec:dvest}.
In the high-dimensional setting, following \cite{tan2024model}, we take $f(X) = h(X)$ which consists of all main effects and two-way interactions of $X$ except those with the fractions of nonzero values less than 46 (i.e. 0.8\% of the sample size 5735). The dimension is $p = m = 1855$, excluding the constant. All regressors are standardized with sample means 0 and variances 1.  The Lasso tuning parameters are selected by fivefold cross-validation over a discrete set $\{\kappa^* / 2^{j/4}: j = 0, 1, \ldots, 24\}$, where $\kappa^*$ is a data-dependent upper limit.

For eMSM sensitivity parameters $(\Lambda_1, \Lambda_2, \delta)$, we consider $\Lambda_1^{-1} = \Lambda_2 = \Lambda \in \{1, 1.2, 1.5, 2\}$ as in \citet{tan2024model} and $\delta\in\{0.2, 0.5, 0.8, 1\}$. For DV sensitivity parameter $\Theta$, we consider $\Theta \in \{\Theta^+ / 2, \Theta^+, 3\Theta^+ / 2\}$,
where $\Theta^+$ in \eqref{eq:this} is used to roughly align eMSM and DV models. Table \ref{tb:theta+} reports $\Theta^+$ values for different $\Lambda$ and $\delta$.
For $\delta = 0.2$, $\Theta^+ / 2$ is less than 1 and hence excluded because $\Theta \ge 1$ by definition and $\Theta=1$ represents no unmeasured confounding.

\begin{table} 
\caption{\footnotesize $\Theta^+$ values given $\Lambda$ and $\delta$.} \label{tb:theta+} \vspace{-4ex}
\begin{center}
\begin{tabular}{lcccc}
\hline
$\delta \backslash \Lambda$ & 1.0 & 1.2 & 1.5 & 2.0 \\
\hline
0.2 & 1.5 & 1.55 & 1.625 & 1.75 \\
0.5 & 3 & 3.2 & 3.5 & 4 \\
0.8 & 9 & 9.8 & 11 & 13 \\
1.0 & $\infty$ & $\infty$ & $\infty$ & $\infty$ \\
\hline
\end{tabular}
\end{center}
\end{table}

\begin{figure}[ht]
    \centering
    \begin{minipage}{0.49\textwidth}
        \centering
        \includegraphics[width=\textwidth]{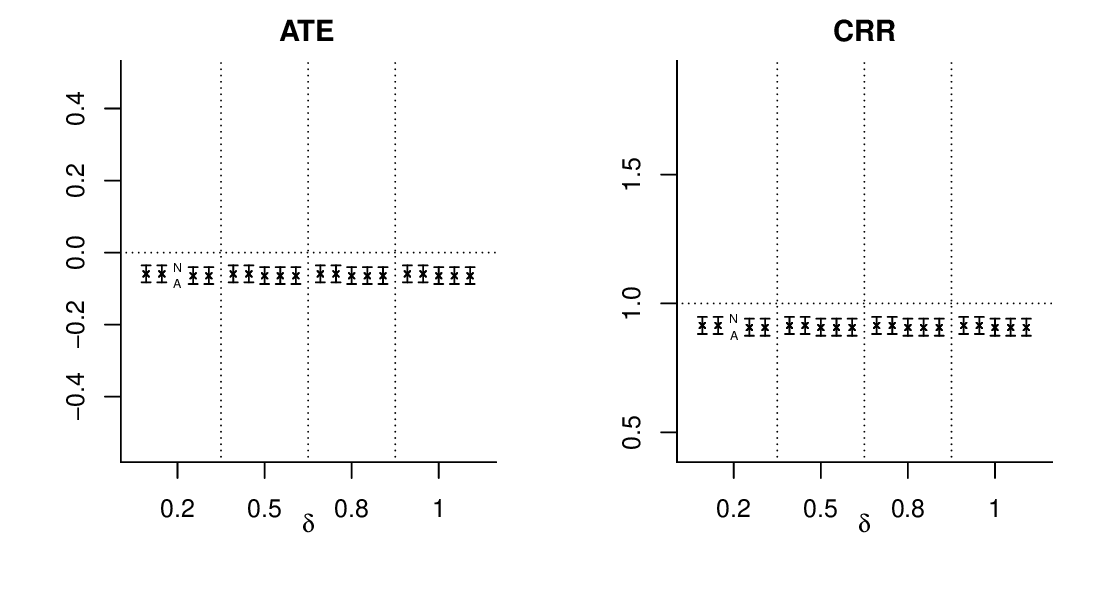} \vspace{-.4in}
        \subcaption{ \small $\Lambda = 1$}
    \end{minipage}
    \hfill
    \begin{minipage}{0.49\textwidth}
        \centering
        \includegraphics[width=\textwidth]{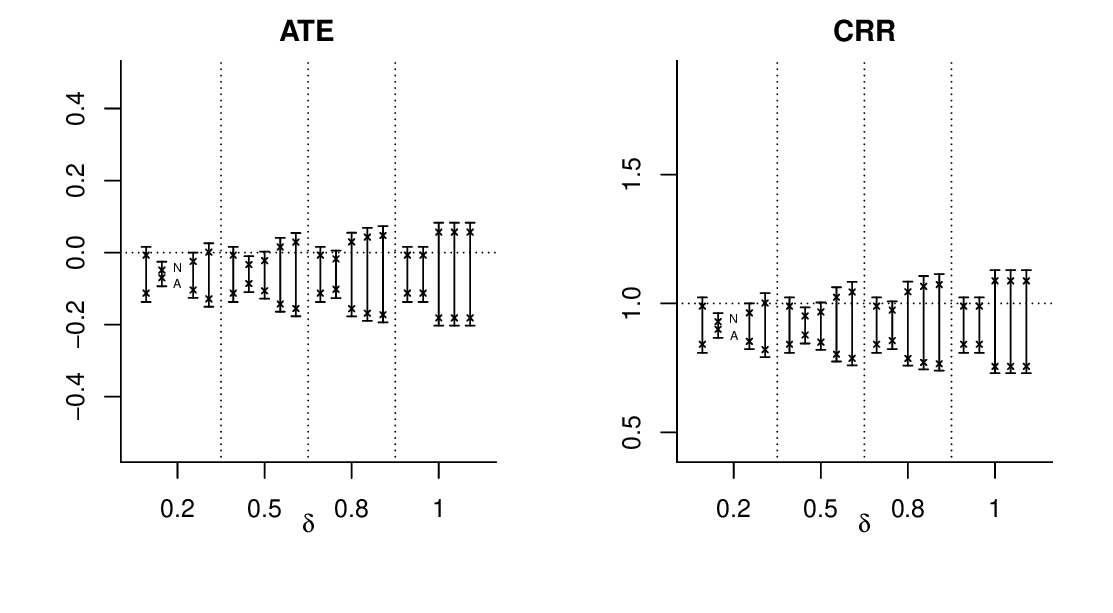} \vspace{-.4in}
        \subcaption{\small $\Lambda = 1.2$}
    \end{minipage}
    
    \vspace{.3in} 

    \begin{minipage}{0.49\textwidth}
        \centering
        \includegraphics[width=\textwidth]{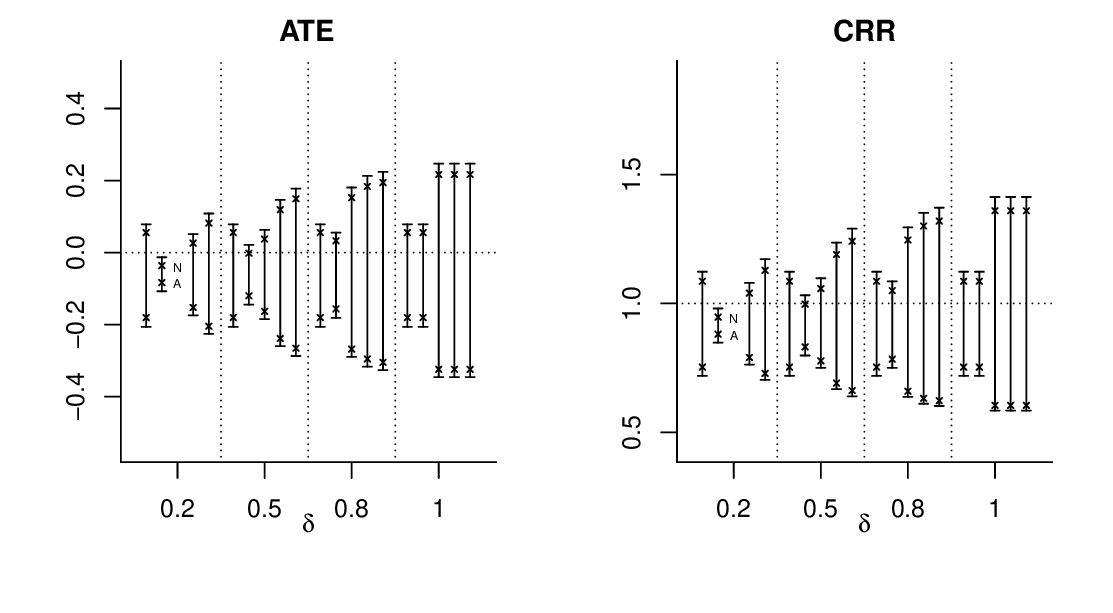} \vspace{-.4in}
        \subcaption{\small $\Lambda = 1.5$}
    \end{minipage}
    \hfill
    \begin{minipage}{0.49\textwidth}
        \centering
        \includegraphics[width=\textwidth]{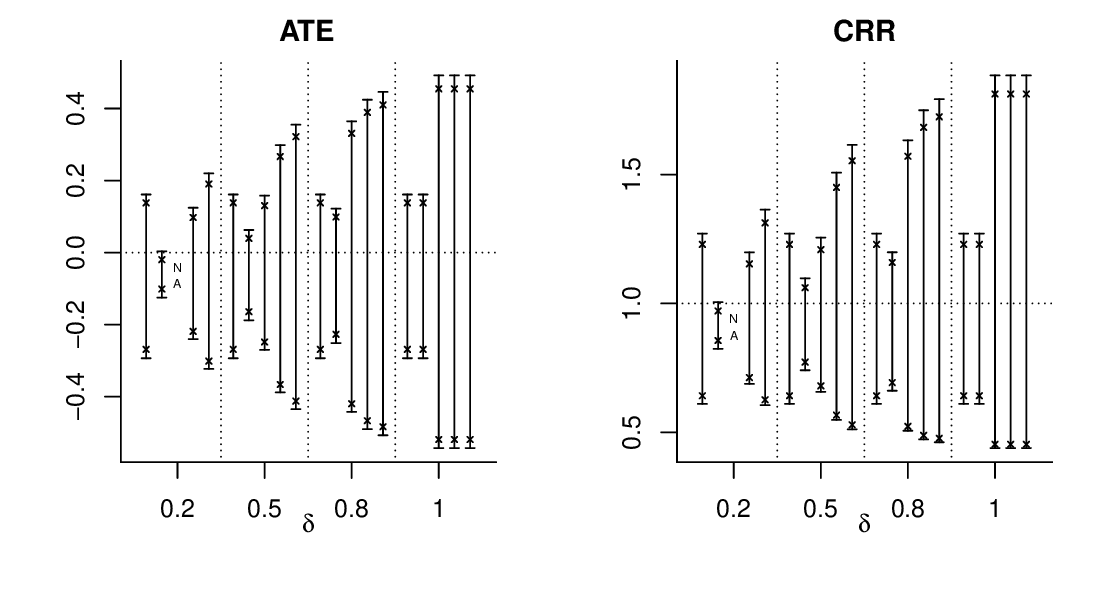} \vspace{-.4in}
        \subcaption{\small $\Lambda = 2$}
    \end{minipage}

    \caption{\small
Point bounds (x) and 90\% confidence intervals (–) on ATE and causal risk ratio, using working models with only main effects and logistic outcome mean regression for CAL estimation in the RHC study.
Each sub-figure corresponds to a choice of $\Lambda \in \{1, 1.2, 1.5, 2\}$, labeled as (a)–(d).
Within each plot, results are presented over four choices of $\delta$, separated by vertical lines. For each $\delta$, results from the following methods are shown (from left to right): CAL under MSM, CAL under eMSM, and DV method with $\Theta = \Theta^+ / 2$, $\Theta^+$, or $3\Theta^+ / 2$,
except that DV is excluded with $\Theta=\Theta^+/2$ when $\delta=0.2$ and shown as ``NA".}
    \label{fig:rhc-lowdim-logit-ate}
\end{figure}

Figure \ref{fig:rhc-lowdim-logit-ate} shows the point bounds and 90\% confidence intervals on ATE and causal risk ratio,
using working models with only main effects and logistic outcome mean regression for CAL estimation. 
Additional results are presented in Supplement Section \ref{sec:additional-numerical},
including sensitivity intervals on 30-day survival probabilities $(\mu_0,\mu_1)$ and 
the results from CAL estimation using linear outcome mean regression and from RCAL estimation using high-dimensional working models with main effects and interactions. The latter results are similar to those in Figure \ref{fig:rhc-lowdim-logit-ate}.

From Figure \ref{fig:rhc-lowdim-logit-ate}, we observe that as expected by design, the eMSM sensitivity intervals are always narrower than
MSM intervals at the same $\Lambda \in \{1.2,1.5,2\}$, 
while the difference in width becomes smaller as $\delta$ increases from $0.2$ to $1$ where eMSM and MSM intervals coincide.
Moreover, the eMSM sensitivity intervals are consistently and considerably narrower than DV intervals with the roughly aligned $\Theta$ choices
(excluding $\Theta=\Theta^+/2$ when $\delta=0.2$ as explained earlier).
In fact, although the DV method accounts for both treatment and outcome unmeasured confounding, its sensitivity intervals are even wider than MSM intervals at $\Theta = \Theta^+$ whenever $\delta\geq 0.5$.

Under unconfoundedness $(\Lambda=1)$, ATE estimation indicates a negative effect (the results from MSM and eMSM at $\Lambda = 1$ are the same for all $\delta$). 
The MSM sensitivity intervals include 0 at all $\Lambda \in \{1.2,1.5,2\}$ as in \cite{tan2024model}. The eMSM intervals are below 0 at $\Lambda = 1.2$ and $\delta\in\{0.2, 0.5\}$ and $(\Lambda = 1.5, \delta = 0.2)$, almost below 0 at $(\Lambda = 2, \delta = 0.2)$ with an upper interval point of 0.0030 (Supplement Table \ref{tb:rhc-lowdim-cal-lam2.0}), but include 0 at the remaining $(\Lambda,\delta)$ choices.
These results suggest that a negative ATE may exist 
while allowing an unmeasured confounder which induces treatment confounding of similar magnitudes $\Lambda$ as in previous MSM analyses
but outcome confounding constrained by a shrinkage factor $\delta \le 0.2$ when $\Lambda \le 2$ or $\delta \le 0.5$ when $\Lambda \le 1.2$.
Overall, a negative ATE seems plausible only under relatively weak unmeasured confounding. 
The same pattern also applies to CRR estimation.

\subsection{Application to NHANES} \label{sec:nhanes}
The outcome $Y$ of this study is the total blood mercury (in $\log_2$), measured in micrograms per liter. The treatment $T$ is coded as 1 if an individual consumed more than 12 servings of fish or shellfish in the preceding month or $0$ otherwise. There are 234 total treated units and 873 control units resulting in a sample size $n = 1107$. There are eight covariates in $X$: gender, age, income, whether income is missing, race, education, ever smoked and number of cigarettes smoked last month. After converting categorical covariates into dummy variables, we obtain 15 covariates.

For MSM and eMSM sensitivity analysis, we apply CAL estimation using only main-effect working models, and RCAL estimation using high-dimensional working models that allow main effects and interactions, similarly as in the RHC application (Section \ref{sec:rhc}). Because DV method is designed for the case of binary outcomes, we do not apply DV method here. In the high-dimensional setting, we take $f(X) = h(X)$ which consists of all main effects and two-way interactions of $X$. The dimension is $p = m = 104$, excluding the constant.

\begin{figure}
\centering
\includegraphics[scale=0.6]{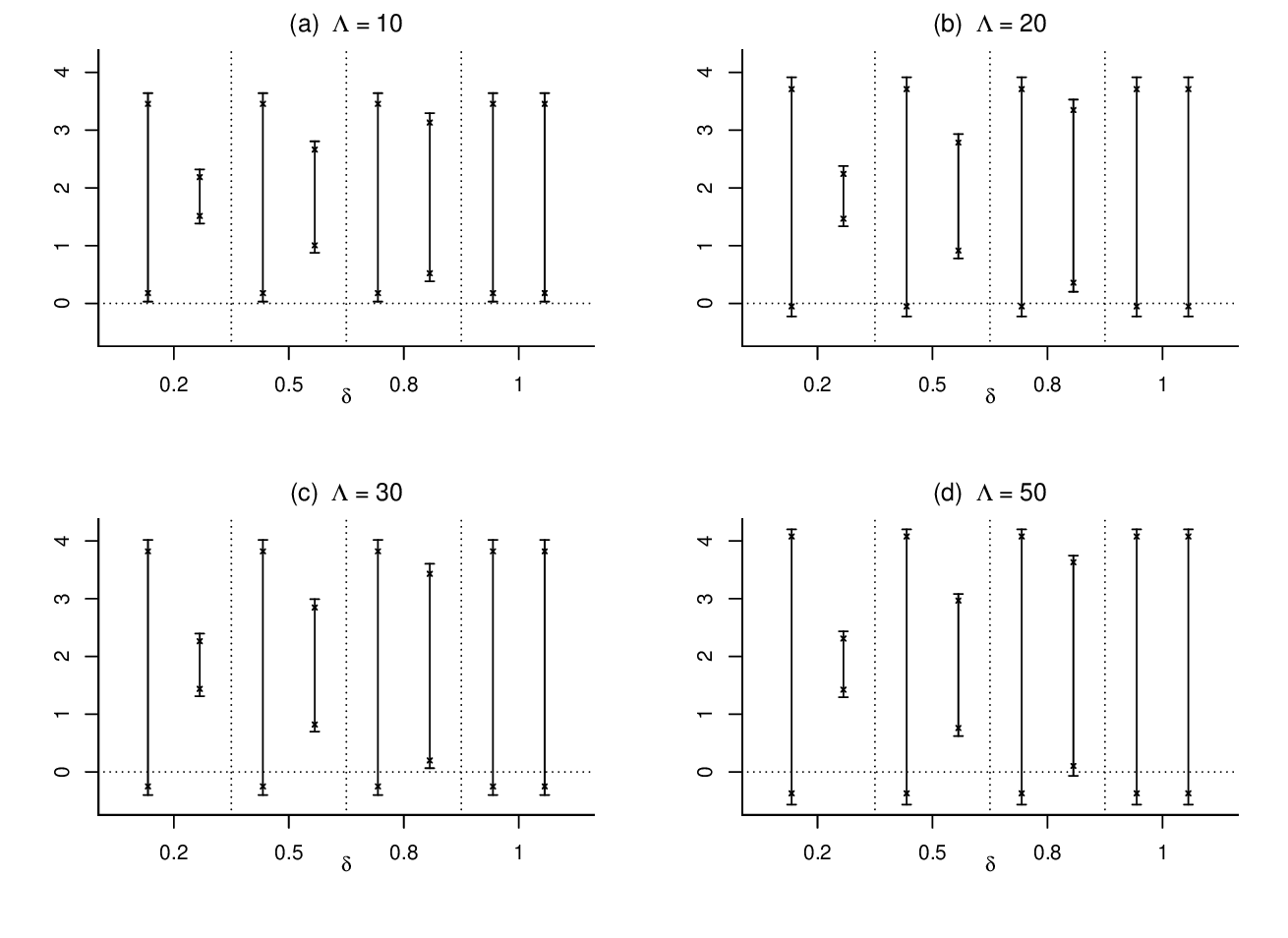} \vspace{-.26in}
\caption{{\small
Point bounds (x) and 90\% confidence intervals (–) on ATE, using working models with main effects and interactions and linear outcome mean regression for RCAL estimation in the NHANES study. Each sub-figure corresponds to a different choice of $\Lambda \in \{10, 20, 30, 50\}$, labeled as (a)–(d).
Within each plot, results are presented over four choices of $\delta$, separated by vertical lines. For each $\delta$, two methods are used (from left to right):
RCAL under MSM and RCAL under eMSM. 
}}
\label{fig:nhanes-highdim-lin-ate}
\end{figure}

We consider sensitivity parameters $\delta\in\{0.2, 0.5, 0.8, 1\}$, same as in Section~\ref{sec:rhc}, and $\Lambda \in \{1, 10, 20, 30, 50\}$,
which are selected for presentation after obtaining results for a denser sequence from 1 to 50.
Figure \ref{fig:nhanes-highdim-lin-ate} shows the point bounds and 90\% confidence intervals on ATE using working models with main effects and interactions and linear outcome mean regression for RCAL estimation for $\Lambda\in\{10, 20, 30, 50\}$. 
The results from MSM and eMSM at $\Lambda = 1$ are the same for all $\delta$ (Supplement Figure \ref{fig:nhanes-highdim-lin}). Additional results are presented in Supplement Section \ref{sec:additional-numerical}, 
including sensitivity intervals on the mean parameters $(\mu_0,\mu_1)$ and the results from CAL estimation using working models with only main effects,
which are briefly discussed below.

Under unconfoundedness $(\Lambda = 1)$, ATE estimation indicates a positive effect (see Supplement Figure \ref{fig:nhanes-highdim-lin}). 
As unmeasured treatment confounding is allowed to increase ($\Lambda$ increases), the MSM sensitivity intervals 
are above 0 at $\Lambda=10$ but include 0 at, for example, $\Lambda =20$.
Furthermore, when additionally accounting for unmeasured outcome confounding, the eMSM sensitivity intervals remain above 0 
for all $\delta\in\{0.2, 0.5, 0.8\}$ at $\Lambda = 30$.
For $\Lambda= 50$ and $\delta=0.8$, the eMSM sensitivity interval only narrowly includes 0.
These results demonstrate the robustness of a positive ATE, which persists 
not only under unmeasured treatment confounding of large magnitudes ($\Lambda=10$) and unconstrained unmeasured outcome confounding as in previous MSM analyses,
but also under unmeasured treatment confounding of even larger magnitudes ($\Lambda=30$ to $50-$)
and unmeasured outcome confounding constrained by a large shrinkage factor ($\delta =0.8$).

Unlike the RHC application, there are substantial differences in point bounds between using working models with only main effects or with both main effects and interactions as shown in Supplement Section \ref{sec:additional-numerical}.
For example, at $(\Lambda = 50, \delta = 1)$, the lower point bound of ATE from CAL and RCAL estimation
is -0.009 and -0.371, leading to a difference of 0.362, but the associated standard errors are 0.065 and 0.118 respectively.
This suggests that working models relying solely on main effects may be misspecified, because otherwise CAL and RCAL would yield similar estimates.

\section{Conclusion}\label{sec:conclusion}
We propose eMSM that incorporates an outcome sensitivity constraint to effectively tighten the sharp bounds from the popular MSM approach. In the case of $\mu^1$, our outcome sensitivity constraint \eqref{eq:UYconstr} controls the $U$-$Y^1$ association by bounding the deviation of $\mg^*(X,U)$ from the reference value, $E(Y|T=1,X)$, under unconfoundedness. For completeness, we point out two alternative sensitivity models with outcome sensitivity constraints that are based on $\mg^*(X,U)$.

The first outcome sensitivity constraint assumes a bound on $\mg^*(X,U)$ in the differences
over possible values of $U$ but without referring to $E(Y|T=1,X)$:
\begin{align}
    \max_{u}\mg^*(X,u)-\min_{u}\mg^*(X,u)\le \Delta(X).\label{eq:eta-diff}
\end{align}
We refer to the model specified by \eqref{eq:msm} and \eqref{eq:eta-diff} as $\rm{dMSM}(\Delta)$, with ``d'' for difference. The sharp bounds of $\mu^1$ under $\rm{dMSM}(\Delta)$ are
\begin{subequations}
    \begin{align}
    \mu_{\mytext{dMSM}}^{1+}(\Delta)=&E\Big(E(Y|T=1,X)+(1-T)\{\Lambda_2(X)-\Lambda_1(X)\}\times\\
        &\min\left[\tau(X)\left\{1-\tau(X)\right\}\Delta(X),E \{\rho_{\tau}(Y,q^*_{1,\tau})|T=1,X \}\right]\Big),\\
        \mu_{\mytext{dMSM}}^{1-}(\Delta)=&E\Big(E(Y|T=1,X)-(1-T)\{\Lambda_2(X)-\Lambda_1(X)\}\times\\
        &\min\left[\tau(X)\left\{1-\tau(X)\right\}\Delta(X),E \{\rho_{1-\tau}(Y,q^*_{1,1-\tau})|T=1,X \}\right]\Big).
    \end{align}
\end{subequations}

The second outcome sensitivity constraint applies to any nonnegative outcome variable $Y$ and bounds $\mg^*(X,U)$ in the ratios over possible values of $U$:
\begin{align}
    \frac{\max_{u}\mg^*(X,u)}{\min_{u}\mg^*(X,u)}\le\Theta(X).\label{eq:eta-ratio}
\end{align}
We refer to the model specified by \eqref{eq:msm} and \eqref{eq:eta-ratio} as $\rm{rMSM}(\Theta)$,
with ``r" for ratio. The sharp bounds of $\mu^1$ under $\rm{rMSM}(\Theta)$ are
\begin{subequations}
    \begin{align}
    \mu_{\mytext{rMSM}}^{1+}(\Theta)=&E\bigg(\{E(Y|T=1,X)+(1-T)\{\Lambda_2(X)-\Lambda_1(X)\}\times\\
        &\min\left[\frac{E(Y|T=1,X)(\Theta(X)-1)}{1/\left\{1-\tau(X)\right\}+\Theta(X)/\tau(X)},E \{\rho_{\tau}(Y,q^*_{1,\tau})|T=1,X \}\right]\bigg),\\
        \mu_{\mytext{rMSM}}^{1-}(\Theta)=&E\bigg(E(Y|T=1,X)-(1-T)\{\Lambda_2(X)-\Lambda_1(X)\}\times\\
        &\min\left[\frac{E(Y|T=1,X)(\Theta(X)-1)}{\Theta(X)/\left\{1-\tau(X)\right\}+1/\tau(X)},E \{\rho_{1-\tau}(Y,q^*_{1,1-\tau})|T=1,X \}\right]\bigg).
    \end{align}
\end{subequations}

Similar to in Proposition \ref{prop:DVcompare}, both dMSM and rMSM are collections of eMSMs with compatible sensitivity parameters, and hence, their sharp bounds are derived by optimizing the eMSM sharp bounds over the compatible parameter sets. See Supplement Section \ref{sec:d-r-MSM} for additional discussion and comparison. These findings demonstrate the fundamental value of eMSM.

\bibliographystyle{apalike}
\bibliography{references.bib}
\newpage
\appendix
\setcounter{page}{1}

\setcounter{section}{0}
\setcounter{equation}{0}

\setcounter{table}{0}
\setcounter{figure}{0}

\setcounter{pro}{0}
\renewcommand{\thepro}{S\arabic{pro}}

\setcounter{lem}{0}
\renewcommand{\thelem}{S\arabic{lem}}

\setcounter{thm}{0}
\renewcommand{\thethm}{S\arabic{thm}}

\setcounter{ass}{0}
\renewcommand{\theass}{S\arabic{ass}}

\renewcommand\thesection{\Roman{section}}

\renewcommand\thesubsection{\thesection.\arabic{subsection}}

\renewcommand\theequation{S\arabic{equation}}

\renewcommand\thetable{S\arabic{table}}
\renewcommand\thefigure{S\arabic{figure}}

\begin{center}
{\Large Supplementary Material for}\\
{\Large``Enhanced Marginal Sensitivity Model and Bounds''}

\vspace{.1in}
Yi Zhang, Wenfu Xu, Zhiqiang Tan
\vspace{.1in}
\end{center}

\section{Additional discussion on MSM}\label{sec:addonMSM}
The MSM \eqref{eq:msm} in the main text is a variation of the original MSM in \citeappend{tan2006distributional} in the following aspects.
First, the bounds $\Lambda_1(X)$ and $\Lambda_2(X)$ in \eqref{eq:msm} are allowed to depend on $X$, unlike the simpler choice of \citeappend{tan2006distributional}, where $\Lambda_1(X) \equiv \Lambda^{-1}$ and $\Lambda_2 (X) \equiv \Lambda$
for a constant $\Lambda \ge 1$.
Second, the original MSM in \citeappend{tan2006distributional} is formulated
in terms of the density ratio of $Y^0$ or $Y^1$ without explicitly involving $U$. Such an MSM with covariate-dependent bounds assumes that
\begin{subequations}\label{eq:msm-Yt}
\vspace{.5em}
\hspace{-1.5em}
\noindent\begin{minipage}{0.5\textwidth}
\begin{equation}\label{eq:msm-Y0}
\frac{\ud P_{Y^0}(\cdot |T=0,X)}{\ud P_{Y^0} (\cdot|T=1,X)} \in [\Lambda_1(X),\Lambda_2(X)],
\end{equation}
    \end{minipage}
    \hspace{-1em}
    \begin{minipage}{0.5\textwidth}
\begin{equation}\label{eq:msm-Y1}
\frac{\ud P_{Y^1} (\cdot|T=0,X)}{\ud P_{Y^1} (\cdot|T=1,X)} \in [\Lambda_1(X),\Lambda_2(X)].
\end{equation}
\end{minipage}\vspace{1em}
\end{subequations}
\\~
To relate to \eqref{eq:msm}, a slightly different MSM assumes that
\begin{align}
 &  \frac{\ud P_{Y^0,Y^1} (\cdot|T=0,X)}{\ud P_{Y^0,Y^1} (\cdot|T=1,X)} \in [\Lambda_1(X),\Lambda_2(X)].  \label{eq:msm-Y0Y1}
\end{align}
The relationships among different MSMs can be summarized as follows.
\begin{lem}(Relationships among $\rm{MSMs}$)\label{lm:msm-variations}
    Fix $\Lambda_1(X)$ and $\Lambda_2(X)$.

    (i) A distribution of $(Y^0,Y^1,T,X,U)$ that satisfies \eqref{eq:msm} also satisfies \eqref{eq:msm-Y0Y1}.

    (ii) For a distribution of $(Y^0,Y^1,T,X)$ that satisfies \eqref{eq:msm-Y0Y1}, there exists an unmeasured variable $U$ that satisfies \eqref{eq:msm-indep}, i.e., $(Y^0,Y^1)\perp T | X,U$, and the distribution of $(Y^0,Y^1,T,X,U)$ satisfies \eqref{eq:msm-bound}.

    (iii) A distribution of $(Y^0,Y^1,T,X)$ satisfying \eqref{eq:msm-Y0Y1}, or a distribution of $(Y^0,Y^1,T,X,U)$ satisfying \eqref{eq:msm}, also satisfies \eqref{eq:msm-Yt}.
\end{lem}

A proof is provided in Section \ref{sec:proof-lm-MSMs}. Similar to Lemma 1 of \citeappend{tan2024model}, Lemma \ref{lm:msm-variations} (i) and (ii) together establish an equivalence between models \eqref{eq:msm} and \eqref{eq:msm-Y0Y1} in the sense that the collections of allowed distributions, when marginalized to $(Y^0,Y^1,T,X)$, are the same.
Lemma \ref{lm:msm-variations} (iii) shows that model \eqref{eq:msm} and \eqref{eq:msm-Y0Y1} are sub-models of model \eqref{eq:msm-Yt} when all distributions considered are marginalized to $(Y^0,Y^1,T,X)$. The converse is generally not true. To see this, consider $Q$ under model \eqref{eq:msm-Yt} such that $Y^0\perp Y^1|T,X$ and at some $(y_0,y_1)$
$$\frac{\ud Q_{Y^1}(y_1|T=0,X)}{\ud Q_{Y^1}(y_1|T=1,X)} =\Lambda_2, \quad
\frac{\ud Q_{Y^0}(y_0|T=0,X)}{\ud Q_{Y^0}(y_0|T=1,X)} >1 .$$
Then the density ratio of $Y^0,Y^1$ between treatment groups exceeds $\Lambda_2$ at $(y_0,y_1)$.

Despite the differences, the sharp population bounds of $\mu^0$ or, respectively, $\mu^1$
under models \eqref{eq:msm}, \eqref{eq:msm-Yt} and \eqref{eq:msm-Y0Y1}
are identical. The sharp bounds under model \eqref{eq:msm}, or equivalently under model \eqref{eq:msm-Y0Y1} by Lemma~\ref{lm:msm-variations}, are derived in \citeappend{dorn2023sharp}, while identical bounds are obtained in \citeappend{tan2024model} with model \eqref{eq:msm-Yt}.
To explain the identical bounds of $\mu^1$, for any distribution $Q$ of $(Y^0,Y^1,T,X)$ allowed in model \eqref{eq:msm-Yt},
we construct $\tilde Q$ as follows:
(a) $\tilde Q_{Y^1,T,X}= Q_{Y^1,T,X}$, (b) $Y^0 \perp T |X$ under $\tilde Q$, and (c) $Y^0\perp Y^1|T,X$ under $\tilde Q$.
Then $E_{\tilde Q} (Y^1) = E_Q (Y^1)$ and $\tilde Q$ is allowed in model \eqref{eq:msm-Y0Y1}.
Hence the sharp bounds of $\mu^1$ are identical under models \eqref{eq:msm-Yt} and \eqref{eq:msm-Y0Y1}.

The sharp bounds of ATE are also identical among the models \eqref{eq:msm}, \eqref{eq:msm-Yt} and \eqref{eq:msm-Y0Y1}, which follows from the simultaneous sharpness of $(\mu^{0-}_{\mytext{MSM}},\mu^{1+}_{\mytext{MSM}})$ and that of $(\mu^{0+}_{\mytext{MSM}},\mu^{1-}_{\mytext{MSM}})$. Formally, two bounds are said to be simultaneously sharp if they are achieved by a single distribution allowed in the respective model.
The simultaneous sharpness in model \eqref{eq:msm-Yt} is immediate: one may separately consider $Q_{Y^0}(\cdot|T,X)$ for the sharp bound of $\mu^0$ and $Q_{Y^1}(\cdot|T,X)$ for that of $\mu^1$, and then multiply the two for a joint distribution on $(Y^0,Y^1,T,X)$ under which $Y^0\perp Y^1|T,X$.
The simultaneous sharpness in model \eqref{eq:msm-Yt} or \eqref{eq:msm} requires a different construction, as
from the preceding discussion of Lemma~\ref{lm:msm-variations} (iii),
the sharp bounds of $(\mu^0, \mu^1)$ under model \eqref{eq:msm-Y0Y1} \textit{cannot}
be simultaneously achieved by a distribution $Q$ under which $Y^0\perp Y^1|T,X$.
Nevertheless, the simultaneous sharpness is demonstrated for model \eqref{eq:msm} by \citeappend{dorn2023sharp}
and hence also valid for model \eqref{eq:msm-Y0Y1} by the equivalence from Lemma~\ref{lm:msm-variations} (i)--(ii).
Moreover, the simultaneous sharpess for bounds of $(\mu^0, \mu^1)$ can be extended to eMSM, as discussed in Section \ref{sec:proof-prop1-cor1} later.

Finally, we comment that although MSMs \eqref{eq:msm} and \eqref{eq:msm-Y0Y1} are equivalent,
there is a subtle difference between their formulations.
While MSM \eqref{eq:msm-Y0Y1} characterizes unmeasured confounding via the density ratio of $(Y^0,Y^1)$ between the treatment groups,
MSM \eqref{eq:msm} does so via
the density ratio of a responsible unmeasured confounder $U$ as in the treatment sensitivity constraint \eqref{eq:msm-bound}.
The introduction of $U$ facilitates the formulation of eMSM by incorporating
the outcome sensitivity constraint \eqref{eq:UYconstr} into MSM \eqref{eq:msm}. It would be difficult to do this by directly extending MSM \eqref{eq:msm-Y0Y1}.

\section{Characterization of unmeasured confounders}\label{sec:Characterize-U}
As mentioned in the main paper, an unmeasured variable $U$ that is responsible for all confounding beyond $X$ needs to be associated with both $T$ and $(Y^0,Y^1)|T$, within some level of $X$.
For completeness, we formally state and prove this result below.

\begin{lem} \label{lem:U-confounding}
Suppose that $ Y^1 \perp T | X, U$.

(i) If $U \perp T | X$, then $Y^1 \perp T |X$.

(ii) The two properties $U \perp Y^1 | T=1, X$ and $U \perp Y^1 | T=0, X$
are equivalent to each other. Moreover, if either property holds, then $Y^1 \perp T | X$.

(iii) The property that $E(Y^1| T=1, X, U)$ is constant in $U$ and the property that $E ( Y^1 | T=0, X, U)$ is constant in $U$ are equivalent to each other. Moreover, if either property holds, then $E(Y^1 | T=0,X) = E(Y^1 | T=1,X)=E(Y|T=1,X)$.
\end{lem}

\begin{prf}
(i) For $t=0,1$, we have
\begin{align*}
P ( T=t | Y^1, X) &= E\{ P(T=t | Y^1,U,X) | Y^1, X\}\\
 & = E\{P( T=t | X, U)|Y^1,X\} \quad  \text{[because  $ Y^1 \perp T | X, U$]} \\
 & = E\{P( T=t | X )|Y^1,X\} \quad  \text{[because $U \perp T | X$]}\\
 &=P( T=t | X ).
\end{align*}
Hence $Y^1 \perp T |X$.

(ii) We show that if $U \perp Y^1 | T=1, X$, then $Y^1 \perp T | X$. In fact,
\begin{align}  \label{eq:U-confounding-prf1}
\begin{split}
\ud P_{Y^1} ( \cdot | X)&=E\left\{\ud P_{Y^1} ( \cdot | X,U)|X\right\}\\
&= E\left\{\ud P_{Y^1} (  \cdot | T=1, X, U) |X\right\}\quad  \text{[because  $ Y^1 \perp T | X, U$]}  \\
 & =  E\left\{\ud P_{Y^1}  ( \cdot | T=1, X)|X\right\} \quad  \text{[because $U \perp Y^1 | T=1, X$]}\\
 &=\ud P_{Y^1}  ( \cdot | T=1, X)
\end{split}
\end{align}
Combining this with the fact that $$ \ud P(Y^1 | X) = \pi^*(X) \ud P(Y^1 | T=1,X) + (1-\pi^*(X)) \ud P(Y^1 | T=0,X),$$ we obtain
$ \ud P(Y^1 | X) = \ud P(Y^1 | T=0,X)$. Hence  $Y^1 \perp T |X$.

Second, we show that if $U \perp Y^1 | T=1, X$, then $U \perp Y^1 | T=0, X$. In fact,
\begin{align*}
 \ud P (Y^1 | T=0, X,U) & = \ud P ( Y^1|T=1, X, U) \quad  \text{[because  $ Y^1 \perp T | X, U$]}  \\
 & = \ud P ( Y^1 | T=1, X) \quad  \text{[because $U \perp Y^1 | T=1, X$]} \\
 & = \ud P ( Y^1 | T=0, X) \quad  \text{[because $Y^1 \perp T |X$],}
\end{align*}
that is, $U \perp Y^1 | T=0, X$.

Finally, by symmetry, it can be shown that if $U \perp Y^1 | T=0, X$, then $U \perp Y^1 | T=1, X$.

(iii) The proof is similar to that in (ii). First, we show that if $E(Y^1| T=1, X,U)$ is constant in $U$, then $E(Y^1|T=0, X) = E(Y^1|T=1,X)$. In fact,
\begin{align}  \label{eq:U-confounding-prf2}
\begin{split}
 E (Y^1 | X,U) & = E ( Y^1 | T=1, X, U) \quad  \text{[because  $ Y^1 \perp T | X, U$]}  \\
 & = E ( Y^1 | T=1, X) \quad  \text{[because $E(Y^1| T=1, X,U)$ is constant in $U$].}
\end{split}
\end{align}
Then $ E(Y^1 | X) = E\{ E (Y^1|X,U) | X\} = E (Y^1 | T=1,X)$.
Combining this with the fact that $$ E (Y^1 | X) = \pi^*(X) E (Y^1 | T=1,X) + (1-\pi^*(X)) E (Y^1 | T=0,X),$$
we obtain
$ E (Y^1 | X) = E (Y^1 | T=0,X) = E (Y^1 | T=1,X)$. The equivalence to $ E (Y |T=1, X)$ follows from the consistency assumption.

Second, we show that if $E(Y^1| T=1, X,U)$ is constant in $U$, then $E(Y^1| T=0, X,U)$ is constant in $U$. In fact,
\begin{align*}
 E (Y^1 | T=0, X,U) & = E ( Y^1 | X, U) \quad  \text{[because  $ Y^1 \perp T | X, U$]}  \\
 & = E ( Y^1 | T=1, X) \quad  \text{[from \eqref{eq:U-confounding-prf2}]} \\
 & = E ( Y^1 | T=0, X) \quad  \text{[as shown above],}
\end{align*}
that is, $E(Y^1| T=0, X,U)$ is constant in $U$.

Finally, by symmetry, it can be shown that if $E(Y^1| T=0, X,U)$ is constant in $U$, then $E(Y^1| T=1, X,U)$ is constant in $U$.
\end{prf}

Conditional on $X$, Lemma \ref{lem:U-confounding} (i) and (ii) show that if $U$ is not associated with $T$ or $Y^1$ in either treatment group, then unconfoundedness holds in $Y^1$. Lemma \ref{lem:U-confounding} (iii) is variation of result (ii), dealing with unconfoundedness in means only and corresponding to the degenerate case $\Delta_1(X) = \Delta_2(X) \equiv 0$ in the proposed outcome sensitivity constraint \eqref{eq:UYconstr}.

For simplicity, Lemma \ref{lem:U-confounding} is stated for unmeasured confounding in $Y^1$, where $U$ is assumed to satisfy $ Y^1 \perp T | X, U$.
Under $(Y^0,Y^1) \perp T | X, U$, Lemma \ref{lem:U-confounding} (i) and (ii) remain valid and extend to the unmeausred confounding in $(Y^0,Y^1)$ by substituting $(Y^0,Y^1)$ for $Y^1$,
and Lemma \ref{lem:U-confounding} (iii) applies to both $Y^0$ and $Y^1$.

\section{eMSM sharp bounds and estimation of $\mu^0$ and ATE}\label{sec:additionalbounds}
The main text presents the outcome sensitivity constraint \eqref{eq:UYconstr} and the eMSM-assisted inference about $\mu^1$. We now extend them to $\mu^0$ and ATE.

Similar to \eqref{eq:UYconstr} on the $U$-$Y^1$ association, the $U$-$Y^0$ association is constrained as
\begin{align}\label{eq:eMSM_Y0}
    E(Y^0|T=0,X,U)-E(Y^0|T=0,X)\in[-\Delta_{1}'(X),\Delta_{2}'(X)],
\end{align}
where $\Delta_{1}'(X),\Delta_{2}'(X)\ge 0$ are prespecified covariate functions. We refer to the model specified by \eqref{eq:msm}, \eqref{eq:UYconstr} and \eqref{eq:eMSM_Y0} as a complete eMSM (C-eMSM), while the model specified by \eqref{eq:msm} and \eqref{eq:UYconstr}, or \eqref{eq:msm} and \eqref{eq:eMSM_Y0}, as a (partial) eMSM.

By symmetry of $T$ in its two labels and Propositions \ref{pro:pop-sol} and \ref{pro:pop-sol-lower}, the sharp population bounds of $\nu^{0}(X)=E(Y^0|T=1,X)$ under the partial eMSM defined by \eqref{eq:msm} and \eqref{eq:eMSM_Y0} are
\begin{align*}
&\nu^{0+}_{\mytext{eMSM}} (X) =E(Y|T=0,X)+\{\Lambda_1^{-1}(X)-\Lambda_2^{-1}(X)\}\psi_{0+}(X)E\{\rho_{\tau'}(Y,q^*_{0,\tau'})|T=0,X\},\\
&\nu^{0-}_{\mytext{eMSM}} (X) = E(Y|T=0,X)-\{\Lambda_1^{-1}(X)-\Lambda_2^{-1}(X)\}\psi_{0-}(X)E\{\rho_{1-\tau'}(Y,q^*_{0,1-\tau'})|T=0,X\},
\end{align*}
where $\tau'(X)=\{\Lambda_1^{-1}(X)-1\}/\{\Lambda_1^{-1}(X)-\Lambda_2^{-1}(X)\}$, $q^*_{0,\tau'}(X)$ is the $\tau'(X)$-quantile of $P_Y(\cdot| T=0,X)$,
$\psi_{0+}(X)=\min \{\delta'_{1+}(X), \delta'_{2+}(X), 1\}$ with
\begin{align*}
\delta'_{1+}(X) = \frac{\tau'(X)\Delta_1'(X)}{E\{\rho_{\tau'}(Y,q^*_{0,\tau'})|T=0,X\}} , \quad
\delta'_{2+}(X) = \frac{\{1-\tau'(X)\} \Delta_2'(X)}{E\{\rho_{\tau'}(Y,q^*_{0,\tau'})|T=0,X\}},
\end{align*}
and $\psi_{0-}(X)=\min \{\delta'_{1-}(X), \delta'_{2-}(X), 1\}$ is $\psi_{0+}(X)$ with $\tau'(X)$ replaced by $1-\tau'(X)$.
The sharp bounds on $\mu^{0}$ can be obtained as
\begin{align}
    \mu^{0+}_{\mytext{eMSM}} = E \{(1-T)Y+T\nu^{0+}_{\mytext{eMSM}} (X) \},\quad \mu^{0-}_{\mytext{eMSM}} = E \{(1-T)Y+T\nu^{0-}_{\mytext{eMSM}} (X) \}.
\end{align}

As discussed in Section \ref{sec:ParaSpec}, we typically specify the sensitivity parameters in a covariate-independent
manner, possibly after some re-parametrization. Following this convention, the recommended specification sets $\Lambda_1(X)\equiv\Lambda_1$, $\Lambda_2(X)\equiv\Lambda_2$, and, if $\tau'=\{\Lambda_1^{-1}-1\}/\{\Lambda_1^{-1}-\Lambda_2^{-1}\}\ge 1/2$, analogously to \eqref{eq:recspec}
\begin{align}
    \Delta_2'(X)=\frac{\delta'}{1-\tau'} E\{\rho_{\tau'}(Y,q^*_{0,\tau'})|T=0,X\},\quad \Delta_1'(X)=\frac{\delta'}{1-\tau'} E\{\rho_{1-\tau'}(Y,q^*_{0,1-\tau'})|T=0,X\},
\end{align}
for some $\delta'\in[0,1]$. If $\tau'<1/2$, we specify
\begin{align}
    \Delta_2'(X)=\frac{\delta'}{\tau'} E\{\rho_{1-\tau'}(Y,q^*_{0,1-\tau'})|T=0,X\},\quad \Delta_1'(X)=\frac{\delta'}{\tau'} E\{\rho_{\tau'}(Y,q^*_{0,\tau'})|T=0,X\}.
\end{align}
In either case, the sharp bounds of $\mu^0$ under the partial eMSM can be simplified as in Proposition \ref{prop:propspecbd} to
\begin{align}
    \mu^{0+}_{\mytext{eMSM}}
    &=E\left[E(Y|T=0,X)+T(\Lambda_1^{-1}-\Lambda_2^{-1})\delta' E\left\{\rho_{\tau'}(Y,q^*_{0,\tau'})|T=0,X\right\}\right]\\
    &=E\left\{\frac{1-T}{1-\pi^*(X)}Y+(1-T)\frac{\pi^*(X)}{1-\pi^*(X)}(\Lambda_1^{-1}-\Lambda_2^{-1})\delta'\rho_{\tau'}(Y,q^*_{0,\tau'})\right\}  ,  \label{eq:mu0up}\\
    \mu^{0-}_{\mytext{eMSM}}
    &=E\left[E(Y|T=0,X)-T(\Lambda_1^{-1}-\Lambda_2^{-1})\delta' E\left\{\rho_{1-\tau'}(Y,q^*_{0,1-\tau'})|T=0,X\right\}\right]\\
    &=E\left\{\frac{1-T}{1-\pi^*(X)}Y+(1-T)\frac{\pi^*(X)}{1-\pi^*(X)}(\Lambda_1^{-1}-\Lambda_2^{-1})\delta'\rho_{1-\tau'}(Y,q^*_{0,1-\tau'})\right\}.\label{eq:mu0low}
\end{align}

C-eMSM is needed for deriving bounds of causal parameters contrasting $\mu^0$ and $\mu^1$, e.g., ATE. Fortunately,
the sharp bounds of $(\nu^0,\nu^1)$ and $(\mu^0,\mu^1)$ under C-eMSM are identical to those from the two partial eMSMs, which is proved in Section \ref{sec:proof-prop1-cor1}. Moreover, the proof shows that the sharp lower bound of $\mu^0$ and the sharp upper bound of $\mu^1$ under respective partial eMSMs (or the sharp upper bound of $\mu^0$ and the sharp lower bound of $\mu^1$) are attained simultaneously by a single distribution $Q$ allowed in C-eMSM. Due to the simultaneous sharpness, the sharp bounds of causal parameters like ATE under C-eMSM can be obtained by contrasting the sharp bounds of $\mu^0$ and $\mu^1$ under respective partial eMSMs. For example, $\mu^{1+}_{\mytext{eMSM}}-\mu^{0-}_{\mytext{eMSM}}$ is the sharp upper bound of ATE under C-eMSM and $\mu^{1-}_{\mytext{eMSM}}-\mu^{0+}_{\mytext{eMSM}}$ the sharp lower bound.

Confidence intervals for $\mu^{0}$ and ATE can be obtained similarly to that of $\mu^1$ as in Section \ref{sec:est}. Consider the same working models \eqref{eq:propensity}--\eqref{eq:outreg} for nuisance parameters, except that $q(X;\beta)$ is used to estimate $q^*_{0,\tau'}(X)$ and $m_+ (X; q(\cdot;\beta),\alpha)$ is used to estimate $E\{\tilde{Y}_+' (q(\cdot;\beta))|T=0,X\}$
where $\tilde{Y}_+^\prime (q) = Y+(\Lambda_1^{-1}-\Lambda_2^{-1})\delta'\rho_{\tau'}(Y,q)$. Given estimates $(\hgamma,\hbeta,\halpha)$, the sample upper bound for $\mu^0$ is
\begin{align*}
    \hat{\mu}^{0+} (\hgamma,\hbeta,\halpha) = \tE \{ \varphi_{0+}(\hgamma,\hbeta,\halpha) \},
\end{align*}
where $\tilde E (\cdot)$ denotes the sample average over the observed data $\{ (Y_i, T_i, X_i) :i=1,\ldots,n\}$, and
\begin{align}
    \varphi_{0+} (\gamma,\beta,\alpha)=&\frac{1-T}{1-\pi(X;\gamma)}Y+(1-T)\frac{\pi(X;\gamma)}{1-\pi(X;\gamma)}(\Lambda_1^{-1}-\Lambda_2^{-1})\delta'\rho_{\tau'}(Y,q(X;\beta))\\
    &
    -\left(\frac{1-T}{1-\pi(X;\gamma)}-1\right)  m_+(X;q(\cdot;\beta),\alpha).
\end{align}
The CAL estimates $(\hgamma_{\mytext{CAL}}',\hbeta_{\mytext{WQ},+}',\halpha_{\mytext{WL},+}')$ can be obtained through
\begin{align}
    \hgamma_{\mytext{CAL}}'&=\arg\min_{\gamma}\tE \left\{(1-T)e^{f^\mytext{T}(X)\gamma}-Tf^\mytext{T}(X)\gamma\right\},\label{eq:gammaest-mu0} \\
    \hbeta_{\mytext{WQ},+}'&=\arg\min_{\beta}\tE\left\{ (1-T)\frac{\pi(X;\hgamma_{\mytext{CAL}}')}{1-\pi(X;\hgamma_{\mytext{CAL}}')}\delta'\rho_{\tau'}\left(Y,h^\mytext{T}(X)\beta\right)\right\},\label{eq:betaest-mu0}\\
   \halpha_{\mytext{WL},+}'&=\arg\min_{\alpha}\tE\left[ (1-T)\frac{\pi(X;\hgamma_{\mytext{CAL}}')}{1-\pi(X;\hgamma_{\mytext{CAL}}')}\left\{\tilde{Y}_+'\left(h^\mytext{T}\hbeta_{\mytext{WQ},+}' \right)-f^\mytext{T}(X)\alpha\right\}^2\right].\label{eq:alphaest-mu0}
\end{align}

Let $\hat{\mu}^{0+}_{\mytext{CAL}} = \hat{\mu}^{0+}(\hgamma_{\mytext{CAL}}',\hbeta_{\mytext{WQ},+}',\halpha_{\mytext{WL},+}')$ and
$\hat V_{\mytext{CAL}}^{0+}= \tE \{\varphi_{0+}(\hgamma_{\mytext{CAL}}',\hbeta_{\mytext{WQ},+}',\halpha_{\mytext{WL},+}') - \hat{\mu}^{0+}_{\mytext{CAL}}\}^2$. The one-sided confidence interval
\begin{equation}\label{eq:CI-mu0-upper}
    \left(-\infty,\hat{\mu}^{0+}_{\mytext{CAL}}+z_{c}\sqrt{ \hat V_{\mytext{CAL}}^{0+}} /n\right]
\end{equation}
is doubly robust with asymptotic level $1-c$ for the sharp bound $\mu^{0+}_{\mytext{eMSM}} $ and the true value $\mu^0$. In other words, if either model \eqref{eq:propensity} or \eqref{eq:outreg} is correctly specified, \eqref{eq:CI-mu0-upper} achieves asymptotic level $1-c$ in containing $\mu^{0+}_{\mytext{eMSM}}$ and $\mu^0$.
If further model \eqref{eq:quantile} is correctly specified, then \eqref{eq:CI-mu0-upper}
is doubly robust with asymptotic size $1-c$ for $\mu^{0+}_{\mytext{eMSM}}$.

The sample lower bound of ATE is $\hat{\mu}^{1-}_{\mytext{CAL}}-\hat{\mu}^{0+}_{\mytext{CAL}}$, and its related doubly robust one-sided confidence interval of asymptotic level $1-c$ is
\begin{equation}\label{eq:CI-ATE-upper}
    \left[\hat{\mu}^{1-}_{\mytext{CAL}}-\hat{\mu}^{0+}_{\mytext{CAL}}-z_{c}\sqrt{ \hat V_{\mytext{CAL}}} /n,+\infty\right),
\end{equation}
where
\begin{align}
    \hat V_{\mytext{CAL}}=\tilde E \left\{\varphi_{1-}(\hgamma_{\mytext{CAL}},\hbeta_{\mytext{WQ},-},\halpha_{\mytext{WL},-})-\varphi_{0+}(\hgamma_{\mytext{CAL}}',\hbeta_{\mytext{WQ},+}',\halpha_{\mytext{WL},+}')\right\}^2 .
\end{align}
Two-sided confidence intervals for $\mu^0$ and ATE can be similarly derived as in \eqref{eq:Twosided-mu1-CI} and are omitted.

\section{Additional discussion on DV}\label{sec:dvest}
The DV approach considered in Section \ref{sec:compare}, as in the main text of \citeappend{ding2016sensitivity}, is conditional on $X$ and on the population level. We summarize DV unconditional bounds and finite sample results, as well as our alternative unconditional bounds.

\noindent\textbf{Conditional bounds and finite sample results}

First, we expand the notations to indicate conditioning on the observed covariates $X$.
As in Section \ref{sec:compare}, we define for $t=0,1$
\begin{equation*}
\RR_{\mytext{UT,t}}(X)=\max_{u}\frac{\ud P_U(u\mid T=1-t,X)}{\ud P_U(u\mid T=t,X)},
\quad\RR_{\mytext{UY,t}}(X)=\frac{\max_u E(Y\mid T=t,X,U=u)}{\min_u E(Y\mid T=t,X,U=u)},
\end{equation*}
which are constrained as
\begin{align}
    \RR_{\mytext{UT,0}}(X)\le1/\Lambda_1(X),\quad \RR_{\mytext{UT,1}}(X)\le\Lambda_2(X),\quad  \RR_{\mytext{UY}}(X)\le \Theta(X),
\end{align}
with $\RR_{\mytext{UY}}(X)=\max\{\RR_{\mytext{UY,0}}(X),\RR_{\mytext{UY,1}}(X)\}$. Let $p_t(X)=E(Y|T=t,X)$ and $\mu^t(X)=E(Y^t|X)$ for $t=0,1$, $\ORR(X)= p_1(X)/p_0(X)$ and $\CRR(X)= E(Y^1|X)/E(Y^0|X)$. With $B(x,y)=xy/(x+y-1),$ the DV bounds of $\mu^1(X)$ and $\mu^0(X)$ are
\begin{align}\label{eq:dv-ab-1}
 &  \mu^{1+}_{\mytext{DV}}(X) =
 p_1(X)\left\{P(T=1|X)+P(T=0|X) B(\Lambda_2(X),\Theta(X))\right\},\\
&  \mu^{1-}_{\mytext{DV}}(X) =
  p_1(X)\left\{P(T=1|X)+\frac{P(T=0|X)}{B(1/\Lambda_1(X),\Theta(X))}\right\},\\
  &  \mu^{0+}_{\mytext{DV}}(X) =
 p_0(X)\left\{P(T=0|X)+P(T=1|X) B(1/\Lambda_1(X),\Theta(X))\right\},\\
 &  \mu^{0-}_{\mytext{DV}}(X) =
  p_0(X)\left\{P(T=0|X)+\frac{P(T=1|X)}{B(\Lambda_2(X),\Theta(X))}\right\}.
\end{align}
The main result of DV can be stated as follows:
\begin{equation}\label{eq:conditionalDV}
    \ORR(X)/B(1/\Lambda_1(X),\Theta(X))\le\CRR(X)\le \ORR(X)\cdot B(\Lambda_2(X),\Theta(X)),
\end{equation}
where the upper bound of $\CRR(X)$ is obtained as the ratio of $\mu^{1+}_{\mytext{DV}}(X)$ over $\mu^{0-}_{\mytext{DV}}(X)$, and the lower bound the ratio of $\mu^{1-}_{\mytext{DV}}(X)$ over $\mu^{0+}_{\mytext{DV}}(X)$.

The bounds in \eqref{eq:conditionalDV} are on the population level, as $\ORR(X)$ is not a sample estimate but a population quantity. In applications, given a confidence interval $[\widehat{\ORR}_{\rm L}(X),\widehat{\ORR}_{\rm U}(X)]$ for $\ORR(X)$, DV used the following interval estimates for $\CRR(X)$:
\begin{equation}
    \left[\widehat{\ORR}_{\rm L}(X)/B(1/\Lambda_1(X),\Theta(X)),\widehat{\ORR}_{\rm U}(X)\cdot B(\Lambda_2(X),\Theta(X))\right].
\end{equation}

\citeappend{ding2016sensitivity} also obtained the following bounds of the conditional average treatment effect ($\rm{CATE}(X)=E(Y^1|X)-E(Y^0|X)$):
\begin{align*}
    &\left\{p_1(X)-p_0(X)B(1/\Lambda_1(X),\Theta(X))\right\}\left\{\pi^*(X)+(1-\pi^*(X))/B(1/\Lambda_1(X),\Theta(X))\right\}\\
   &\hspace{12em}\le \rm{CATE}(X)\le\\
   &\left\{p_1(X)-p_0(X)/B(\Lambda_2(X),\Theta(X))\right\}\left\{\pi^*(X)+(1-\pi^*(X))B(\Lambda_2(X),\Theta(X))\right\}\nonumber.
\end{align*}
 Given sample estimates $\hat{p}_t(X)$ with $t=0,1$, $\hat{\pi}(X)$, and their standard errors $s_t(X)$, $s(X)$, DV obtained the variance of the lower bound of $\rm{CATE}(X)$ as
\begin{align}\label{eq:condVarL}
    &\left\{s_1^2(X)+s_0^2(X)\cdot B(1/\Lambda_1(X),\Theta(X))^2\right\}\left\{\hat{\pi}(X)+\frac{1-\hat{\pi}(X)}{B(1/\Lambda_1(X),\Theta(X))}\right\}^2+\nonumber\\
    &\hspace{5em}\left\{\hat{p}_1(X)-\hat{p}_0(X)B(1/\Lambda_1(X),\Theta(X))\right\}^2\left\{1-B(1/\Lambda_1(X),\Theta(X))^{-1}\right\}^2s^2(X),
\end{align}
and the variance of the upper bound as
\begin{align}\label{eq:condVarU}
    &\left\{s_1^2(X)\cdot B(\Lambda_2(X),\Theta(X))^2+s_0^2(X)\right\}\left\{1-\hat{\pi}(X)+\frac{\hat{\pi}(X)}{B(\Lambda_2(X),\Theta(X))}\right\}^2+\nonumber\\
    &\hspace{5em}\left\{\hat{p}_1(X)B(\Lambda_2(X),\Theta(X))-\hat{p}_0(X)\right\}^2\left\{1-B(\Lambda_2(X),\Theta(X))^{-1}\right\}^2s^2(X).
\end{align}
A sensitivity interval for $\rm{CATE}(X)$ is then obtained with standard normal approximation.

\noindent\textbf{Unconditional bounds}

For unconditional causal relative risk, $\CRR=E(Y^1)/E(Y^0)$, DV proposed a relaxed bound:
\begin{align}
    \min_{X}\ORR(X)/ B(1/\Lambda_1(X),\Theta(X))\le\CRR\le \max_{X}\ORR(X)\cdot  B(\Lambda_2(X),\Theta(X)).
\end{align}
However, DV did not address how sampling uncertainty can be properly incorporated into these bounds. One naive approach is to take the union of all conditional intervals over $X$, but the resulting interval may often be too wide.

Alternatively, as in our numerical study, $E(Y^1)$ and $E(Y^0)$ are bounded through iterated expectations:
\begin{align}\label{eq:unconditional1}
   &E\left[p_1(X)\left\{\pi^*(X)+(1-\pi^*(X))/B(1/\Lambda_1(X),\Theta(X))\right\}\right]\nonumber\\
   &\hspace{8em}\le E(Y^1)\le\\
   &E\left[p_1(X)\left\{\pi^*(X)+(1-\pi^*(X))B(\Lambda_2(X),\Theta(X))\right\}\right],\nonumber\\
    &E\left[p_0(X)\left\{\pi^*(X)/B(\Lambda_2(X),\Theta(X))+1-\pi^*(X)\right\}\right]\nonumber\\
   &\hspace{8em}\le E(Y^0)\le\\
   &E\left[p_0(X)\left\{\pi^*(X)B(1/\Lambda_1(X),\Theta(X))+1-\pi^*(X)\right\}\right]\nonumber.
\end{align}
It follows that the unconditional CRR are bounded by
\begin{align}\label{eq:uncondCRR}
&\frac{E\left[p_1(X)\left\{\pi^*(X)+(1-\pi^*(X))/B(1/\Lambda_1(X),\Theta(X))\right\}\right]}{E\left[p_0(X)\left\{\pi^*(X)B(1/\Lambda_1(X),\Theta(X))+1-\pi^*(X)\right\}\right]}
     &\nonumber\\
   &\hspace{8em}\le \CRR\le\\
   &\frac{E\left[p_1(X)\left\{\pi^*(X)+(1-\pi^*(X))B(\Lambda_2(X),\Theta(X))\right\}\right]}{E\left[p_0(X)\left\{\pi^*(X)/B(\Lambda_2(X),\Theta(X))+1-\pi^*(X)\right\}\right]}\nonumber,
\end{align}
and the average treatment effect, $\rm{ATE}=E(Y^1)- E(Y^0)$, by
\begin{align}\label{eq:uncondCRD}
&E\left[\left\{p_1(X)-p_0(X)B(1/\Lambda_1(X),\Theta(X))\right\}\left\{\pi^*(X)+(1-\pi^*(X))/B(1/\Lambda_1(X),\Theta(X))\right\}\right]\nonumber\\
   &\hspace{12em}\le \rm{ATE}\le\\
   &E\left[\left\{p_1(X)-p_0(X)/B(\Lambda_2(X),\Theta(X))\right\}\left\{\pi^*(X)+(1-\pi^*(X))B(\Lambda_2(X),\Theta(X))\right\}\right]\nonumber.
\end{align}
Sample versions of the point bounds \eqref{eq:uncondCRR} and \eqref{eq:uncondCRD} can be constructed as the corresponding sample averages,
with $p_t(X)$ and $\pi^*(X)$ replaced by estimates $\hat{p}_t(X)$ and $\hat{\pi}(X)$.
For non-regularized estimation using low-dimensional regression models,
bootstrapping can be used to obtain confidence intervals for the point bounds. 
In our numerical applications (Section~\ref{sec:numerical}), ML estimation is used and bootstrapping is implemented with 1000 bootstrap samples.

\section{Additional discussion on dMSM and rMSM}\label{sec:d-r-MSM}
Under the outcome sensitivity constraint of dMSM, $\mg^*(X,U)$ is bounded in the differences
over possible values of $U$:
\begin{align}
    \max_{u}\mg^*(X,u)-\min_{u}\mg^*(X,u)\le \Delta(X).
\end{align}
When $\Delta(X)\equiv0$, unconfoundedness holds in means, i.e., $E(Y^1 | T=0,X) = E(Y^1 | T=1,X)$ for all $X$.

Similar as in Proposition \ref{prop:DVcompare}, $\rm{dMSM}(\Delta)$ is the collection of all $\rm{eMSM}s$ with $\Delta_1(X)+\Delta_2(X)\le\Delta(X)$.
Consequently, the sharp bounds under $\rm{dMSM}(\Delta)$ are no tighter than those under individual $\rm{eMSM}$ with $\Delta_1(X)+\Delta_2(X)\le\Delta(X)$, and can be derived from those under eMSM with proper choice of $\Delta_1(X)$ and $\Delta_2(X)$. Specifically, the sharp bounds of $\mu^1$ under $\rm{dMSM}(\Delta)$ are
\begin{subequations}
    \begin{align}
    \mu_{\mytext{dMSM}}^{1+}(\Delta)=&E\Big(E(Y|T=1,X)+(1-T)\{\Lambda_2(X)-\Lambda_1(X)\}\times\\
        &\min\left[\tau(X)\left\{1-\tau(X)\right\}\Delta(X),E \{\rho_{\tau}(Y,q^*_{1,\tau})|T=1,X \}\right]\Big),\label{eq:dMSM-up}\\
    \mu_{\mytext{dMSM}}^{1-}(\Delta)=&E\Big(E(Y|T=1,X)-(1-T)\{\Lambda_2(X)-\Lambda_1(X)\}\times\\
        &\min\left[\tau(X)\left\{1-\tau(X)\right\}\Delta(X),E \{\rho_{1-\tau}(Y,q^*_{1,1-\tau})|T=1,X \}\right]\Big)\label{eq:dMSM-low}.
    \end{align}
\end{subequations}
The sharp upper bound $\mu_{\mytext{dMSM}}^{1+}(\Delta)$ coincides with that of the eMSM with $\Delta_1(X)=\{1-\tau(X)\}\Delta(X)$ and $\Delta_2(X)=\tau(X)\Delta(X)$, or equivalently
$\delta_{1+}(X)=\delta_{2+}(X) = \tau(X)\{1-\tau(X)\}\Delta(X) / E \{\rho_{\tau}(Y,q^*_{1,\tau})|T=1,X \}$. Conversely, for a constant $\delta \in [0,1]$, taking $\delta_{1+}(X) = \delta_{2+}(X) \equiv \delta$ as in the specification \eqref{eq:upspec} leads to
\begin{align}\label{eq:dMSM-spec-up}
   \Delta (X) = \delta \frac{E \{\rho_{\tau}(Y,q^*_{1,\tau})|T=1,X \}} {\tau(X)\{1-\tau(X)\} }.
\end{align}
The sharp lower bound $\mu_{\mytext{dMSM}}^{1-}(\Delta)$ coincides with that of another eMSM with $\Delta_1(X)=\tau(X)\Delta(X)$ and $\Delta_2(X)=\{1-\tau(X)\}\Delta(X)$
or equivalently
$\delta_{1-}(X) =\delta_{2-}(X) = \tau(X)\{1-\tau(X)\}\Delta(X) /E \{\rho_{1-\tau}(Y,q^*_{1,1-\tau})|T=1,X \}$. Conversely, taking $\delta_{1-}(X) = \delta_{2-}(X) \equiv \delta$ as in the specification \eqref{eq:lowspec} leads to
\begin{align}\label{eq:dMSM-spec-low}
    \Delta (X) = \delta \frac{E \{\rho_{\tau}(Y,q^*_{1,1-\tau})|T=1,X \}} {\tau(X)\{1-\tau(X)\} }.
\end{align}
In general, unless $Y|T=1,X$ is symmetric around its median, the above two ``constant'' dMSM specifications, \eqref{eq:dMSM-spec-up} and \eqref{eq:dMSM-spec-low},
are not compatible and cannot be combined similarly as in the ``constant'' $(\Delta_1,\Delta_2)$-specification \eqref{eq:recspec} for eMSM.
Therefore, eMSM seems to be more flexible and interpretable than dMSM, due to the use of two parameters $(\Delta_1,\Delta_2)$ instead of $\Delta$.

Under the outcome sensitivity constraint of rMSM, $\mg^*(X,U)$ is bounded in the ratios over possible values of $U$:
\begin{align}
    \frac{\max_{u}\mg^*(X,u)}{\min_{u}\mg^*(X,u)}\le\Theta(X).
\end{align}
When $\Theta(X)\equiv1$, unconfoundedness holds in means. $\rm{rMSM}$ extends the DV model discussed in Section \ref{sec:compare} from binary to nonnegative outcomes. Following a similar argument as for Proposition \ref{prop:DVcompare}, $\rm{rMSM}(\Theta)$ is a collection of $\rm{eMSM} (\Delta_{1},\Delta_{2})$ with $\theta(\Delta_{1},\Delta_{2})\le\Theta$. Therefore, the sharp bounds of $\mu^1$ under $\rm{rMSM}(\Theta)$ are no tighter than those of individual $\rm{eMSM}$, and are
\begin{subequations}
    \begin{align}
    \mu_{\mytext{rMSM}}^{1+}(\Theta)=&E\bigg(E(Y|T=1,X)+(1-T)\{\Lambda_2(X)-\Lambda_1(X)\}\times\\
        &\min\left[\frac{E(Y|T=1,X)(\Theta(X)-1)}{1/\left\{1-\tau(X)\right\}+\Theta(X)/\tau(X)},E \{\rho_{\tau}(Y,q^*_{1,\tau})|T=1,X \}\right]\bigg),\label{eq:rMSM-up}\\
        \mu_{\mytext{rMSM}}^{1-}(\Theta)=&E\bigg(E(Y|T=1,X)-(1-T)\{\Lambda_2(X)-\Lambda_1(X)\}\times\\
        &\min\left[\frac{E(Y|T=1,X)(\Theta(X)-1)}{\Theta(X)/\left\{1-\tau(X)\right\}+1/\tau(X)},E \{\rho_{1-\tau}(Y,q^*_{1,1-\tau})|T=1,X \}\right]\bigg)\label{eq:rMSM-low}.
    \end{align}
\end{subequations}
Both bounds mimic the respective sharp DV bounds in Proposition \ref{prop:DVcompare}, with $p_1$ replaced by $E(Y|T=1,X)$ and $X$ integrated out. 
\section{Proofs of main results}\label{sec:proofsmain}
\subsection{Proof of equation (\ref{eq:Expected-Y}) }  \label{sec:proof-eq-EQY}
Under any full data distribution $Q$ that is compatible with the observed-data distribution on $(Y,T,X)$,
we have
\begin{align*}
    E_Q(Y^1) & =P(T=1) E_Q(Y^1|T=1)+P(T=0)E_Q(Y^1|T=0)\\
    & =P(T=1) E(Y|T=1)+P(T=0)E\left\{E_Q(Y^1|T=0,X)|T=0\right\}\\
   & =E \left\{TY+(1-T) E_Q (Y^1|T=0,X) \right\}.
\end{align*}
Moreover,
\begin{align*}
    E_Q (Y^1|T=0,X)&=E_Q\left\{ E_Q (Y^1|T=0,X,U)|T=0,X\right\}\\
    &=\int \mg_Q(X,u)\ud Q_U(u|T=0,X)\\
    &=\int \mg_Q(X,u)\lambda_Q(X,u) \ud Q_U(u|T=1,X),
\end{align*}
where we used \eqref{eq:Q-indep} in the second equality and the definition of $\lambda_Q$ in the third.

\subsection{Proof of inequality (\ref{eq:g-inherentNP})}\label{sec:proof-g-inherentNP}
Throughout the proof, all expressions are implicitly conditional on $T=1$ and $X$. For example, $\mg_Q(X,u)=E_Q(Y|T=1,X,U=u)$ is simplified to $\mg_Q(u)=E_Q(Y|U=u)$. It suffices to show that if $Q_U(u)=c$ for some constant $c\in(0,1]$, then $\mg_Q(u)$ satisfies
\begin{align}\label{eq:g-inherentNP-simp}
E \left\{ Y - c^{-1} \rho_{c} \left(Y, q^*_{c}\right) \right\}\le\mg_Q(u) \le E \left\{ Y + c^{-1} \rho_{1-c} \left(Y, q^*_{1-c}\right)\right\},
\end{align}
where $q^*_{c}$ is the $c$-quantile of $P_Y$, the observed distribution of $Y$, and $\rho_c(Y,q)$ is the check function.

Let $(\Omega,\mathcal{F},Q)$ be the probability space on which $Y$ and $U$ are defined. It follows from the definition of conditional expectation that
\begin{align}\label{eq:cond-exp-bound-up}
    E_Q(Y|U=u)=\frac{E\left(Y_Q\mathbf{1}_{U=u}\right)}{Q_U(u)}\le\sup_{A\in\mathcal{F},Q(A)= c}E_Q\left(c^{-1}Y\mathbf{1}_A\right),
\end{align}
where $\mathbf{1}_A$ is the indicator function of $A$. By relaxing $c^{-1}\mathbf{1}_A$ to a random variable $D$ that takes value in $[0,c^{-1}]$, \eqref{eq:cond-exp-bound-up} is then bounded from above by
\begin{align}\label{eq:rGNP}
    &\sup_{D} E_Q\left( YD\right),\\
    &\hspace{-3em}\text{subject to}\quad D\in[0,c^{-1}],\\
    &\hspace{1.5em}\quad E_Q(D)= 1. 
\end{align}
If $D$ is restricted to a function of $Y$, \eqref{eq:rGNP} reduces to the optimization for the MSM sharp upper bound of $\nu^{1}$ with $(\Lambda_1,\Lambda_2)=(0,c^{-1})$, which is solved as a generalized Neyman--Pearson lemma in \citeappend{francis1969some} and \citetappend{tan2024model}. From Property 2.2 of \citeappend{francis1969some} and Lemma S2 of \citetappend{tan2024model}, the latter problem admits an optimizer  
\begin{align}\label{eq:Y-Decision}
    \bar D=\begin{cases}
        c^{-1}\quad &\text{if}\quad Y>q^*_{1-c}\\
        0\quad &\text{if}\quad Y<q^*_{1-c}\\
        c' \quad&\text{if}\quad Y=q^*_{1-c}
    \end{cases},
\end{align}
where $c'\in[0,c^{-1}]$ is a constant such that $E_Q(\bar D)=1$. In fact, \eqref{eq:Y-Decision} is also an optimizer for \eqref{eq:rGNP}. This is because for any $D$ feasible in \eqref{eq:rGNP}, it can be replaced by a function of $Y$, namely $E_Q(D|Y)$, without changing the evaluation of objective function or violating the constraints.

With $\bar D$ in \eqref{eq:Y-Decision} as a function of $Y$, $E_Q$ can be evaluated as an expectation with respect to $Q_Y$, which is identical to $P_Y$ by assumption. Then the objective function is determined as
\begin{align}
    &E( Y\bar D)=E\left\{c^{-1}\left(Y-q^*_{1-c}\right)_+\right\}+q^*_{1-c}\\
    &=E\left\{\left(Y-q^*_{1-c}\right)_+-\left(q^*_{1-c}-Y\right)_+\right\}+q^*_{1-c}+c^{-1}E\left\{(1-c)\left(Y-q^*_{1-c}\right)_++c\left(q^*_{1-c}-Y\right)_+\right\}\\
    &=E \left\{ Y + c^{-1} \rho_{1-c} \left(Y, q^*_{1-c}\right)\right\},\label{eq:eta-simp-up}
\end{align}
which combined with \eqref{eq:cond-exp-bound-up} establishes the upper bound in \eqref{eq:g-inherentNP-simp}. The lower bound in \eqref{eq:g-inherentNP-simp} follows from a similar argument by replacing $Y$ with $-Y$ in \eqref{eq:cond-exp-bound-up} and \eqref{eq:eta-simp-up}.

\subsection{Proof of Propositions \ref{pro:pop-sol}, \ref{pro:pop-sol-lower} and Corollary \ref{cor:eMSM-Q}}\label{sec:proof-prop1-cor1}
As outlined in Section \ref{sec:sharpbounds} and \ref{sec:additionalbounds}, we prove as follows:
    \begin{itemize}\addtolength{\itemsep}{-.1in}
        \item[(a)] Solve optimization \eqref{eq:condtightup-b} for $\nu^{1+}_{\mytext{e}} (X)$ and a similar optimization that defines $\nu^{0-}_{\mytext{e}} (X)$;
        \item[(b)] Show that $\min \{\nu^{1+}_{\mytext{MSM}} (X), \nu^{1+}_{\mytext{e}} (X)\}$, which is a valid upper bound of $\nu^1$ under the partial eMSM for $Y^1$, and $\max \{\nu^{0-}_{\mytext{MSM}} (X), \nu^{0-}_{\mytext{e}} (X)\}$, which is a valid lower bound of $\nu^0$ under the partial eMSM for $Y^0$, are simultaneously sharp under C-eMSM specified by \eqref{eq:msm}, \eqref{eq:UYconstr} and \eqref{eq:eMSM_Y0}, which is a common sub-model to both partial eMSMs.
    \end{itemize}
If valid bounds under parent models (i.e., partial eMSMs) are sharp in a common sub-model (i.e., C-eMSM), then they are also sharp in their respective parent models. This leads to Propositions \ref{pro:pop-sol} and \ref{pro:pop-sol-lower}. In proving point (b) by construction, Corollary \ref{cor:eMSM-Q} is also proved.
Moreover, as mentioned in Section \ref{sec:additionalbounds}, the proof also establishes the sharpness, for example, of $\mu^{1+}_{\mytext{eMSM}}-\mu^{0-}_{\mytext{eMSM}}$ as the upper bound of ATE under C-eMSM. 

To simplify notations, we make the conditioning on $X$ implicit.

\noindent\textbf{Solution for upper bound $\nu^{1+}_{\mytext{e}}$}

Recall optimization \eqref{eq:condtightup-b}:
\begin{align}\label{eq:V_e-simp}
\nu^{1+}_{\mytext{e}}  =\sup_{Q_U(\cdot|T=1),\mg,\lambda}\;
\int \mg(u) \lambda (u)\ud Q_U(u|T=1),
\end{align}
where the two measurable functions $\lambda$ (nonnegative)
and $\mg$ of $U|T=1$ are subject to the following constraints
\begin{align}
&\lambda(U) \in [\Lambda_1,\Lambda_2],\label{eq:msm-simp}\\
    &\mg(U) - E(Y|T=1)\in[-\Delta_1, \Delta_2],\label{eq:emsm-simp}\\
    &\int\lambda(u)\ud Q_U(u|T=1)=1,\label{eq:lambda-simp}\\
    &\int \mg(u) \ud Q_U(u|T=1) = E (Y|T=1)\label{eq:eta-simp}.
\end{align}

Fix any $(Q_U(\cdot|T=1),\mg)$ allowed in \eqref{eq:V_e-simp}, it follows from Property 2.2 of \citeappend{francis1969some}
(related to Proposition 2 of \citeappend{dorn2023sharp} and Proposition 1 of \citeappend{tan2024model})
that an optimal $\lambda$ solving \eqref{eq:V_e-simp} subject to \eqref{eq:msm-simp} and \eqref{eq:lambda-simp}, regardless of 
\eqref{eq:emsm-simp} and \eqref{eq:eta-simp}, is of the form
\begin{align}\label{eq:lambda-optimal}
\lambda(u)=\begin{cases}
        \Lambda_2\quad &\text{if}\quad \mg(u)>q_{\mg,\tau}\\
        \Lambda_1\quad &\text{if}\quad \mg(u)<q_{\mg,\tau}\\
    \end{cases},
\end{align}
where $q_{\mg,\tau}$ is the $\tau$-quantile of $\mg(U)$ and $\tau=(\Lambda_2-1)/(\Lambda_2-\Lambda_1)$. From this property, it suffices to consider $\lambda$ in \eqref{eq:lambda-optimal}. Moreover, as stated in Lemma \ref{lm:BinaryU} (whose proof is provided in Section \ref{sec:proof-lm-BinaryU}),
optimization \eqref{eq:V_e-simp} can be further simplified by only considering the class of $(Q_U(\cdot|T=1),\mg,\lambda)$ that corresponds to some binary $U|T=1$.
\begin{lem}{}\label{lm:BinaryU}
    For any $(Q_{U'}(\cdot|T=1),\mg',\lambda')$ allowed in \eqref{eq:V_e-simp} with $\lambda'$ as in \eqref{eq:lambda-optimal}, there exist another allowed triple $(Q_U(\cdot|T=1),\mg,\lambda)$ corresponding to a binary $U$ such that $\int \mg(u)\lambda(u)\ud Q_U(u|T=1)=\int \mg'(u)\lambda'(u)\ud Q_{U'}(u|T=1)$. Moreover, $Q_U(0|T=1)=1-Q_U(1|T=1)=\tau$ and
\begin{align}\label{eq:lambda-binary}
    \lambda(u)=\begin{cases}
        \Lambda_2&\textit{if}\quad u=1\\
        \Lambda_1&\textit{if}\quad u=0\\
    \end{cases}.
\end{align}
\end{lem}

Hereafter, we assume that $(Q_U(\cdot|T=1),\mg,\lambda)$ is as in Lemma \ref{lm:BinaryU}. Then
the objective function can be evaluated as
\begin{align}
    \int \mg(u) \lambda (u)\ud Q_U(u|T=1)&=\mg(1)\Lambda_2(1-\tau)+\mg(0)\Lambda_1\tau\\
    &=\mg(1)\Lambda_2(1-\tau)+\left\{E(Y|T=1)-\mg(1)(1-\tau)\right\}\hspace{1.6em} [\text{by \eqref{eq:eta-simp}}]\\
    &=E(Y|T=1)+(1-\Lambda_1)\left\{\mg(1)-E(Y|T=1)\right\},\label{eq:nu1-e-simp}
\end{align}
which is maximized over $\mg(1)$ under constraints \eqref{eq:emsm-simp} and \eqref{eq:eta-simp}. With $\mg(0)=\{E(Y|T=1)-\mg(1)(1-\tau)\}/\tau$ and $\mg(1)$, \eqref{eq:eta-simp} is automatically satisfied. Then requiring both $\mg(0)$ and $\mg(1)$ to satisfy \eqref{eq:emsm-simp} is equivalent to the condition:
\begin{align}\label{eq:nu-binary-bounds}
 \max\left\{-\Delta_1,-\frac{\tau}{1-\tau}\Delta_2\right\}   \le \mg(1)-E(Y|T=1)\le\min\left\{\frac{\tau}{1-\tau}\Delta_1,\Delta_2\right\}.
\end{align}
Substitute $\mg(1)$ in \eqref{eq:nu1-e-simp} with its upper bound in \eqref{eq:nu-binary-bounds},
$$\nu^{1+}_{\mytext{e}}= E(Y|T=1) +(\Lambda_2-\Lambda_1) \min\big\{ \tau \Delta_1, (1-\tau) \Delta_2  \big\},$$
which solves optimization \eqref{eq:V_e-simp}.

\noindent\textbf{Solution for lower bound $\nu_e^{0-}$}

The bound $\nu_e^{1-}$ defined as the infimum of the objective function in \eqref{eq:V_e-simp} under the same set of constrains \eqref{eq:msm-simp}--\eqref{eq:eta-simp}, can be equivalently expressed as
\begin{align}\label{eq:V_e-low}
-\nu^{1-}_{\mytext{e}}  =\sup_{Q_U(\cdot|T=1),-\mg,\lambda}\;
\int -\mg(u) \lambda (u)\ud Q_U(u|T=1),
\end{align}
where the two measurable functions $\lambda$ (nonnegative)
and $-\mg$ of $U|T=1$ are constrained by
\begin{align}
&\lambda(U) \in [\Lambda_1,\Lambda_2],\\
    &-\mg(U) -\{- E(Y|T=1)\}\in[-\Delta_2, \Delta_1],\\
    &\int\lambda(u)\ud Q_U(u|T=1)=1,\\
    &\int -\mg(u) \ud Q_U(u|T=1) = -E (Y|T=1).
\end{align}
From the discussion of $\nu^{1+}_{\mytext{e}}$, replacing $\mg(1)$ in \eqref{eq:nu1-e-simp} with its lower bound in \eqref{eq:nu-binary-bounds}
gives $\nu^{1-}_{\mytext{e}}= E(Y|T=1) -(\Lambda_2-\Lambda_1) \min\big\{ (1-\tau) \Delta_1,\tau \Delta_2  \big\}$.

To bound $\nu^0=E(Y^0|T=1)$ under the partial eMSM specified by \eqref{eq:msm} and \eqref{eq:eMSM_Y0}, we define a similar relaxed upper bound for $\nu^0$ via optimization:
\begin{align}\label{eq:V_e-0-low}
\nu^{0+}_{\mytext{e}}  =\sup_{Q_U(\cdot|T=0),\mg,\lambda}\;
\int \mg(u) \lambda (u)\ud Q_U(u|T=0),
\end{align}
where the two measurable functions $\lambda$ (nonnegative)
and $\mg$ of $U|T=0$ are constrained by
\begin{align}
&\lambda(U) \in [\Lambda_2^{-1},\Lambda_1^{-1}],\\
    &\mg(U) -E(Y|T=1)\in[-\Delta_1', \Delta_2'],\\
    &\int\lambda(u)\ud Q_U(u|T=0)=1,\\
    &\int \mg(u) \ud Q_U(u|T=0) = E (Y|T=0).
\end{align}
From the discussion of $\nu^{1+}_{\mytext{e}}$,
$$\nu^{0+}_{\mytext{e}}= E(Y|T=0) +(\Lambda_1^{-1}-\Lambda_2^{-1}) \min\big\{ \tau' \Delta_1', (1-\tau') \Delta_2'  \big\},$$
where $\tau'=(\Lambda_1^{-1}-1)/(\Lambda_1^{-1}-\Lambda_2^{-1})=(\Lambda_2-\Lambda_1\Lambda_2)/(\Lambda_2-\Lambda_1)$. Similarly, $$\nu^{0-}_{\mytext{e}}= E(Y|T=0) -(\Lambda_1^{-1}-\Lambda_2^{-1}) \min\big\{  (1-\tau') \Delta_1',\tau' \Delta_2'  \big\}.$$

\noindent\textbf{Simultaneous sharpness under C-eMSM}

By definition, \eqref{eq:nu-1-up-min} and \eqref{eq:nu-1-low-max} below are respectively valid upper bound and lower bound for $\nu^1$ under the partial eMSM specified \eqref{eq:msm} and \eqref{eq:UYconstr}.
\begin{align}
   &\min \left\{\nu^{1+}_{\mytext{MSM}}, \nu^{1+}_{\mytext{e}}\right\}=E(Y|T=1)+\\
   &\hspace{1.5in}(\Lambda_2-\Lambda_1)\min\big[ \tau \Delta_1, (1-\tau) \Delta_2,E\{\rho_\tau(Y,q^*_{1,\tau})|T=1\} \big],\label{eq:nu-1-up-min}\\
   &\max \left\{\nu^{1-}_{\mytext{MSM}}, \nu^{1-}_{\mytext{e}}\right\}=E(Y|T=1)-\\
   &\hspace{1.5in}(\Lambda_2-\Lambda_1)\min\big[ (1-\tau) \Delta_1,\tau \Delta_2, E\{\rho_{1-\tau}(Y,q^*_{1,1-\tau})|T=1\} \big].\label{eq:nu-1-low-max}
\end{align}
Similarly, \eqref{eq:nu-0-up-min} and \eqref{eq:nu-0-low-max} below are respectively valid upper bound and lower bound for $\nu^0$ under the partial eMSM specified by \eqref{eq:msm} and \eqref{eq:eMSM_Y0}.
\begin{align}
   &\min \left\{\nu^{0+}_{\mytext{MSM}}, \nu^{0+}_{\mytext{e}}\right\}=E(Y|T=0) +\\
   &\hspace{1.5in}(\Lambda_1^{-1}-\Lambda_2^{-1}) \min\big\{ \tau' \Delta_1', (1-\tau') \Delta_2',E\{\rho_{\tau'}(Y,q^*_{0,\tau'})|T=0\} \big],\label{eq:nu-0-up-min}\\
   &\max \left\{\nu^{0-}_{\mytext{MSM}}, \nu^{0-}_{\mytext{e}}\right\}=E(Y|T=0)-\\
   &\hspace{1.5in}(\Lambda_1^{-1}-\Lambda_2^{-1}) \min\big[  (1-\tau') \Delta_1', \tau' \Delta_2',E\{\rho_{1-\tau'}(Y,q^*_{0,1-\tau'})|T=0\} \big].\label{eq:nu-0-low-max}
\end{align}

We construct a $Q$ on $(Y^0,Y^1,T,U)|X$, with $Q_X=P_X$ implicitly assumed, so that
\begin{itemize}\addtolength{\itemsep}{-.1in}
    \item[(i)] $Q$ is compatible with the observed-data distribution with $Y=TY^1+(1-T)Y^0$;
    \item[(ii)] $Q$ satisfies the C-eMSM constraints \eqref{eq:msm}, \eqref{eq:UYconstr} and \eqref{eq:eMSM_Y0};
    \item[(iii)] $Q$ attains both \eqref{eq:nu-1-up-min} and \eqref{eq:nu-0-low-max}.
\end{itemize}
To facilitate the construction, we first define an auxiliary variable $U$ conditionally on $T$ as follows:
\begin{align}
   U|(T=1)=&\mathbf {1}(Y> q^*_{1,\tau}|T=1)\times Bernoulli\left(1-\tau+\tau\psi_{1+}\right)+\nonumber\\
    &\mathbf {1}(Y< q^*_{1,\tau}|T=1)\times Bernoulli\left\{1-\tau-(1-\tau)\psi_{1+}\right\}+\nonumber\\
    &\mathbf {1}(Y= q^*_{1,\tau}|T=1)\times Bernoulli\left\{ 1-\tau-(1-\tau)\psi_{1+} +\frac{1-\tau-p_{1,+}}{p_{1,\sim}}\psi_{1+}\right\},\label{eq:U|T=1}\\
    U|(T=0)=&\mathbf {1}(Y> q^*_{0,1-\tau'}|T=0)\times Bernoulli\left\{\tau'+(1-\tau')\psi_{0-}\right\}+\\
    &\mathbf {1}(Y< q^*_{0,1-\tau'}|T=0)\times Bernoulli\left(\tau'-\tau'\psi_{0-}\right)+\\
    &\mathbf {1}(Y= q^*_{0,1-\tau'}|T=0)\times Bernoulli\left\{ \tau'-\tau'\psi_{0-}+\frac{\tau'-p_{0,+}}{p_{0,\sim}}\psi_{0-}\right\}.\label{eq:U|T=0}
\end{align}
Here $p_{1,+}=P(Y> q^*_{1,\tau}|T=1)$, $p_{1,-}=P(Y< q^*_{1,\tau}|T=1)$, $p_{1,\sim}=P(Y= q^*_{1,\tau}|T=1)$, and $\psi_{1+}=\min \{ \delta_{1+}, \delta_{2+}, 1\}$ with
\begin{align*}
 \delta_{1+}  = \frac{\tau \Delta_1}{E\{\rho_{\tau}(Y,q^*_{1,\tau})|T=1\}},\quad
 \delta_{2+} = \frac{(1-\tau)\Delta_2}{E\{\rho_{\tau}(Y,q^*_{1,\tau})|T=1\}}.
\end{align*}
When $p_{1,\sim}=0$, necessarily $1-\tau-p_{1,+}=0$, in which case $(1-\tau-p_{1,+})/p_{1,\sim}$ is set to $0$ in \eqref{eq:U|T=1}. Similarly, 
$p_{0,+}=P(Y> q^*_{0,1-\tau'}|T=0)$, $p_{0,-}=P(Y< q^*_{0,1-\tau'}|T=0)$, $p_{0,\sim}=P(Y= q^*_{0,1-\tau'}|T=0)$, and $\psi_{0-}=\min \{\delta'_{1-}, \delta'_{2-}, 1\}$ with
\begin{align*}
 \delta'_{1-}  = \frac{(1-\tau)' \Delta_1'}{E\{\rho_{1-\tau'}(Y,q^*_{0,1-\tau'})|T=0\}},\quad
 \delta'_{2-} = \frac{\tau'\Delta_2'}{E\{\rho_{1-\tau'}(Y,q^*_{0,1-\tau'})|T=0\}}.
\end{align*}
When $p_{0,\sim}=0$, necessarily $\tau'-p_{0,+}=0$, in which case  $(\tau'-p_{0,+})/p_{0,\sim}$ is set to $0$ in \eqref{eq:U|T=0}.
The independent Bernoulli variables individually randomize the value of their corresponding indicator functions of $Y|T$ with known probabilities based on the observed-data distribution $P_{Y,T}$.

By construction, $U|T=1$ and $U|T=0$ are randomized indicator functions of $Y|T$. Formally, $(Y,T,U)$ are defined on an augmented sample space obtained from the original sample space by incorporating the independent Bernoulli variables.
The joint distribution of $(Y,T,U)$, denoted as $P_{Y,T,U}$, is fully determined from $P_{Y,T}$ and independent Bernoulli variables. We caution that $U$ serves only as an auxiliary variable for the construction of $Q$ and is, in general, not the unmeasurable confounder in the true distribution $P$ under eMSM. 

With $U$ defined above, $Q$ is then constructed step by step as follows:
\begin{subequations}\label{eq:Q-construct}
    \begin{align}
        &Q_{T,U}(\cdot)=P_{T,U}(\cdot), \label{eq:T-U-Q}\\
        &Q_{Y^1}(\cdot|T=0,U=u)=Q_{Y^1}(\cdot|T=1,U=u)=P_Y(\cdot|T=1,U=u),\label{eq:Y1-Q}\\
        &Q_{Y^0}(\cdot|T=1,U=u)=Q_{Y^0}(\cdot|T=0,U=u)=P_Y(\cdot|T=0,U=u),\label{eq:Y0-Q}\\
&Q_{Y^1,Y^0}(\cdot,\cdot|T,U)=Q_{Y^1}(\cdot|T,U)Q_{Y^0}(\cdot|T,U).\label{eq:Y0Y1-indep-Q}
    \end{align}
\end{subequations}
By enforcing $Y=TY^1+(1-T)Y^0$ through the consistency assumption, \eqref{eq:T-U-Q}, \eqref{eq:Y1-Q} and \eqref{eq:Y0-Q} together ensure that $Q_{Y,T}(\cdot,t)=Q_{Y^t,T}(\cdot,t)=P_{Y,T}(\cdot, t)$ for $t=0,1$. This proves claim (i). Moreover, for any two events $A_0$ and $A_1$,
\begin{align}
    &Q_{Y^0,Y^1}(A_0,A_1|T=1,U)=Q_{Y^0}(A_0|T=1,U)Q_{Y^1}(A_1|T=1,U)\quad  \text{[from \eqref{eq:Y0Y1-indep-Q}]}\\
    &=Q_{Y^0}(A_0|T=0,U)Q_{Y^1}(A_1|T=0,U)\quad  \text{[from \eqref{eq:Y1-Q} and \eqref{eq:Y0-Q}]}\\
    &=Q_{Y^0,Y^1}(A_0,A_1|T=0,U),\quad  \text{[from \eqref{eq:Y0Y1-indep-Q}]}
\end{align}
which proves \eqref{eq:msm-indep}, i.e., $(Y^0,Y^1)\perp T|U$, for claim (ii).

Next, we derive relevant quantities to prove claims (ii) and (iii).
From \eqref{eq:U|T=1}, \eqref{eq:U|T=0} and \eqref{eq:T-U-Q},
\begin{align}
    Q_U(1|T=1)&=P_{U}(1|T=1)=p_{1,+}(1-\tau+\tau\psi_{1+})+p_{1,-}\{1-\tau-(1-\tau)\psi_{1+}\}+\\
    &\hspace{1.2em}p_{1,\sim}\left(1-\tau-(1-\tau)\psi_{1+} +\frac{1-\tau-p_{1,+}}{p_{1,\sim}}\psi_{1+}\right)\\
    &=1-\tau+\left\{p_{1,+}\tau-p_{1,-}(1-\tau)-p_{1,\sim}(1-\tau)+1-\tau-p_{1,+}\right\}\psi_{1+}\\
    &=1-\tau,\label{eq:U1-prob}\\
    Q_U(1|T=0)&=P_{U}(1|T=0)=p_{0,+}\{\tau'+(1-\tau')\psi_{0-}\}+p_{0,-}(\tau'-\tau'\psi_{0-}\}+\\
    &\hspace{1.2em}p_{0,\sim}\left(\tau'-\tau'\psi_{0-} +\frac{\tau'-p_{0,+}}{p_{0,\sim}}\psi_{0-}\right)\\
    &=\tau'+\left\{p_{0,+}(1-\tau')-p_{0,-}\tau'-p_{0,\sim}\tau'+\tau'-p_{0,+}\right\}\psi_{0-}\\
    &=\tau'.\label{eq:U0-prob}
\end{align}
Therefore,
\begin{align}\label{eq:lam-Q}
    \lambda_Q(u)=\frac{Q_U(u|T=0)}{Q_U(u|T=1)}=\begin{cases}
        \tau'/(1-\tau)=\Lambda_2\quad &\text{if}\quad u=1\\
        (1-\tau')/\tau=\Lambda_1\quad &\text{if}\quad u=0
    \end{cases}.
\end{align}
Moreover,
\begin{align}
    \ud Q_{Y^1}(y|T,U=1)&\stackrel{\text{\eqref{eq:Y1-Q}}}{=\joinrel=}\ud P_{Y}(y|T=1,U=1)\\
    &=\frac{P_{U}(1|T=1,Y=y)\ud P_Y(y|T=1)}{P_{U}(1|T=1)}\\
    &=\begin{cases}
        \left(1+\frac{\tau}{1-\tau}\psi_{1+}\right)\ud P_Y (y | T=1) &\text{if}\quad y>q^*_{1,\tau} \\
        \left(1-\psi_{1+}\right)\ud P_Y (y | T=1) &\text{if}\quad y<q^*_{1,\tau} \\
   \end{cases},\label{eq:Y|U-1-dis-Q}\\
    \ud Q_{Y^1}(y|T,U=0)&\stackrel{\text{\eqref{eq:Y1-Q}}}{=\joinrel=}\ud P_{Y} (y |T=1,U=0) \\
    & =\frac{P_{U}(0|T=1,Y=y)\ud P_Y(y|T=1)}{P_{U}(0|T=1)} \\&=\begin{cases}
        \left(1-\psi_{1+}\right)\ud P_Y (y | T=1)&\text{if}\quad y>q^*_{1,\tau} \\
        \left(1+\frac{1-\tau}{\tau}\psi_{1+}\right)\ud P_Y (y | T=1) &\text{if}\quad y<q^*_{1,\tau}
   \end{cases},
\end{align}
and if $p_{1,\sim}=P(Y=q^*_{1,\tau}|T=1)>0$,
\begin{align}\label{eq:Y|U-1-point-Q}
&Q_{Y^1}(q^*_{1,\tau}|T,U=1)=P_{Y}(q^*_{1,\tau}|T=1,U=1)= (1-\psi_{1+}) p_{1,\sim} +\left(1-\frac{p_{1,+}}{1-\tau}\right)\psi_{1+},\\
&Q_{Y^1}(q^*_{1,\tau}|T,U=0)=P_{Y} (q^*_{1,\tau} | T=1,U=0)=\left(1-\psi_{1+}\right) p_{1,\sim}+\left(1-\frac{p_{1,-}}{\tau}\right)\psi_{1+}.
\end{align}
Similarly,
\begin{align}
    \ud Q_{Y^0}(y|T,U=1)&\stackrel{\text{\eqref{eq:Y0-Q}}}{=\joinrel=}\begin{cases}
        \left(1+\frac{1-\tau'}{\tau'}\psi_{0-}\right)\ud P_Y (y | T=0) &\text{if}\quad y> q^*_{0,1-\tau'} \\
        \left(1-\psi_{0-}\right)\ud P_Y (y | T=0) &\text{if}\quad y< q^*_{0,1-\tau'} \\
   \end{cases},\label{eq:Y|U-0-dis-Q}\\
    \ud Q_{Y^0}(y|T,U=0)&\stackrel{\text{\eqref{eq:Y0-Q}}}{=\joinrel=}\begin{cases}
        \left(1-\psi_{0-}\right)\ud P_Y (y | T=0)&\text{if}\quad y> q^*_{0,1-\tau'} \\
        \left(1+\frac{\tau'}{1-\tau'}\psi_{0-}\right)\ud P_Y (y | T=0) &\text{if}\quad y< q^*_{0,1-\tau'}
   \end{cases},
\end{align}
and if $p_{0,\sim}=P(Y= q^*_{0,1-\tau'}|T=0)>0$,
\begin{align}
&Q_{Y^)}(q^*_{0,1-\tau'}|T,U=1)=P_{Y}(q^*_{0,1-\tau'}|T=0,U=1)= (1-\psi_{0-}) p_{0,\sim} +\left(1-\frac{p_{0,+}}{\tau'}\right)\psi_{0-},\label{eq:Y|U-0-point-Q}\\
&Q_{Y^)}(q^*_{0,1-\tau'}|T,U=0)=P_{Y} (q^*_{0,1-\tau'} | T=0,U=0)=\left(1+\frac{\tau'}{1-\tau'}\psi_{0-}\right) p_{1,\sim}-\left(\frac{1-p_{0,+}}{1-\tau'}-1\right)\psi_{0-}.\nonumber
\end{align}
By \eqref{eq:Y|U-1-dis-Q} and \eqref{eq:Y|U-1-point-Q},
\begin{align}
    &\mg_Q(1)=E_Q(Y^1|T,U=1)=E(Y|T=1,U=1)\\
    &=\left(1-\psi_{1+}\right)\left[p_{1,-}\cdot q^*_{1,\tau}-E\left\{(q^*_{1,\tau}-Y)_+|T=1\right\}\right]+\left\{(1-\psi_{1+}) p_{1,\sim} +\left(1-\frac{p_{1,+}}{1-\tau}\right)\psi_{1+}\right\}q^*_{1,\tau}\\
    &\hspace{2em}+\left(1+\frac{\tau}{1-\tau}\psi_{1+}\right)\left[E\left\{(Y-q^*_{1,\tau})_+|T=1\right\}+p_{1,+}\cdot q^*_{1,\tau}\right]\\
    &=E(Y|T=1)+\frac{\psi_{1+}E\left\{\rho_\tau(Y,q^*_{1,\tau})|T=1\right\}}{1-\tau}-\psi_{1+} q^*_{1,\tau}\left(p_{1,+}+p_{1,-}+p_{1,\sim}-1\right)\\
    &=E(Y|T=1)+\frac{\min\big[ \tau \Delta_1, (1-\tau) \Delta_2,E\{\rho_\tau(Y,q^*_{1,\tau})|T=1\} \big]}{1-\tau},\label{eq:eta-1-Q}
\end{align}
It follows from the law of iterated expectations that
\begin{align}
    &\mg_Q(0)=E_Q(Y^1|T,U=0)=E(Y|T=1,U=0)\\
    &=\frac{E(Y|T=1)-E(Y|T=1,U=1)P_{U}(1|T=1)}{P_{U}(0|T=1)}\\
    &=E(Y|T=1)-\frac{\min\big[ \tau \Delta_1, (1-\tau) \Delta_2,E\{\rho_\tau(Y,q^*_{1,\tau})|T=1\} \big]}{\tau}.\label{eq:eta-0-Q}
\end{align}
Similarly,
\begin{align}
    &E_Q(Y^0|T,U=1)
    =E(Y|T=0)+\frac{\min\big[  (1-\tau') \Delta_1', \tau' \Delta_2',E\{\rho_{1-\tau'}(Y,q^*_{0,1-\tau'})|T=0\} \big]}{\tau'},\label{eq:EY0-U1-Q}
\end{align}
and
\begin{align}
    E_Q(Y^0|T,U=0)=E(Y|T=0)-\frac{\min\big[  (1-\tau') \Delta_1', \tau' \Delta_2',E\{\rho_{1-\tau'}(Y,q^*_{0,1-\tau'})|T=0\} \big]}{1-\tau'}.\label{eq:EY0-U0-Q}
\end{align}

Finally, we are ready to complete the proof.
As shown earlier, $Q$ satisfies \eqref{eq:msm-indep} by construction. From \eqref{eq:lam-Q}, \eqref{eq:msm-bound} is satisfied. \eqref{eq:UYconstr} follows from \eqref{eq:eta-1-Q} and \eqref{eq:eta-0-Q}, while \eqref{eq:eMSM_Y0} follows from \eqref{eq:EY0-U1-Q} and \eqref{eq:EY0-U0-Q}. This completes the proof of claim (ii).

From \eqref{eq:U0-prob}, \eqref{eq:eta-1-Q} and \eqref{eq:eta-0-Q}, we have
\begin{align}
    &\nu_Q^1=E_Q(Y^1|T=0)=\mg_Q(1)\tau'+\mg_Q(0)(1-\tau')=\mg_Q(1)(1-\tau)\Lambda_2+\mg_Q(0)\tau\Lambda_1\\
    &=E(Y|T=1)+(\Lambda_2-\Lambda_1)\min\big[ \tau \Delta_1, (1-\tau) \Delta_2,E\{\rho_\tau(Y,q^*_{1,\tau})|T=1\} \big]\equiv\eqref{eq:nu-1-up-min}.
\end{align}
From \eqref{eq:U1-prob}, \eqref{eq:EY0-U1-Q} and \eqref{eq:EY0-U0-Q},
\begin{align}
    &\nu_Q^0
    &=E(Y|T=0)-(\Lambda_1^{-1}-\Lambda_2^{-1})\min\big[  (1-\tau') \Delta_1', \tau' \Delta_2',E\{\rho_{1-\tau'}(Y,q^*_{0,1-\tau'})|T=0\} \big]\equiv\eqref{eq:nu-0-low-max}.
\end{align}
This completes the proof of claim (iii) and consequently Proposition \ref{pro:pop-sol}.

Proposition \ref{pro:pop-sol-lower} follows from the proof for $\nu_Q^0$ by flipping the label of $T$, which leads to $\lambda_Q$ being replaced with its reciprocal and $(\Lambda_1,\Lambda_2)$ replaced with $(\Lambda_2^{-1},\Lambda_1^{-1})$.

Corollary \ref{cor:eMSM-Q} follows from the construction of $Q$.
Note that the construction of $Q$ does not seem to be unique when $\psi_{1+}$ or $\psi_{0-}$ is strictly less than 1. For example, the construction of $U|T$ may be modified by varying the Bernoulli probabilities in \eqref{eq:U|T=1} or \eqref{eq:U|T=0} along with the quantile points used to partition $P_{Y}(\cdot|T)$.

\subsection{Proof of Proposition \ref{prop:propspecbd}}\label{sec:proof-propspecbd}
The main text only covers the recommended specification for $\tau\ge1/2$, which is
\begin{align}\label{eq:recspec-tau-ge.5}
\begin{split}
& \Delta_2(X)= \delta /(1-\tau) \cdot E\{ \rho_{\tau}(Y,q^*_{1,\tau})|T=1,X\}, \\
& \Delta_1(X)= \delta /(1-\tau) \cdot E\{ \rho_{1-\tau}(Y,q^*_{1,1-\tau})|T=1,X\},
\end{split}
\end{align}
where $\delta\in[0,1]$.
The recommended specification for $\tau<1/2$ is
\begin{align}\label{eq:recspec-tau-l.5}
\begin{split}
& \Delta_1(X)= \delta /\tau \cdot E\{ \rho_{\tau}(Y,q^*_{1,\tau})|T=1,X\}, \\
& \Delta_2(X)= \delta /\tau \cdot E\{ \rho_{1-\tau}(Y,q^*_{1,1-\tau})|T=1,X\}.
\end{split}
\end{align}

By Propositions \ref{pro:pop-sol} and \ref{pro:pop-sol-lower},
\begin{align}
     \nu^{1+}_{\mytext{eMSM}}(X) &=E (Y|T=1,X) +(\Lambda_2-\Lambda_1)\psi_{1+} E\left\{\rho_{\tau}(Y,q^*_{1,\tau})|T=1,X\right\}\label{eq:nu-1-up-psi}\\
     \nu^{1-}_{\mytext{eMSM}}(X) &=E (Y|T=1,X) -(\Lambda_2-\Lambda_1)\psi_{1-} E\left\{\rho_{1-\tau}(Y,q^*_{1,1-\tau})|T=1,X\right\}\label{eq:nu-1-low-psi}
\end{align}
where $\psi_{1+}(X)=\min \{ \delta_{1+}(X), \delta_{2+}(X), 1\}$ with
\begin{align*}
 \delta_{1+} (X) = \frac{\tau(X) \Delta_1(X)}{E\{\rho_{\tau}(Y,q^*_{1,\tau})|T=1,X\}},\quad
 \delta_{2+}(X) = \frac{\{1-\tau(X)\}\Delta_2(X)}{E\{\rho_{\tau}(Y,q^*_{1,\tau})|T=1,X\}},
\end{align*}
and
$\psi_{1-}(X)=\min \{\delta_{1-}(X), \delta_{2-}(X), 1\}$ with
\begin{align*}
\delta_{1-}(X) = \frac{\{1-\tau(X)\}\Delta_1(X)}{E\{\rho_{1-\tau}(Y,q^*_{1,1-\tau})|T=1,X\}} , \quad
\delta_{2-}(X) = \frac{\tau(X) \Delta_2(X)}{E\{\rho_{1-\tau}(Y,q^*_{1,1-\tau})|T=1,X\}}.
\end{align*}

It suffices to show that $\psi_{1+}(X)=\psi_{1-}(X)=\delta$ in \eqref{eq:nu-1-up-psi} and \eqref{eq:nu-1-low-psi} under the eMSM with recommended specification \eqref{eq:recspec-tau-ge.5} if $\tau\ge1/2$, or \eqref{eq:recspec-tau-l.5} if $\tau<1/2$.

\noindent\textbf{Case $\tau\ge1/2$}

For $(\Delta_1,\Delta_2)$-specification \eqref{eq:recspec-tau-ge.5}, $ \delta_{2+}(X)=\delta_{1-}(X)\equiv\delta$, and
\begin{align}
    \delta_{1+}(X)= \delta\frac{\tau}{1-\tau}\frac{E\{\rho_{1-\tau}(Y,q^*_{1,1-\tau})|T=1,X\}}{ E\{\rho_\tau(Y,q^*_{1,\tau})|T=1,X\}},\quad\delta_{2-}(X)= \delta\frac{\tau}{1-\tau}\frac{E\{\rho_\tau(Y,q^*_{1,\tau})|T=1,X\}}{ E\{\rho_{1-\tau}(Y,q^*_{1,1-\tau})|T=1,X\}}.
\end{align}
To show that $\psi_{1+}(X)=\psi_{1-}(X)=\delta$ in \eqref{eq:nu-1-up-psi} and \eqref{eq:nu-1-low-psi}, it suffices to prove
\begin{align}\label{eq:rhoratio}
     \frac{1-\tau}{\tau}\le\frac{E\{\rho_\tau(Y,q^*_{1,\tau})|T=1,X\}}{ E\{\rho_{1-\tau}(Y,q^*_{1,1-\tau})|T=1,X\}}\le\frac{\tau}{1-\tau}.
 \end{align}
For a covariate function $q$, let $A(q)=E\{ (Y-q)_{+}|T=1,X\}$ and $B(q)= E\{(q-Y)_{+}|T=1,X\}$. As $\tau\ge1/2\ge1-\tau$, we have
\begin{align*}
    E\{\rho_{\tau}(Y,q^*_{1,\tau})|T=1,X\}\le E\{\rho_{\tau}(Y,q)|T=1,X\}
    &=\tau A(q)+(1-\tau) B(q)\\&\le \frac{\tau}{1-\tau}\left\{(1-\tau) A(q)+\tau B(q)\right\}\\
    &=\frac{\tau}{1-\tau}E\{\rho_{1-\tau}(Y,q)|T=1,X\}.
\end{align*}
Setting $q=q^*_{1,1-\tau}$ gives the upper bound in \eqref{eq:rhoratio}.
Similarly,
\begin{align*}
    E\{\rho_{\tau}(Y,q)|T=1,X\}=\tau A(q)+(1-\tau) B(q)&\ge \frac{1-\tau}{\tau}\left\{(1-\tau) A(q)+\tau B(q)\right\}\\
    &=\frac{1-\tau}{\tau}E\{\rho_{1-\tau}(Y,q)|T=1,X\}\\
    &\ge \frac{1-\tau}{\tau}E\{\rho_{1-\tau}(Y,q^*_{1,1-\tau})|T=1,X\}.
\end{align*}
Setting $q=q^*_{1,\tau}$ gives the lower bound in \eqref{eq:rhoratio}.

\noindent\textbf{Case $\tau<1/2$}

For $(\Delta_1,\Delta_2)$-specification \eqref{eq:recspec-tau-l.5}, $ \delta_{1+}(X)=\delta_{2-}(X)\equiv\delta$, and
\begin{equation*}
    \delta_{2+}(X)= \delta\frac{1-\tau}{\tau}\frac{E\{\rho_{1-\tau}(Y,q^*_{1,1-\tau})|T=1,X\}}{ E\{\rho_\tau(Y,q^*_{1,\tau})|T=1,X\}}, \quad \delta_{1-}(X)= \delta\frac{1-\tau}{\tau}\frac{E\{\rho_\tau(Y,q^*_{1,\tau})|T=1,X\}}{ E\{\rho_{1-\tau}(Y,q^*_{1,1-\tau})|T=1,X\}}.
\end{equation*}
It suffices to show that \begin{equation}\label{eq:rhoratio2}
     \frac{\tau}{1-\tau}\le\frac{E\{\rho_\tau(Y,q^*_{1,\tau})|T=1,X\}}{ E\{\rho_{1-\tau}(Y,q^*_{1,1-\tau})|T=1,X\}}\le\frac{1-\tau}{\tau}.
 \end{equation}
Since $\tau<1/2<1-\tau$, the inequality follows from \eqref{eq:rhoratio} by switching $\tau$ and $1-\tau$ in the bounds.

\subsection{Proof of Proposition \ref{prop:doublyrobust}}\label{sec:proof-prop-dbrb}
 \noindent\textbf{Double Robustness}

 Recall that
\begin{align}
    \varphi_{1+} (\pi,q, m_+ )&=\frac{T}{\pi(X)}Y+T\frac{1-\pi(X)}{\pi(X)}(\Lambda_2-\Lambda_1)\delta\rho_{\tau}(Y,q)
    -\left(\frac{T}{\pi(X)}-1\right)  m_+(X;q)\label{eq:dbpropensity}\\
    &=TY+(1-T)m_+(X;q)+T\frac{1-\pi(X)}{\pi(X)}\left\{\tilde{Y}(q,X)-m_+(X;q)\right\}\label{eq:dbmean},
\end{align}
where $\tilde{Y}_+(q)=Y+(\Lambda_2-\Lambda_1)\delta\rho_\tau(Y,q)$.

If $\pi=\pi^*$,
\begin{align}
   E \left\{  -\left(\frac{T}{\pi(X)}-1\right)  m_+(X;q)\right\} =  E \left\{  -E\left(\frac{T}{\pi^*(X)}-1\bigg|X\right)  m_+(X;q)\right\}=0.
\end{align}
By \eqref{eq:dbpropensity}, $E \{ \varphi_{1+}(\pi,q,m_+) \} =\mu^{1+}_{\mytext{eMSM}} (q)$.

If $m_+(\cdot;q)= m_+^*(\cdot;q)=E\{\tilde{Y}_+(q)|T=1,X\}$,
\begin{align}
   &E\left\{T\frac{1-\pi(X)}{\pi(X)}\left[\tilde{Y}(q,X)-E\{\tilde{Y}_+(q)|T=1,X\}\right] \right\}\\
   &=  E \left\{\frac{1-\pi(X)}{\pi(X)}E\left[\tilde{Y}(q,X)-E\{\tilde{Y}_+(q)|T=1,X\}\bigg|T=1,X\right] \right\}=0.
\end{align}
By \eqref{eq:dbmean}, $E \{ \varphi_{1+}(\pi,q,m_+) \}=\mu^{1+}_{\mytext{eMSM}} (q)$.

\noindent\textbf{Minimization at true quantile}

The minimization of  $\mu^{1+}_{\mytext{eMSM}} (q)$ at $q^*_{1,\tau}$ follows from that of the conditional quantile loss $E\{\rho_\tau(Y,q)|T=1,X\}$ at $q^*_{1,\tau}(X)$. See \citeappend{koenker1978regression} for details.

\subsection{Proof of Proposition \ref{prop:DVcompare}}\label{sec:proof-dv}

We state a stronger result that includes the simultaneous sharp bounds of $(\mu^0,\mu^1)$ and prove the stronger result instead. Assume, as in the main text, that all statements are implicitly conditional on $X$, and extend the following notations:
\begin{itemize}\setlength\itemsep{-.5em}
\item Denote as $\rm{eMSM}(\underline\Delta)$ the C-eMSM specified by \eqref{eq:msm}, \eqref{eq:UYconstr} and \eqref{eq:eMSM_Y0} with outcome sensitivity parameters $\underline \Delta=(\Delta_1,\Delta_2,\Delta_1',\Delta_2')$.
\item Denote as $\mu^{t+}_{\mytext{eMSM}} (\underline \Delta)$ and
$\mu^{t-}_{\mytext{eMSM}} (\underline\Delta)$ the C-eMSM sharp bounds of $\mu^t$ for $t=0,1$.
\item Denote as $\mu^{1+}_{\mytext{eMSM}} (\Delta_1,\Delta_2)$, $\mu^{1-}_{\mytext{eMSM}} (\Delta_1,\Delta_2)$ the sharp bounds of $\mu^1$ under partial eMSM specified by \eqref{eq:msm} and \eqref{eq:UYconstr}; and $\mu^{0+}_{\mytext{eMSM}} (\Delta_1',\Delta_2')$, $\mu^{0-}_{\mytext{eMSM}} (\Delta_1',\Delta_2')$ the sharp bounds of $\mu^0$ under partial eMSM specified by \eqref{eq:msm} and \eqref{eq:eMSM_Y0}.
\item Denote as $\rm{DV}(\Theta)$ the DV model specified by \eqref{eq:UT0}, \eqref{eq:UT1}, \eqref{eq:UY0} and \eqref{eq:UY1}, with $0<\Lambda_1\le1\le\Lambda_2$ common among models and made implicit.
\item Denote as $\mu^{t+}_{\mytext{DV}}(\Theta)$ the sharp upper bound of $\mu^t$ under $\rm{DV}(\Theta)$  and $\mu^{t-}_{\mytext{DV}}(\Theta)$ the sharp lower bound of $\mu^t$ under $\rm{DV}(\Theta)$, where $t=0,1$.
\item $\mathcal{D} (\Theta) = \{\underline \Delta:\theta_1 (\Delta_{1},\Delta_{2})\le \Theta, \theta_0 (\Delta_{1}',\Delta_{2}')\le \Theta \}$,
where $\theta_t (x,y)=\min\{p_t +y,1\}/\max\{p_t -x,0\}$ for $t=0,1$.
\end{itemize}

\begin{pro}[Sharp bounds for DV model]\label{prop:dvsharp}
For the model $\rm{DV}(\Theta)$, the following results hold.

(i) $\bigcup_{\underline \Delta\in\mathcal{D}(\Theta)}\rm{eMSM}(\underline\Delta)=\rm{DV}(\Theta)$;

(ii) The sharp bounds of $\mu^1$ and $\mu^0$ under $\rm{DV}(\Theta)$ are determined as
\begin{align*}
    &\mu^{1+}_{\mytext{DV}}(\Theta)=\max_{\underline\Delta\in\mathcal{D}(\Theta)} \; \mu^{1+}_{\mytext{eMSM}} (\underline\Delta)=\mu^{1+}_{\mytext{eMSM}} (\underline{\Delta}^+),\\
    &\mu^{0-}_{\mytext{DV}}(\Theta)=\min_{\underline\Delta\in\mathcal{D}(\Theta)} \; \mu^{0-}_{\mytext{eMSM}}(\underline\Delta)=\mu^{0-}_{\mytext{eMSM}} (\underline{\Delta}^+),
\end{align*}
where $\underline{\Delta}^+\in\D(\Theta)$ is given by
\begin{align}
\hspace{-1em}    \Delta_{1}=\frac{(\Theta-1)p_1}{\odds(\tau)+ \Theta}, \Delta_{2}=\frac{(\Theta-1)p_1}{1+ \odds(1-\tau)\Theta}, \Delta_{1}'=\frac{(\Theta-1)p_0}{\odds(1-\tau')+ \Theta}, \Delta_{2}'=\frac{(\Theta-1)p_0}{1+ \odds(\tau')\Theta}, \label{eq:Delta+}
\end{align}
with $\tau'=(\Lambda_1^{-1}-1)/(\Lambda_1^{-1}-\Lambda_2^{-1})$ and $\odds(x)=x/(1-x)$. Moreover, $\mu^{1+}_{\mytext{DV}}(\Theta)$ and $\mu^{0-}_{\mytext{DV}}(\Theta)$ are simultaneously sharp, i.e., they can be obtained by a single distribution under $\rm{DV}(\Theta)$. Similarly,
$\mu^{1-}_{\mytext{DV}}(\Theta)=\mu^{1-}_{\mytext{eMSM}} (\underline{\Delta}^-)$  and $\mu^{0+}_{\mytext{DV}}(\Theta)=\mu^{0+}_{\mytext{eMSM}} (\underline{\Delta}^-)$ are simultaneously sharp, where $\underline{\Delta}^-\in\D(\Theta)$ is given by
\begin{align}
 \hspace{-1em}   \Delta_{1}=\frac{(\Theta-1)p_1}{\odds(1-\tau)+ \Theta}, \Delta_{2}=\frac{(\Theta-1)p_1}{1+ \odds(\tau)\Theta}, \Delta_{1}'=\frac{(\Theta-1)p_0}{\odds(\tau')+ \Theta}, \Delta_{2}'=\frac{(\Theta-1)p_0}{1+ \odds(1-\tau')\Theta}.\label{eq:Delta-}
\end{align}

(iii) The sharp bounds of $\mu^1$ and $\mu^0$ in (ii) can be expressed as
\begin{align}
\mu^{1+}_{\mytext{DV}} (\Theta) &=p_1+P(T=0)(\Lambda_2-\Lambda_1)\min\left[\frac{\tau(\Theta-1)p_1}{\odds(\tau)+\Theta},E\{\rho_\tau(Y,q^*_{1,\tau})|T=1\}\right],\label{eq:dvsharp-mu1+}\\
    \mu^{1-}_{\mytext{DV}} (\Theta) &=p_1-P(T=0)(\Lambda_2-\Lambda_1)\min\left[\frac{(1-\tau)(\Theta-1)p_1}{\odds(1-\tau)+\Theta},E\{\rho_{1-\tau}(Y,q^*_{1,1-\tau})|T=1\}\right],\label{eq:dvsharp-mu1-}\\
    \mu^{0+}_{\mytext{DV}} (\Theta) &= p_0+P(T=1)(\Lambda_1^{-1}-\Lambda_2^{-1})\min\left[\frac{\tau'(\Theta-1)p_0}{\odds(\tau')+\Theta},E\{\rho_{\tau'}(Y,q^*_{0,\tau'})|T=0\}\right],\label{eq:dvsharp-mu0+}\\
    \mu^{0-}_{\mytext{DV}} (\Theta) &= p_0-P(T=1)(\Lambda_1^{-1}-\Lambda_2^{-1})\min\left[\frac{(1-\tau')(\Theta-1)p_0}{\odds(1-\tau')+\Theta},E\{\rho_{1-\tau'}(Y,q^*_{0,1-\tau'})|T=0\}\right].\label{eq:dvsharp-mu0-}
\end{align}
\end{pro}

\noindent\textit{Proof of Proposition \ref{prop:dvsharp} (i)}:
For any $Q\in\rm{eMSM}(\underline\Delta)$ such that $\underline\Delta\in\D(\Theta)$, we have
\begin{align}\label{eq:UYQ}
  \RR_{\mytext{UY,t}}(Q)= \frac{max_{u}E_Q(Y|U=u,T=t)}{\min_{u}E_Q(Y|U=u,T=t)}\le  \max\{\theta_1(\Delta_{1},\Delta_{2}),\theta_0(\Delta_{1}',\Delta_{2}')\}\le\Theta.
\end{align}
Therefore,
$\bigcup_{\underline \Delta\in\mathcal{D}(\Theta)}\rm{eMSM}(\underline\Delta)\subset\rm{DV}(\Theta)$.

Conversely, for $Q\in\rm{DV}(\Theta)$, $Q$
is, by definition, an element of $\rm{eMSM}(\underline\Delta^Q)$ with $\underline\Delta^Q$ given by
\begin{align}
    &\Delta_{1}=\max_{u}p_1-E_Q(Y^1|U=u,T=1),\quad\Delta_{2}=\max_{u}E_Q(Y^1|U=u,T=1)-p_1,\label{eq:}\\
    &\Delta_{1}'=\max_{u}p_0-E_Q(Y|U=u,T=0)),\quad \Delta_{2}'=\max_{u}E_Q(Y^1|U=u,T=0)-p_0.
\end{align}
By construction, $\underline\Delta^Q\in\D(\Theta)$ and $\rm{DV}(\Theta)\subset\bigcup_{\underline \Delta\in\mathcal{D}(\Theta)}\rm{eMSM}(\underline\Delta)$.

\noindent\textit{Proof of Proposition \ref{prop:dvsharp} (ii)}:
Consider $\mu^{1+}_{\mytext{DV}}(\Theta)$ and $\mu^{0-}_{\mytext{DV}}(\Theta)$.
By Proposition \ref{prop:dvsharp} (i),
\begin{align}
    &\mu^{1+}_{\mytext{DV}}(\Theta)=\max_{\underline\Delta\in\mathcal{D}(\Theta)}\;\mu^{1+}_{\mytext{eMSM}} (\underline\Delta)=\max_{(\Delta_1,\Delta_2):\theta_1(\Delta_1,\Delta_2)\le\Theta}\mu^{1+}_{\mytext{eMSM}} (\Delta_1,\Delta_2),\label{eq:C-to-P1+}\\
    &\mu^{0-}_{\mytext{DV}}(\Theta)=\min_{\underline\Delta\in\mathcal{D}(\Theta)} \; \mu^{0-}_{\mytext{eMSM}}(\underline\Delta)=\min_{(\Delta_1',\Delta_2'):\theta_0(\Delta_1',\Delta_2')\le\Theta}\mu^{0-}_{\mytext{eMSM}} (\Delta_1',\Delta_2'),\label{eq:C-to-P0-}
\end{align}
where the second equality in \eqref{eq:C-to-P1+} and that in \eqref{eq:C-to-P0-} follow from our discussion in Section \ref{sec:additionalbounds}, that C-eMSM shares the same sharp bounds of $\mu^1$ with the partial eMSM for $Y^1$ and the same sharp bound of $\mu^0$ with the partial eMSM for $Y^0$. As a result,
combining $(\Delta_1,\Delta_2)$ of the maximizing partial eMSM for $\mu^{1+}$ and $(\Delta_1',\Delta_2')$ of the minimizing partial eMSM for $\mu^{0-}$
leads to an C-eMSM that attains both $\mu^{1+}_{\mytext{DV}}(\Theta)$ and $\mu^{0-}_{\mytext{DV}}(\Theta)$. 

\noindent\textbf{Simultaneous sharpness}

With the C-eMSM that attains both $\mu^{1+}_{\mytext{DV}}(\Theta)$ and $\mu^{0-}_{\mytext{DV}}(\Theta)$, the simultaneous sharpness of DV bounds is an immediate consequence of that of C-eMSM bounds. The proofs for $\mu^{1-}_{\mytext{DV}}(\Theta)$ and $\mu^{0+}_{\mytext{DV}}(\Theta)$ are similar.

\noindent\textbf{Optimal $(\Delta_1,\Delta_2)$ for sharp upper bound of $\mu^1$}

By Proposition \ref{pro:pop-sol}, optimal $(\Delta_1,\Delta_2)$ in \eqref{eq:C-to-P1+} can be obtained by solving the following optimization
\begin{subequations}\label{eq:dvsharpopt}
\begin{align}
&\hspace{1.2in}\arg\max_{(\delta_{1+},\delta_{2+})}\min \{ \delta_{1+}, \delta_{2+}, 1\}\label{eq:dvobj}\\
&\text{subject to}\quad \frac{\min\{\delta_{2+}/(1-\tau)E(\rho_\tau(Y,q^*_{1,\tau})|T=1)+p_1,1\}}{\max\{-\delta_{1+}/\tau E(\rho_\tau(Y,q^*_{1,\tau})|T=1)+p_1,0\}}\le\Theta,\label{eq:frac-constr}
\end{align}
\end{subequations}
and then setting
\begin{align*}
\Delta_1 = \frac{\delta_{1+}E\{\rho_{\tau}(Y,q^*_{1,\tau})|T=1\} }{\tau},\quad
 \Delta_2 = \frac{\delta_{2+}E\{\rho_{\tau}(Y,q^*_{1,\tau})|T=1\}}{1-\tau}.
\end{align*}

Since $E\{\rho_\tau(Y,q^*_{1,\tau})|T=1\}=\min\{(1-\tau)(1-p_1),\tau p_1\}$, the first term in the numerator of \eqref{eq:frac-constr} is no larger than one if and only if
\begin{align}\label{eq:condless1}
   \begin{cases}
    &\delta_{2+}\le 1\quad\textit{when}\quad p_1\ge 1-\tau\\
    &\delta_{2+}\le\odds(1-\tau)/\odds(p_1)\quad\textit{when}\quad p_1< 1-\tau
    \end{cases}.
\end{align}
For any pair $(\delta_{1+},\delta_{2+})$ that satisfies \eqref{eq:frac-constr} but violates \eqref{eq:condless1},
resetting $\delta_{2+}$ to 1 does not change the objective function value or violate the constraint \eqref{eq:frac-constr}, which is a non-decreasing function of $\delta_2$. Henceforth we assume \eqref{eq:condless1} is satisfied.

The denominator of \eqref{eq:frac-constr} is greater than 0 if and only if
\begin{align}\label{eq:condgreater0}
   \begin{cases}
    &\delta_{1+}<\odds(p_1)/\odds(1-\tau))\quad\textit{when}\quad p_1\ge 1-\tau\\
    &\delta_{1+}< 1 \quad\textit{when}\quad p_1< 1-\tau\
    \end{cases}.
\end{align}
Consider first $(\delta_{1+},\delta_{2+})$ satisfying \eqref{eq:condgreater0}. The constraint \eqref{eq:frac-constr} is then equivalent to
\begin{align}\label{eq:linear-constr}
    \frac{\Theta E(\rho_{\tau}(Y,q^*_{1,\tau})|T=1)}{\tau}\delta_{1+}+\frac{E(\rho_{\tau}(Y,q^*_{1,\tau})|T=1)}{1-\tau}\delta_{2+}\le (\Theta-1)p_1.
\end{align}
Under this linear constraint, \eqref{eq:dvobj} is maximized by setting $\delta_{1+}=\delta_{2+}$ such that $\Theta$ is reached in \eqref{eq:frac-constr}, or equivalently $(\Theta-1)p_1$ is met in \eqref{eq:linear-constr}, which leads to
\begin{equation}\label{eq:sharpr}
\delta_{1+}=\delta_{2+}=\frac{(\Theta-1)p_1}{E(\rho_{\tau}(Y,q^*_{1,\tau})|T=1)\{1/(1-\tau)+\Theta/\tau\}}.
\end{equation}
When $(\delta_{1+},\delta_{2+})$ is allowed to differ from \eqref{eq:condgreater0}, necessarily $\Theta=\infty$. In this case, \eqref{eq:sharpr} is larger than 1, which gives a trivial but valid solution.
Therefore, \eqref{eq:sharpr} is a solution to \eqref{eq:dvsharpopt}, under which
\begin{align}\label{eq:Del-mu1+}
\Delta_1 = \frac{(\Theta-1)p_1 }{\odds(\tau)+\Theta},\quad
\Delta_2 = \frac{(\Theta-1)p_1 }{1+\odds(1-\tau)\Theta}.
\end{align}

\noindent\textbf{Optimal $(\Delta_1,\Delta_2)$ for sharp lower bound of $\mu^1$}

Similar to how sharp lower bound is related to sharp upper bound under eMSM,
switching $\tau$ with $1-\tau$ in \eqref{eq:Del-mu1+} gives the solution as in \eqref{eq:Delta-}.

\noindent\textbf{Optimal $(\Delta_1',\Delta_2')$ for sharp bounds of $\mu^0$}

By the symmetry of $T$ in its two labels, replacing $\tau$ with $\tau'$ and $p_1$ with $p_0$ in respective bounds of $\mu^1$ gives the desired results.

\noindent\textit{Proof of Proposition \ref{prop:dvsharp} (iii)}:
Consider $\mu^{1+}_{\mytext{DV}}(\Theta)$. By \eqref{eq:Del-mu1+},
\begin{align}\label{eq:eMSM-min-simp}
   \min\{ \tau \Delta_1, (1-\tau) \Delta_2 \} = \frac{\tau(\Theta-1)p_1}{\odds(\tau)+\Theta}.
\end{align}
$\mu^{1+}_{\mytext{DV}}(\Theta)$ is then obtained by plugging \eqref{eq:eMSM-min-simp} into \eqref{eq:binary-emsm-up}. The other bounds under $\rm{DV}(\Theta)$ can be similarly obtained.

\subsection{Proof of the sharp bounds under dMSM}\label{sec:proof-dMSM}
We first show $\rm{dMSM}(\Delta)$ is the collection of $\rm{eMSM}(\Delta_1,\Delta_2)$ with $\Delta_1(X)+\Delta_2(X)\le\Delta(X)$, and hence
$\mu_{\mytext{dMSM}}^{1+}(\Delta)=\max_{\Delta_1+\Delta_2\le\Delta}\mu_{\mytext{eMSM}}^{1+}(\Delta_1,\Delta_2)$.

For any full data distribution $Q\in\rm{eMSM}(\Delta_1,\Delta_2)$ with $\Delta_1(X)+\Delta_2(X)\le\Delta(X)$,
\begin{align}
    &\max_{u}\mg_Q(X,u)-\min_u\mg_Q(X,u)\\
    &=\max_{u}\mg_Q(X,u)-E(Y|T=1,X)-\left\{\min_{u}\mg_Q(X,u)-E(Y|T=1,X)\right\}\\
    &=\Delta_2(X)+\Delta_1(X)\leq\Delta(X),
\end{align}
which shows that $\bigcup_{\Delta_1+\Delta_2\le\Delta}\rm{eMSM}(\Delta_1,\Delta_2)\subset\rm{dMSM}(\Delta)$.

Conversely, for any $Q\in\rm{dMSM}(\Delta)$, $Q\in\rm{eMSM}(\Delta_m,\Delta_M)$, where $\Delta_M(X)=\max_{u}\eta_Q(X,u)-E(Y|T=1,X)$, $\Delta_m(X)=E(Y|T=1,X)-\min_{u}\eta_Q(X,u)$ and by construction $\Delta_m(X)+\Delta_M(X)\le \Delta(X)$. Therefore, $\rm{dMSM}(\Delta)\subset\bigcup_{\Delta_1+\Delta_2\le\Delta}\rm{eMSM}(\Delta_1,\Delta_2)$.
This proves
$\rm{dMSM}(\Delta)=\bigcup_{\Delta_1+\Delta_2\le\Delta}\rm{eMSM}(\Delta_1,\Delta_2)$
and consequently, $\mu_{\mytext{dMSM}}^{1+}(\Delta)=\max_{\Delta_1+\Delta_2\le\Delta}\mu_{\mytext{eMSM}}^{1+}(\Delta_1,\Delta_2)$.

Second, we solve $\max_{\Delta_1+\Delta_2\le\Delta}\mu_{\mytext{eMSM}}^{1+}(\Delta_1,\Delta_2)$ for $\mu_{\mytext{dMSM}}^{1+}(\Delta)$. By Proposition \ref{pro:pop-sol}, the optimal $(\Delta_1,\Delta_2)$ also solves
\begin{align}
&\max_{\Delta_{1},\Delta_{2}}\min\big[ \tau(X) \Delta_1(X), \{1-\tau(X)\} \Delta_2 (X), E \{\rho_{\tau}(Y,q^*_{1,\tau})|T=1,X \} \big]\label{eq:dMSMobj}\\
&\text{subject to}\quad \Delta_1(X)+\Delta_2(X)\le\Delta(X).\label{eq:dMSM-constr}
\end{align}

As $E \{\rho_{\tau}(Y,q^*_{1,\tau})|T=1,X \}$ does not depend on $\Delta_1,\Delta_2$, it suffices to consider
\begin{subequations}\label{eq:dMSM2}
\begin{align}
&\max_{\Delta_{1},\Delta_{2}}\min\big[ \tau(X) \Delta_1(X), \{1-\tau(X)\} \Delta_2 (X)\big]\label{eq:dMSMobj2}\\
&\text{subject to}\quad \Delta_1(X)+\Delta_2(X)\le\Delta(X) . \label{eq:dMSM-constr2}
\end{align}
\end{subequations}
This is solved when $\tau(X) \Delta_1(X)=\{1-\tau(X)\}\Delta_2 (X)$ and $\Delta_1(X)+\Delta_2(X)=\Delta(X)$. Therefore, $$\Delta_1=\{1-\tau(X)\}\Delta (X),\text{and}\quad\Delta_2=\tau(X)\Delta (X),$$
from which
\begin{align}
     \mu_{\mytext{dMSM}}^{1+}(\Delta)=&E\left\{E(Y|T=1,X)+(1-T)\{\Lambda_2(X)-\Lambda_1(X)\}\times\right.\\
        &\left.\min\left[\tau(X)\left\{1-\tau(X)\right\}\Delta(X),E \{\rho_{\tau}(Y,q^*_{1,\tau})|T=1,X \}\right]\right\}.
\end{align}
The expression for $\mu_{\mytext{dMSM}}^{1-}(\Delta)$ follows similarly.

\section{Proofs of technical lemmas}\label{sec:prooflm}
\subsection{Proof of Lemma \ref{lm:msm-variations}} \label{sec:proof-lm-MSMs}
\textit{Proof of Lemma \ref{lm:msm-variations} (i):}
By \eqref{eq:msm-indep} and \eqref{eq:msm-bound}, we have
\begin{align}
    \frac{\ud P_{Y^0,Y^1,U}(\cdot|T=0,X)}{\ud P_{Y^0,Y^1,U}(\cdot|T=1,X)}&=\frac{\ud P_{Y^0,Y^1}(\cdot|T=0,X,U)\ud P_U(\cdot|T=0,X)}{\ud P_{Y^0,Y^1}(\cdot|T=1,X,U)\ud P_U(\cdot|T=1,X)}\\
    &=\frac{\ud P_U(\cdot|T=0,X)}{\ud P_U(\cdot|T=1,X)}\in [\Lambda_1(X),\Lambda_2(X)].
\end{align}
Then for some dominating measure $Q$,
\begin{align}
     \Lambda_1(X)\frac{\ud P_{Y^0,Y^1,U}(\cdot|T=1,X)}{\ud Q}\le\frac{\ud P_{Y^0,Y^1,U}(\cdot|T=0,X)}{\ud Q}\le \Lambda_2(X)\frac{\ud P_{Y^0,Y^1,U}(\cdot|T=1,X)}{\ud Q}.\label{eq:msms-equiv}
\end{align}
Therefore, \eqref{eq:msm-Y0Y1} follows from \eqref{eq:msms-equiv} by marginalizing over $U$.

\noindent\textit{Proof of Lemma \ref{lm:msm-variations} (ii):} The result follows by setting $U=(Y^0,Y^1)$.

\noindent\textit{Proof of Lemma \ref{lm:msm-variations} (iii):} By (i), it suffices to consider distribution $P$ on $(Y^0,Y^1,T,X)$ that satisfies \eqref{eq:msm-Y0Y1}.
From \eqref{eq:msms-equiv}, marginalizing over $Y^1$ or $Y^0$ respectively gives \eqref{eq:msm-Y0} or \eqref{eq:msm-Y1}.

\subsection{Proof of Lemma \ref{lm:BinaryU}}\label{sec:proof-lm-BinaryU}
To further simplify the notation, the conditioning on $T=1$ is made implicit.
   We consider two cases (a) $Q(\mg'(U')=q_{\mg',\tau})=0$, and (b) $Q(\mg'(U')=q_{\mg',\tau})>0$.

In case (a), $\lambda'$ as in \eqref{eq:lambda-optimal} assumes binary values $\Lambda_1$ and $\Lambda_2$ $Q$-almost surely. Define $U$ as the indicator function of $\lambda'(U')=\Lambda_2$, i.e., $U=\mathbf{1}_{\lambda'(U')=\Lambda_2}$.
Then $Q_U(1)=Q(\lambda'(U')=\Lambda_2)=Q(\mg'(U')\ge q_{\mg',\tau})=1-\tau$ and $Q_U(0)=Q(\lambda'(U')=\Lambda_1)=Q(\mg'(U')<q_{\mg',\tau})=\tau$. Define $\lambda$ as in \eqref{eq:lambda-binary}, then $\lambda(U)$ satisfies \eqref{eq:msm-simp} and \eqref{eq:lambda-simp}. Define
\begin{align}
  \mg(1)=E\{\mg'(U')|U=1\},\quad \mg(0)=E\{\mg'(U')|U=0\}.
\end{align}
As $\mg'(U')$ satisfies \eqref{eq:emsm-simp} and \eqref{eq:eta-simp}, so does $\mg(U)$. Finally, $E_Q\{\mg(U)\lambda(U)\}=E_Q\{\mg'(U')\lambda'(U')\}$ by construction.

For case (b), we first construct an intermediate triple from $(Q_{U'},\mg', \lambda')$ that randomizes on $\mg'(U')=q_{\mg',\tau}$ and then apply a similar strategy as in case (a). Specifically, we define $U'_{\mytext{Z}}=(U',Z)$ by augmenting $U'$ with $Z$, where
\begin{align}
Z|(U'=u)=
\begin{cases}
      1 \quad\text{with probability } p_r,  &\text{if } u\in\{u:\mg'(U')=q_{\mg',\tau}\} \\
      -1 \quad\text{with probability } 1-p_r, & \text{if } u\in\{u:\mg'(U')=q_{\mg',\tau}\}\\
      0, & \text{if } u\notin \{u:\mg'(U')=q_{\mg',\tau}\}\\
    \end{cases},
\end{align}
and $p_r=\{1-\tau-Q(\mg'(U')>q_{\mg',\tau})\}/Q(\mg'(U')=q_{\mg',\tau})$. By definition,
\begin{align}
   \ud Q_{U'_Z}(U'=u,Z=z)=\begin{cases}
      p_r\ud Q_{U'}(u) &\text{if } u\in\{u:\mg'(U')=q_{\mg',\tau}\}, z=1 \\
      (1-p_r)\ud Q_{U'}(u) &\text{if } u\in\{u:\mg'(U')=q_{\mg',\tau}\}, z=-1\\
      0, & \text{if } u\in\{u:\mg'(U')=q_{\mg',\tau}\}, z=0
    \end{cases},
\end{align}
and
\begin{align}
   \ud Q_{U'_Z}(U'=u,Z=z)=\begin{cases}
      \ud Q_{U'}(u) &\text{if } u\notin\{u:\mg'(U')=q_{\mg',\tau}\}, z=0 \\
      0 &\text{if } u\notin\{u:\mg'(U')=q_{\mg',\tau}\}, z\neq 0
    \end{cases}.
\end{align}
We define  $\lambda'_Z$ and $\mg'_Z$ for $U'_Z$ as follows:
\begin{align}
    \lambda'_Z(U'=u,Z=z)=\begin{cases}
      \lambda'(u) &\text{if } z=0 \\
      \Lambda_2 &\text{if } z= 1\\
      \Lambda_1 &\text{if } z= -1
    \end{cases},
\end{align}
and $\mg'_Z(U'=u,Z=z)=\mg'(u)$. It follows immediately that $\mg'_Z(U'_Z)$ satisfies \eqref{eq:emsm-simp} and \eqref{eq:eta-simp}, and that $\lambda'_Z(U'_Z)$ satisfies \eqref{eq:msm-simp}. We show \eqref{eq:lambda-simp} holds with $(Q_{U'_Z},\lambda'_Z)$:
\begin{align}
    &\int \lambda'_Z(u,z)\ud Q_{U'_Z}=\int_{\mg'(U')\neq q_{\mg',\tau}} \lambda'(u)\ud Q_{U'}+\left\{\Lambda_2 p_r+\Lambda_1 (1-p_r)\right\}Q\left(\mg'(U')= q_{\mg',\tau}\right)\\
    &=\int_{\mg'(U')\neq q_{\mg',\tau}} \lambda'(u)\ud Q_{U'}+\Lambda_2(1-\tau)-\Lambda_2 Q(\mg'(U')>q_{\mg',\tau})+\Lambda_1\tau-\Lambda_1 Q(\mg'(U')<q_{\mg',\tau})\\
    &=\Lambda_2(1-\tau)+\Lambda_1\tau=1.\label{eq:lamb-Z-1}
\end{align}
Next, we show that the objective function in \eqref{eq:V_e-simp} has the same evaluation under $(Q_{U'_Z},\mg'_Z,\lambda'_Z)$ and $(Q_{U'},\mg',\lambda')$:
\begin{align}
    &\int \mg'_Z(u,z)\lambda'_Z(u,z)\ud Q_{U'_Z}\\
    &=\int_{\mg'(U')\neq q_{\mg',\tau}}\mg'(u) \lambda'(u)\ud Q_{U'}+\left\{\Lambda_2 p_r+\Lambda_1 (1-p_r)\right\}Q\left(\mg'(U')= q_{\mg',\tau}\right)q_{\mg',\tau}\\
    &=\int_{\mg'(U')\neq q_{\mg',\tau}}\mg'(u) \lambda'(u)\ud Q_{U'}+\left(1-\int_{\mg'(U')\neq q_{\mg',\tau}} \lambda'(u)\ud Q_{U'}\right)q_{\mg',\tau}\quad\text{[By \eqref{eq:lamb-Z-1}]}\\
    &=\int\mg'(u)\lambda'(u)\ud Q_{U'}.
\end{align}
To complete the proof for case (b), we set $U=\mathbf{1}_{\lambda'_Z(U'_Z)=\Lambda_2}$. Define $\lambda(U)$ as in \eqref{eq:lambda-binary}, and $\mg(U)$ by
\begin{align}
  &\mg(1)=E\{\mg'_Z(U'_Z)|U=1\},\quad \mg(0)=E\{\mg'_Z(U'_Z)|U=0\}.
\end{align}
The rest follows similarly as in case (a). 
\section{Additional results from empirical applications} \label{sec:additional-numerical}
We provide additional results for the empirical applications to RHC and NHANES.

\subsection{Additional numerical result for RHC} \label{sec:additional-numerical-rhc}
\begin{figure}[h]
    \centering
    \begin{minipage}{0.49\textwidth}
        \centering
        \includegraphics[width=\textwidth]{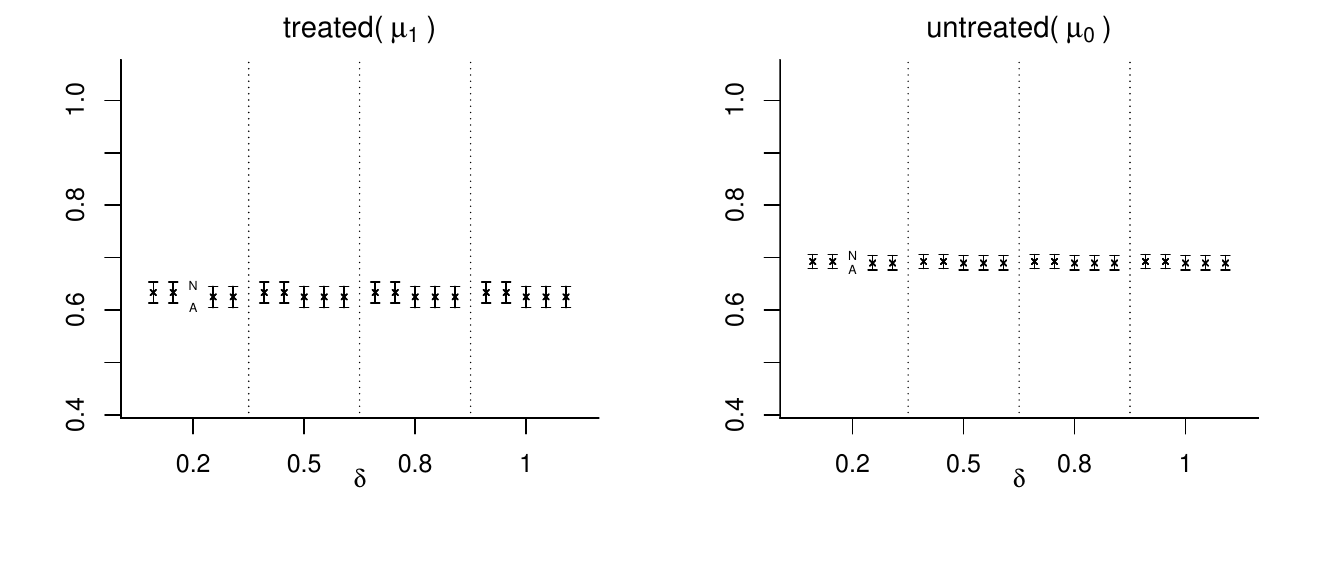} \vspace{-.4in}
        \subcaption{ \small $\Lambda = 1$}
    \end{minipage}
    \hfill
    \begin{minipage}{0.49\textwidth}
        \centering
        \includegraphics[width=\textwidth]{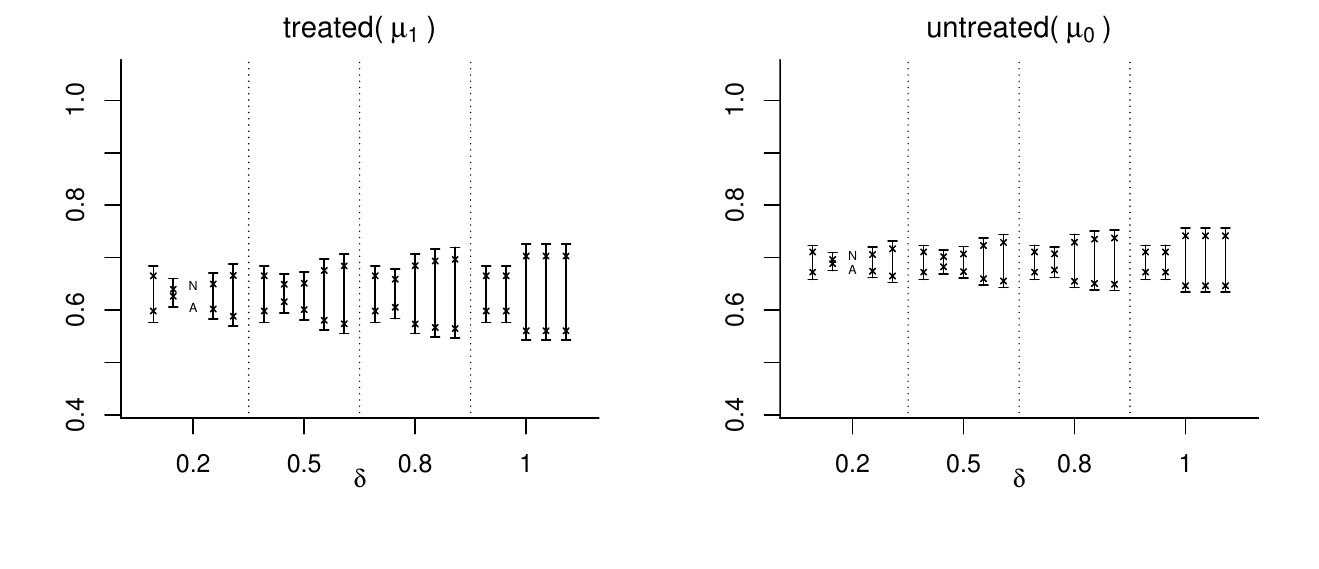} \vspace{-.4in}
        \subcaption{\small $\Lambda = 1.2$}
    \end{minipage}
    
    \vspace{0.3in} 

    \begin{minipage}{0.49\textwidth}
        \centering
        \includegraphics[width=\textwidth]{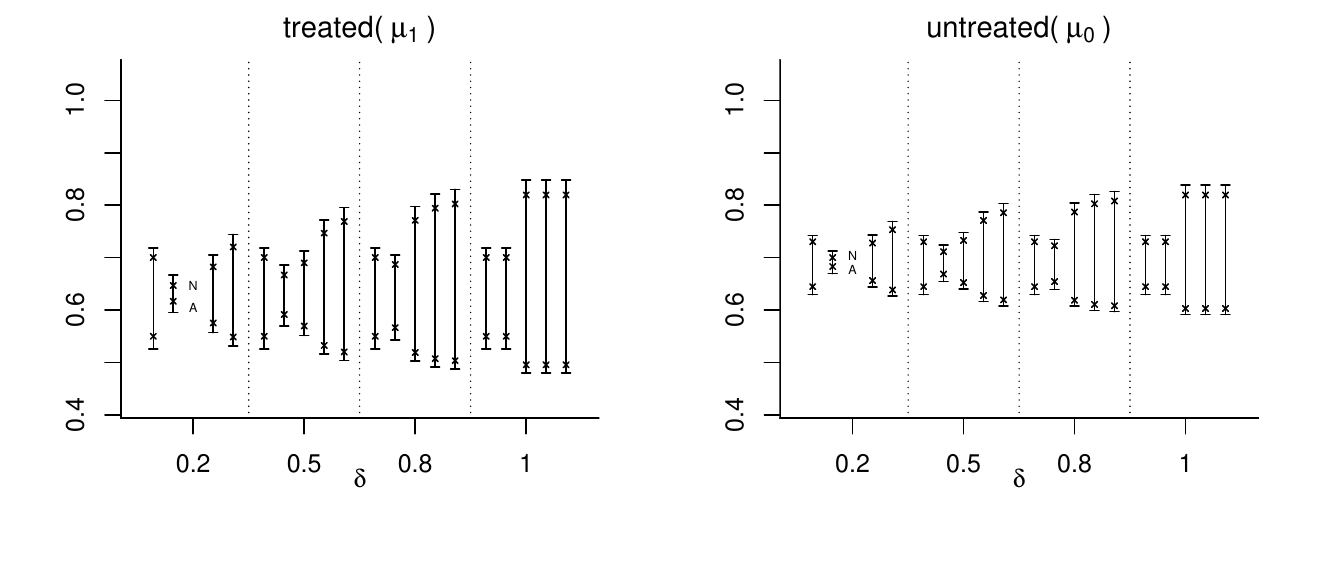} \vspace{-.4in}
        \subcaption{\small $\Lambda = 1.5$}
    \end{minipage}
    \hfill
    \begin{minipage}{0.49\textwidth}
        \centering
        \includegraphics[width=\textwidth]{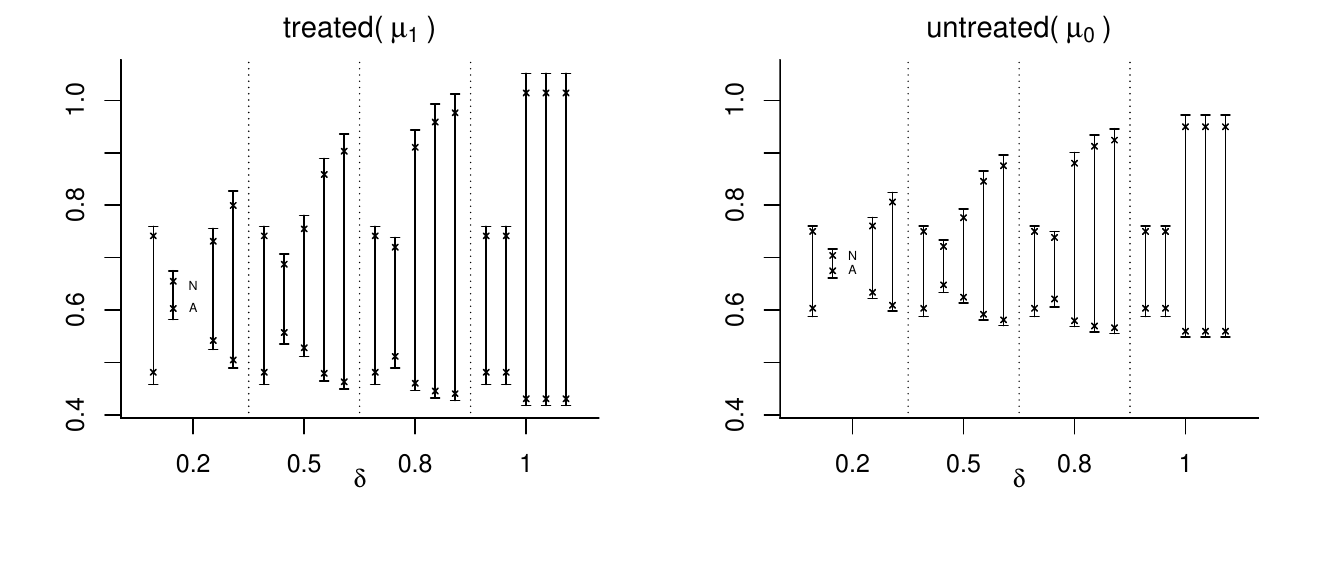} \vspace{-.4in}
        \subcaption{\small $\Lambda = 2$}
    \end{minipage} \vspace{-.1in}

    \caption{\small
Point bounds (x) and 90\% confidence intervals (–) on 30-day survival probabilities in the RHC study, similarly as Figure \ref{fig:rhc-lowdim-logit-ate}.}
    \label{fig:rhc-lowdim-logit-mu}
\end{figure}

\begin{figure} 
\centering
\begin{minipage}{\textwidth}
        \centering
        \includegraphics[width=\textwidth]{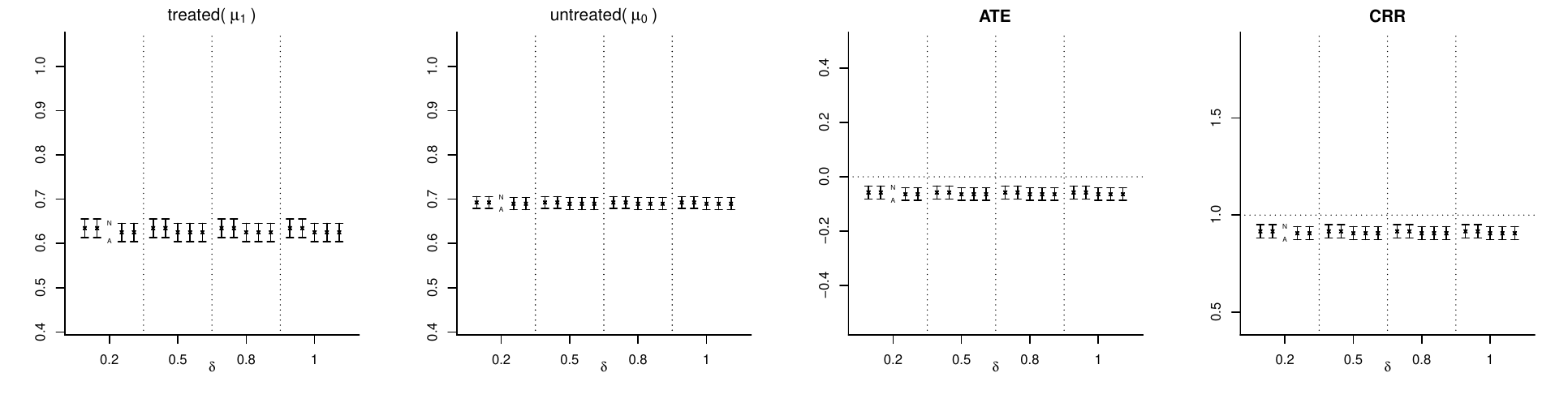} \vspace{-.4in}
        \subcaption{$\Lambda = 1$}
    \end{minipage}
    
    \vspace{.3in}
    
  \begin{minipage}{\textwidth}
        \centering
        \includegraphics[width=\textwidth]{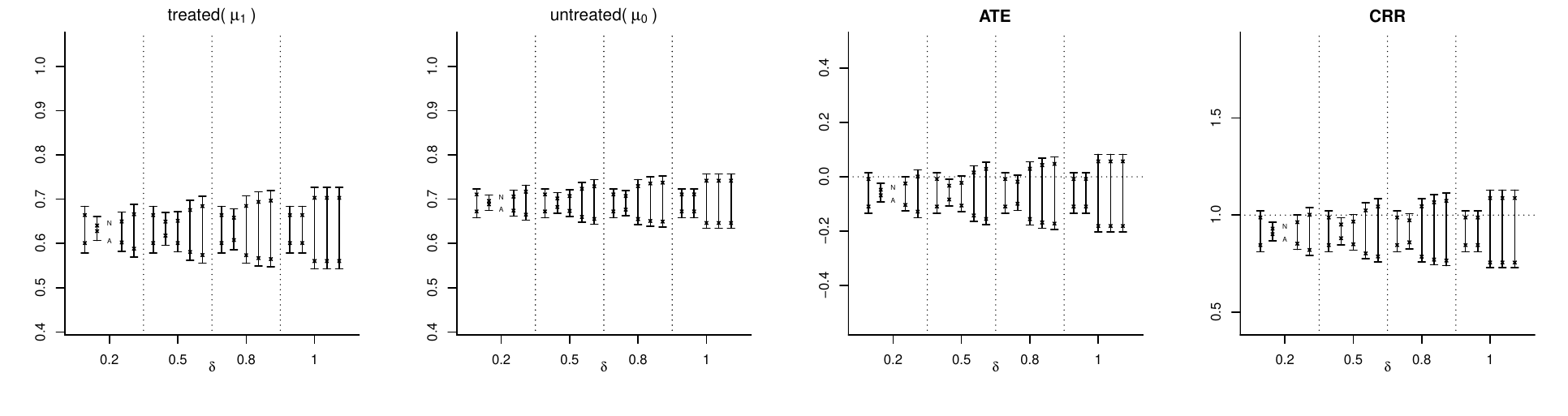} \vspace{-.4in}
        \subcaption{$\Lambda = 1.2$}
    \end{minipage}
    
    \vspace{.3in}
    
    \begin{minipage}{\textwidth}
        \centering
        \includegraphics[width=\textwidth]{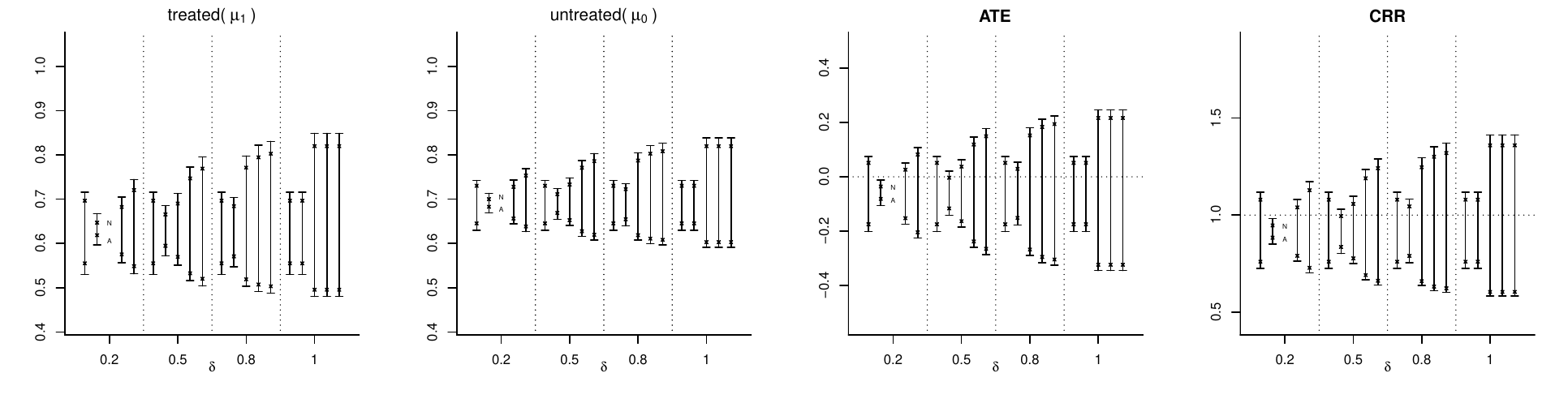} \vspace{-.4in}
        \subcaption{$\Lambda = 1.5$}
    \end{minipage}
    
    \vspace{.3in}
    
    \begin{minipage}{\textwidth}
        \centering
        \includegraphics[width=\textwidth]{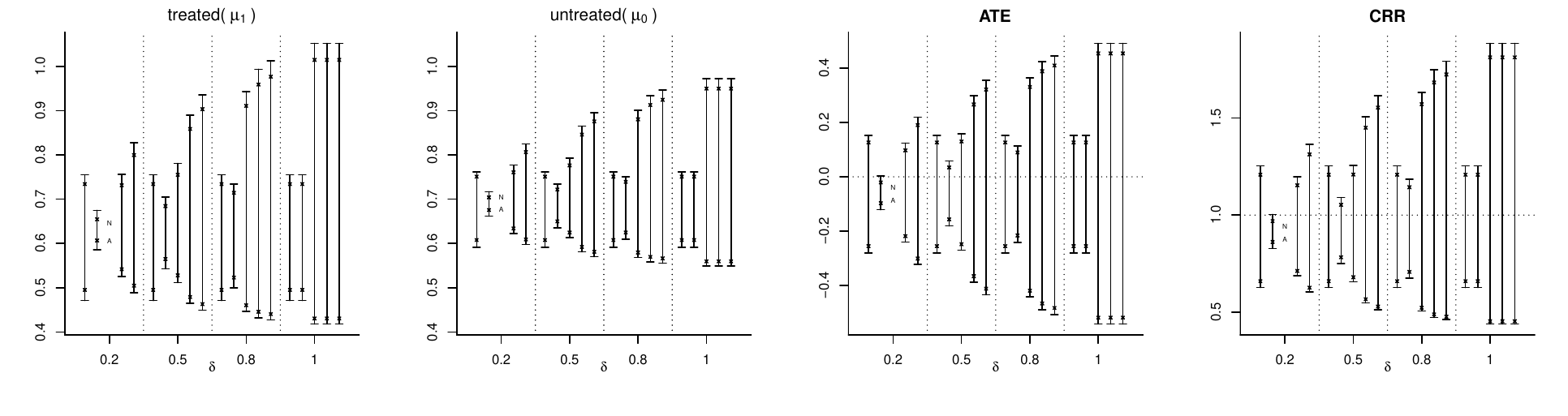} \vspace{-.4in}
        \subcaption{$\Lambda = 2$}
    \end{minipage} \vspace{-.1in}
\caption{\small
Point bounds (x) and 90\% confidence intervals (–) on 30-day survival probabilities, ATE and CRR in the RHC study, similarly as Figure \ref{fig:rhc-lowdim-logit-ate}, but using linear outcome mean regression for CAL under MSM and eMSM.}
\label{fig:rhc-lowdim-lin}
\end{figure}

\begin{figure} 
\centering
\begin{minipage}{\textwidth}
        \centering
        \includegraphics[width=\textwidth]{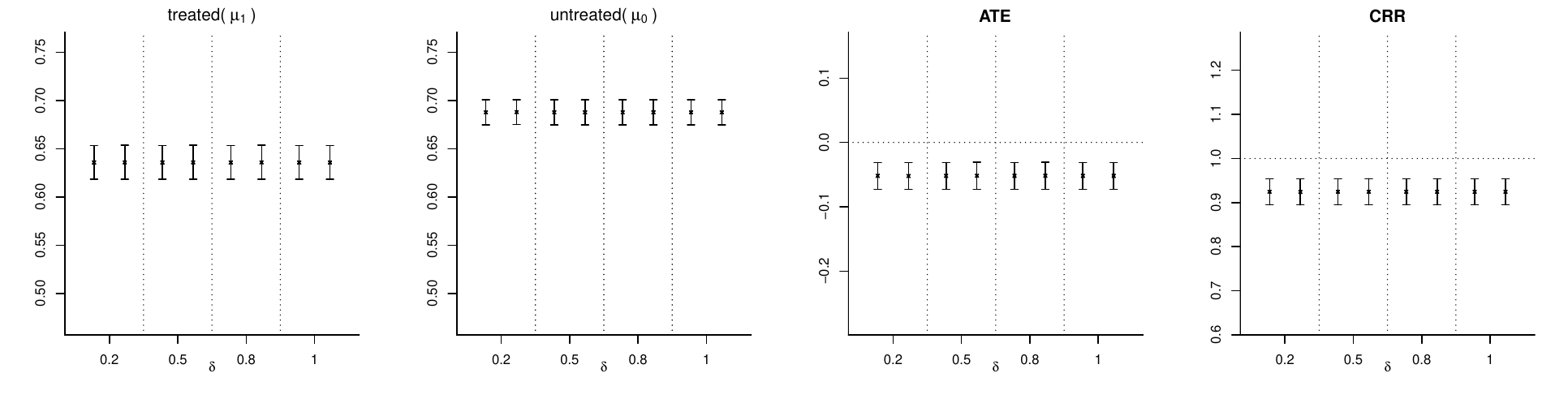} \vspace{-.4in}
        \subcaption{$\Lambda = 1$}
    \end{minipage}
    
    \vspace{.3in}
    
  \begin{minipage}{\textwidth}
        \centering
        \includegraphics[width=\textwidth]{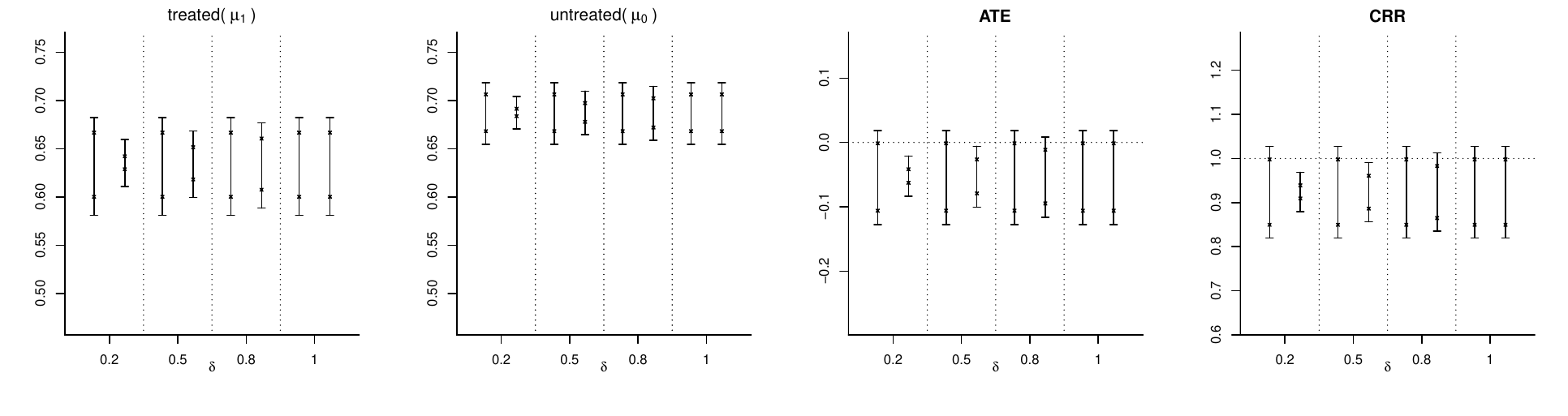} \vspace{-.4in}
        \subcaption{$\Lambda = 1.2$}
    \end{minipage}
    
    \vspace{.3in}
    
    \begin{minipage}{\textwidth}
        \centering
        \includegraphics[width=\textwidth]{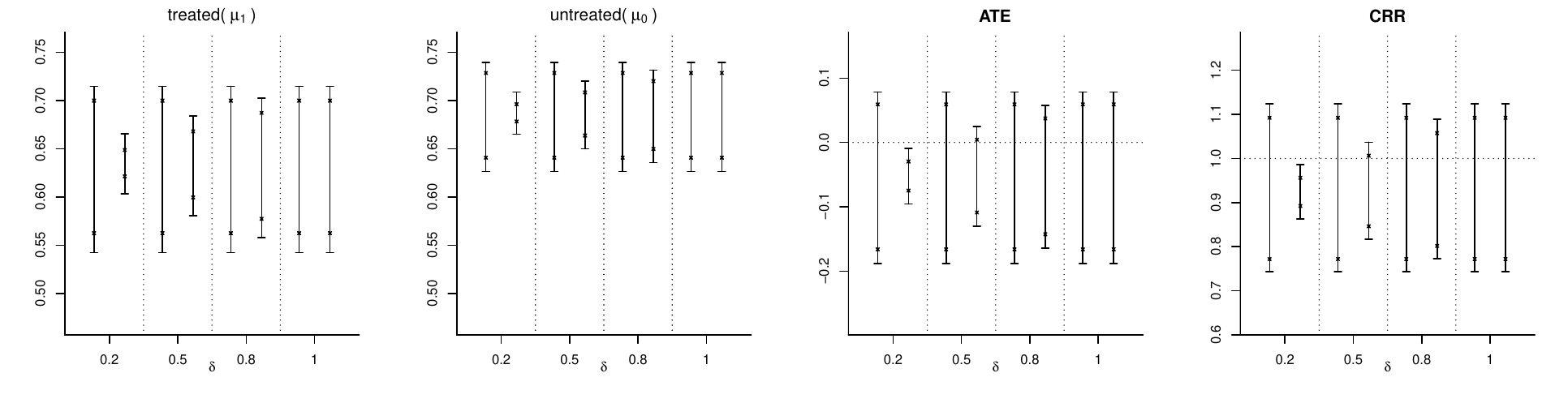} \vspace{-.4in}
        \subcaption{$\Lambda = 1.5$}
    \end{minipage}
    
    \vspace{.3in}
    
    \begin{minipage}{\textwidth}
        \centering
        \includegraphics[width=\textwidth]{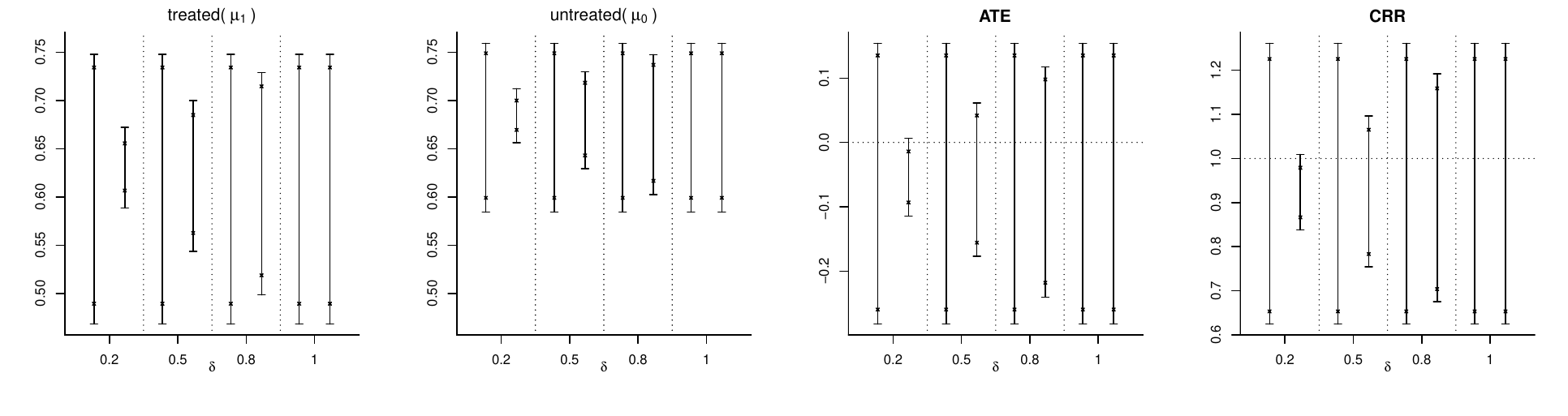} \vspace{-.4in}
        \subcaption{$\Lambda = 2$}
    \end{minipage} \vspace{-.1in}
\caption{\small
Point bounds (x) and 90\% confidence intervals (–) on 30-day survival probabilities, ATE and CRR in the RHC study, similarly as Figure \ref{fig:rhc-lowdim-logit-ate}, but using working models with main effects and interactions.}
\label{fig:rhc-highdim-logit}
\end{figure}

\begin{figure} 
\centering
\begin{minipage}{\textwidth}
        \centering
        \includegraphics[width=\textwidth]{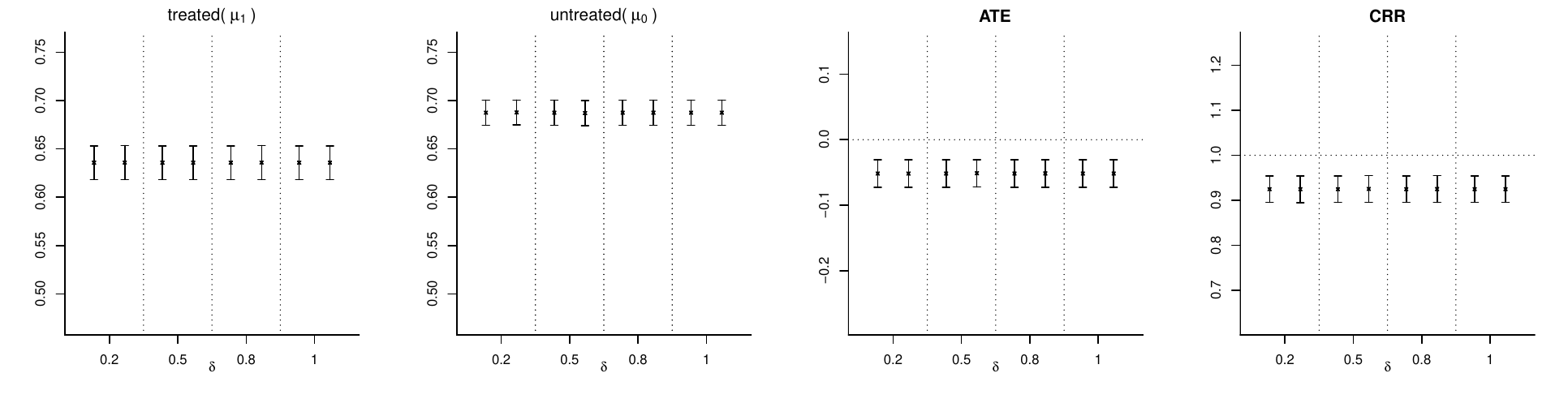} \vspace{-.4in}
        \subcaption{$\Lambda = 1$}
    \end{minipage}
    
    \vspace{.3in}
    
  \begin{minipage}{\textwidth}
        \centering
        \includegraphics[width=\textwidth]{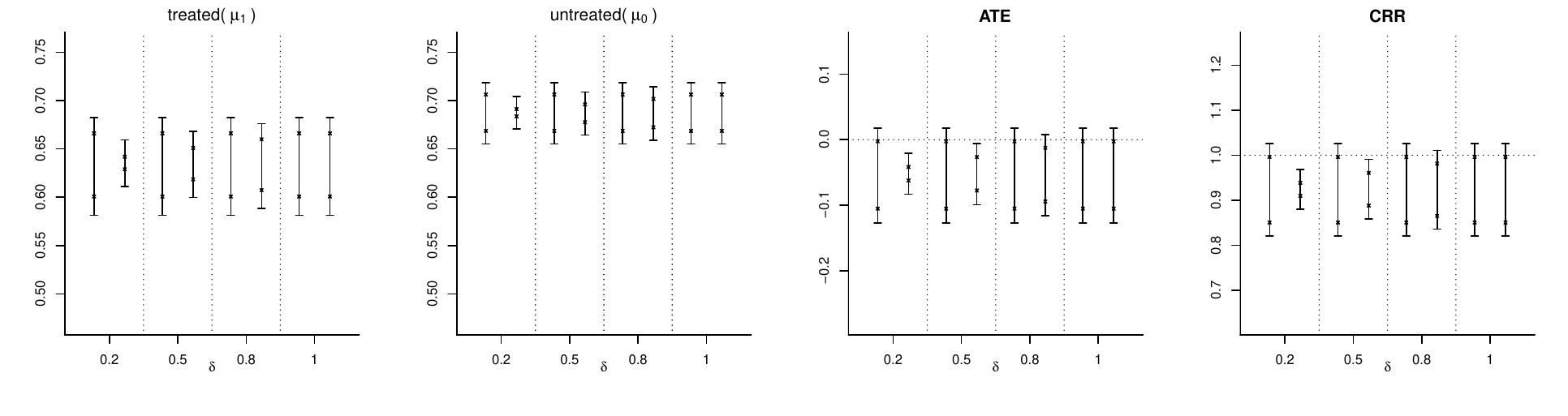} \vspace{-.4in}
        \subcaption{$\Lambda = 1.2$}
    \end{minipage}
    
    \vspace{.3in}
    
    \begin{minipage}{\textwidth}
        \centering
        \includegraphics[width=\textwidth]{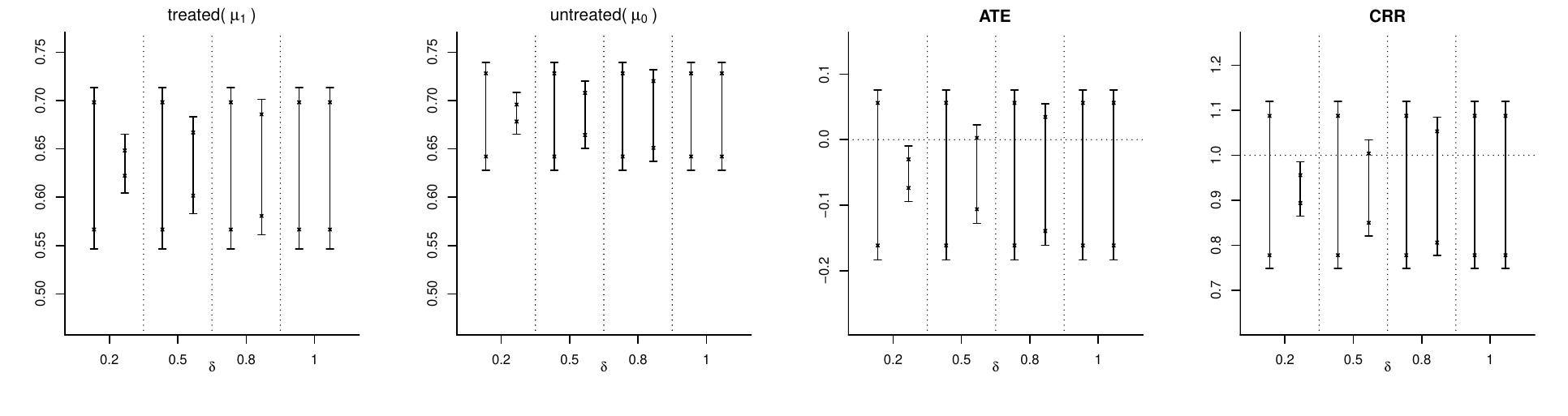} \vspace{-.4in}
        \subcaption{$\Lambda = 1.5$}
    \end{minipage}
    
    \vspace{.3in}
    
    \begin{minipage}{\textwidth}
        \centering
        \includegraphics[width=\textwidth]{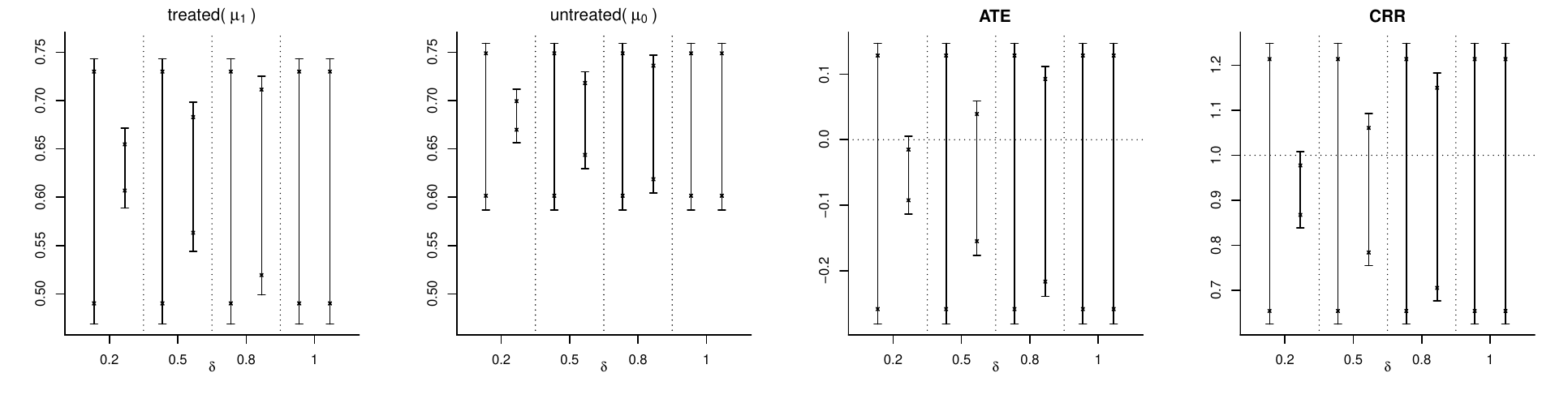} \vspace{-.4in}
        \subcaption{$\Lambda = 2$}
    \end{minipage} \vspace{-.1in}
\caption{\small
Point bounds (x) and 90\% confidence intervals (–) on 30-day survival probabilities, ATE and CRR in the RHC study, similarly as Figure \ref{fig:rhc-highdim-logit}, but using linear outcome mean regression for RCAL under MSM and eMSM.}
\label{fig:rhc-highdim-lin}
\end{figure}

Figure \ref{fig:rhc-lowdim-logit-mu} presents the point bounds and 90\% confidence intervals on 30-day survival probabilities, similarly as Figure \ref{fig:rhc-lowdim-logit-ate}.
Figure \ref{fig:rhc-lowdim-lin} presents the point bounds and 90\% confidence intervals using working models with only main effects and linear outcome mean regression for CAL under MSM and eMSM. Figures \ref{fig:rhc-highdim-logit} and \ref{fig:rhc-highdim-lin} present the point bounds and 90\% confidence intervals using working models with main effects and interactions, with logistic and linear outcome mean regression applied to RCAL under MSM and eMSM, respectively. Tables \ref{tb:rhc-lowdim-cal-lam1.0}--\ref{tb:rhc-lowdim-dv-lam2.0} show the numerical estimates and standard errors using working models with only main effects. Tables \ref{tb:rhc-highdim-cal-lam1.0}--\ref{tb:rhc-highdim-cal-lam2.0} show the numerical estimates and standard errors using workings models with main effects and interactions.

Comparing Figures \ref{fig:rhc-lowdim-logit-ate} and \ref{fig:rhc-lowdim-lin}, we find that the conclusions in Section \ref{sec:rhc} remain valid for CAL under MSM and eMSM when using working models with linear outcome mean regression. By comparing numerical estimates and standard errors between the low-dimensional setting (Tables \ref{tb:rhc-lowdim-cal-lam1.0}–\ref{tb:rhc-lowdim-dv-lam2.0}) and the high-dimensional setting (Tables \ref{tb:rhc-highdim-cal-lam1.0}–\ref{tb:rhc-highdim-cal-lam2.0}), we observe that, in the high-dimensional setting, the point bounds for ATE are overall similar, while the standard errors of the point bounds are consistently smaller due to regularized estimation with more flexible working models, as observed in Tan (2024a).

\begin{table}
\caption{\footnotesize Point bounds and SEs for CAL under MSM and eMSM with $\Lambda = 1$, using working models with only main effects in the RHC study.} \label{tb:rhc-lowdim-cal-lam1.0} \vspace{-4ex}
\begin{center}
\renewcommand{\arraystretch}{0.6}
\small
\resizebox{\textwidth}{!}{\begin{tabular}{lccccccccccc}
\hline
& \multicolumn{2}{c}{$\delta = 0.2$} & & \multicolumn{2}{c}{$\delta = 0.5$} & & \multicolumn{2}{c}{$\delta = 0.8$} & & \multicolumn{2}{c}{$\delta = 1$} \\
& Bounds & SEs & & Bounds & SEs & & Bounds & SEs & & Bounds & SEs \\
\hline
& \multicolumn{11}{c}{Lower bounds and SEs for $\mu^1$} \\
MSM logit &  0.634 &  0.012 &  &  0.634 &  0.012 &  &  0.634 &  0.012 &  &  0.634 &  0.012\\
MSM lin &  0.635 &  0.013 &  &  0.635 &  0.013 &  &  0.635 &  0.013 &  &  0.635 &  0.013\\
eMSM logit &  0.634 &  0.012 &  &  0.634 &  0.012 &  &  0.634 &  0.012 &  &  0.634 &  0.012\\
eMSM lin &  0.635 &  0.013 &  &  0.635 &  0.013 &  &  0.635 &  0.013 &  &  0.635 &  0.013\\
& \multicolumn{11}{c}{Upper bounds and SEs for $\mu^1$} \\
MSM logit &  0.634 &  0.012 &  &  0.634 &  0.012 &  &  0.634 &  0.012 &  &  0.634 &  0.012\\
MSM lin &  0.635 &  0.013 &  &  0.635 &  0.013 &  &  0.635 &  0.013 &  &  0.635 &  0.013\\
eMSM logit &  0.634 &  0.012 &  &  0.634 &  0.012 &  &  0.634 &  0.012 &  &  0.634 &  0.012\\
eMSM lin &  0.635 &  0.013 &  &  0.635 &  0.013 &  &  0.635 &  0.013 &  &  0.635 &  0.013\\
& \multicolumn{11}{c}{Lower bounds and SEs for $\mu^0$} \\
MSM logit &  0.693 &  0.008 &  &  0.693 &  0.008 &  &  0.693 &  0.008 &  &  0.693 &  0.008\\
MSM lin &  0.692 &  0.008 &  &  0.692 &  0.008 &  &  0.692 &  0.008 &  &  0.692 &  0.008\\
eMSM logit &  0.693 &  0.008 &  &  0.693 &  0.008 &  &  0.693 &  0.008 &  &  0.693 &  0.008\\
eMSM lin &  0.692 &  0.008 &  &  0.692 &  0.008 &  &  0.692 &  0.008 &  &  0.692 &  0.008\\
& \multicolumn{11}{c}{Upper bounds and SEs for $\mu^0$} \\
MSM logit &  0.693 &  0.008 &  &  0.693 &  0.008 &  &  0.693 &  0.008 &  &  0.693 &  0.008\\
MSM lin &  0.692 &  0.008 &  &  0.692 &  0.008 &  &  0.692 &  0.008 &  &  0.692 &  0.008\\
eMSM logit &  0.693 &  0.008 &  &  0.693 &  0.008 &  &  0.693 &  0.008 &  &  0.693 &  0.008\\
eMSM lin &  0.692 &  0.008 &  &  0.692 &  0.008 &  &  0.692 &  0.008 &  &  0.692 &  0.008\\
& \multicolumn{11}{c}{Lower bounds and SEs for $\mu^1 - \mu^0$} \\
MSM logit &  -0.059 &   0.014 &  &  -0.059 &   0.014 &  &  -0.059 &   0.014 &  &  -0.059 &   0.014\\
MSM lin &  -0.058 &   0.015 &  &  -0.058 &   0.015 &  &  -0.058 &   0.015 &  &  -0.058 &   0.015\\
eMSM logit &  -0.059 &   0.014 &  &  -0.059 &   0.014 &  &  -0.059 &   0.014 &  &  -0.059 &   0.014\\
eMSM lin &  -0.058 &   0.015 &  &  -0.058 &   0.015 &  &  -0.058 &   0.015 &  &  -0.058 &   0.015\\
& \multicolumn{11}{c}{Upper bounds and SEs for $\mu^1 - \mu^0$} \\
MSM logit &  -0.059 &   0.014 &  &  -0.059 &   0.014 &  &  -0.059 &   0.014 &  &  -0.059 &   0.014\\
MSM lin &  -0.058 &   0.015 &  &  -0.058 &   0.015 &  &  -0.058 &   0.015 &  &  -0.058 &   0.015\\
eMSM logit &  -0.059 &   0.014 &  &  -0.059 &   0.014 &  &  -0.059 &   0.014 &  &  -0.059 &   0.014\\
eMSM lin &  -0.058 &   0.015 &  &  -0.058 &   0.015 &  &  -0.058 &   0.015 &  &  -0.058 &   0.015\\
& \multicolumn{11}{c}{Lower bounds and SEs for $\mu^1 / \mu^0$} \\
MSM logit &  0.915 &  0.020 &  &  0.915 &  0.020 &  &  0.915 &  0.020 &  &  0.915 &  0.020\\
MSM lin &  0.916 &  0.021 &  &  0.916 &  0.021 &  &  0.916 &  0.021 &  &  0.916 &  0.021\\
eMSM logit &  0.915 &  0.020 &  &  0.915 &  0.020 &  &  0.915 &  0.020 &  &  0.915 &  0.020\\
eMSM lin &  0.916 &  0.021 &  &  0.916 &  0.021 &  &  0.916 &  0.021 &  &  0.916 &  0.021\\
& \multicolumn{11}{c}{Upper bounds and SEs for $\mu^1 / \mu^0$} \\
MSM logit &  0.915 &  0.020 &  &  0.915 &  0.020 &  &  0.915 &  0.020 &  &  0.915 &  0.020\\
MSM lin &  0.916 &  0.021 &  &  0.916 &  0.021 &  &  0.916 &  0.021 &  &  0.916 &  0.021\\
eMSM logit &  0.915 &  0.020 &  &  0.915 &  0.020 &  &  0.915 &  0.020 &  &  0.915 &  0.020\\
eMSM lin &  0.916 &  0.021 &  &  0.916 &  0.021 &  &  0.916 &  0.021 &  &  0.916 &  0.021\\
\hline
\end{tabular}}
\end{center}
\setlength{\baselineskip}{0.5\baselineskip}
\vspace{-.15in}\noindent{\tiny
\textbf{Note}: MSM logit and eMSM logit denote CAL under MSM and eMSM using logistic outcome mean regression. MSM lin and eMSM lin denote CAL under MSM and eMSM using linear outcome mean regression.}
\vspace{-.2in}
\end{table}

\begin{table}
\caption{\footnotesize Point bounds and SEs for DV method under DV model with $\Lambda = 1$, using working models with only main effects in the RHC study.} \label{tb:rhc-lowdim-dv-lam1.0} \vspace{-4ex}
\begin{center}
\renewcommand{\arraystretch}{0.6}
\small
\resizebox{\textwidth}{!}{\begin{tabular}{lccccccccccc}
\hline
& \multicolumn{2}{c}{$\delta = 0.2$} & & \multicolumn{2}{c}{$\delta = 0.5$} & & \multicolumn{2}{c}{$\delta = 0.8$} & & \multicolumn{2}{c}{$\delta = 1$} \\
& Bounds & SEs & & Bounds & SEs & & Bounds & SEs & & Bounds & SEs \\
\hline
& \multicolumn{11}{c}{Lower bounds and SEs for $\mu^1$} \\
DV $\Theta^+/2$ &  0.625 &  0.012 &  &  0.625 &  0.013 &  &  0.625 &  0.013 &  &  0.625 &  0.013\\
DV $\Theta^+$ &  0.625 &  0.013 &  &  0.625 &  0.013 &  &  0.625 &  0.013 &  &  0.625 &  0.013\\
DV $3\Theta^+/2$ &  0.625 &  0.013 &  &  0.625 &  0.013 &  &  0.625 &  0.013 &  &  0.625 &  0.013\\
& \multicolumn{11}{c}{Upper bounds and SEs for $\mu^1$} \\
DV $\Theta^+/2$ &  0.625 &  0.012 &  &  0.625 &  0.012 &  &  0.625 &  0.012 &  &  0.625 &  0.012\\
DV $\Theta^+$ &  0.625 &  0.012 &  &  0.625 &  0.012 &  &  0.625 &  0.012 &  &  0.625 &  0.012\\
DV $3\Theta^+/2$ &  0.625 &  0.012 &  &  0.625 &  0.012 &  &  0.625 &  0.012 &  &  0.625 &  0.012\\
& \multicolumn{11}{c}{Lower bounds and SEs for $\mu^0$} \\
DV $\Theta^+/2$ &  0.690 &  0.008 &  &  0.690 &  0.008 &  &  0.690 &  0.008 &  &  0.690 &  0.008\\
DV $\Theta^+$ &  0.690 &  0.008 &  &  0.690 &  0.008 &  &  0.690 &  0.008 &  &  0.690 &  0.008\\
DV $3\Theta^+/2$ &  0.690 &  0.008 &  &  0.690 &  0.008 &  &  0.690 &  0.008 &  &  0.690 &  0.008\\
& \multicolumn{11}{c}{Upper bounds and SEs for $\mu^0$} \\
DV $\Theta^+/2$ &  0.690 &  0.009 &  &  0.690 &  0.009 &  &  0.690 &  0.009 &  &  0.690 &  0.009\\
DV $\Theta^+$ &  0.690 &  0.009 &  &  0.690 &  0.009 &  &  0.690 &  0.009 &  &  0.690 &  0.009\\
DV $3\Theta^+/2$ &  0.690 &  0.009 &  &  0.690 &  0.009 &  &  0.690 &  0.009 &  &  0.690 &  0.009\\
& \multicolumn{11}{c}{Lower bounds and SEs for $\mu^1 - \mu^0$} \\
DV $\Theta^+/2$ &  -0.064 &   0.014 &  &  -0.064 &   0.014 &  &  -0.064 &   0.014 &  &  -0.064 &   0.014\\
DV $\Theta^+$ &  -0.064 &   0.014 &  &  -0.064 &   0.014 &  &  -0.064 &   0.014 &  &  -0.064 &   0.014\\
DV $3\Theta^+/2$ &  -0.064 &   0.014 &  &  -0.064 &   0.014 &  &  -0.064 &   0.014 &  &  -0.064 &   0.014\\
& \multicolumn{11}{c}{Upper bounds and SEs for $\mu^1 - \mu^0$} \\
DV $\Theta^+/2$ &  -0.064 &   0.015 &  &  -0.064 &   0.015 &  &  -0.064 &   0.015 &  &  -0.064 &   0.015\\
DV $\Theta^+$ &  -0.064 &   0.015 &  &  -0.064 &   0.015 &  &  -0.064 &   0.015 &  &  -0.064 &   0.015\\
DV $3\Theta^+/2$ &  -0.064 &   0.015 &  &  -0.064 &   0.015 &  &  -0.064 &   0.015 &  &  -0.064 &   0.015\\
& \multicolumn{11}{c}{Lower bounds and SEs for $\mu^1 / \mu^0$} \\
DV $\Theta^+/2$ &  0.907 &  0.019 &  &  0.907 &  0.019 &  &  0.907 &  0.019 &  &  0.907 &  0.019\\
DV $\Theta^+$ &  0.907 &  0.019 &  &  0.907 &  0.019 &  &  0.907 &  0.019 &  &  0.907 &  0.019\\
DV $3\Theta^+/2$ &  0.907 &  0.019 &  &  0.907 &  0.019 &  &  0.907 &  0.019 &  &  0.907 &  0.019\\
& \multicolumn{11}{c}{Upper bounds and SEs for $\mu^1 / \mu^0$} \\
DV $\Theta^+/2$ &  0.907 &  0.021 &  &  0.907 &  0.021 &  &  0.907 &  0.021 &  &  0.907 &  0.021\\
DV $\Theta^+$ &  0.907 &  0.021 &  &  0.907 &  0.021 &  &  0.907 &  0.021 &  &  0.907 &  0.021\\
DV $3\Theta^+/2$ &  0.907 &  0.021 &  &  0.907 &  0.021 &  &  0.907 &  0.021 &  &  0.907 &  0.021\\
\hline
\end{tabular}}
\end{center}
\setlength{\baselineskip}{0.5\baselineskip}
\vspace{-.15in}\noindent{\tiny
\textbf{Note}: DV $\Theta^+ / 2$,  DV $\Theta^+$ and DV $3\Theta^+ / 2$ denote DV method under DV model with $\Theta = \Theta^+ / 2$, $\Theta = \Theta^+$ and $\Theta = 3\Theta^+ / 2$. SEs for DV method are computed by matching the lower and upper points of the 90\% sensitivity interval with 
``lower point bound$- z_{0.05}$ SE'' and ``upper point bound$+ z_{0.05}$ SE'' respectively, where $z_{0.05}$ is the 95\% quantile of the standard normal distribution.}
\vspace{-.2in}
\end{table}

\begin{table}
\caption{\footnotesize Point bounds and SEs for CAL under MSM and eMSM with $\Lambda = 1.2$, using working models with only main effects in the RHC study.} \label{tb:rhc-lowdim-cal-lam1.2} \vspace{-4ex}
\begin{center}
\renewcommand{\arraystretch}{0.6}
\small
\resizebox{\textwidth}{!}{\begin{tabular}{lccccccccccc}
\hline
& \multicolumn{2}{c}{$\delta = 0.2$} & & \multicolumn{2}{c}{$\delta = 0.5$} & & \multicolumn{2}{c}{$\delta = 0.8$} & & \multicolumn{2}{c}{$\delta = 1$} \\
& Bounds & SEs & & Bounds & SEs & & Bounds & SEs & & Bounds & SEs \\
\hline
& \multicolumn{11}{c}{Lower bounds and SEs for $\mu^1$} \\
MSM logit &  0.598 &  0.013 &  &  0.598 &  0.013 &  &  0.598 &  0.013 &  &  0.598 &  0.013\\
MSM lin &  0.601 &  0.014 &  &  0.601 &  0.014 &  &  0.601 &  0.014 &  &  0.601 &  0.014\\
eMSM logit &  0.626 &  0.013 &  &  0.616 &  0.013 &  &  0.605 &  0.013 &  &  0.598 &  0.013\\
eMSM lin &  0.628 &  0.013 &  &  0.618 &  0.013 &  &  0.608 &  0.014 &  &  0.601 &  0.014\\
& \multicolumn{11}{c}{Upper bounds and SEs for $\mu^1$} \\
MSM logit &  0.665 &  0.012 &  &  0.665 &  0.012 &  &  0.665 &  0.012 &  &  0.665 &  0.012\\
MSM lin &  0.664 &  0.012 &  &  0.664 &  0.012 &  &  0.664 &  0.012 &  &  0.664 &  0.012\\
eMSM logit &  0.640 &  0.012 &  &  0.649 &  0.012 &  &  0.659 &  0.012 &  &  0.665 &  0.012\\
eMSM lin &  0.640 &  0.013 &  &  0.649 &  0.012 &  &  0.658 &  0.012 &  &  0.664 &  0.012\\
& \multicolumn{11}{c}{Lower bounds and SEs for $\mu^0$} \\
MSM logit &  0.672 &  0.009 &  &  0.672 &  0.009 &  &  0.672 &  0.009 &  &  0.672 &  0.009\\
MSM lin &  0.672 &  0.009 &  &  0.672 &  0.009 &  &  0.672 &  0.009 &  &  0.672 &  0.009\\
eMSM logit &  0.689 &  0.008 &  &  0.683 &  0.008 &  &  0.676 &  0.008 &  &  0.672 &  0.009\\
eMSM lin &  0.688 &  0.008 &  &  0.682 &  0.008 &  &  0.676 &  0.009 &  &  0.672 &  0.009\\
& \multicolumn{11}{c}{Upper bounds and SEs for $\mu^0$} \\
MSM logit &  0.711 &  0.008 &  &  0.711 &  0.008 &  &  0.711 &  0.008 &  &  0.711 &  0.008\\
MSM lin &  0.711 &  0.008 &  &  0.711 &  0.008 &  &  0.711 &  0.008 &  &  0.711 &  0.008\\
eMSM logit &  0.696 &  0.008 &  &  0.702 &  0.008 &  &  0.707 &  0.008 &  &  0.711 &  0.008\\
eMSM lin &  0.696 &  0.008 &  &  0.701 &  0.008 &  &  0.707 &  0.008 &  &  0.711 &  0.008\\
& \multicolumn{11}{c}{Lower bounds and SEs for $\mu^1 - \mu^0$} \\
MSM logit &  -0.113 &   0.015 &  &  -0.113 &   0.015 &  &  -0.113 &   0.015 &  &  -0.113 &   0.015\\
MSM lin &  -0.110 &   0.015 &  &  -0.110 &   0.015 &  &  -0.110 &   0.015 &  &  -0.110 &   0.015\\
eMSM logit &  -0.070 &   0.014 &  &  -0.086 &   0.015 &  &  -0.102 &   0.015 &  &  -0.113 &   0.015\\
eMSM lin &  -0.068 &   0.015 &  &  -0.084 &   0.015 &  &  -0.099 &   0.015 &  &  -0.110 &   0.015\\
& \multicolumn{11}{c}{Upper bounds and SEs for $\mu^1 - \mu^0$} \\
MSM logit &  -0.007 &   0.014 &  &  -0.007 &   0.014 &  &  -0.007 &   0.014 &  &  -0.007 &   0.014\\
MSM lin &  -0.008 &   0.014 &  &  -0.008 &   0.014 &  &  -0.008 &   0.014 &  &  -0.008 &   0.014\\
eMSM logit &  -0.049 &   0.014 &  &  -0.033 &   0.014 &  &  -0.018 &   0.014 &  &  -0.007 &   0.014\\
eMSM lin &  -0.048 &   0.014 &  &  -0.033 &   0.014 &  &  -0.018 &   0.014 &  &  -0.008 &   0.014\\
& \multicolumn{11}{c}{Lower bounds and SEs for $\mu^1 / \mu^0$} \\
MSM logit &  0.841 &  0.020 &  &  0.841 &  0.020 &  &  0.841 &  0.020 &  &  0.841 &  0.020\\
MSM lin &  0.846 &  0.021 &  &  0.846 &  0.021 &  &  0.846 &  0.021 &  &  0.846 &  0.021\\
eMSM logit &  0.900 &  0.020 &  &  0.878 &  0.020 &  &  0.856 &  0.020 &  &  0.841 &  0.020\\
eMSM lin &  0.902 &  0.021 &  &  0.881 &  0.021 &  &  0.860 &  0.021 &  &  0.846 &  0.021\\
& \multicolumn{11}{c}{Upper bounds and SEs for $\mu^1 / \mu^0$} \\
MSM logit &  0.989 &  0.021 &  &  0.989 &  0.021 &  &  0.989 &  0.021 &  &  0.989 &  0.021\\
MSM lin &  0.988 &  0.021 &  &  0.988 &  0.021 &  &  0.988 &  0.021 &  &  0.988 &  0.021\\
eMSM logit &  0.929 &  0.020 &  &  0.951 &  0.020 &  &  0.974 &  0.020 &  &  0.989 &  0.021\\
eMSM lin &  0.930 &  0.021 &  &  0.952 &  0.021 &  &  0.973 &  0.021 &  &  0.988 &  0.021\\
\hline
\end{tabular}}
\end{center}
\setlength{\baselineskip}{0.5\baselineskip}
\vspace{-.15in}\noindent{\tiny
\textbf{Note}: MSM logit and eMSM logit denote CAL under MSM and eMSM using logistic outcome mean regression. MSM lin and eMSM lin denote CAL under MSM and eMSM using linear outcome mean regression.}
\vspace{-.2in}
\end{table}

\begin{table}
\caption{\footnotesize Point bounds and SEs for DV method under DV model with $\Lambda = 1.2$, using working models with only main effects in the RHC study.} \label{tb:rhc-lowdim-dv-lam1.2} \vspace{-4ex}
\begin{center}
\renewcommand{\arraystretch}{0.6}
\small
\resizebox{\textwidth}{!}{\begin{tabular}{lccccccccccc}
\hline
& \multicolumn{2}{c}{$\delta = 0.2$} & & \multicolumn{2}{c}{$\delta = 0.5$} & & \multicolumn{2}{c}{$\delta = 0.8$} & & \multicolumn{2}{c}{$\delta = 1$} \\
& Bounds & SEs & & Bounds & SEs & & Bounds & SEs & & Bounds & SEs \\
\hline
& \multicolumn{11}{c}{Lower bounds and SEs for $\mu^1$} \\
DV $\Theta^+/2$ &  0.625 &  0.012 &  &  0.601 &  0.012 &  &  0.574 &  0.011 &  &  0.560 &  0.011\\
DV $\Theta^+$ &  0.602 &  0.012 &  &  0.581 &  0.011 &  &  0.567 &  0.011 &  &  0.560 &  0.011\\
DV $3\Theta^+/2$ &  0.588 &  0.012 &  &  0.574 &  0.011 &  &  0.565 &  0.011 &  &  0.560 &  0.011\\
& \multicolumn{11}{c}{Upper bounds and SEs for $\mu^1$} \\
DV $\Theta^+/2$ &  0.625 &  0.012 &  &  0.651 &  0.013 &  &  0.685 &  0.014 &  &  0.703 &  0.014\\
DV $\Theta^+$ &  0.650 &  0.013 &  &  0.676 &  0.013 &  &  0.694 &  0.014 &  &  0.703 &  0.014\\
DV $3\Theta^+/2$ &  0.666 &  0.013 &  &  0.684 &  0.014 &  &  0.697 &  0.014 &  &  0.703 &  0.014\\
& \multicolumn{11}{c}{Lower bounds and SEs for $\mu^0$} \\
DV $\Theta^+/2$ &  0.690 &  0.008 &  &  0.673 &  0.008 &  &  0.655 &  0.007 &  &  0.646 &  0.007\\
DV $\Theta^+$ &  0.674 &  0.008 &  &  0.660 &  0.007 &  &  0.651 &  0.007 &  &  0.646 &  0.007\\
DV $3\Theta^+/2$ &  0.665 &  0.007 &  &  0.655 &  0.007 &  &  0.649 &  0.007 &  &  0.646 &  0.007\\
& \multicolumn{11}{c}{Upper bounds and SEs for $\mu^0$} \\
DV $\Theta^+/2$ &  0.690 &  0.009 &  &  0.707 &  0.009 &  &  0.729 &  0.009 &  &  0.742 &  0.009\\
DV $\Theta^+$ &  0.706 &  0.009 &  &  0.723 &  0.009 &  &  0.735 &  0.009 &  &  0.742 &  0.009\\
DV $3\Theta^+/2$ &  0.717 &  0.009 &  &  0.729 &  0.009 &  &  0.737 &  0.009 &  &  0.742 &  0.009\\
& \multicolumn{11}{c}{Lower bounds and SEs for $\mu^1 - \mu^0$} \\
DV $\Theta^+/2$ &  -0.064 &   0.014 &  &  -0.106 &   0.013 &  &  -0.156 &   0.013 &  &  -0.181 &   0.013\\
DV $\Theta^+$ &  -0.104 &   0.013 &  &  -0.143 &   0.013 &  &  -0.168 &   0.013 &  &  -0.181 &   0.013\\
DV $3\Theta^+/2$ &  -0.129 &   0.013 &  &  -0.155 &   0.013 &  &  -0.173 &   0.013 &  &  -0.181 &   0.013\\
& \multicolumn{11}{c}{Upper bounds and SEs for $\mu^1 - \mu^0$} \\
DV $\Theta^+/2$ &  -0.064 &   0.015 &  &  -0.022 &   0.015 &  &   0.030 &   0.015 &  &   0.057 &   0.016\\
DV $\Theta^+$ &  -0.025 &   0.015 &  &   0.016 &   0.015 &  &   0.043 &   0.016 &  &   0.057 &   0.016\\
DV $3\Theta^+/2$ &   0.001 &   0.015 &  &   0.029 &   0.015 &  &   0.048 &   0.016 &  &   0.057 &   0.016\\
& \multicolumn{11}{c}{Lower bounds and SEs for $\mu^1 / \mu^0$} \\
DV $\Theta^+/2$ &  0.907 &  0.019 &  &  0.850 &  0.018 &  &  0.786 &  0.017 &  &  0.756 &  0.016\\
DV $\Theta^+$ &  0.853 &  0.018 &  &  0.803 &  0.017 &  &  0.771 &  0.016 &  &  0.756 &  0.016\\
DV $3\Theta^+/2$ &  0.821 &  0.017 &  &  0.787 &  0.017 &  &  0.766 &  0.016 &  &  0.756 &  0.016\\
& \multicolumn{11}{c}{Upper bounds and SEs for $\mu^1 / \mu^0$} \\
DV $\Theta^+/2$ &  0.907 &  0.021 &  &  0.967 &  0.022 &  &  1.045 &  0.024 &  &  1.088 &  0.025\\
DV $\Theta^+$ &  0.964 &  0.022 &  &  1.024 &  0.024 &  &  1.066 &  0.025 &  &  1.088 &  0.025\\
DV $3\Theta^+/2$ &  1.002 &  0.023 &  &  1.044 &  0.024 &  &  1.073 &  0.025 &  &  1.088 &  0.025\\
\hline
\end{tabular}}
\end{center}
\setlength{\baselineskip}{0.5\baselineskip}
\vspace{-.15in}\noindent{\tiny
\textbf{Note}: DV $\Theta^+ / 2$,  DV $\Theta^+$ and DV $3\Theta^+ / 2$ denote DV method under DV model with $\Theta = \Theta^+ / 2$, $\Theta = \Theta^+$ and $\Theta = 3\Theta^+ / 2$. SEs for DV method are computed by matching the lower and upper points of the 90\% sensitivity interval with 
``lower point bound$- z_{0.05}$ SE'' and ``upper point bound$+ z_{0.05}$ SE'' respectively, where $z_{0.05}$ is the 95\% quantile of the standard normal distribution.}
\vspace{-.2in}
\end{table}

\begin{table}
\caption{\footnotesize Point bounds and SEs for CAL under MSM and eMSM with $\Lambda = 1.5$, using working models with only main effects in the RHC study.} \label{tb:rhc-lowdim-cal-lam1.5} \vspace{-4ex}
\begin{center}
\renewcommand{\arraystretch}{0.6}
\small
\resizebox{\textwidth}{!}{\begin{tabular}{lccccccccccc}
\hline
& \multicolumn{2}{c}{$\delta = 0.2$} & & \multicolumn{2}{c}{$\delta = 0.5$} & & \multicolumn{2}{c}{$\delta = 0.8$} & & \multicolumn{2}{c}{$\delta = 1$} \\
& Bounds & SEs & & Bounds & SEs & & Bounds & SEs & & Bounds & SEs \\
\hline
& \multicolumn{11}{c}{Lower bounds and SEs for $\mu^1$} \\
MSM logit &  0.550 &  0.015 &  &  0.550 &  0.015 &  &  0.550 &  0.015 &  &  0.550 &  0.015\\
MSM lin &  0.555 &  0.015 &  &  0.555 &  0.015 &  &  0.555 &  0.015 &  &  0.555 &  0.015\\
eMSM logit &  0.617 &  0.013 &  &  0.592 &  0.013 &  &  0.567 &  0.014 &  &  0.550 &  0.015\\
eMSM lin &  0.619 &  0.013 &  &  0.595 &  0.014 &  &  0.571 &  0.015 &  &  0.555 &  0.015\\
& \multicolumn{11}{c}{Upper bounds and SEs for $\mu^1$} \\
MSM logit &  0.700 &  0.011 &  &  0.700 &  0.011 &  &  0.700 &  0.011 &  &  0.700 &  0.011\\
MSM lin &  0.696 &  0.012 &  &  0.696 &  0.012 &  &  0.696 &  0.012 &  &  0.696 &  0.012\\
eMSM logit &  0.647 &  0.012 &  &  0.667 &  0.012 &  &  0.687 &  0.011 &  &  0.700 &  0.011\\
eMSM lin &  0.647 &  0.012 &  &  0.666 &  0.012 &  &  0.684 &  0.012 &  &  0.696 &  0.012\\
& \multicolumn{11}{c}{Lower bounds and SEs for $\mu^0$} \\
MSM logit &  0.645 &  0.009 &  &  0.645 &  0.009 &  &  0.645 &  0.009 &  &  0.645 &  0.009\\
MSM lin &  0.645 &  0.010 &  &  0.645 &  0.010 &  &  0.645 &  0.010 &  &  0.645 &  0.010\\
eMSM logit &  0.683 &  0.008 &  &  0.669 &  0.009 &  &  0.654 &  0.009 &  &  0.645 &  0.009\\
eMSM lin &  0.683 &  0.008 &  &  0.669 &  0.009 &  &  0.655 &  0.009 &  &  0.645 &  0.010\\
& \multicolumn{11}{c}{Upper bounds and SEs for $\mu^0$} \\
MSM logit &  0.730 &  0.007 &  &  0.730 &  0.007 &  &  0.730 &  0.007 &  &  0.730 &  0.007\\
MSM lin &  0.730 &  0.007 &  &  0.730 &  0.007 &  &  0.730 &  0.007 &  &  0.730 &  0.007\\
eMSM logit &  0.700 &  0.008 &  &  0.711 &  0.008 &  &  0.723 &  0.007 &  &  0.730 &  0.007\\
eMSM lin &  0.700 &  0.008 &  &  0.711 &  0.008 &  &  0.723 &  0.007 &  &  0.730 &  0.007\\
& \multicolumn{11}{c}{Lower bounds and SEs for $\mu^1 - \mu^0$} \\
MSM logit &  -0.180 &   0.016 &  &  -0.180 &   0.016 &  &  -0.180 &   0.016 &  &  -0.180 &   0.016\\
MSM lin &  -0.175 &   0.016 &  &  -0.175 &   0.016 &  &  -0.175 &   0.016 &  &  -0.175 &   0.016\\
eMSM logit &  -0.084 &   0.015 &  &  -0.120 &   0.015 &  &  -0.156 &   0.015 &  &  -0.180 &   0.016\\
eMSM lin &  -0.081 &   0.015 &  &  -0.117 &   0.015 &  &  -0.152 &   0.016 &  &  -0.175 &   0.016\\
& \multicolumn{11}{c}{Upper bounds and SEs for $\mu^1 - \mu^0$} \\
MSM logit &   0.056 &   0.014 &  &   0.056 &   0.014 &  &   0.056 &   0.014 &  &   0.056 &   0.014\\
MSM lin &   0.051 &   0.015 &  &   0.051 &   0.015 &  &   0.051 &   0.015 &  &   0.051 &   0.015\\
eMSM logit &  -0.036 &   0.014 &  &  -0.002 &   0.014 &  &   0.033 &   0.014 &  &   0.056 &   0.014\\
eMSM lin &  -0.036 &   0.014 &  &  -0.003 &   0.014 &  &   0.029 &   0.014 &  &   0.051 &   0.015\\
& \multicolumn{11}{c}{Lower bounds and SEs for $\mu^1 / \mu^0$} \\
MSM logit &  0.753 &  0.021 &  &  0.753 &  0.021 &  &  0.753 &  0.021 &  &  0.753 &  0.021\\
MSM lin &  0.760 &  0.022 &  &  0.760 &  0.022 &  &  0.760 &  0.022 &  &  0.760 &  0.022\\
eMSM logit &  0.881 &  0.020 &  &  0.831 &  0.020 &  &  0.784 &  0.020 &  &  0.753 &  0.021\\
eMSM lin &  0.884 &  0.021 &  &  0.836 &  0.021 &  &  0.790 &  0.021 &  &  0.760 &  0.022\\
& \multicolumn{11}{c}{Upper bounds and SEs for $\mu^1 / \mu^0$} \\
MSM logit &  1.086 &  0.022 &  &  1.086 &  0.022 &  &  1.086 &  0.022 &  &  1.086 &  0.022\\
MSM lin &  1.079 &  0.023 &  &  1.079 &  0.023 &  &  1.079 &  0.023 &  &  1.079 &  0.023\\
eMSM logit &  0.947 &  0.020 &  &  0.997 &  0.021 &  &  1.050 &  0.022 &  &  1.086 &  0.022\\
eMSM lin &  0.947 &  0.021 &  &  0.995 &  0.021 &  &  1.045 &  0.022 &  &  1.079 &  0.023\\
\hline
\end{tabular}}
\end{center}
\setlength{\baselineskip}{0.5\baselineskip}
\vspace{-.15in}\noindent{\tiny
\textbf{Note}: MSM logit and eMSM logit denote CAL under MSM and eMSM using logistic outcome mean regression. MSM lin and eMSM lin denote CAL under MSM and eMSM using linear outcome mean regression.}
\vspace{-.2in}
\end{table}

\begin{table}
\caption{\footnotesize Point bounds and SEs for DV method under DV model with $\Lambda = 1.5$, using working models with only main effects in the RHC study.} \label{tb:rhc-lowdim-dv-lam1.5} \vspace{-4ex}
\begin{center}
\renewcommand{\arraystretch}{0.6}
\small
\resizebox{\textwidth}{!}{\begin{tabular}{lccccccccccc}
\hline
& \multicolumn{2}{c}{$\delta = 0.2$} & & \multicolumn{2}{c}{$\delta = 0.5$} & & \multicolumn{2}{c}{$\delta = 0.8$} & & \multicolumn{2}{c}{$\delta = 1$} \\
& Bounds & SEs & & Bounds & SEs & & Bounds & SEs & & Bounds & SEs \\
\hline
& \multicolumn{11}{c}{Lower bounds and SEs for $\mu^1$} \\
DV $\Theta^+/2$ &  0.625 &  0.012 &  &  0.570 &  0.011 &  &  0.519 &  0.010 &  &  0.495 &  0.009\\
DV $\Theta^+$ &  0.575 &  0.011 &  &  0.533 &  0.010 &  &  0.507 &  0.010 &  &  0.495 &  0.009\\
DV $3\Theta^+/2$ &  0.549 &  0.010 &  &  0.520 &  0.010 &  &  0.503 &  0.009 &  &  0.495 &  0.009\\
& \multicolumn{11}{c}{Upper bounds and SEs for $\mu^1$} \\
DV $\Theta^+/2$ &  0.625 &  0.012 &  &  0.690 &  0.014 &  &  0.771 &  0.016 &  &  0.820 &  0.017\\
DV $\Theta^+$ &  0.682 &  0.014 &  &  0.747 &  0.015 &  &  0.794 &  0.017 &  &  0.820 &  0.017\\
DV $3\Theta^+/2$ &  0.720 &  0.015 &  &  0.769 &  0.016 &  &  0.803 &  0.017 &  &  0.820 &  0.017\\
& \multicolumn{11}{c}{Lower bounds and SEs for $\mu^0$} \\
DV $\Theta^+/2$ &  0.690 &  0.008 &  &  0.652 &  0.007 &  &  0.619 &  0.007 &  &  0.603 &  0.007\\
DV $\Theta^+$ &  0.656 &  0.007 &  &  0.628 &  0.007 &  &  0.611 &  0.007 &  &  0.603 &  0.007\\
DV $3\Theta^+/2$ &  0.639 &  0.007 &  &  0.619 &  0.007 &  &  0.608 &  0.007 &  &  0.603 &  0.007\\
& \multicolumn{11}{c}{Upper bounds and SEs for $\mu^0$} \\
DV $\Theta^+/2$ &  0.690 &  0.009 &  &  0.733 &  0.009 &  &  0.787 &  0.011 &  &  0.820 &  0.011\\
DV $\Theta^+$ &  0.728 &  0.009 &  &  0.771 &  0.010 &  &  0.803 &  0.011 &  &  0.820 &  0.011\\
DV $3\Theta^+/2$ &  0.753 &  0.010 &  &  0.786 &  0.011 &  &  0.808 &  0.011 &  &  0.820 &  0.011\\
& \multicolumn{11}{c}{Lower bounds and SEs for $\mu^1 - \mu^0$} \\
DV $\Theta^+/2$ &  -0.064 &   0.014 &  &  -0.163 &   0.013 &  &  -0.268 &   0.013 &  &  -0.324 &   0.013\\
DV $\Theta^+$ &  -0.153 &   0.013 &  &  -0.238 &   0.013 &  &  -0.295 &   0.013 &  &  -0.324 &   0.013\\
DV $3\Theta^+/2$ &  -0.205 &   0.013 &  &  -0.265 &   0.013 &  &  -0.305 &   0.013 &  &  -0.324 &   0.013\\
& \multicolumn{11}{c}{Upper bounds and SEs for $\mu^1 - \mu^0$} \\
DV $\Theta^+/2$ &  -0.064 &   0.015 &  &   0.038 &   0.016 &  &   0.152 &   0.017 &  &   0.217 &   0.018\\
DV $\Theta^+$ &   0.026 &   0.015 &  &   0.119 &   0.017 &  &   0.184 &   0.018 &  &   0.217 &   0.018\\
DV $3\Theta^+/2$ &   0.082 &   0.016 &  &   0.150 &   0.017 &  &   0.194 &   0.018 &  &   0.217 &   0.018\\
& \multicolumn{11}{c}{Lower bounds and SEs for $\mu^1 / \mu^0$} \\
DV $\Theta^+/2$ &  0.907 &  0.019 &  &  0.777 &  0.016 &  &  0.659 &  0.014 &  &  0.605 &  0.012\\
DV $\Theta^+$ &  0.790 &  0.017 &  &  0.691 &  0.014 &  &  0.632 &  0.013 &  &  0.605 &  0.012\\
DV $3\Theta^+/2$ &  0.728 &  0.015 &  &  0.662 &  0.014 &  &  0.623 &  0.013 &  &  0.605 &  0.012\\
& \multicolumn{11}{c}{Upper bounds and SEs for $\mu^1 / \mu^0$} \\
DV $\Theta^+/2$ &  0.907 &  0.021 &  &  1.058 &  0.024 &  &  1.246 &  0.030 &  &  1.360 &  0.033\\
DV $\Theta^+$ &  1.040 &  0.024 &  &  1.190 &  0.028 &  &  1.300 &  0.031 &  &  1.360 &  0.033\\
DV $3\Theta^+/2$ &  1.128 &  0.026 &  &  1.241 &  0.029 &  &  1.320 &  0.031 &  &  1.360 &  0.033\\
\hline
\end{tabular}}
\end{center}
\setlength{\baselineskip}{0.5\baselineskip}
\vspace{-.15in}\noindent{\tiny
\textbf{Note}: DV $\Theta^+ / 2$,  DV $\Theta^+$ and DV $3\Theta^+ / 2$ denote DV method under DV model with $\Theta = \Theta^+ / 2$, $\Theta = \Theta^+$ and $\Theta = 3\Theta^+ / 2$. SEs for DV method are computed by matching the lower and upper points of the 90\% sensitivity interval with 
``lower point bound$- z_{0.05}$ SE'' and ``upper point bound$+ z_{0.05}$ SE'' respectively, where $z_{0.05}$ is the 95\% quantile of the standard normal distribution.}
\vspace{-.2in}
\end{table}

\begin{table}
\caption{\footnotesize Point bounds and SEs for CAL under MSM and eMSM with $\Lambda = 2$, using working models with only main effects in the RHC study.} \label{tb:rhc-lowdim-cal-lam2.0} \vspace{-4ex}
\begin{center}
\renewcommand{\arraystretch}{0.6}
\small
\resizebox{\textwidth}{!}{\begin{tabular}{lccccccccccc}
\hline
& \multicolumn{2}{c}{$\delta = 0.2$} & & \multicolumn{2}{c}{$\delta = 0.5$} & & \multicolumn{2}{c}{$\delta = 0.8$} & & \multicolumn{2}{c}{$\delta = 1$} \\
& Bounds & SEs & & Bounds & SEs & & Bounds & SEs & & Bounds & SEs \\
\hline
& \multicolumn{11}{c}{Lower bounds and SEs for $\mu^1$} \\
MSM logit &  0.481 &  0.014 &  &  0.481 &  0.014 &  &  0.481 &  0.014 &  &  0.481 &  0.014\\
MSM lin &  0.495 &  0.015 &  &  0.495 &  0.015 &  &  0.495 &  0.015 &  &  0.495 &  0.015\\
eMSM logit &  0.603 &  0.013 &  &  0.557 &  0.013 &  &  0.512 &  0.014 &  &  0.481 &  0.014\\
eMSM lin &  0.607 &  0.013 &  &  0.565 &  0.014 &  &  0.523 &  0.014 &  &  0.495 &  0.015\\
& \multicolumn{11}{c}{Upper bounds and SEs for $\mu^1$} \\
MSM logit &  0.742 &  0.011 &  &  0.742 &  0.011 &  &  0.742 &  0.011 &  &  0.742 &  0.011\\
MSM lin &  0.734 &  0.013 &  &  0.734 &  0.013 &  &  0.734 &  0.013 &  &  0.734 &  0.013\\
eMSM logit &  0.655 &  0.012 &  &  0.688 &  0.012 &  &  0.720 &  0.011 &  &  0.742 &  0.011\\
eMSM lin &  0.654 &  0.012 &  &  0.684 &  0.012 &  &  0.714 &  0.012 &  &  0.734 &  0.013\\
& \multicolumn{11}{c}{Lower bounds and SEs for $\mu^0$} \\
MSM logit &  0.603 &  0.010 &  &  0.603 &  0.010 &  &  0.603 &  0.010 &  &  0.603 &  0.010\\
MSM lin &  0.608 &  0.010 &  &  0.608 &  0.010 &  &  0.608 &  0.010 &  &  0.608 &  0.010\\
eMSM logit &  0.675 &  0.008 &  &  0.648 &  0.009 &  &  0.621 &  0.009 &  &  0.603 &  0.010\\
eMSM lin &  0.675 &  0.008 &  &  0.650 &  0.009 &  &  0.625 &  0.009 &  &  0.608 &  0.010\\
& \multicolumn{11}{c}{Upper bounds and SEs for $\mu^0$} \\
MSM logit &  0.750 &  0.007 &  &  0.750 &  0.007 &  &  0.750 &  0.007 &  &  0.750 &  0.007\\
MSM lin &  0.751 &  0.007 &  &  0.751 &  0.007 &  &  0.751 &  0.007 &  &  0.751 &  0.007\\
eMSM logit &  0.704 &  0.008 &  &  0.721 &  0.007 &  &  0.739 &  0.007 &  &  0.750 &  0.007\\
eMSM lin &  0.704 &  0.008 &  &  0.722 &  0.007 &  &  0.739 &  0.007 &  &  0.751 &  0.007\\
& \multicolumn{11}{c}{Lower bounds and SEs for $\mu^1 - \mu^0$} \\
MSM logit &  -0.269 &   0.015 &  &  -0.269 &   0.015 &  &  -0.269 &   0.015 &  &  -0.269 &   0.015\\
MSM lin &  -0.256 &   0.016 &  &  -0.256 &   0.016 &  &  -0.256 &   0.016 &  &  -0.256 &   0.016\\
eMSM logit &  -0.101 &   0.014 &  &  -0.164 &   0.015 &  &  -0.227 &   0.015 &  &  -0.269 &   0.015\\
eMSM lin &  -0.097 &   0.015 &  &  -0.157 &   0.015 &  &  -0.216 &   0.015 &  &  -0.256 &   0.016\\
& \multicolumn{11}{c}{Upper bounds and SEs for $\mu^1 - \mu^0$} \\
MSM logit &   0.138 &   0.014 &  &   0.138 &   0.014 &  &   0.138 &   0.014 &  &   0.138 &   0.014\\
MSM lin &   0.127 &   0.015 &  &   0.127 &   0.015 &  &   0.127 &   0.015 &  &   0.127 &   0.015\\
eMSM logit &  -0.020 &   0.014 &  &   0.040 &   0.014 &  &   0.099 &   0.014 &  &   0.138 &   0.014\\
eMSM lin &  -0.021 &   0.014 &  &   0.034 &   0.015 &  &   0.090 &   0.015 &  &   0.127 &   0.015\\
& \multicolumn{11}{c}{Lower bounds and SEs for $\mu^1 / \mu^0$} \\
MSM logit &  0.642 &  0.019 &  &  0.642 &  0.019 &  &  0.642 &  0.019 &  &  0.642 &  0.019\\
MSM lin &  0.659 &  0.020 &  &  0.659 &  0.020 &  &  0.659 &  0.020 &  &  0.659 &  0.020\\
eMSM logit &  0.856 &  0.020 &  &  0.772 &  0.019 &  &  0.693 &  0.019 &  &  0.642 &  0.019\\
eMSM lin &  0.862 &  0.020 &  &  0.783 &  0.020 &  &  0.708 &  0.020 &  &  0.659 &  0.020\\
& \multicolumn{11}{c}{Upper bounds and SEs for $\mu^1 / \mu^0$} \\
MSM logit &  1.229 &  0.026 &  &  1.229 &  0.026 &  &  1.229 &  0.026 &  &  1.229 &  0.026\\
MSM lin &  1.208 &  0.027 &  &  1.208 &  0.027 &  &  1.208 &  0.027 &  &  1.208 &  0.027\\
eMSM logit &  0.971 &  0.021 &  &  1.061 &  0.022 &  &  1.159 &  0.024 &  &  1.229 &  0.026\\
eMSM lin &  0.969 &  0.021 &  &  1.053 &  0.023 &  &  1.144 &  0.025 &  &  1.208 &  0.027\\
\hline
\end{tabular}}
\end{center}
\setlength{\baselineskip}{0.5\baselineskip}
\vspace{-.15in}\noindent{\tiny
\textbf{Note}: MSM logit and eMSM logit denote CAL under MSM and eMSM using logistic outcome mean regression. MSM lin and eMSM lin denote CAL under MSM and eMSM using linear outcome mean regression. The upper or lower point of the 90\% sensitivity interval is calculated as ``upper point bound$ + z_{0.05}$ SE'' or ``lower point bound$ -  z_{0.05}$ SE'' respectively. For example, as mentioned in the main text, the upper interval point for $\mu^1 - \mu^0$ from eMSM logit at $(\Lambda = 2, \delta = 0.2)$ is $-0.020 +  z_{0.05}\times 0.014 \approx 0.0030$.}
\vspace{-.2in}
\end{table}

\begin{table}
\caption{\footnotesize Point bounds and SEs for DV method under DV model with $\Lambda = 2$, using working models with only main effects in the RHC study.} \label{tb:rhc-lowdim-dv-lam2.0} \vspace{-4ex}
\begin{center}
\renewcommand{\arraystretch}{0.6}
\small
\resizebox{\textwidth}{!}{\begin{tabular}{lccccccccccc}
\hline
& \multicolumn{2}{c}{$\delta = 0.2$} & & \multicolumn{2}{c}{$\delta = 0.5$} & & \multicolumn{2}{c}{$\delta = 0.8$} & & \multicolumn{2}{c}{$\delta = 1$} \\
& Bounds & SEs & & Bounds & SEs & & Bounds & SEs & & Bounds & SEs \\
\hline
& \multicolumn{11}{c}{Lower bounds and SEs for $\mu^1$} \\
DV $\Theta^+/2$ &  0.625 &  0.012 &  &  0.528 &  0.010 &  &  0.461 &  0.008 &  &  0.431 &  0.008\\
DV $\Theta^+$ &  0.542 &  0.010 &  &  0.479 &  0.009 &  &  0.446 &  0.008 &  &  0.431 &  0.008\\
DV $3\Theta^+/2$ &  0.505 &  0.009 &  &  0.463 &  0.008 &  &  0.441 &  0.008 &  &  0.431 &  0.008\\
& \multicolumn{11}{c}{Upper bounds and SEs for $\mu^1$} \\
DV $\Theta^+/2$ &  0.625 &  0.012 &  &  0.755 &  0.016 &  &  0.911 &  0.020 &  &  1.014 &  0.023\\
DV $\Theta^+$ &  0.731 &  0.015 &  &  0.859 &  0.019 &  &  0.959 &  0.021 &  &  1.014 &  0.023\\
DV $3\Theta^+/2$ &  0.800 &  0.017 &  &  0.903 &  0.020 &  &  0.976 &  0.022 &  &  1.014 &  0.023\\
& \multicolumn{11}{c}{Lower bounds and SEs for $\mu^0$} \\
DV $\Theta^+/2$ &  0.690 &  0.008 &  &  0.625 &  0.007 &  &  0.580 &  0.007 &  &  0.560 &  0.007\\
DV $\Theta^+$ &  0.634 &  0.007 &  &  0.592 &  0.007 &  &  0.570 &  0.007 &  &  0.560 &  0.007\\
DV $3\Theta^+/2$ &  0.609 &  0.007 &  &  0.581 &  0.007 &  &  0.566 &  0.007 &  &  0.560 &  0.007\\
 & \multicolumn{11}{c}{Upper bounds and SEs for $\mu^0$} \\
DV $\Theta^+/2$ &  0.690 &  0.009 &  &  0.776 &  0.010 &  &  0.880 &  0.012 &  &  0.950 &  0.014\\
DV $\Theta^+$ &  0.761 &  0.010 &  &  0.846 &  0.012 &  &  0.912 &  0.013 &  &  0.950 &  0.014\\
DV $3\Theta^+/2$ &  0.806 &  0.011 &  &  0.875 &  0.012 &  &  0.924 &  0.013 &  &  0.950 &  0.014\\
& \multicolumn{11}{c}{Lower bounds and SEs for $\mu^1 - \mu^0$} \\
DV $\Theta^+/2$ &  -0.064 &   0.014 &  &  -0.248 &   0.013 &  &  -0.420 &   0.014 &  &  -0.519 &   0.015\\
DV $\Theta^+$ &  -0.219 &   0.013 &  &  -0.366 &   0.013 &  &  -0.467 &   0.014 &  &  -0.519 &   0.015\\
DV $3\Theta^+/2$ &  -0.301 &   0.013 &  &  -0.412 &   0.014 &  &  -0.484 &   0.014 &  &  -0.519 &   0.015\\
& \multicolumn{11}{c}{Upper bounds and SEs for $\mu^1 - \mu^0$} \\
DV $\Theta^+/2$ &  -0.064 &   0.015 &  &   0.130 &   0.017 &  &   0.331 &   0.020 &  &   0.455 &   0.023\\
DV $\Theta^+$ &   0.097 &   0.016 &  &   0.267 &   0.019 &  &   0.389 &   0.021 &  &   0.455 &   0.023\\
DV $3\Theta^+/2$ &   0.190 &   0.018 &  &   0.322 &   0.020 &  &   0.410 &   0.022 &  &   0.455 &   0.023\\
& \multicolumn{11}{c}{Lower bounds and SEs for $\mu^1 / \mu^0$} \\
DV $\Theta^+/2$ &  0.907 &  0.019 &  &  0.680 &  0.014 &  &  0.523 &  0.010 &  &  0.453 &  0.009\\
DV $\Theta^+$ &  0.712 &  0.015 &  &  0.567 &  0.011 &  &  0.488 &  0.010 &  &  0.453 &  0.009\\
DV $3\Theta^+/2$ &  0.626 &  0.013 &  &  0.529 &  0.011 &  &  0.477 &  0.009 &  &  0.453 &  0.009\\
& \multicolumn{11}{c}{Upper bounds and SEs for $\mu^1 / \mu^0$} \\
DV $\Theta^+/2$ &  0.907 &  0.021 &  &  1.209 &  0.029 &  &  1.571 &  0.038 &  &  1.812 &  0.044\\
DV $\Theta^+$ &  1.154 &  0.027 &  &  1.450 &  0.035 &  &  1.683 &  0.040 &  &  1.812 &  0.044\\
DV $3\Theta^+/2$ &  1.313 &  0.031 &  &  1.554 &  0.037 &  &  1.724 &  0.042 &  &  1.812 &  0.044\\
\hline
\end{tabular}}
\end{center}
\setlength{\baselineskip}{0.5\baselineskip}
\vspace{-.15in}\noindent{\tiny
\textbf{Note}: DV $\Theta^+ / 2$,  DV $\Theta^+$ and DV $3\Theta^+ / 2$ denote DV method under DV model with $\Theta = \Theta^+ / 2$, $\Theta = \Theta^+$ and $\Theta = 3\Theta^+ / 2$. SEs for DV method are computed by matching the lower and upper points of the 90\% sensitivity interval with 
``lower point bound$- z_{0.05}$ SE'' and ``upper point bound$+ z_{0.05}$ SE'' respectively, where $z_{0.05}$ is the 95\% quantile of the standard normal distribution.}
\vspace{-.2in}
\end{table}

\begin{table}
\caption{\footnotesize Point bounds and SEs for RCAL under MSM and eMSM with $\Lambda = 1$, using working models with main effects and interactions in the RHC study.} \label{tb:rhc-highdim-cal-lam1.0} \vspace{-4ex}
\begin{center}
\renewcommand{\arraystretch}{0.6}
\small
\resizebox{\textwidth}{!}{\begin{tabular}{lccccccccccc}
\hline
& \multicolumn{2}{c}{$\delta = 0.2$} & & \multicolumn{2}{c}{$\delta = 0.5$} & & \multicolumn{2}{c}{$\delta = 0.8$} & & \multicolumn{2}{c}{$\delta = 1$} \\
& Bounds & SEs & & Bounds & SEs & & Bounds & SEs & & Bounds & SEs \\
\hline
& \multicolumn{11}{c}{Lower bounds and SEs for $\mu^1$} \\
MSM logit &  0.636 &  0.011 &  &  0.636 &  0.011 &  &  0.636 &  0.011 &  &  0.636 &  0.011\\
MSM lin &  0.636 &  0.011 &  &  0.636 &  0.011 &  &  0.636 &  0.011 &  &  0.636 &  0.011\\
eMSM logit &  0.636 &  0.011 &  &  0.636 &  0.011 &  &  0.636 &  0.011 &  &  0.636 &  0.011\\
eMSM lin &  0.636 &  0.011 &  &  0.636 &  0.011 &  &  0.636 &  0.011 &  &  0.636 &  0.011\\
& \multicolumn{11}{c}{Upper bounds and SEs for $\mu^1$} \\
MSM logit &  0.636 &  0.011 &  &  0.636 &  0.011 &  &  0.636 &  0.011 &  &  0.636 &  0.011\\
MSM lin &  0.636 &  0.011 &  &  0.636 &  0.011 &  &  0.636 &  0.011 &  &  0.636 &  0.011\\
eMSM logit &  0.636 &  0.011 &  &  0.636 &  0.011 &  &  0.636 &  0.011 &  &  0.636 &  0.011\\
eMSM lin &  0.636 &  0.011 &  &  0.636 &  0.011 &  &  0.636 &  0.011 &  &  0.636 &  0.011\\
& \multicolumn{11}{c}{Lower bounds and SEs for $\mu^0$} \\
MSM logit &  0.688 &  0.008 &  &  0.688 &  0.008 &  &  0.688 &  0.008 &  &  0.688 &  0.008\\
MSM lin &  0.687 &  0.008 &  &  0.687 &  0.008 &  &  0.687 &  0.008 &  &  0.687 &  0.008\\
eMSM logit &  0.688 &  0.008 &  &  0.688 &  0.008 &  &  0.688 &  0.008 &  &  0.688 &  0.008\\
eMSM lin &  0.688 &  0.008 &  &  0.687 &  0.008 &  &  0.687 &  0.008 &  &  0.687 &  0.008\\
& \multicolumn{11}{c}{Upper bounds and SEs for $\mu^0$} \\
MSM logit &  0.688 &  0.008 &  &  0.688 &  0.008 &  &  0.688 &  0.008 &  &  0.688 &  0.008\\
MSM lin &  0.687 &  0.008 &  &  0.687 &  0.008 &  &  0.687 &  0.008 &  &  0.687 &  0.008\\
eMSM logit &  0.688 &  0.008 &  &  0.688 &  0.008 &  &  0.688 &  0.008 &  &  0.688 &  0.008\\
eMSM lin &  0.688 &  0.008 &  &  0.687 &  0.008 &  &  0.687 &  0.008 &  &  0.687 &  0.008\\
& \multicolumn{11}{c}{Lower bounds and SEs for $\mu^1 - \mu^0$} \\
MSM logit &  -0.052 &   0.013 &  &  -0.052 &   0.013 &  &  -0.052 &   0.013 &  &  -0.052 &   0.013\\
MSM lin &  -0.052 &   0.013 &  &  -0.052 &   0.013 &  &  -0.052 &   0.013 &  &  -0.052 &   0.013\\
eMSM logit &  -0.052 &   0.013 &  &  -0.052 &   0.013 &  &  -0.052 &   0.013 &  &  -0.052 &   0.013\\
eMSM lin &  -0.052 &   0.013 &  &  -0.051 &   0.013 &  &  -0.052 &   0.013 &  &  -0.052 &   0.013\\
& \multicolumn{11}{c}{Upper bounds and SEs for $\mu^1 - \mu^0$} \\
MSM logit &  -0.052 &   0.013 &  &  -0.052 &   0.013 &  &  -0.052 &   0.013 &  &  -0.052 &   0.013\\
MSM lin &  -0.052 &   0.013 &  &  -0.052 &   0.013 &  &  -0.052 &   0.013 &  &  -0.052 &   0.013\\
eMSM logit &  -0.052 &   0.013 &  &  -0.052 &   0.013 &  &  -0.052 &   0.013 &  &  -0.052 &   0.013\\
eMSM lin &  -0.052 &   0.013 &  &  -0.051 &   0.013 &  &  -0.052 &   0.013 &  &  -0.052 &   0.013\\
& \multicolumn{11}{c}{Lower bounds and SEs for $\mu^1 / \mu^0$} \\
MSM logit &  0.924 &  0.018 &  &  0.924 &  0.018 &  &  0.924 &  0.018 &  &  0.924 &  0.018\\
MSM lin &  0.925 &  0.018 &  &  0.925 &  0.018 &  &  0.925 &  0.018 &  &  0.925 &  0.018\\
eMSM logit &  0.924 &  0.018 &  &  0.925 &  0.018 &  &  0.925 &  0.018 &  &  0.924 &  0.018\\
eMSM lin &  0.925 &  0.018 &  &  0.925 &  0.018 &  &  0.925 &  0.018 &  &  0.925 &  0.018\\
& \multicolumn{11}{c}{Upper bounds and SEs for $\mu^1 / \mu^0$} \\
MSM logit &  0.924 &  0.018 &  &  0.924 &  0.018 &  &  0.924 &  0.018 &  &  0.924 &  0.018\\
MSM lin &  0.925 &  0.018 &  &  0.925 &  0.018 &  &  0.925 &  0.018 &  &  0.925 &  0.018\\
eMSM logit &  0.924 &  0.018 &  &  0.925 &  0.018 &  &  0.925 &  0.018 &  &  0.924 &  0.018\\
eMSM lin &  0.925 &  0.018 &  &  0.925 &  0.018 &  &  0.925 &  0.018 &  &  0.925 &  0.018\\
\hline
\end{tabular}}
\end{center}
\setlength{\baselineskip}{0.5\baselineskip}
\vspace{-.15in}\noindent{\tiny
\textbf{Note}: MSM logit and eMSM logit denote RCAL under MSM and eMSM using logistic outcome mean regression. MSM lin and eMSM lin denote RCAL under MSM and eMSM using linear outcome mean regression.}
\vspace{-.2in}
\end{table}

\begin{table}
\caption{\footnotesize Point bounds and SEs for RCAL under MSM and eMSM with $\Lambda = 1.2$, using working models with main effects and interactions in the RHC study.} \label{tb:rhc-highdim-cal-lam1.2} \vspace{-4ex}
\begin{center}
\renewcommand{\arraystretch}{0.6}
\small
\resizebox{\textwidth}{!}{\begin{tabular}{lccccccccccc}
\hline
& \multicolumn{2}{c}{$\delta = 0.2$} & & \multicolumn{2}{c}{$\delta = 0.5$} & & \multicolumn{2}{c}{$\delta = 0.8$} & & \multicolumn{2}{c}{$\delta = 1$} \\
& Bounds & SEs & & Bounds & SEs & & Bounds & SEs & & Bounds & SEs \\
\hline
& \multicolumn{11}{c}{Lower bounds and SEs for $\mu^1$} \\
MSM logit &  0.600 &  0.012 &  &  0.600 &  0.012 &  &  0.600 &  0.012 &  &  0.600 &  0.012\\
MSM lin &  0.601 &  0.012 &  &  0.601 &  0.012 &  &  0.601 &  0.012 &  &  0.601 &  0.012\\
eMSM logit &  0.629 &  0.011 &  &  0.618 &  0.011 &  &  0.608 &  0.012 &  &  0.600 &  0.012\\
eMSM lin &  0.629 &  0.011 &  &  0.618 &  0.011 &  &  0.607 &  0.011 &  &  0.601 &  0.012\\
& \multicolumn{11}{c}{Upper bounds and SEs for $\mu^1$} \\
MSM logit &  0.667 &  0.010 &  &  0.667 &  0.010 &  &  0.667 &  0.010 &  &  0.667 &  0.010\\
MSM lin &  0.666 &  0.010 &  &  0.666 &  0.010 &  &  0.666 &  0.010 &  &  0.666 &  0.010\\
eMSM logit &  0.642 &  0.011 &  &  0.652 &  0.010 &  &  0.661 &  0.010 &  &  0.667 &  0.010\\
eMSM lin &  0.642 &  0.011 &  &  0.651 &  0.010 &  &  0.660 &  0.010 &  &  0.666 &  0.010\\
& \multicolumn{11}{c}{Lower bounds and SEs for $\mu^0$} \\
MSM logit &  0.668 &  0.008 &  &  0.668 &  0.008 &  &  0.668 &  0.008 &  &  0.668 &  0.008\\
MSM lin &  0.669 &  0.008 &  &  0.669 &  0.008 &  &  0.669 &  0.008 &  &  0.669 &  0.008\\
eMSM logit &  0.684 &  0.008 &  &  0.678 &  0.008 &  &  0.672 &  0.008 &  &  0.668 &  0.008\\
eMSM lin &  0.684 &  0.008 &  &  0.678 &  0.008 &  &  0.672 &  0.008 &  &  0.669 &  0.008\\
& \multicolumn{11}{c}{Upper bounds and SEs for $\mu^0$} \\
MSM logit &  0.706 &  0.007 &  &  0.706 &  0.007 &  &  0.706 &  0.007 &  &  0.706 &  0.007\\
MSM lin &  0.706 &  0.007 &  &  0.706 &  0.007 &  &  0.706 &  0.007 &  &  0.706 &  0.007\\
eMSM logit &  0.691 &  0.008 &  &  0.697 &  0.008 &  &  0.702 &  0.008 &  &  0.706 &  0.007\\
eMSM lin &  0.691 &  0.008 &  &  0.696 &  0.008 &  &  0.702 &  0.008 &  &  0.706 &  0.007\\
& \multicolumn{11}{c}{Lower bounds and SEs for $\mu^1 - \mu^0$} \\
MSM logit &  -0.106 &   0.013 &  &  -0.106 &   0.013 &  &  -0.106 &   0.013 &  &  -0.106 &   0.013\\
MSM lin &  -0.105 &   0.013 &  &  -0.105 &   0.013 &  &  -0.105 &   0.013 &  &  -0.105 &   0.013\\
eMSM logit &  -0.063 &   0.013 &  &  -0.079 &   0.013 &  &  -0.095 &   0.013 &  &  -0.106 &   0.013\\
eMSM lin &  -0.062 &   0.013 &  &  -0.078 &   0.013 &  &  -0.094 &   0.013 &  &  -0.105 &   0.013\\
& \multicolumn{11}{c}{Upper bounds and SEs for $\mu^1 - \mu^0$} \\
MSM logit &  -0.001 &   0.012 &  &  -0.001 &   0.012 &  &  -0.001 &   0.012 &  &  -0.001 &   0.012\\
MSM lin &  -0.003 &   0.012 &  &  -0.003 &   0.012 &  &  -0.003 &   0.012 &  &  -0.003 &   0.012\\
eMSM logit &  -0.042 &   0.013 &  &  -0.026 &   0.012 &  &  -0.012 &   0.012 &  &  -0.001 &   0.012\\
eMSM lin &  -0.042 &   0.013 &  &  -0.027 &   0.013 &  &  -0.012 &   0.012 &  &  -0.003 &   0.012\\
& \multicolumn{11}{c}{Lower bounds and SEs for $\mu^1 / \mu^0$} \\
MSM logit &  0.850 &  0.018 &  &  0.850 &  0.018 &  &  0.850 &  0.018 &  &  0.850 &  0.018\\
MSM lin &  0.851 &  0.018 &  &  0.851 &  0.018 &  &  0.851 &  0.018 &  &  0.851 &  0.018\\
eMSM logit &  0.909 &  0.018 &  &  0.886 &  0.018 &  &  0.865 &  0.018 &  &  0.850 &  0.018\\
eMSM lin &  0.910 &  0.018 &  &  0.888 &  0.018 &  &  0.865 &  0.018 &  &  0.851 &  0.018\\
& \multicolumn{11}{c}{Upper bounds and SEs for $\mu^1 / \mu^0$} \\
MSM logit &  0.998 &  0.018 &  &  0.998 &  0.018 &  &  0.998 &  0.018 &  &  0.998 &  0.018\\
MSM lin &  0.996 &  0.018 &  &  0.996 &  0.018 &  &  0.996 &  0.018 &  &  0.996 &  0.018\\
eMSM logit &  0.939 &  0.018 &  &  0.961 &  0.018 &  &  0.983 &  0.018 &  &  0.998 &  0.018\\
eMSM lin &  0.939 &  0.018 &  &  0.961 &  0.018 &  &  0.981 &  0.018 &  &  0.996 &  0.018\\
\hline
\end{tabular}}
\end{center}
\setlength{\baselineskip}{0.5\baselineskip}
\vspace{-.15in}\noindent{\tiny
\textbf{Note}: MSM logit and eMSM logit denote RCAL under MSM and eMSM using logistic outcome mean regression. MSM lin and eMSM lin denote RCAL under MSM and eMSM using linear outcome mean regression.}
\vspace{-.2in}
\end{table}

\begin{table}
\caption{\footnotesize Point bounds and SEs for RCAL under MSM and eMSM with $\Lambda = 1.5$, using working models with main effects and interactions in the RHC study.} \label{tb:rhc-highdim-cal-lam1.5} \vspace{-4ex}
\begin{center}
\renewcommand{\arraystretch}{0.6}
\small
\resizebox{\textwidth}{!}{\begin{tabular}{lccccccccccc}
\hline
& \multicolumn{2}{c}{$\delta = 0.2$} & & \multicolumn{2}{c}{$\delta = 0.5$} & & \multicolumn{2}{c}{$\delta = 0.8$} & & \multicolumn{2}{c}{$\delta = 1$} \\
& Bounds & SEs & & Bounds & SEs & & Bounds & SEs & & Bounds & SEs \\
\hline
& \multicolumn{11}{c}{Lower bounds and SEs for $\mu^1$} \\
MSM logit &  0.562 &  0.012 &  &  0.562 &  0.012 &  &  0.562 &  0.012 &  &  0.562 &  0.012\\
MSM lin &  0.567 &  0.012 &  &  0.567 &  0.012 &  &  0.567 &  0.012 &  &  0.567 &  0.012\\
eMSM logit &  0.621 &  0.011 &  &  0.600 &  0.011 &  &  0.577 &  0.012 &  &  0.562 &  0.012\\
eMSM lin &  0.622 &  0.011 &  &  0.602 &  0.011 &  &  0.581 &  0.012 &  &  0.567 &  0.012\\
& \multicolumn{11}{c}{Upper bounds and SEs for $\mu^1$} \\
MSM logit &  0.700 &  0.009 &  &  0.700 &  0.009 &  &  0.700 &  0.009 &  &  0.700 &  0.009\\
MSM lin &  0.698 &  0.009 &  &  0.698 &  0.009 &  &  0.698 &  0.009 &  &  0.698 &  0.009\\
eMSM logit &  0.649 &  0.010 &  &  0.668 &  0.010 &  &  0.687 &  0.009 &  &  0.700 &  0.009\\
eMSM lin &  0.648 &  0.010 &  &  0.667 &  0.010 &  &  0.686 &  0.009 &  &  0.698 &  0.009\\
& \multicolumn{11}{c}{Lower bounds and SEs for $\mu^0$} \\
MSM logit &  0.641 &  0.009 &  &  0.641 &  0.009 &  &  0.641 &  0.009 &  &  0.641 &  0.009\\
MSM lin &  0.642 &  0.009 &  &  0.642 &  0.009 &  &  0.642 &  0.009 &  &  0.642 &  0.009\\
eMSM logit &  0.678 &  0.008 &  &  0.664 &  0.008 &  &  0.650 &  0.009 &  &  0.641 &  0.009\\
eMSM lin &  0.678 &  0.008 &  &  0.664 &  0.008 &  &  0.651 &  0.009 &  &  0.642 &  0.009\\
& \multicolumn{11}{c}{Upper bounds and SEs for $\mu^0$} \\
MSM logit &  0.729 &  0.007 &  &  0.729 &  0.007 &  &  0.729 &  0.007 &  &  0.729 &  0.007\\
MSM lin &  0.728 &  0.007 &  &  0.728 &  0.007 &  &  0.728 &  0.007 &  &  0.728 &  0.007\\
eMSM logit &  0.696 &  0.008 &  &  0.708 &  0.007 &  &  0.720 &  0.007 &  &  0.729 &  0.007\\
eMSM lin &  0.696 &  0.008 &  &  0.708 &  0.007 &  &  0.720 &  0.007 &  &  0.728 &  0.007\\
& \multicolumn{11}{c}{Lower bounds and SEs for $\mu^1 - \mu^0$} \\
MSM logit &  -0.166 &   0.013 &  &  -0.166 &   0.013 &  &  -0.166 &   0.013 &  &  -0.166 &   0.013\\
MSM lin &  -0.162 &   0.013 &  &  -0.162 &   0.013 &  &  -0.162 &   0.013 &  &  -0.162 &   0.013\\
eMSM logit &  -0.075 &   0.013 &  &  -0.109 &   0.013 &  &  -0.143 &   0.013 &  &  -0.166 &   0.013\\
eMSM lin &  -0.074 &   0.013 &  &  -0.106 &   0.013 &  &  -0.140 &   0.013 &  &  -0.162 &   0.013\\
& \multicolumn{11}{c}{Upper bounds and SEs for $\mu^1 - \mu^0$} \\
MSM logit &   0.059 &   0.012 &  &   0.059 &   0.012 &  &   0.059 &   0.012 &  &   0.059 &   0.012\\
MSM lin &   0.056 &   0.012 &  &   0.056 &   0.012 &  &   0.056 &   0.012 &  &   0.056 &   0.012\\
eMSM logit &  -0.030 &   0.012 &  &   0.004 &   0.012 &  &   0.037 &   0.012 &  &   0.059 &   0.012\\
eMSM lin &  -0.030 &   0.012 &  &   0.003 &   0.012 &  &   0.035 &   0.012 &  &   0.056 &   0.012\\
& \multicolumn{11}{c}{Lower bounds and SEs for $\mu^1 / \mu^0$} \\
MSM logit &  0.772 &  0.018 &  &  0.772 &  0.018 &  &  0.772 &  0.018 &  &  0.772 &  0.018\\
MSM lin &  0.778 &  0.018 &  &  0.778 &  0.018 &  &  0.778 &  0.018 &  &  0.778 &  0.018\\
eMSM logit &  0.892 &  0.018 &  &  0.846 &  0.018 &  &  0.802 &  0.018 &  &  0.772 &  0.018\\
eMSM lin &  0.894 &  0.018 &  &  0.850 &  0.018 &  &  0.806 &  0.018 &  &  0.778 &  0.018\\
& \multicolumn{11}{c}{Upper bounds and SEs for $\mu^1 / \mu^0$} \\
MSM logit &  1.092 &  0.019 &  &  1.092 &  0.019 &  &  1.092 &  0.019 &  &  1.092 &  0.019\\
MSM lin &  1.087 &  0.020 &  &  1.087 &  0.020 &  &  1.087 &  0.020 &  &  1.087 &  0.020\\
eMSM logit &  0.956 &  0.018 &  &  1.006 &  0.019 &  &  1.058 &  0.019 &  &  1.092 &  0.019\\
eMSM lin &  0.956 &  0.018 &  &  1.004 &  0.019 &  &  1.053 &  0.019 &  &  1.087 &  0.020\\
\hline
\end{tabular}}
\end{center}
\setlength{\baselineskip}{0.5\baselineskip}
\vspace{-.15in}\noindent{\tiny
\textbf{Note}: MSM logit and eMSM logit denote RCAL under MSM and eMSM using logistic outcome mean regression. MSM lin and eMSM lin denote RCAL under MSM and eMSM using linear outcome mean regression.}
\vspace{-.2in}
\end{table}

\begin{table}
\caption{\footnotesize Point bounds and SEs for RCAL under MSM and eMSM with $\Lambda = 2$, using working models with main effects and interactions in the RHC study.} \label{tb:rhc-highdim-cal-lam2.0} \vspace{-4ex}
\begin{center}
\renewcommand{\arraystretch}{0.6} 
\small
\resizebox{\textwidth}{!}{\begin{tabular}{lccccccccccc}
\hline
& \multicolumn{2}{c}{$\delta = 0.2$} & & \multicolumn{2}{c}{$\delta = 0.5$} & & \multicolumn{2}{c}{$\delta = 0.8$} & & \multicolumn{2}{c}{$\delta = 1$} \\
& Bounds & SEs & & Bounds & SEs & & Bounds & SEs & & Bounds & SEs \\
\hline
& \multicolumn{11}{c}{Lower bounds and SEs for $\mu^1$} \\
MSM logit &  0.489 &  0.013 &  &  0.489 &  0.013 &  &  0.489 &  0.013 &  &  0.489 &  0.013\\
MSM lin &  0.490 &  0.013 &  &  0.490 &  0.013 &  &  0.490 &  0.013 &  &  0.490 &  0.013\\
eMSM logit &  0.607 &  0.011 &  &  0.563 &  0.012 &  &  0.519 &  0.012 &  &  0.489 &  0.013\\
eMSM lin &  0.607 &  0.011 &  &  0.563 &  0.012 &  &  0.519 &  0.012 &  &  0.490 &  0.013\\
& \multicolumn{11}{c}{Upper bounds and SEs for $\mu^1$} \\
MSM logit &  0.734 &  0.008 &  &  0.734 &  0.008 &  &  0.734 &  0.008 &  &  0.734 &  0.008\\
MSM lin &  0.730 &  0.008 &  &  0.730 &  0.008 &  &  0.730 &  0.008 &  &  0.730 &  0.008\\
eMSM logit &  0.656 &  0.010 &  &  0.685 &  0.009 &  &  0.715 &  0.009 &  &  0.734 &  0.008\\
eMSM lin &  0.655 &  0.010 &  &  0.683 &  0.009 &  &  0.711 &  0.009 &  &  0.730 &  0.008\\
& \multicolumn{11}{c}{Lower bounds and SEs for $\mu^0$} \\
MSM logit &  0.599 &  0.009 &  &  0.599 &  0.009 &  &  0.599 &  0.009 &  &  0.599 &  0.009\\
MSM lin &  0.602 &  0.009 &  &  0.602 &  0.009 &  &  0.602 &  0.009 &  &  0.602 &  0.009\\
eMSM logit &  0.670 &  0.008 &  &  0.643 &  0.008 &  &  0.617 &  0.009 &  &  0.599 &  0.009\\
eMSM lin &  0.670 &  0.008 &  &  0.644 &  0.009 &  &  0.619 &  0.009 &  &  0.602 &  0.009\\
& \multicolumn{11}{c}{Upper bounds and SEs for $\mu^0$} \\
MSM logit &  0.749 &  0.006 &  &  0.749 &  0.006 &  &  0.749 &  0.006 &  &  0.749 &  0.006\\
MSM lin &  0.749 &  0.006 &  &  0.749 &  0.006 &  &  0.749 &  0.006 &  &  0.749 &  0.006\\
eMSM logit &  0.700 &  0.008 &  &  0.718 &  0.007 &  &  0.737 &  0.007 &  &  0.749 &  0.006\\
eMSM lin &  0.699 &  0.008 &  &  0.718 &  0.007 &  &  0.736 &  0.007 &  &  0.749 &  0.006\\
& \multicolumn{11}{c}{Lower bounds and SEs for $\mu^1 - \mu^0$} \\
MSM logit &  -0.260 &   0.014 &  &  -0.260 &   0.014 &  &  -0.260 &   0.014 &  &  -0.260 &   0.014\\
MSM lin &  -0.259 &   0.014 &  &  -0.259 &   0.014 &  &  -0.259 &   0.014 &  &  -0.259 &   0.014\\
eMSM logit &  -0.093 &   0.013 &  &  -0.156 &   0.013 &  &  -0.218 &   0.013 &  &  -0.260 &   0.014\\
eMSM lin &  -0.093 &   0.013 &  &  -0.155 &   0.013 &  &  -0.217 &   0.014 &  &  -0.259 &   0.014\\
& \multicolumn{11}{c}{Upper bounds and SEs for $\mu^1 - \mu^0$} \\
MSM logit &   0.135 &   0.011 &  &   0.135 &   0.011 &  &   0.135 &   0.011 &  &   0.135 &   0.011\\
MSM lin &   0.128 &   0.011 &  &   0.128 &   0.011 &  &   0.128 &   0.011 &  &   0.128 &   0.011\\
eMSM logit &  -0.014 &   0.012 &  &   0.042 &   0.012 &  &   0.098 &   0.012 &  &   0.135 &   0.011\\
eMSM lin &  -0.015 &   0.012 &  &   0.039 &   0.012 &  &   0.093 &   0.012 &  &   0.128 &   0.011\\
& \multicolumn{11}{c}{Lower bounds and SEs for $\mu^1 / \mu^0$} \\
MSM logit &  0.653 &  0.018 &  &  0.653 &  0.018 &  &  0.653 &  0.018 &  &  0.653 &  0.018\\
MSM lin &  0.655 &  0.018 &  &  0.655 &  0.018 &  &  0.655 &  0.018 &  &  0.655 &  0.018\\
eMSM logit &  0.867 &  0.018 &  &  0.783 &  0.017 &  &  0.704 &  0.017 &  &  0.653 &  0.018\\
eMSM lin &  0.868 &  0.018 &  &  0.784 &  0.017 &  &  0.706 &  0.018 &  &  0.655 &  0.018\\
& \multicolumn{11}{c}{Upper bounds and SEs for $\mu^1 / \mu^0$} \\
MSM logit &  1.226 &  0.022 &  &  1.226 &  0.022 &  &  1.226 &  0.022 &  &  1.226 &  0.022\\
MSM lin &  1.213 &  0.021 &  &  1.213 &  0.021 &  &  1.213 &  0.021 &  &  1.213 &  0.021\\
eMSM logit &  0.979 &  0.018 &  &  1.065 &  0.019 &  &  1.159 &  0.020 &  &  1.226 &  0.022\\
eMSM lin &  0.977 &  0.018 &  &  1.061 &  0.019 &  &  1.150 &  0.020 &  &  1.213 &  0.021\\
\hline
\end{tabular}}
\end{center}
\setlength{\baselineskip}{0.5\baselineskip}
\vspace{-.15in}\noindent{\tiny
\textbf{Note}: MSM logit and eMSM logit denote RCAL under MSM and eMSM using logistic outcome mean regression. MSM lin and eMSM lin denote RCAL under MSM and eMSM using linear outcome mean regression.}
\vspace{-.2in}
\end{table}
\clearpage
\subsection{Additional numerical result for NHANES} \label{sec:additional-numerical-nhanes}
\begin{figure} 
\centering
\begin{minipage}{\textwidth}
        \centering
        \includegraphics[scale=0.525]{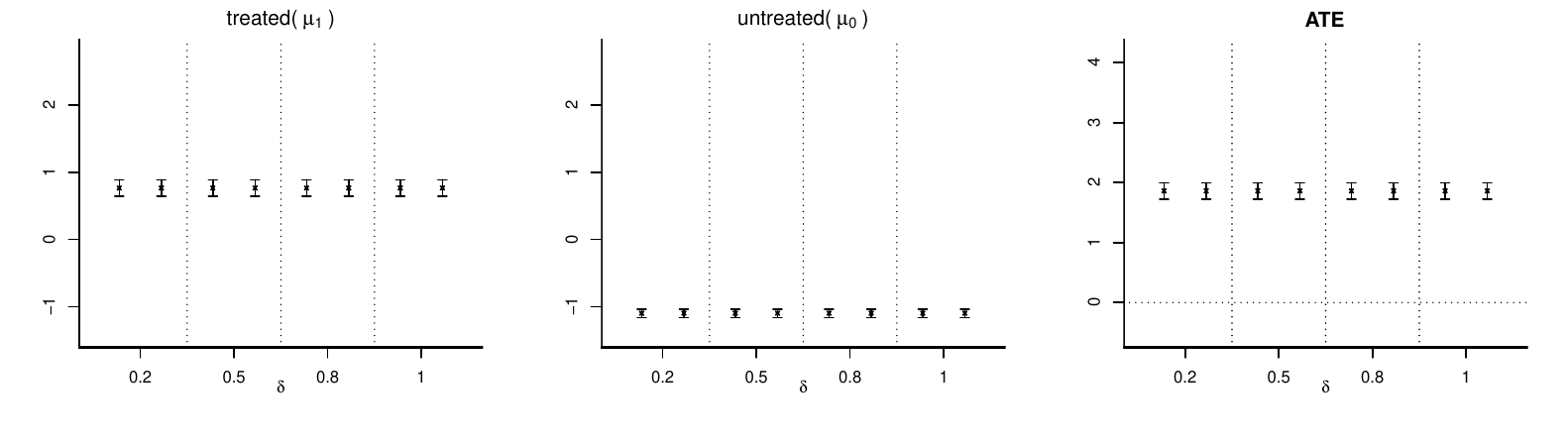} \vspace{-.25in}
        \subcaption{$\Lambda = 1$}
    \end{minipage}
    
     \vspace{.18in}
     
  \begin{minipage}{\textwidth}
        \centering
        \includegraphics[scale=0.525]{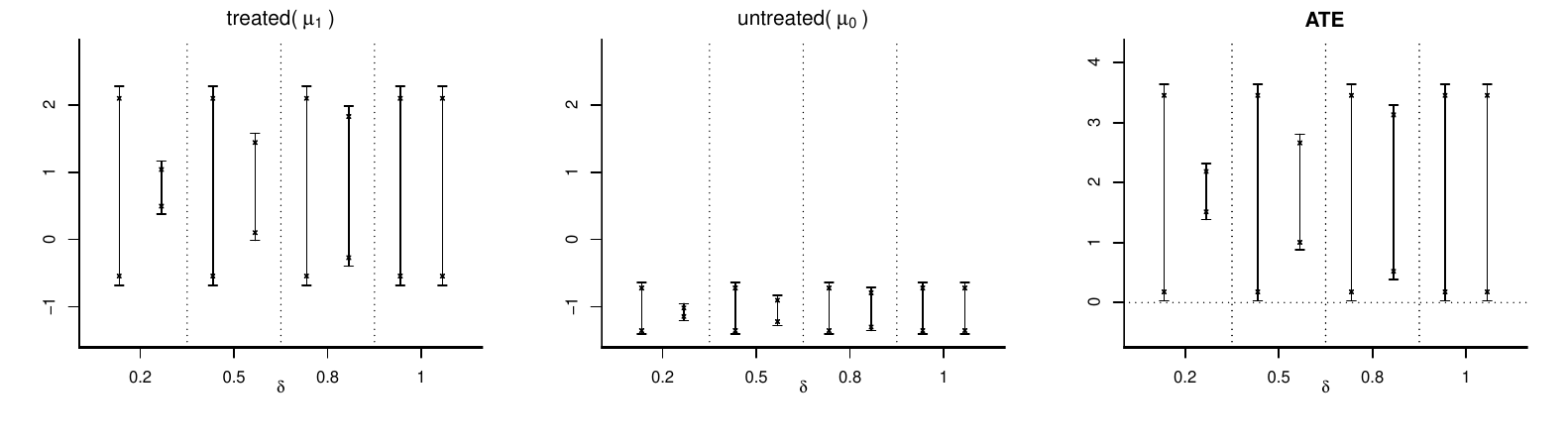} \vspace{-.25in}
        \subcaption{$\Lambda = 10$}
    \end{minipage}
    
     \vspace{.18in}
     
    \begin{minipage}{\textwidth}
        \centering
        \includegraphics[scale=0.525]{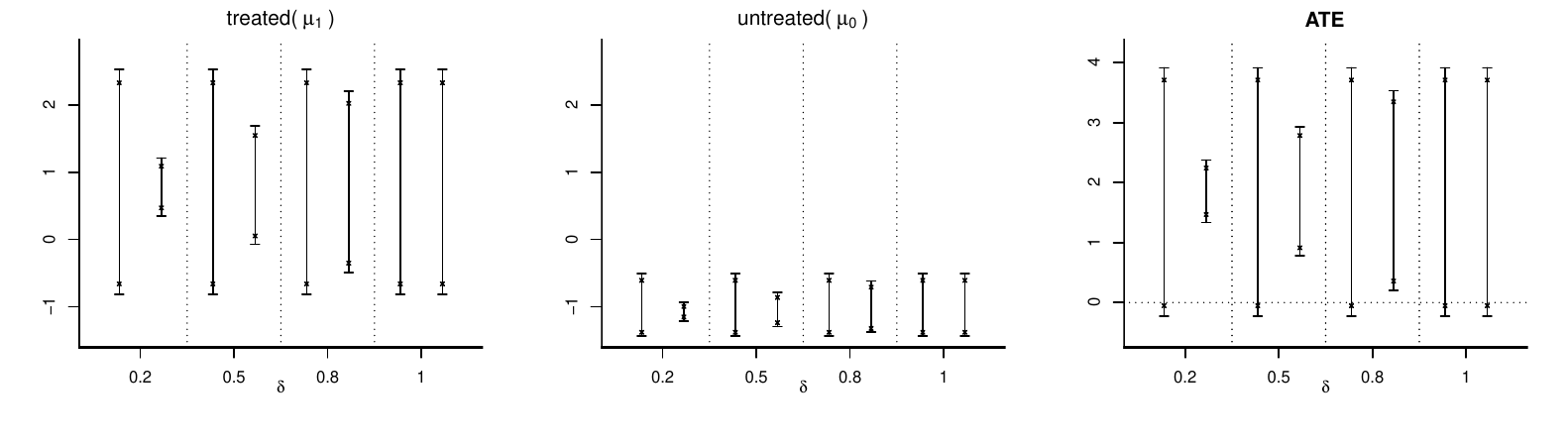} \vspace{-.25in}
        \subcaption{$\Lambda = 20$}
    \end{minipage}
    
     \vspace{.18in}
     
    \begin{minipage}{\textwidth}
        \centering
        \includegraphics[scale=0.525]{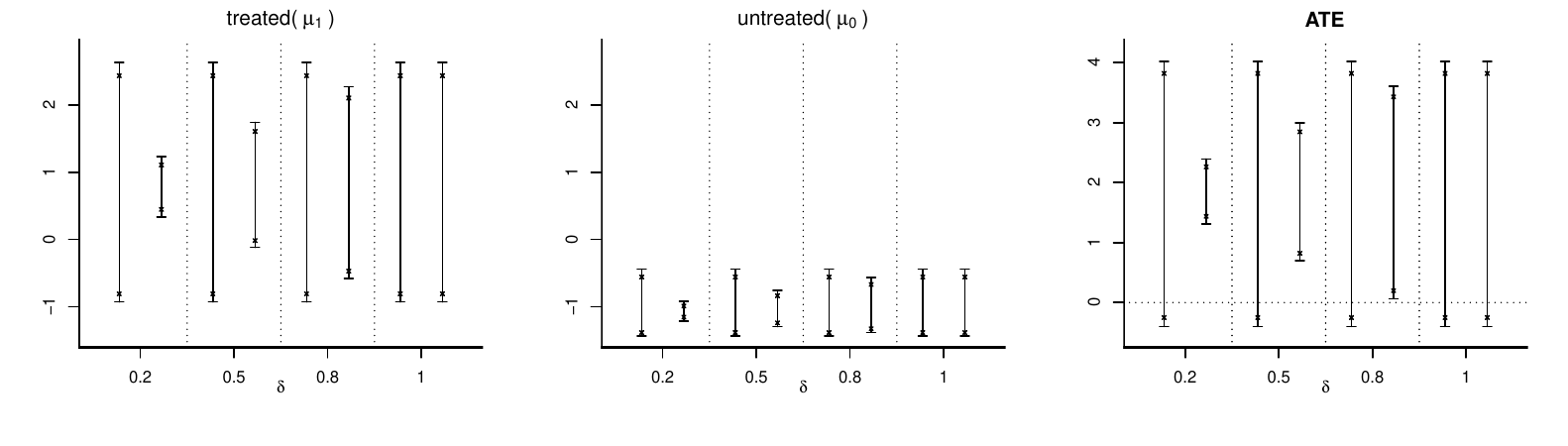}  \vspace{-.25in}
        \subcaption{$\Lambda = 30$}
    \end{minipage}
    
     \vspace{.18in}
     
    \begin{minipage}{\textwidth}
        \centering
        \includegraphics[scale=0.525]{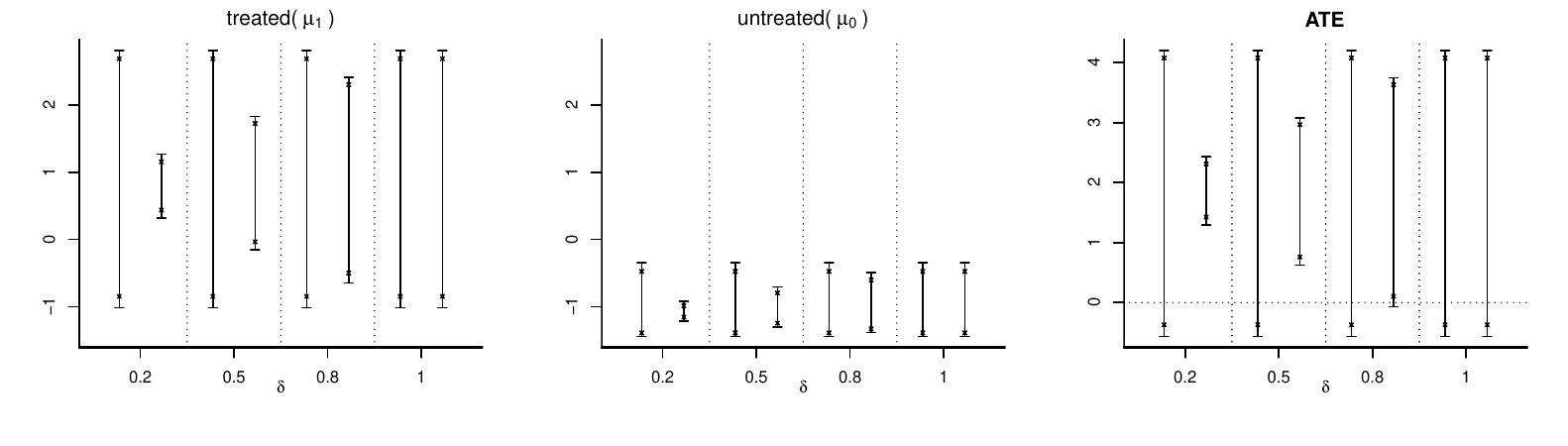} \vspace{-.25in}
        \subcaption{$\Lambda = 50$}
    \end{minipage} \vspace{-.1in}
\caption{\small
Point bounds (x) and 90\% confidence intervals (–) on mean blood mercury and ATE in the NHANES study, similarly as Figure \ref{fig:nhanes-highdim-lin-ate}.}
\label{fig:nhanes-highdim-lin}
\end{figure}

\begin{figure} 
\centering
\begin{minipage}{\textwidth}
        \centering
        \includegraphics[scale=0.525]{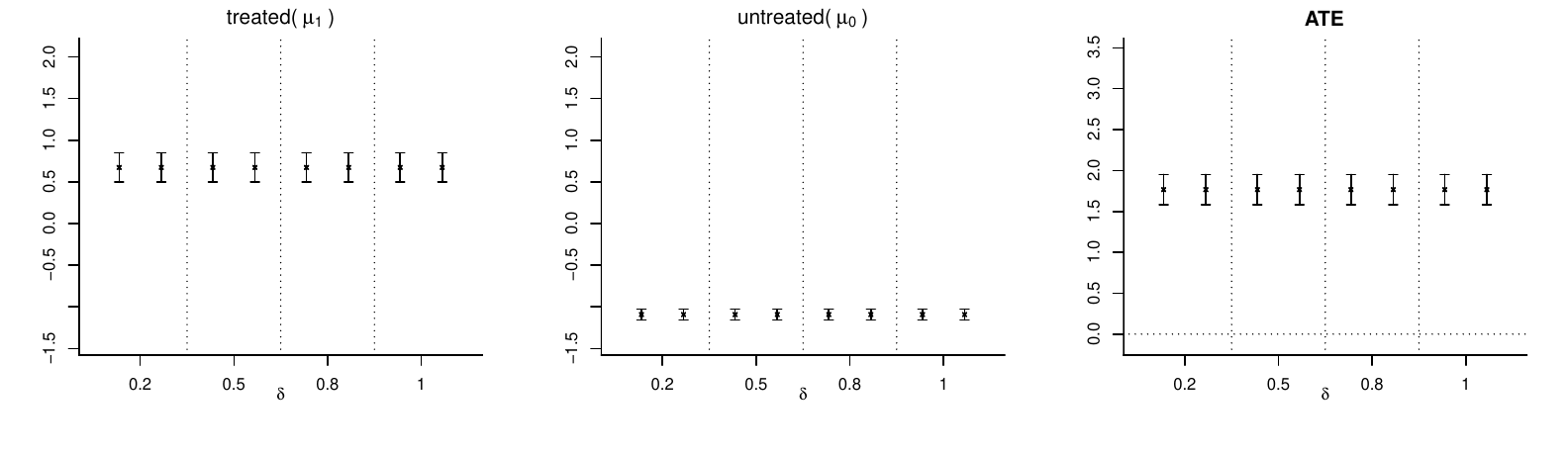} \vspace{-.25in}
        \subcaption{$\Lambda = 1$}
    \end{minipage}
    
     \vspace{.18in}
     
  \begin{minipage}{\textwidth}
        \centering
        \includegraphics[scale=0.525]{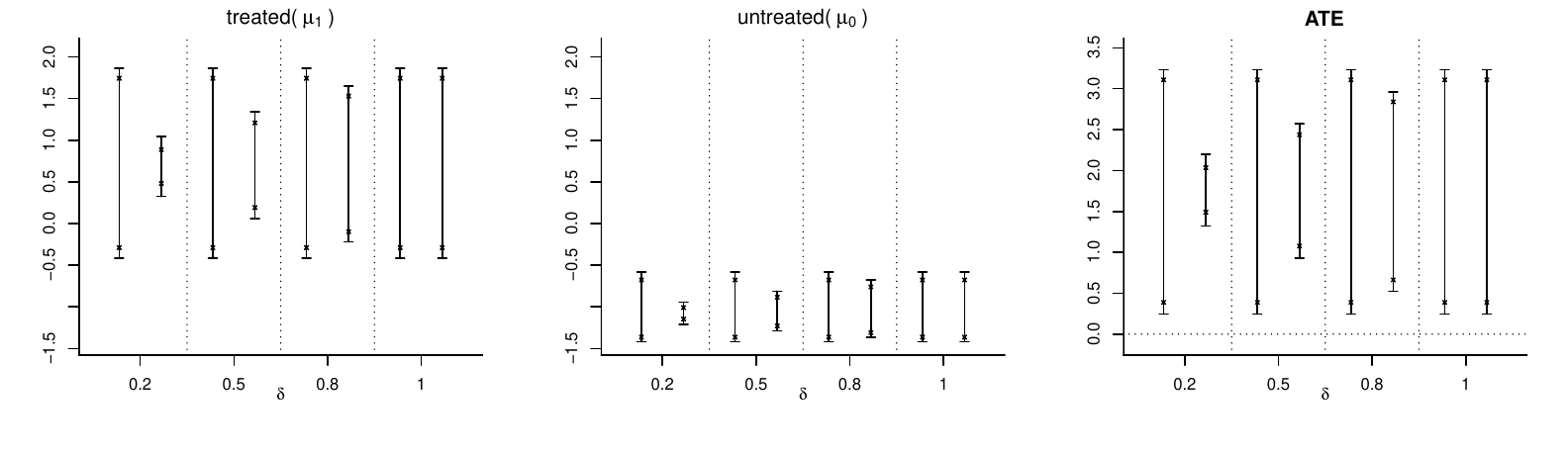} \vspace{-.25in}
        \subcaption{$\Lambda = 10$}
    \end{minipage}
    
     \vspace{.18in}
     
    \begin{minipage}{\textwidth}
        \centering
        \includegraphics[scale=0.525]{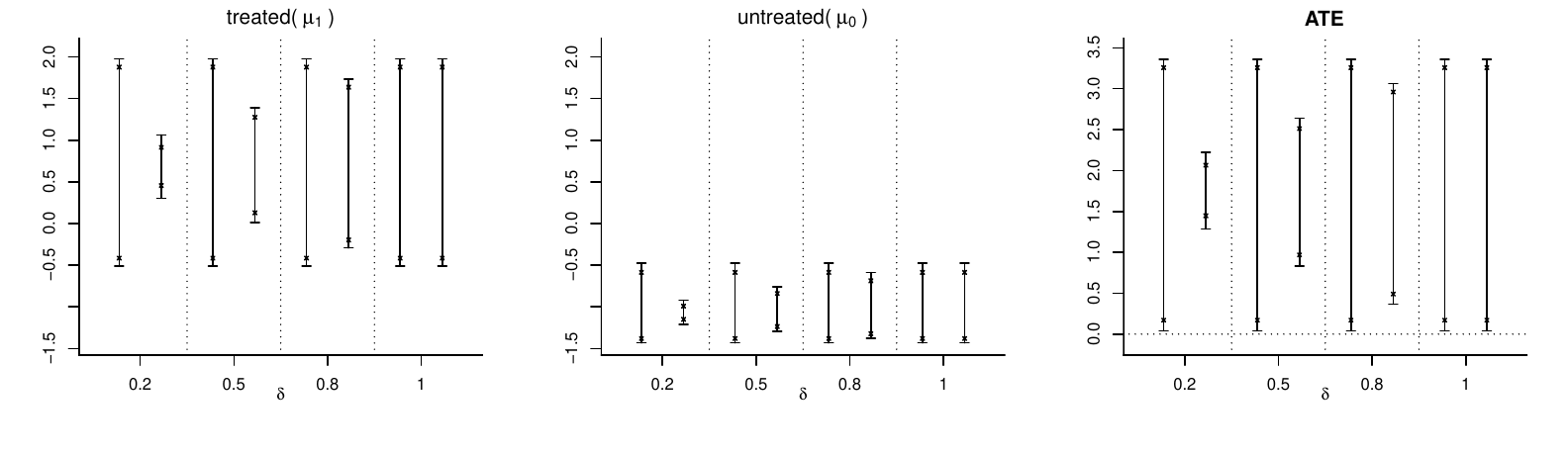} \vspace{-.25in}
        \subcaption{$\Lambda = 20$}
    \end{minipage}
    
     \vspace{.18in}
     
    \begin{minipage}{\textwidth}
        \centering
        \includegraphics[scale=0.525]{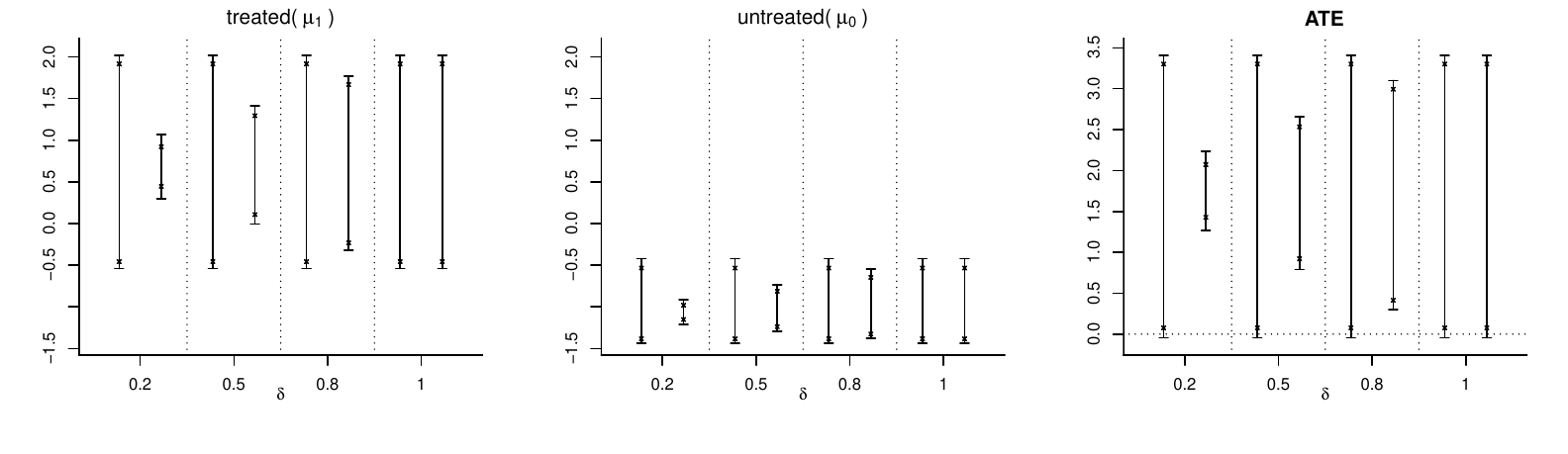}  \vspace{-.25in}
        \subcaption{$\Lambda = 30$}
    \end{minipage}
    
     \vspace{.18in}
     
    \begin{minipage}{\textwidth}
        \centering
        \includegraphics[scale=0.525]{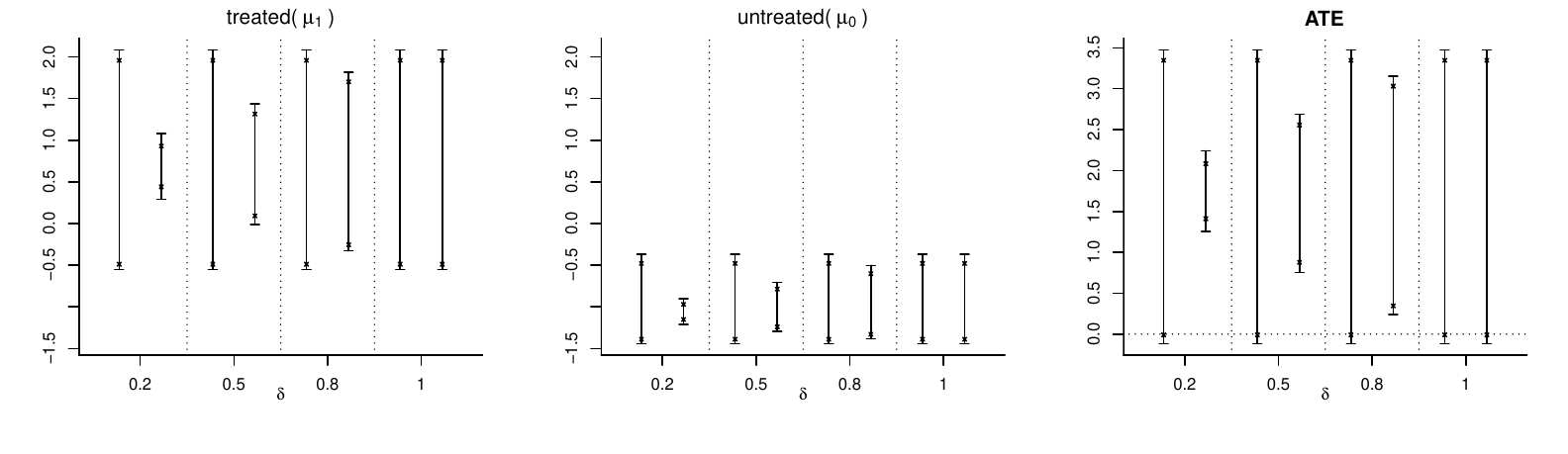} \vspace{-.25in}
        \subcaption{$\Lambda = 50$}
    \end{minipage} \vspace{-.1in}
\caption{\small
Point bounds (x) and 90\% confidence intervals (–) on mean blood mercury and ATE in the NHANES study, similarly as Figure \ref{fig:nhanes-highdim-lin-ate}, but using working models with main effects only.}
\label{fig:nhanes-lowdim-lin}
\end{figure}

Figure \ref{fig:nhanes-highdim-lin} presents the point bounds and 90\% confidence intervals on mean blood mercury and average treatment effect, similarly as Figure \ref{fig:nhanes-highdim-lin-ate}.
Figure \ref{fig:nhanes-lowdim-lin} presents the point bounds and 90\% confidence intervals using working models with only main effects and linear outcome mean regression for CAL under MSM and eMSM. Tables \ref{tb:nhanes-lowdim-cal-lam1.0}--\ref{tb:nhanes-lowdim-cal-lam50} show numerical estimates and standard errors using working models with only main effects and Tables \ref{tb:nhanes-highdim-cal-lam1.0}--\ref{tb:nhanes-highdim-cal-lam50} show numerical estimates and standard errors using working models with main effects and interactions.

Comparing Figures \ref{fig:nhanes-highdim-lin-ate} and \ref{fig:nhanes-lowdim-lin}, we see notable differences in point bounds between using
 working models with only main effects or with both main effects and interactions, as mentioned in the main text.
In fact, the sensitivity intervals from RCAL under MSM begin to include zero at $\Lambda = 20$, whereas those from CAL under MSM do not include zero until $\Lambda = 30$. Furthermore, the sensitivity intervals from RCAL under eMSM begin to include zero at $(\Lambda = 50, \delta = 0.8)$, whereas those from CAL under eMSM still do not include zero at the same $(\Lambda, \delta)$ values. 
 A possible explanation is that working models relying solely on main effects may be misspecified.


    
    
    
    
    
    
    
    


\begin{table}
\caption{\footnotesize Point bounds and SEs for CAL under MSM and eMSM with $\Lambda = 1$, using working models with only main effects in the NHANES study.} \label{tb:nhanes-lowdim-cal-lam1.0} \vspace{-4ex}
\begin{center}
\small
\resizebox{\textwidth}{!}{\begin{tabular}{lccccccccccc}
\hline
& \multicolumn{2}{c}{$\delta = 0.2$} & & \multicolumn{2}{c}{$\delta = 0.5$} & & \multicolumn{2}{c}{$\delta = 0.8$} & & \multicolumn{2}{c}{$\delta = 1$} \\
& Bounds & SEs & & Bounds & SEs & & Bounds & SEs & & Bounds & SEs \\
\hline
& \multicolumn{11}{c}{Lower bounds and SEs for $\mu^1$} \\
MSM lin &  0.673 &  0.108 &  &  0.673 &  0.108 &  &  0.673 &  0.108 &  &  0.673 &  0.108\\
eMSM lin &  0.673 &  0.108 &  &  0.673 &  0.108 &  &  0.673 &  0.108 &  &  0.673 &  0.108\\
& \multicolumn{11}{c}{Upper bounds and SEs for $\mu^1$} \\
MSM lin &  0.673 &  0.108 &  &  0.673 &  0.108 &  &  0.673 &  0.108 &  &  0.673 &  0.108\\
eMSM lin &  0.673 &  0.108 &  &  0.673 &  0.108 &  &  0.673 &  0.108 &  &  0.673 &  0.108\\
& \multicolumn{11}{c}{Lower bounds and SEs for $\mu^0$} \\
MSM lin &  -1.093 &   0.040 &  &  -1.093 &   0.040 &  &  -1.093 &   0.040 &  &  -1.093 &   0.040\\
eMSM lin &  -1.093 &   0.040 &  &  -1.093 &   0.040 &  &  -1.093 &   0.040 &  &  -1.093 &   0.040\\
& \multicolumn{11}{c}{Upper bounds and SEs for $\mu^0$} \\
MSM lin &  -1.093 &   0.040 &  &  -1.093 &   0.040 &  &  -1.093 &   0.040 &  &  -1.093 &   0.040\\
eMSM lin &  -1.093 &   0.040 &  &  -1.093 &   0.040 &  &  -1.093 &   0.040 &  &  -1.093 &   0.040\\
& \multicolumn{11}{c}{Lower bounds and SEs for $\mu^1 - \mu^0$} \\
MSM lin &  1.766 &  0.113 &  &  1.766 &  0.113 &  &  1.766 &  0.113 &  &  1.766 &  0.113\\
eMSM lin &  1.766 &  0.113 &  &  1.766 &  0.113 &  &  1.766 &  0.113 &  &  1.766 &  0.113\\
& \multicolumn{11}{c}{Upper bounds and SEs for $\mu^1 - \mu^0$} \\
MSM lin &  1.766 &  0.113 &  &  1.766 &  0.113 &  &  1.766 &  0.113 &  &  1.766 &  0.113\\
eMSM lin &  1.766 &  0.113 &  &  1.766 &  0.113 &  &  1.766 &  0.113 &  &  1.766 &  0.113\\
\hline
\end{tabular}}
\end{center}
\setlength{\baselineskip}{0.5\baselineskip}
\vspace{-.15in}\noindent{\tiny
\textbf{Note}: MSM lin and eMSM lin denote CAL under MSM and eMSM using linear outcome mean regression.}
\vspace{-.2in}
\end{table}

\begin{table}
\caption{\footnotesize Point bounds and SEs for CAL under MSM and eMSM with $\Lambda = 10$, using working models with only main effects in the NHANES study.} \label{tb:nhanes-lowdim-cal-lam10} \vspace{-4ex}
\begin{center}
\small
\resizebox{\textwidth}{!}{\begin{tabular}{lccccccccccc}
\hline
& \multicolumn{2}{c}{$\delta = 0.2$} & & \multicolumn{2}{c}{$\delta = 0.5$} & & \multicolumn{2}{c}{$\delta = 0.8$} & & \multicolumn{2}{c}{$\delta = 1$} \\
& Bounds & SEs & & Bounds & SEs & & Bounds & SEs & & Bounds & SEs \\
\hline
& \multicolumn{11}{c}{Lower bounds and SEs for $\mu^1$} \\
MSM lin &  -0.290 &   0.076 &  &  -0.290 &   0.076 &  &  -0.290 &   0.076 &  &  -0.290 &   0.076\\
eMSM lin &   0.480 &   0.095 &  &   0.192 &   0.080 &  &  -0.097 &   0.074 &  &  -0.290 &   0.076\\
& \multicolumn{11}{c}{Upper bounds and SEs for $\mu^1$} \\
MSM lin &  1.744 &  0.072 &  &  1.744 &  0.072 &  &  1.744 &  0.072 &  &  1.744 &  0.072\\
eMSM lin &  0.887 &  0.095 &  &  1.208 &  0.079 &  &  1.530 &  0.071 &  &  1.744 &  0.072\\
& \multicolumn{11}{c}{Lower bounds and SEs for $\mu^0$} \\
MSM lin &  -1.363 &   0.033 &  &  -1.363 &   0.033 &  &  -1.363 &   0.033 &  &  -1.363 &   0.033\\
eMSM lin &  -1.147 &   0.038 &  &  -1.228 &   0.035 &  &  -1.309 &   0.033 &  &  -1.363 &   0.033\\
& \multicolumn{11}{c}{Upper bounds and SEs for $\mu^0$} \\
MSM lin &  -0.677 &   0.056 &  &  -0.677 &   0.056 &  &  -0.677 &   0.056 &  &  -0.677 &   0.056\\
eMSM lin &  -1.010 &   0.041 &  &  -0.885 &   0.045 &  &  -0.760 &   0.051 &  &  -0.677 &   0.056\\
& \multicolumn{11}{c}{Lower bounds and SEs for $\mu^1 - \mu^0$} \\
MSM lin &  0.388 &  0.087 &  &  0.388 &  0.087 &  &  0.388 &  0.087 &  &  0.388 &  0.087\\
eMSM lin &  1.490 &  0.101 &  &  1.077 &  0.088 &  &  0.663 &  0.084 &  &  0.388 &  0.087\\
& \multicolumn{11}{c}{Upper bounds and SEs for $\mu^1 - \mu^0$} \\
MSM lin &  3.107 &  0.075 &  &  3.107 &  0.075 &  &  3.107 &  0.075 &  &  3.107 &  0.075\\
eMSM lin &  2.034 &  0.100 &  &  2.437 &  0.084 &  &  2.839 &  0.075 &  &  3.107 &  0.075\\
\hline
\end{tabular}}
\end{center}
\setlength{\baselineskip}{0.5\baselineskip}
\vspace{-.15in}\noindent{\tiny
\textbf{Note}: MSM lin and eMSM lin denote CAL under MSM and eMSM using linear outcome mean regression.}
\vspace{-.2in}
\end{table}

\begin{table}
\caption{\footnotesize Point bounds and SEs for CAL under MSM and eMSM with $\Lambda = 20$, using working models with only main effects in the NHANES study.} \label{tb:nhanes-lowdim-cal-lam20} \vspace{-4ex}
\begin{center}
\small
\resizebox{\textwidth}{!}{\begin{tabular}{lccccccccccc}
\hline
& \multicolumn{2}{c}{$\delta = 0.2$} & & \multicolumn{2}{c}{$\delta = 0.5$} & & \multicolumn{2}{c}{$\delta = 0.8$} & & \multicolumn{2}{c}{$\delta = 1$} \\
& Bounds & SEs & & Bounds & SEs & & Bounds & SEs & & Bounds & SEs \\
\hline
& \multicolumn{11}{c}{Lower bounds and SEs for $\mu^1$} \\
MSM lin &  -0.414 &   0.057 &  &  -0.414 &   0.057 &  &  -0.414 &   0.057 &  &  -0.414 &   0.057\\
eMSM lin &   0.455 &   0.092 &  &   0.129 &   0.072 &  &  -0.197 &   0.059 &  &  -0.414 &   0.057\\
& \multicolumn{11}{c}{Upper bounds and SEs for $\mu^1$} \\
MSM lin &  1.878 &  0.059 &  &  1.878 &  0.059 &  &  1.878 &  0.059 &  &  1.878 &  0.059\\
eMSM lin &  0.914 &  0.092 &  &  1.275 &  0.071 &  &  1.637 &  0.059 &  &  1.878 &  0.059\\
& \multicolumn{11}{c}{Lower bounds and SEs for $\mu^0$} \\
MSM lin &  -1.380 &   0.032 &  &  -1.380 &   0.032 &  &  -1.380 &   0.032 &  &  -1.380 &   0.032\\
eMSM lin &  -1.150 &   0.037 &  &  -1.236 &   0.035 &  &  -1.322 &   0.033 &  &  -1.380 &   0.032\\
& \multicolumn{11}{c}{Upper bounds and SEs for $\mu^0$} \\
MSM lin &  -0.586 &   0.068 &  &  -0.586 &   0.068 &  &  -0.586 &   0.068 &  &  -0.586 &   0.068\\
eMSM lin &  -0.991 &   0.042 &  &  -0.839 &   0.050 &  &  -0.687 &   0.060 &  &  -0.586 &   0.068\\
& \multicolumn{11}{c}{Lower bounds and SEs for $\mu^1 - \mu^0$} \\
MSM lin &  0.171 &  0.079 &  &  0.171 &  0.079 &  &  0.171 &  0.079 &  &  0.171 &  0.079\\
eMSM lin &  1.447 &  0.099 &  &  0.969 &  0.082 &  &  0.490 &  0.076 &  &  0.171 &  0.079\\
& \multicolumn{11}{c}{Upper bounds and SEs for $\mu^1 - \mu^0$} \\
MSM lin &  3.257 &  0.060 &  &  3.257 &  0.060 &  &  3.257 &  0.060 &  &  3.257 &  0.060\\
eMSM lin &  2.064 &  0.097 &  &  2.511 &  0.076 &  &  2.959 &  0.063 &  &  3.257 &  0.060\\
\hline
\end{tabular}}
\end{center}
\setlength{\baselineskip}{0.5\baselineskip}
\vspace{-.15in}\noindent{\tiny
\textbf{Note}: MSM lin and eMSM lin denote CAL under MSM and eMSM using linear outcome mean regression.}
\vspace{-.2in}
\end{table}

\begin{table}
\caption{\footnotesize Point bounds and SEs for CAL under MSM and eMSM with $\Lambda = 30$, using working models with only main effects in the NHANES study.} \label{tb:nhanes-lowdim-cal-lam30} \vspace{-4ex}
\begin{center}
\small
\resizebox{\textwidth}{!}{\begin{tabular}{lccccccccccc}
\hline
& \multicolumn{2}{c}{$\delta = 0.2$} & & \multicolumn{2}{c}{$\delta = 0.5$} & & \multicolumn{2}{c}{$\delta = 0.8$} & & \multicolumn{2}{c}{$\delta = 1$} \\
& Bounds & SEs & & Bounds & SEs & & Bounds & SEs & & Bounds & SEs \\
\hline
& \multicolumn{11}{c}{Lower bounds and SEs for $\mu^1$} \\
MSM lin &  -0.458 &   0.050 &  &  -0.458 &   0.050 &  &  -0.458 &   0.050 &  &  -0.458 &   0.050\\
eMSM lin &   0.447 &   0.090 &  &   0.108 &   0.068 &  &  -0.232 &   0.052 &  &  -0.458 &   0.050\\
& \multicolumn{11}{c}{Upper bounds and SEs for $\mu^1$} \\
MSM lin &  1.916 &  0.064 &  &  1.916 &  0.064 &  &  1.916 &  0.064 &  &  1.916 &  0.064\\
eMSM lin &  0.921 &  0.091 &  &  1.294 &  0.072 &  &  1.667 &  0.062 &  &  1.916 &  0.064\\
& \multicolumn{11}{c}{Lower bounds and SEs for $\mu^0$} \\
MSM lin &  -1.385 &   0.032 &  &  -1.385 &   0.032 &  &  -1.385 &   0.032 &  &  -1.385 &   0.032\\
eMSM lin &  -1.151 &   0.037 &  &  -1.239 &   0.034 &  &  -1.326 &   0.033 &  &  -1.385 &   0.032\\
& \multicolumn{11}{c}{Upper bounds and SEs for $\mu^0$} \\
MSM lin &  -0.534 &   0.069 &  &  -0.534 &   0.069 &  &  -0.534 &   0.069 &  &  -0.534 &   0.069\\
eMSM lin &  -0.981 &   0.042 &  &  -0.814 &   0.049 &  &  -0.646 &   0.060 &  &  -0.534 &   0.069\\
& \multicolumn{11}{c}{Lower bounds and SEs for $\mu^1 - \mu^0$} \\
MSM lin &  0.077 &  0.073 &  &  0.077 &  0.073 &  &  0.077 &  0.073 &  &  0.077 &  0.073\\
eMSM lin &  1.428 &  0.097 &  &  0.921 &  0.078 &  &  0.414 &  0.071 &  &  0.077 &  0.073\\
& \multicolumn{11}{c}{Upper bounds and SEs for $\mu^1 - \mu^0$} \\
MSM lin &  3.300 &  0.065 &  &  3.300 &  0.065 &  &  3.300 &  0.065 &  &  3.300 &  0.065\\
eMSM lin &  2.073 &  0.097 &  &  2.533 &  0.076 &  &  2.993 &  0.065 &  &  3.300 &  0.065\\
\hline
\end{tabular}}
\end{center}
\setlength{\baselineskip}{0.5\baselineskip}
\vspace{-.15in}\noindent{\tiny
\textbf{Note}: MSM lin and eMSM lin denote CAL under MSM and eMSM using linear outcome mean regression.}
\vspace{-.2in}
\end{table}

\begin{table}
\caption{\footnotesize Point bounds and SEs for CAL under MSM and eMSM with $\Lambda = 50$, using working models with only main effects in the NHANES study.} \label{tb:nhanes-lowdim-cal-lam50} \vspace{-4ex}
\begin{center}
\small
\resizebox{\textwidth}{!}{\begin{tabular}{lccccccccccc}
\hline
& \multicolumn{2}{c}{$\delta = 0.2$} & & \multicolumn{2}{c}{$\delta = 0.5$} & & \multicolumn{2}{c}{$\delta = 0.8$} & & \multicolumn{2}{c}{$\delta = 1$} \\
& Bounds & SEs & & Bounds & SEs & & Bounds & SEs & & Bounds & SEs \\
\hline
& \multicolumn{11}{c}{Lower bounds and SEs for $\mu^1$} \\
MSM lin &  -0.488 &   0.039 &  &  -0.488 &   0.039 &  &  -0.488 &   0.039 &  &  -0.488 &   0.039\\
eMSM lin &   0.441 &   0.090 &  &   0.092 &   0.064 &  &  -0.256 &   0.044 &  &  -0.488 &   0.039\\
& \multicolumn{11}{c}{Upper bounds and SEs for $\mu^1$} \\
MSM lin &  1.958 &  0.077 &  &  1.958 &  0.077 &  &  1.958 &  0.077 &  &  1.958 &  0.077\\
eMSM lin &  0.930 &  0.092 &  &  1.315 &  0.075 &  &  1.701 &  0.071 &  &  1.958 &  0.077\\
& \multicolumn{11}{c}{Lower bounds and SEs for $\mu^0$} \\
MSM lin &  -1.389 &   0.032 &  &  -1.389 &   0.032 &  &  -1.389 &   0.032 &  &  -1.389 &   0.032\\
eMSM lin &  -1.152 &   0.037 &  &  -1.241 &   0.034 &  &  -1.329 &   0.033 &  &  -1.389 &   0.032\\
& \multicolumn{11}{c}{Upper bounds and SEs for $\mu^0$} \\
MSM lin &  -0.479 &   0.069 &  &  -0.479 &   0.069 &  &  -0.479 &   0.069 &  &  -0.479 &   0.069\\
eMSM lin &  -0.970 &   0.041 &  &  -0.786 &   0.049 &  &  -0.602 &   0.060 &  &  -0.479 &   0.069\\
& \multicolumn{11}{c}{Lower bounds and SEs for $\mu^1 - \mu^0$} \\
MSM lin &  -0.009 &   0.065 &  &  -0.009 &   0.065 &  &  -0.009 &   0.065 &  &  -0.009 &   0.065\\
eMSM lin &   1.411 &   0.096 &  &   0.878 &   0.075 &  &   0.346 &   0.064 &  &  -0.009 &   0.065\\
& \multicolumn{11}{c}{Upper bounds and SEs for $\mu^1 - \mu^0$} \\
MSM lin &  3.347 &  0.077 &  &  3.347 &  0.077 &  &  3.347 &  0.077 &  &  3.347 &  0.077\\
eMSM lin &  2.082 &  0.097 &  &  2.556 &  0.079 &  &  3.030 &  0.073 &  &  3.347 &  0.077\\
\hline
\end{tabular}}
\end{center}
\setlength{\baselineskip}{0.5\baselineskip}
\vspace{-.15in}\noindent{\tiny
\textbf{Note}: MSM lin and eMSM lin denote CAL under MSM and eMSM using linear outcome mean regression.}
\vspace{-.2in}
\end{table}

\begin{table}
\caption{\footnotesize Point bounds and SEs for RCAL under MSM and eMSM with $\Lambda = 1$, using working models with main effects and interactions in the NHANES study.} \label{tb:nhanes-highdim-cal-lam1.0} \vspace{-4ex}
\begin{center}
\small
\resizebox{\textwidth}{!}{\begin{tabular}{lccccccccccc}
\hline
& \multicolumn{2}{c}{$\delta = 0.2$} & & \multicolumn{2}{c}{$\delta = 0.5$} & & \multicolumn{2}{c}{$\delta = 0.8$} & & \multicolumn{2}{c}{$\delta = 1$} \\
& Bounds & SEs & & Bounds & SEs & & Bounds & SEs & & Bounds & SEs \\
\hline
& \multicolumn{11}{c}{Lower bounds and SEs for $\mu^1$} \\
MSM lin &  0.765 &  0.075 &  &  0.765 &  0.075 &  &  0.765 &  0.075 &  &  0.765 &  0.075\\
eMSM lin &  0.765 &  0.075 &  &  0.765 &  0.075 &  &  0.765 &  0.075 &  &  0.765 &  0.075\\
& \multicolumn{11}{c}{Upper bounds and SEs for $\mu^1$} \\
MSM lin &  0.765 &  0.075 &  &  0.765 &  0.075 &  &  0.765 &  0.075 &  &  0.765 &  0.075\\
eMSM lin &  0.765 &  0.075 &  &  0.765 &  0.075 &  &  0.765 &  0.075 &  &  0.765 &  0.075\\
& \multicolumn{11}{c}{Lower bounds and SEs for $\mu^0$} \\
MSM lin &  -1.097 &   0.038 &  &  -1.097 &   0.038 &  &  -1.097 &   0.038 &  &  -1.097 &   0.038\\
eMSM lin &  -1.097 &   0.038 &  &  -1.097 &   0.038 &  &  -1.097 &   0.038 &  &  -1.097 &   0.038\\
& \multicolumn{11}{c}{Upper bounds and SEs for $\mu^0$} \\
MSM lin &  -1.097 &   0.038 &  &  -1.097 &   0.038 &  &  -1.097 &   0.038 &  &  -1.097 &   0.038\\
eMSM lin &  -1.097 &   0.038 &  &  -1.097 &   0.038 &  &  -1.097 &   0.038 &  &  -1.097 &   0.038\\
& \multicolumn{11}{c}{Lower bounds and SEs for $\mu^1 - \mu^0$} \\
MSM lin &  1.862 &  0.082 &  &  1.862 &  0.082 &  &  1.862 &  0.082 &  &  1.862 &  0.082\\
eMSM lin &  1.862 &  0.082 &  &  1.862 &  0.082 &  &  1.862 &  0.082 &  &  1.862 &  0.082\\
& \multicolumn{11}{c}{Upper bounds and SEs for $\mu^1 - \mu^0$} \\
MSM lin &  1.862 &  0.082 &  &  1.862 &  0.082 &  &  1.862 &  0.082 &  &  1.862 &  0.082\\
eMSM lin &  1.862 &  0.082 &  &  1.862 &  0.082 &  &  1.862 &  0.082 &  &  1.862 &  0.082\\
\hline
\end{tabular}}
\end{center}
\setlength{\baselineskip}{0.5\baselineskip}
\vspace{-.15in}\noindent{\tiny
\textbf{Note}: MSM lin and eMSM lin denote RCAL under MSM and eMSM using linear outcome mean regression.}
\vspace{-.2in}
\end{table}

\begin{table}
\caption{\footnotesize Point bounds and SEs for RCAL under MSM and eMSM with $\Lambda = 10$, using working models with main effects and interactions in the NHANES study.} \label{tb:nhanes-highdim-cal-lam10} \vspace{-4ex}
\begin{center}
\small
\resizebox{\textwidth}{!}{\begin{tabular}{lccccccccccc}
\hline
& \multicolumn{2}{c}{$\delta = 0.2$} & & \multicolumn{2}{c}{$\delta = 0.5$} & & \multicolumn{2}{c}{$\delta = 0.8$} & & \multicolumn{2}{c}{$\delta = 1$} \\
& Bounds & SEs & & Bounds & SEs & & Bounds & SEs & & Bounds & SEs \\
\hline
& \multicolumn{11}{c}{Lower bounds and SEs for $\mu^1$} \\
MSM lin &  -0.544 &   0.083 &  &  -0.544 &   0.083 &  &  -0.544 &   0.083 &  &  -0.544 &   0.083\\
eMSM lin &   0.495 &   0.072 &  &   0.102 &   0.071 &  &  -0.270 &   0.077 &  &  -0.544 &   0.083\\
& \multicolumn{11}{c}{Upper bounds and SEs for $\mu^1$} \\
MSM lin &  2.099 &  0.110 &  &  2.099 &  0.110 &  &  2.099 &  0.110 &  &  2.099 &  0.110\\
eMSM lin &  1.040 &  0.075 &  &  1.441 &  0.082 &  &  1.828 &  0.096 &  &  2.099 &  0.110\\
& \multicolumn{11}{c}{Lower bounds and SEs for $\mu^0$} \\
MSM lin &  -1.355 &   0.032 &  &  -1.355 &   0.032 &  &  -1.355 &   0.032 &  &  -1.355 &   0.032\\
eMSM lin &  -1.145 &   0.036 &  &  -1.221 &   0.034 &  &  -1.302 &   0.033 &  &  -1.355 &   0.032\\
& \multicolumn{11}{c}{Upper bounds and SEs for $\mu^0$} \\
MSM lin &  -0.722 &   0.052 &  &  -0.722 &   0.052 &  &  -0.722 &   0.052 &  &  -0.722 &   0.052\\
eMSM lin &  -1.018 &   0.039 &  &  -0.903 &   0.043 &  &  -0.792 &   0.048 &  &  -0.722 &   0.052\\
& \multicolumn{11}{c}{Lower bounds and SEs for $\mu^1 - \mu^0$} \\
MSM lin &  0.178 &  0.090 &  &  0.178 &  0.090 &  &  0.178 &  0.090 &  &  0.178 &  0.090\\
eMSM lin &  1.513 &  0.079 &  &  1.005 &  0.078 &  &  0.522 &  0.084 &  &  0.178 &  0.090\\
& \multicolumn{11}{c}{Upper bounds and SEs for $\mu^1 - \mu^0$} \\
MSM lin &  3.454 &  0.113 &  &  3.454 &  0.113 &  &  3.454 &  0.113 &  &  3.454 &  0.113\\
eMSM lin &  2.185 &  0.082 &  &  2.663 &  0.087 &  &  3.130 &  0.100 &  &  3.454 &  0.113\\
\hline
\end{tabular}}
\end{center}
\setlength{\baselineskip}{0.5\baselineskip}
\vspace{-.15in}\noindent{\tiny
\textbf{Note}: MSM lin and eMSM lin denote RCAL under MSM and eMSM using linear outcome mean regression.}
\vspace{-.2in}
\end{table}

\begin{table}
\caption{\footnotesize Point bounds and SEs for RCAL under MSM and eMSM with $\Lambda = 20$, using working models with main effects and interactions in the NHANES study.} \label{tb:nhanes-highdim-cal-lam20} \vspace{-4ex}
\begin{center}
\small
\resizebox{\textwidth}{!}{\begin{tabular}{lccccccccccc}
\hline
& \multicolumn{2}{c}{$\delta = 0.2$} & & \multicolumn{2}{c}{$\delta = 0.5$} & & \multicolumn{2}{c}{$\delta = 0.8$} & & \multicolumn{2}{c}{$\delta = 1$} \\
& Bounds & SEs & & Bounds & SEs & & Bounds & SEs & & Bounds & SEs \\
\hline
& \multicolumn{11}{c}{Lower bounds and SEs for $\mu^1$} \\
MSM lin &  -0.660 &   0.095 &  &  -0.660 &   0.095 &  &  -0.660 &   0.095 &  &  -0.660 &   0.095\\
eMSM lin &   0.469 &   0.072 &  &   0.052 &   0.074 &  &  -0.352 &   0.085 &  &  -0.660 &   0.095\\
& \multicolumn{11}{c}{Upper bounds and SEs for $\mu^1$} \\
MSM lin &  2.329 &  0.124 &  &  2.329 &  0.124 &  &  2.329 &  0.124 &  &  2.329 &  0.124\\
eMSM lin &  1.089 &  0.076 &  &  1.546 &  0.085 &  &  2.024 &  0.108 &  &  2.329 &  0.124\\
& \multicolumn{11}{c}{Lower bounds and SEs for $\mu^0$} \\
MSM lin &  -1.380 &   0.032 &  &  -1.380 &   0.032 &  &  -1.380 &   0.032 &  &  -1.380 &   0.032\\
eMSM lin &  -1.153 &   0.036 &  &  -1.238 &   0.034 &  &  -1.323 &   0.033 &  &  -1.380 &   0.032\\
& \multicolumn{11}{c}{Upper bounds and SEs for $\mu^0$} \\
MSM lin &  -0.608 &   0.064 &  &  -0.608 &   0.064 &  &  -0.608 &   0.064 &  &  -0.608 &   0.064\\
eMSM lin &  -0.997 &   0.041 &  &  -0.862 &   0.048 &  &  -0.710 &   0.057 &  &  -0.608 &   0.064\\
& \multicolumn{11}{c}{Lower bounds and SEs for $\mu^1 - \mu^0$} \\
MSM lin &  -0.052 &   0.107 &  &  -0.052 &   0.107 &  &  -0.052 &   0.107 &  &  -0.052 &   0.107\\
eMSM lin &   1.466 &   0.079 &  &   0.914 &   0.083 &  &   0.358 &   0.096 &  &  -0.052 &   0.107\\
& \multicolumn{11}{c}{Upper bounds and SEs for $\mu^1 - \mu^0$} \\
MSM lin &  3.709 &  0.126 &  &  3.709 &  0.126 &  &  3.709 &  0.126 &  &  3.709 &  0.126\\
eMSM lin &  2.242 &  0.082 &  &  2.784 &  0.090 &  &  3.347 &  0.111 &  &  3.709 &  0.126\\
\hline
\end{tabular}}
\end{center}
\setlength{\baselineskip}{0.5\baselineskip}
\vspace{-.15in}\noindent{\tiny
\textbf{Note}: MSM lin and eMSM lin denote RCAL under MSM and eMSM using linear outcome mean regression.}
\vspace{-.2in}
\end{table}

\begin{table}
\caption{\footnotesize Point bounds and SEs for RCAL under MSM and eMSM with $\Lambda = 30$, using working models with main effects and interactions in the NHANES study.} \label{tb:nhanes-highdim-cal-lam30} \vspace{-4ex}
\begin{center}
\small
\resizebox{\textwidth}{!}{\begin{tabular}{lccccccccccc}
\hline
& \multicolumn{2}{c}{$\delta = 0.2$} & & \multicolumn{2}{c}{$\delta = 0.5$} & & \multicolumn{2}{c}{$\delta = 0.8$} & & \multicolumn{2}{c}{$\delta = 1$} \\
& Bounds & SEs & & Bounds & SEs & & Bounds & SEs & & Bounds & SEs \\
\hline
& \multicolumn{11}{c}{Lower bounds and SEs for $\mu^1$} \\
MSM lin &  -0.808 &   0.074 &  &  -0.808 &   0.074 &  &  -0.808 &   0.074 &  &  -0.808 &   0.074\\
eMSM lin &   0.447 &   0.068 &  &  -0.016 &   0.062 &  &  -0.471 &   0.065 &  &  -0.808 &   0.074\\
& \multicolumn{11}{c}{Upper bounds and SEs for $\mu^1$} \\
MSM lin &  2.435 &  0.118 &  &  2.435 &  0.118 &  &  2.435 &  0.118 &  &  2.435 &  0.118\\
eMSM lin &  1.108 &  0.074 &  &  1.607 &  0.082 &  &  2.103 &  0.102 &  &  2.435 &  0.118\\
& \multicolumn{11}{c}{Lower bounds and SEs for $\mu^0$} \\
MSM lin &  -1.385 &   0.032 &  &  -1.385 &   0.032 &  &  -1.385 &   0.032 &  &  -1.385 &   0.032\\
eMSM lin &  -1.154 &   0.036 &  &  -1.241 &   0.034 &  &  -1.327 &   0.033 &  &  -1.385 &   0.032\\
& \multicolumn{11}{c}{Upper bounds and SEs for $\mu^0$} \\
MSM lin &  -0.558 &   0.071 &  &  -0.558 &   0.071 &  &  -0.558 &   0.071 &  &  -0.558 &   0.071\\
eMSM lin &  -0.991 &   0.041 &  &  -0.837 &   0.050 &  &  -0.670 &   0.062 &  &  -0.558 &   0.071\\
& \multicolumn{11}{c}{Lower bounds and SEs for $\mu^1 - \mu^0$} \\
MSM lin &  -0.250 &   0.091 &  &  -0.250 &   0.091 &  &  -0.250 &   0.091 &  &  -0.250 &   0.091\\
eMSM lin &   1.438 &   0.076 &  &   0.821 &   0.074 &  &   0.199 &   0.081 &  &  -0.250 &   0.091\\
& \multicolumn{11}{c}{Upper bounds and SEs for $\mu^1 - \mu^0$} \\
MSM lin &  3.820 &  0.120 &  &  3.820 &  0.120 &  &  3.820 &  0.120 &  &  3.820 &  0.120\\
eMSM lin &  2.263 &  0.081 &  &  2.847 &  0.088 &  &  3.430 &  0.105 &  &  3.820 &  0.120\\
\hline
\end{tabular}}
\end{center}
\setlength{\baselineskip}{0.5\baselineskip}
\vspace{-.15in}\noindent{\tiny
\textbf{Note}: MSM lin and eMSM lin denote RCAL under MSM and eMSM using linear outcome mean regression.}
\vspace{-.2in}
\end{table}

\begin{table}
\caption{\footnotesize Point bounds and SEs for RCAL under MSM and eMSM with $\Lambda = 50$, using working models with main effects and interactions in the NHANES study.} \label{tb:nhanes-highdim-cal-lam50} \vspace{-4ex}
\begin{center}
\small
\resizebox{\textwidth}{!}{\begin{tabular}{lccccccccccc}
\hline
& \multicolumn{2}{c}{$\delta = 0.2$} & & \multicolumn{2}{c}{$\delta = 0.5$} & & \multicolumn{2}{c}{$\delta = 0.8$} & & \multicolumn{2}{c}{$\delta = 1$} \\
& Bounds & SEs & & Bounds & SEs & & Bounds & SEs & & Bounds & SEs \\
\hline
& \multicolumn{11}{c}{Lower bounds and SEs for $\mu^1$} \\
MSM lin &  -0.846 &   0.101 &  &  -0.846 &   0.101 &  &  -0.846 &   0.101 &  &  -0.846 &   0.101\\
eMSM lin &   0.438 &   0.070 &  &  -0.035 &   0.073 &  &  -0.499 &   0.089 &  &  -0.846 &   0.101\\
& \multicolumn{11}{c}{Upper bounds and SEs for $\mu^1$} \\
MSM lin &  2.688 &  0.073 &  &  2.688 &  0.073 &  &  2.688 &  0.073 &  &  2.688 &  0.073\\
eMSM lin &  1.155 &  0.068 &  &  1.725 &  0.062 &  &  2.301 &  0.066 &  &  2.688 &  0.073\\
& \multicolumn{11}{c}{Lower bounds and SEs for $\mu^0$} \\
MSM lin &  -1.389 &   0.032 &  &  -1.389 &   0.032 &  &  -1.389 &   0.032 &  &  -1.389 &   0.032\\
eMSM lin &  -1.155 &   0.036 &  &  -1.243 &   0.034 &  &  -1.330 &   0.033 &  &  -1.389 &   0.032\\
& \multicolumn{11}{c}{Upper bounds and SEs for $\mu^0$} \\
MSM lin &  -0.475 &   0.079 &  &  -0.475 &   0.079 &  &  -0.475 &   0.079 &  &  -0.475 &   0.079\\
eMSM lin &  -0.987 &   0.042 &  &  -0.795 &   0.053 &  &  -0.603 &   0.068 &  &  -0.475 &   0.079\\
& \multicolumn{11}{c}{Lower bounds and SEs for $\mu^1 - \mu^0$} \\
MSM lin &  -0.371 &   0.118 &  &  -0.371 &   0.118 &  &  -0.371 &   0.118 &  &  -0.371 &   0.118\\
eMSM lin &   1.425 &   0.079 &  &   0.761 &   0.084 &  &   0.104 &   0.104 &  &  -0.371 &   0.118\\
& \multicolumn{11}{c}{Upper bounds and SEs for $\mu^1 - \mu^0$} \\
MSM lin &  4.076 &  0.075 &  &  4.076 &  0.075 &  &  4.076 &  0.075 &  &  4.076 &  0.075\\
eMSM lin &  2.310 &  0.075 &  &  2.967 &  0.069 &  &  3.631 &  0.070 &  &  4.076 &  0.075\\
\hline
\end{tabular}}
\end{center}
\setlength{\baselineskip}{0.5\baselineskip}
\vspace{-.15in}\noindent{\tiny
\textbf{Note}: MSM lin and eMSM lin denote RCAL under MSM and eMSM using linear outcome mean regression.}
\vspace{-.2in}
\end{table}

\bibliographystyleappend{apalike}
\bibliographyappend{references.bib}



\end{document}